\documentclass{aa} 
\usepackage{natbib}
\bibpunct{(}{)}{;}{a}{}{,} 
 
\usepackage{subfig}

\usepackage{graphicx}
\usepackage{txfonts}
\begin{document}

   \title{Spectroscopic study of the HII regions in the  NGC~1232 galaxy} 
   

   \author{F.~Lima-Costa, 
          \inst{1}
          L.~P.~Martins,
          \inst{1}
          A.~Rodr\'iguez-Ardila
          \inst{2} 
          \and
          L.~Fraga
          \inst{2}
          }
       
   \institute{NAT - Universidade Cidade de São Paulo/Universidade Cruzeiro do Sul,  Rua Galv\~ao Bueno 868, S\~ao Paulo-SP, 01506-000. BR\\
       \email{limacosta.fabiana@gmail.com}
        \and
         Laborat\'orio Nacional de Astrof\'isica, Rua Estados Unidos 154, Itajub\'a- MG, 37530-000. BR\\}

\authorrunning{Lima-Costa, Martins, Rodr\'iguez-Ardila, Fraga} 

   \date{Received September xx, 2019; accepted March xx, 2019}

 
  \abstract
   {NGC~1232 is a face-on spiral galaxy that serves as an excellent laboratory for the study of star formation due 
   to its proximity. Recent studies have revealed interesting features about this galaxy: X-ray observations suggest that it recently collided with a 
dwarf galaxy, however, no apparent remnant is observed.}
  {In this study, we search for evidence of this possible collision.}
 {We used long-slit optical 
  spectra of the galaxy in two different positions obtained with the Goodman spectrograph at the SOAR telescope.}
   {We detected 18 HII regions in the north-south direction and 22 HII regions in the east-west direction
   and a background galaxy, NGC~1232B, for which we present the first redshift measurement and spectral analysis. 
   We used the stellar population fitting technique to study the underlying stellar population of NGC~1232 and 
NGC~1232B 
 and to subtract it from the spectra to measure the emission lines. The emission lines were used to determine
the extinction, electron density, chemical abundance, and the star-formation rate gradient of NGC~1232. }
   { As is common in spiral galaxies, we found a stellar population gradient with older populations at the
central regions and younger ones towards the outskirts, along with a negative oxygen abundance gradient of 
-0.16 dex/r$_e$. Due to the difficulty of measuring important emission lines, the number of objects for the abundance
gradient is small, but there is a hint that this galaxy has a broken gradient profile, with a drop towards the
center. Some authors have explained this effect as the result of a satellite collision, but observations of a large
sample of spiral galaxies shows evidence that goes against such a mechanism. 
If the collision caused any disturbance 
in the galaxy, we believe it would be small and hard to detect with a limited number of 
objects.
From all the other measurements, we found no deviations from a typical spiral galaxy and no 
significant difference between different directions in the galaxy. 
The stellar population and emission line analysis of NGC~1232B suggest that it is a starburst galaxy. 
}
   {}


\maketitle

\section{Introduction}

Spiral galaxies are the most numerous type of galaxy in the Local Universe and
are where most star formation occurs 
\citep{brinchmann+04}. Understanding the formation and 
evolution of these systems is crucial for understanding how stars form
in galaxies. Chemo-dynamical models have made important progress in
understanding these complex systems, agreeing, for example, on a local 
dependence of the star formation law, despite the possibility of
other factors affecting the star formation rate (SFR)
over galactic scales \citep[e.g.][]{schimidt59,
dopita85, kennicutt89,wyse+89,dopita+94,prantzos+95}.
It has been proposed that an inside-out growth of the disc as a
result of the increased timescales of gas infall with radius
leads to a radial dependence on the SFR \citep{matteucci+89, boissier+99}.
The importance 
of other processes, such as metal-rich outflows
\citep[e.g.][]{maclow+99} or radial gas flows 
\citep[e.g.][]{edmunds+95, lacey+85} 
is more controversial. 
In a recent review, \citet{sanchez+20} constructed the largest
dataset of resolved nearby galaxies. By combining his analysis with
others published in recent papers, he concluded that the evolution of galaxies
is mainly shaped by local properties but it is also influenced to some extent by global ones.
The star formation and chemical enrichment
in galaxies, along with their enhancement and quenching, follow local evolutionary laws
that are also verified at kiloparsec scales. He argues that recycling, inflows, and outflows
that influence how the SFR shapes the chemical enrichment happens more on local rather
than global scales.
One of the strongest constraints for the main 
parameters in chemical evolution models is the
spatial distribution of chemical abundances
\citep{koeppen94, edmunds+95, Tsujimoto+95, Molla+96, 
molla+97, prantzos+00, chiappini+01, molla+05, fu+09,
pilkington+12}.

Nebular emission lines
have been one of the main tools used to study the complex physical processes at play in 
spiral galaxies
and their connection with the galaxy's chemical abundances.
In particular, the emission line spectra of extragalactic HII regions trace the young, massive
star components and they have become fundamental to investigate
abundances in galaxies.

While statistical studies with large
sample of galaxies can help us to understand the role and importance of many mechanisms in general, 
the detailed study of individual, nearby objects can reveal important information about the mechanisms
at play. 

In this sense, NGC~1232 is an almost face-on (i $\approx$ 30$^\circ$) gas-rich spiral galaxy 
morphologically classified as SAB(rc)c. It has
a weak bar at its core, a small bulge, and long arms that disperse to the outer regions,
producing a remarkable number of thin arms. 
These arms produce
a well-defined, though somewhat unusual spiral pattern, as they do not bend smoothly as might be expected
of such a galaxy. They bend more abruptly, which \citet{1982ApJ...263...54A}
suggested to be a distortion
caused by a previous interaction with another galaxy.
Due to its frontal position, its proximity 
(19.8 Mpc\footnote{Distance from NED (NASA/IPAC Extragalactic Database).}) 
and because it has numerous star-forming regions,
NGC~1232 is considered an excellent
laboratory for the study of star formation.

Quite nearby in the sky projection, there is another galaxy, NGC~1232A.
However, because of the strong difference in redshift between them
(z=0.00535 for NGC~1232 and z=0.02201 for 
NGC~232A\footnote{Redshifts obtained from NED (NASA/IPAC Extragalactic Database)}), they are not
currently interacting and have probably never interacted before.
Radio studies indicate that NGC~1232 has a large neutral gas envelope that extends far beyond
the optical limit of the galaxy \citep {van1999neutral}.
The nuclear region is dominated by an older stellar population \citep{martins+13}, while the spiral arms are populated
by numerous star-forming regions. 

Diffuse X-ray observations from the Chandra Space Telescope suggest that 
NGC~1232 has recently collided with a dwarf 
galaxy, however, there is no apparent remnant of this event.
According to \cite{garmire+13}, the shock wave produced by this collision may have caused the formation of massive 
bright stars. 
High-resolution imaging studies in H$\alpha$ employed to measure the star formation rate (SFR) in NGC~1232, 
reinforce this hypothesis. According to \citet{2018AJ....155..234A}, there is a significant number 
of luminous 
HII regions in the northern and eastern regions as well as an excess of star formation in the northeast 
region of the galaxy.
This finding is important because the frequency of the collisions between large spirals and dwarf 
galaxies is difficult to estimate and most galaxies show no signs of such collisions. At earlier
times, these collisions were probably very frequent, and contributed significantly to galaxy
growth. If the X-ray excess is confirmed as an evidence of this type of collision, this
may be a way to estimate the frequency of collision with dwarf galaxies and
how much such collisions contribute to the galaxy growth and evolution in the current
epoch \citep{garmire+13}.

In this work, we present the most detailed study of the HII regions of NGC~1232 using long-slit spectroscopy,
with slits positioned along the
north-south and east-west directions. This allows us to carry out a study of the spatial variation of their properties
along these two axis of the galaxy.  

This paper is organised as follows: in \S~2, we describe the observations,
data reduction, and HII regions detections. In \S~3, we use the spectral fitting technique to remove the underlying stellar population
in order to measure the emission lines. The spectral fitting also gives us information about the 
stellar population gradient of the galaxy. In \S~4, we describe the emission line measurements and
in \S~5, we use these emission lines to characterise the HII regions in the galaxy
and describe results for the background galaxy NGC~1232B (first spectroscopic observation
of the galaxy). 
In \S~6, we determine the chemical abundance of the HII regions and in \S~7, the
star formation
gradient of the galaxy. In \S~8, we present our discussion
and conclusions.


\section{Observations, data reduction, and HII region detection}

\subsection{Observations and data reduction}

In order to study the properties of the HII regions of NGC~1232,
we observed the galaxy with the the Goodman spectrograph at SOAR telescope on 
the night of 2 October 2017 and 5 February 2018. We used the 0.84$"$ wide slit mode and
diffraction grating of 400l/mm, with the slit oriented in two different positions:
north-south (N-S  $-$ blue camera) and east-west (E-W  $-$ red camera). 
The charege-coupled device (CCD) that we used has a spatial scale of
0.15 arcsec/pixel, and provides spectral coverage from 3700 to 7200 \AA. The spectral resolution R is 
$\approx $1000 at 5500\AA. 
The CCD binning was set to 1x1 yielding a spatial scale of 0.15 arcsec/pixel.
Three individual on-source integrations were carried out for each
slit, with exposure times of 20~min each. 
The seeing was 0.6$-$0.8” on 2 October and 0.8 $-$1.0” on 5 February, both with photometric skies.

The data were reduced in a standard manner using the
Image Reduction and Analysis Facility (IRAF) software \footnote{IRAF is distributed and maintained by National
Optical Astronomy (NOAO), which are operated by the Association of Universities
for Research in Astronomy, Inc., under a cooperative agreement with
the National Science Foundation.}.
It includes bias subtraction,  division by a normalised flat field, wavelength calibration using a 
HeNeAr lamp for the dispersion solution, and flux calibration. In this last step, observation of a 
\citet{Stone+85} standard star, taken the same night, was employed. Moreover, 
atmospheric extinction correction was applied to each spectrum using the CTIO coefficients available 
in IRAF using the tasks SEN-S FUNC and CALIBRATE.

\subsection {HII region detection}

The spatial coverage of long slits allows for the detection of HII regions from the center of the galaxy to its edges.
Through the identification of H$\alpha$ emission peaks along the slit, 22 HII regions were detected
in the E-W  direction and 18 in the N-S  direction,
plus a background galaxy called NGC~1232B by
\citet{1982ApJ...263...54A}.
Table~\ref{tablinhas} presents basic information about the apertures used for the extractions 
with their respective distances from the center of the galaxy.
In the N-S  slit, apertures 1 to 7 are nuclear apertures, 8 to 12 are in the north direction,
and 13 to 18 in the south direction. Aperture 19 refers to NGC~1232B.
In the E-W  slit, the apertures 1 to 4 and 14 to 16 are in the nuclear region, 
17 to 21 refer to the eastern region, and 5 to 13 to the western region of galaxy. 
It is important to mention that in the nuclear apertures in both directions, 
the emission corresponds to
the emission of the nucleus and the bulge, and it is not possible to disentangle individual
HII regions there. We ensured that the number of pixels summed in each extraction window (from 20
to 50 pixels) corresponded to a window much larger than the seeing, which gives a 
physical meaning to each extraction.

\begin{table}[]
\centering
\footnotesize
\caption {Detected HII Regions and NGC~1232B}
\begin{tabular}{|cccc|}
\hline 
\multicolumn{4}{|c|}{N-S }                        \\ \hline 
Aperture   & distance (pixels) & distance(kpc) & distance(")   \\ 
Nucleus    &                 &             &               \\ \hline
1          & 0               & 0.00           & 0             \\ 
2          & 30              & 0.43         & 5             \\ 
3          & 70              & 1.01        & 11            \\ 
4          & 110             & 1.58        & 17            \\ 
5          & -30             & 0.43         & -5            \\ 
6          & -70             & 0.10        & -11           \\ 
7          & -110            & 1.58        & -17           \\ \hline
RHII North &                 &             &               \\ \hline
8          & 215             & 3.10        & 32            \\
9          & 348             & 5.01        & 52            \\
10         & 448             & 6.45        & 67            \\
11         & 494             & 7.12        & 74            \\
12         & 533             & 7.68        & 80            \\ \hline
RHII South &                 &             &               \\ \hline
13         & -201            & 2.90        & -30           \\
14         & -240            & 3.46        & -36           \\
15         & -350            & 5.04        & -53           \\  
16         & -405            & 5.83        & -61           \\ 
17         & -455            & 6.55        & -68           \\
18         & -680            & 9.79        & -102          \\
19 (NGC1232B)  & 614         & 8.84        & 92            \\\hline  \hline
\multicolumn{4}{|c|}{E-W }                                     \\ \hline 
Aperture   & distance (pixels) & distance (kpc) & distance(") \\ 
Nucleus    &                  &            &              \\ \hline      
1          & 0               & 0.00           & 0            \\
2          & 20              & 0.29         & 3            \\
3          & 40              & 0.58         & 6            \\ 
4          & 60              & 0.86         & 9            \\
14         & -21             & -0.30        & -3           \\
15         & -41             & -0.59        & -6           \\
16         & -61             & -0.88        & -9           \\\hline 
RHII East  &                 &             &              \\\hline 
5          & 157             & 2.27        & 24           \\
6          & 187             & 2.70        & 28           \\
7          & 280             & 4.04        & 42           \\
8          & 291             & 4.20        & 44           \\
9          & 313             & 4.51        & 47           \\
10         & 447             & 6.44        & 67           \\
11         & 484             & 6.96        & 73           \\
12         & 672             & 9.68        & 101          \\
13         & 706             & 10.16       & 106          \\ \hline
RHII West  &                 &             &              \\ \hline
17         & -135            & -1.94       & -20          \\
18         & -175            & -2.52       & -26          \\
19         & -228            & -3.28       & -34          \\
20         & -250            & -3.60       & -38          \\ 
21         & -305            & -4.39       & -46         \\ 
22         & -566            & -8.15       & -85         \\
\hline
\end{tabular}
\label{tablinhas}
\end{table}

\begin{figure*}[!h]
\centering
\subfloat[]{
\includegraphics[scale=0.36]{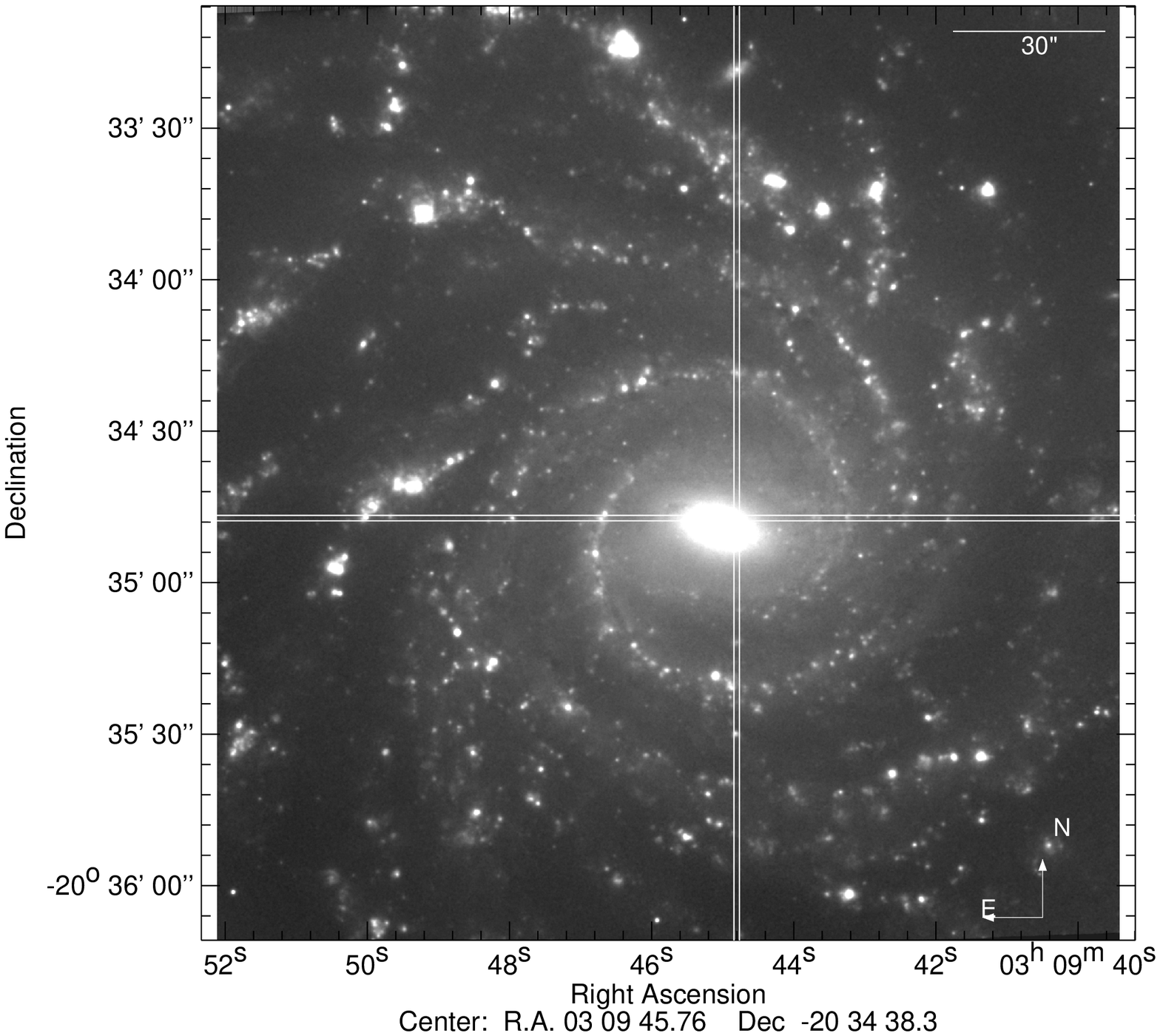}
}
\subfloat[]{
\includegraphics[scale=0.27]{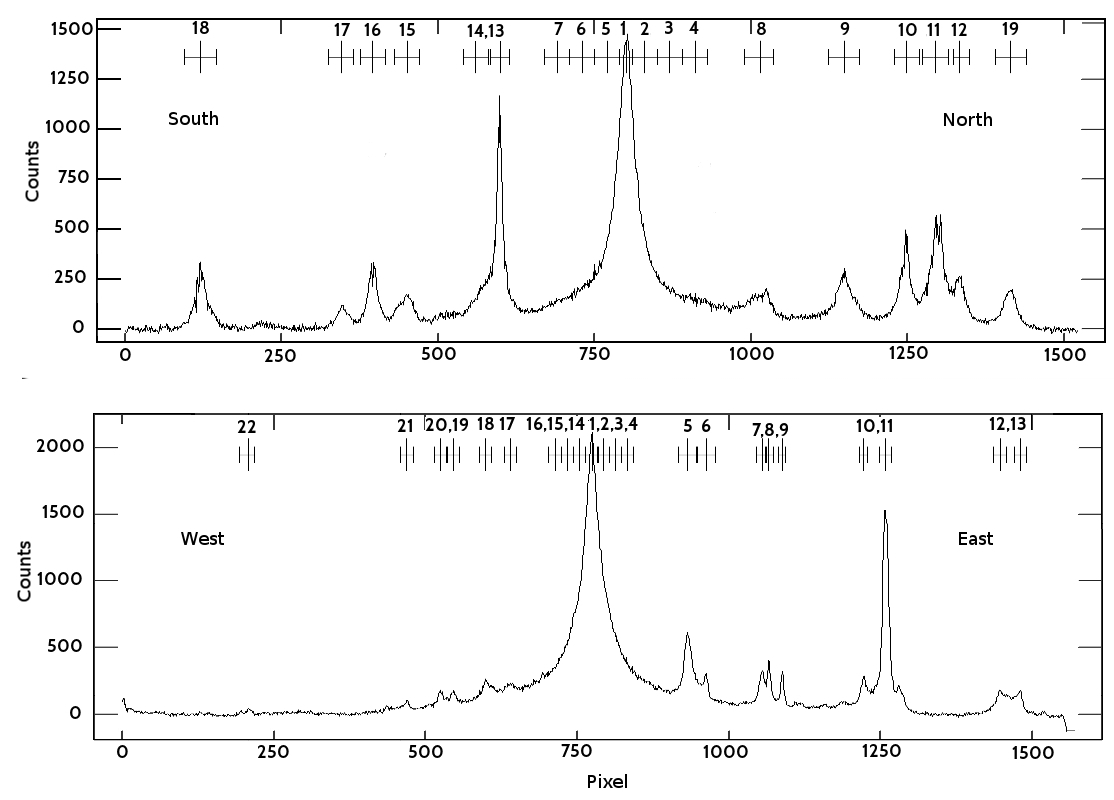}
}
\caption{(a) H$\alpha$ image of NGC~1232 from \citet{2018AJ....155..234A}.
The white lines represent the slits from where the HII regions where extracted 
along the N-S  and E-W  directions. (b) Spatial distribution of flux in H$\alpha$ of the slits N-S  (top)
and E-W  (bottom), showing the peaks which correspond to each HII region (and NGC~1232B) extracted. The numbers on top correspond to their names in Table~\ref{tablinhas} and the size of the extraction of each
HII regions is shown by the ranges below these numbers.}
\label{slits}
\end{figure*}

\begin{figure*}
\begin{center}
\includegraphics [scale=0.6]{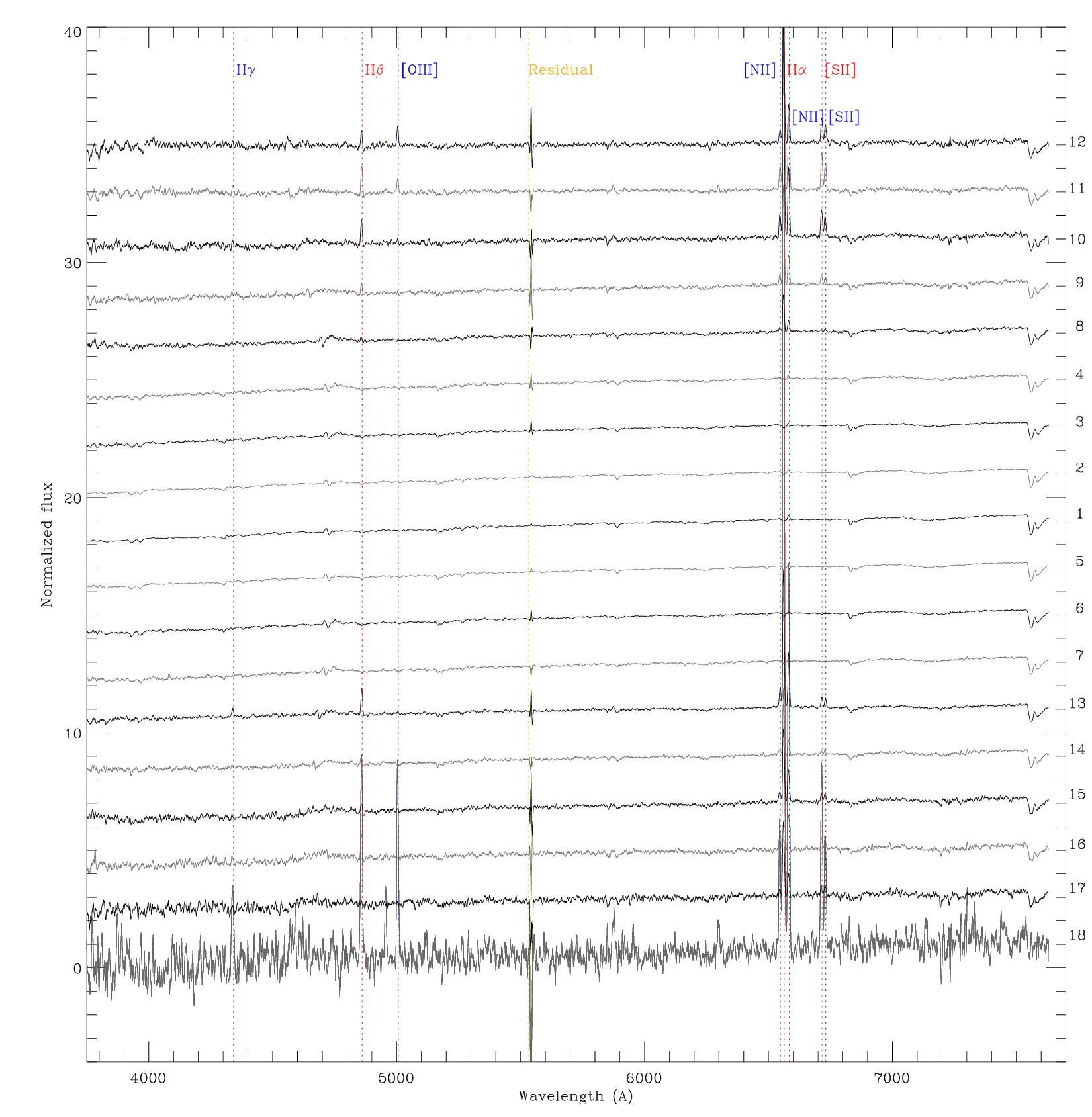} 
\end{center}
\caption {Spectra of HII regions from the N-S slit.
The dotted lines in red and 
blue mark the main emission lines detected. Dotted orange lines mark artificial effects due to
to instrumental problems. The numbers on the right side of the plot identify the apertures.} 
\label{spectra_NS}
\end{figure*}

\begin{figure*}
\begin{center}
\includegraphics [scale=0.6]{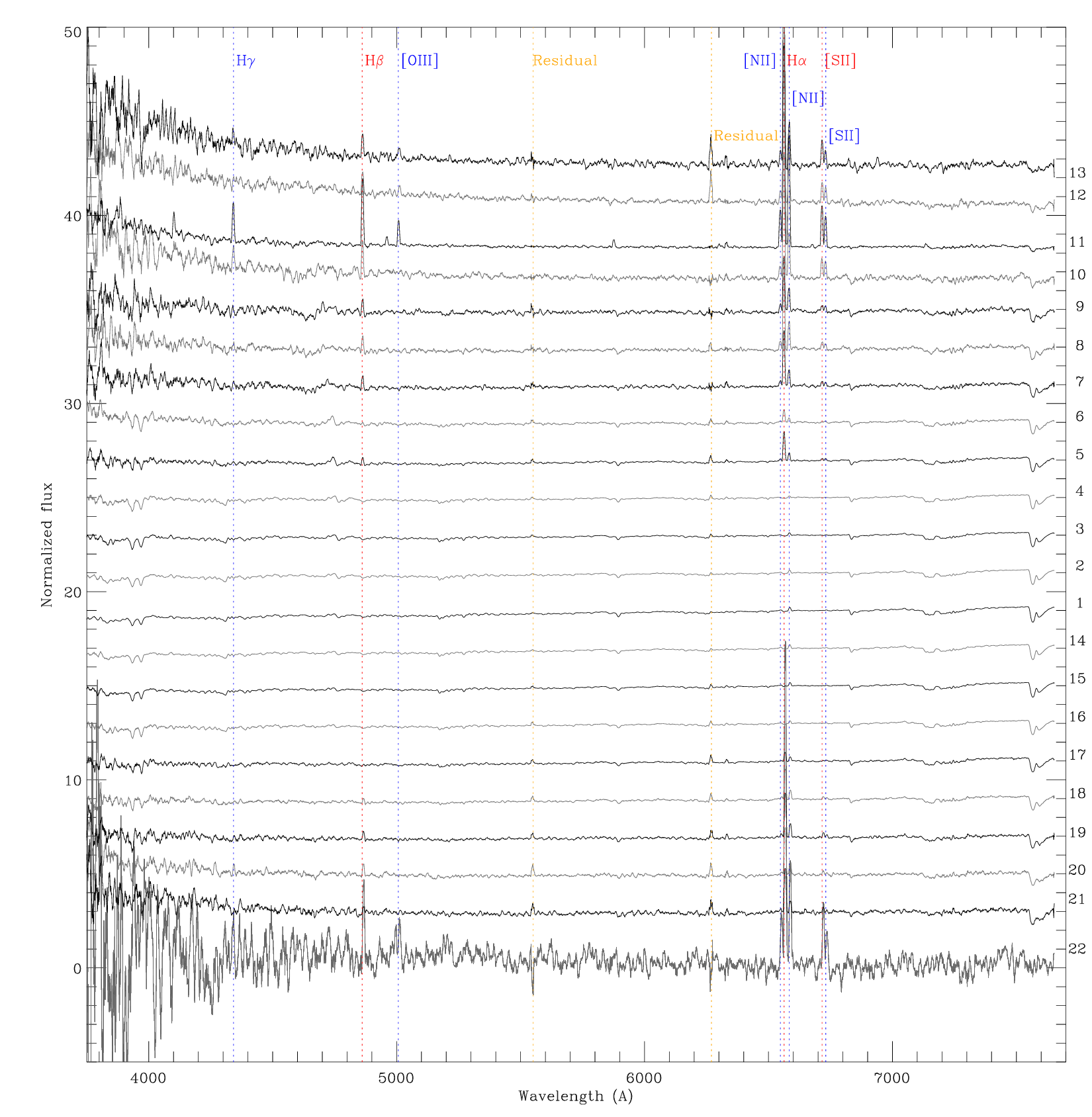} 
\end{center}
\caption {Spectra of HII regions from the E-W slit. 
The dotted lines in red and
blue mark the main emission lines detected. Dotted orange lines mark artificial effects due to
to instrumental problems. The numbers to the right side of the plot identify the apertures.} 
\label{spectra_EW}
\end{figure*}

Figure~\ref{slits}(a) shows an H$\alpha$ image of NGC~1232 \citep{2018AJ....155..234A}, 
where the N-S  and E-W  apertures are represented by 
white lines. Figure~\ref{slits}(b)
shows the spatial distribution of light in each slit and the peaks where the HII regions
and NGC~1232B were extracted. The spectra of 
these extractions are shown in Figure~\ref{spectra_NS} for the N-S slit and
in Figure~\ref{spectra_EW} for the E-W slit. The spectrum of NGC~1232B is shown and
discussed in~\S 5.

\section{Stellar population analysis and subtraction}

The typical spectrum of an HII region is dominated by intense emission lines, 
with very weak or no underlying stellar continuum.
However, our observations were made using a long slit that did not
necessarily encompass strictly the HII regions. To make sure that contamination from the 
underlying stellar population does not alter the emission line measurements we used the spectral fitting 
technique to remove its contribution. 

\begin{figure*}[!h]
\centering
\subfloat[]{
\includegraphics[scale=0.38]{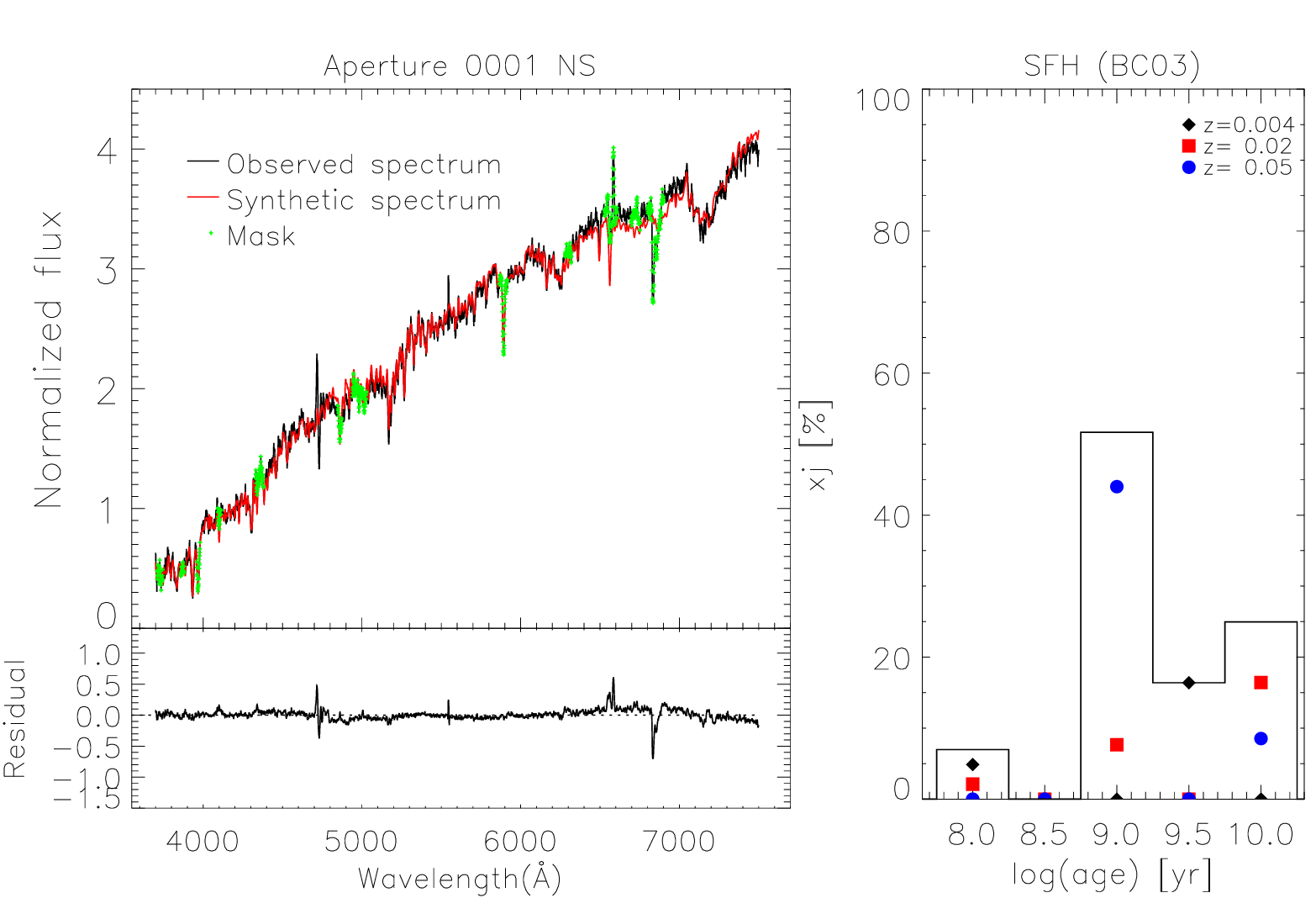}
}
\subfloat[]{
\includegraphics[scale=0.38]{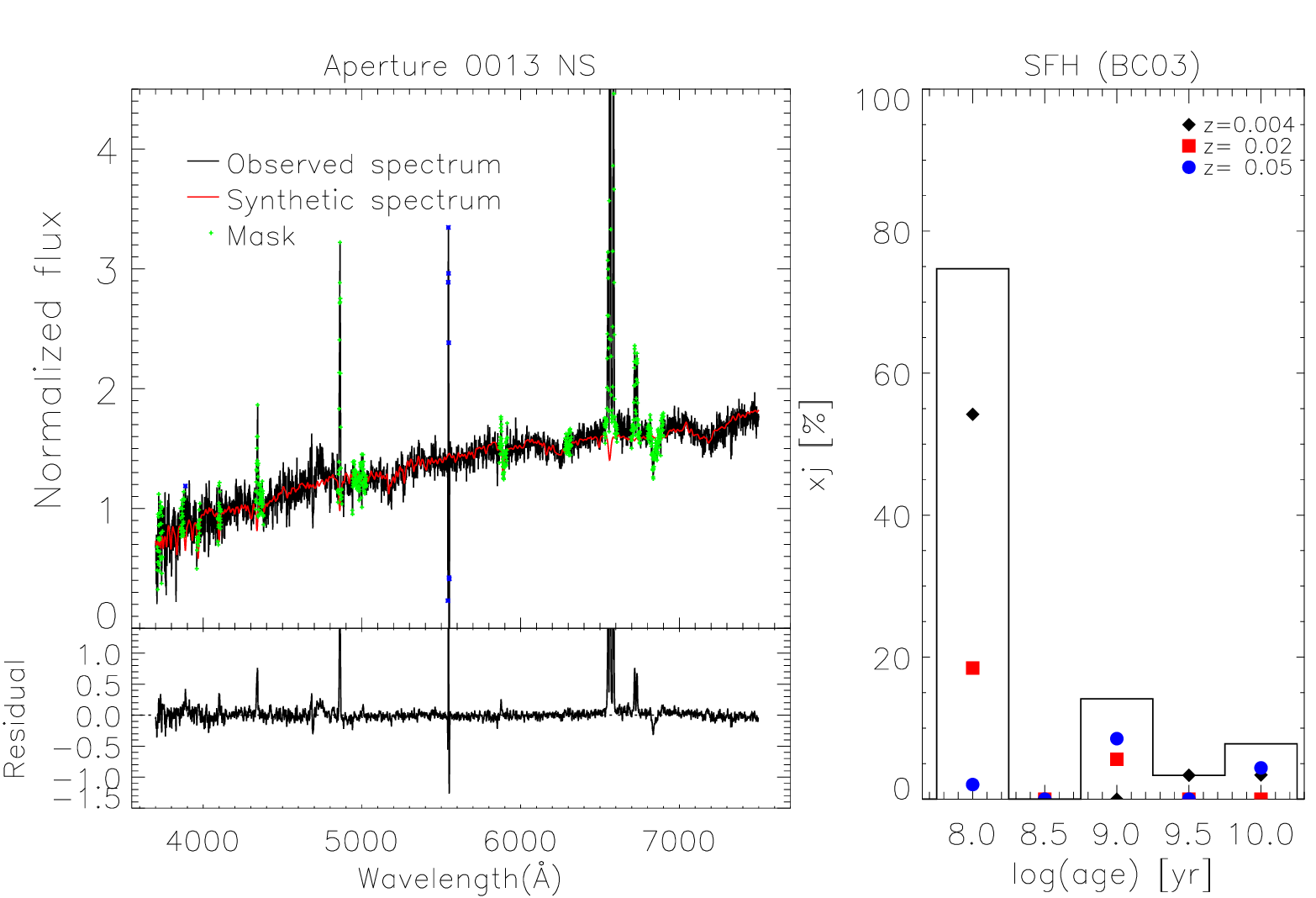}
}
\\
\subfloat[]{
\includegraphics[scale=0.38]{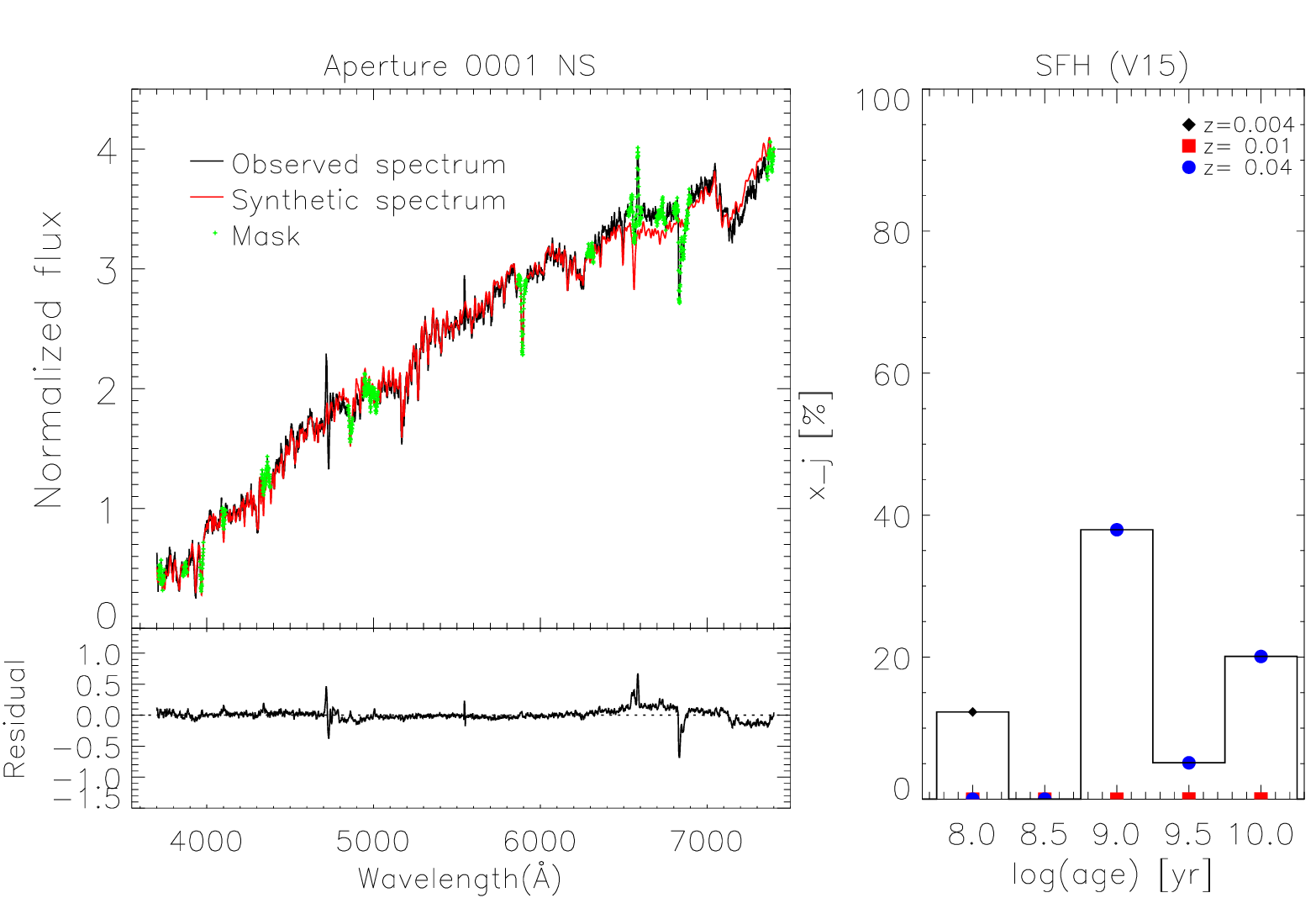}
}
\subfloat[]{
\includegraphics[scale=0.38]{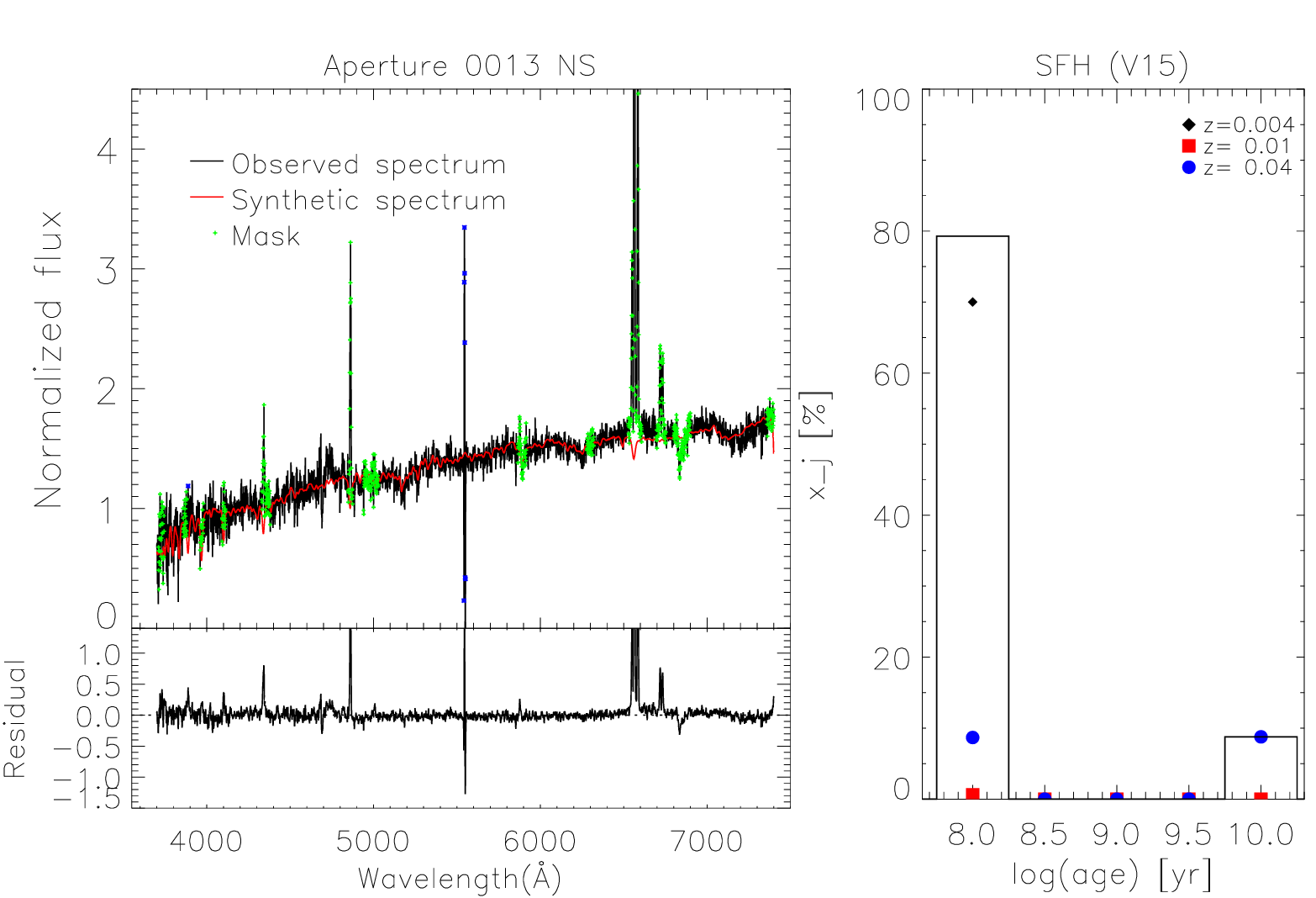}
}
\caption {Example of the stellar population spectral fitting. 
Figures (a) and (c) correspond to aperture 01 (nucleus of the galaxy, N-S  slit) fitted 
with BC03 and V15 models respectively and Figures (b) and (d)
corresponds to aperture 13 (N-S slit), also fitted with
BC03 and V15 models, respectively. 
In each figure, the top left panel shows the observed spectrum in black, and the
fitted synthetic spectrum in red. The regions marked in green
represent masked intervals, ignored by STARLIGHT because
of the presence of emission lines.  The bottom left panel shows the residual of the
difference between the two. The right panel shows the star formation history (SFH), 
which means the contribution
of each population age and metallicity to the final synthetic spectrum. 
} 
\label{popstellar01}
\end{figure*}

For this purpose, we used the code STARLIGHT \citep{cid+04,cid+05,mateus+06,asari+07}. 
This code fits the observed integrated spectrum
of a given stellar population with a combination, 
in different proportions, of single stellar populations with different ages and metallicities. 
To be sure that the patterns found here are not the result of a particular choice of models, 
we tested two different bases of models:  
\citet[][hereafter BC03]{2003MNRAS.344.1000B}, and \citet[][hereafter V15]{Vazdekis+15}. 
Using STARLIGHT, the internal kinematics is determined simultaneously with the population 
parameters. Extinction is also modeled by STARLIGHT as due to a dust screen and parameterised by 
the V-band extinction (A$_V$). We use the  \citet{cardelli+89} extinction law. 
The emission lines have been masked for the fit.
Figure~\ref{popstellar01} shows, as an example of the stellar population fitting, 
the results for apertures 01 (nuclear) and 13 (southern region), both from the N-S  slit. 
Results for all the apertures can be found in Appendix~A.

Although the main purpose of the spectral fitting in this paper is the
removal of the underlying stellar population for the measurement of the emission lines, 
there is information in the fit that deserves some analysis. 
Here, it is important to point out that it has been shown that results for the
stellar population fitting technique can be trusted only in spectra with S/N $=$ 10 or
larger 
 \citep{cid+05}. Because of this, we excluded from this analysis all the spectra 
with a signal-to-noise ratio (S/N) below 10. For the E-W  direction this was more problematic, as all of the
outer spectra are very noisy.

To perform this analysis, we used the condensed population vector as defined by \citet{cid2005stellar}, 
which combines the stellar populations taking into account noise effects
between similar spectral components.
The stellar population is binned into categories and represented by the vectors 
x$_Y$: $\it{t}$ $\leq$5$\times $10$^{7}$ years (young population), x$_I$: 
1$\times$10$^{8}$ $\leq$ $\it{t}$ $\leq$ 2$\times$10$^{9}$ years (intermediate age population) and x$_O$: 
$\it{t}$ $\geq$ 2$\times$10$^{9}$ years (old population), where $\it{t}$ is the age of the stellar population.
These results are presented in Tables~\ref{synthesisNS} 
(N-S slit) and ~\ref{synthesisEW} (E-W slit).
Additional results, namely the extinction value A$_V$, the mean age $\langle log(t_{av}) \rangle, $
and mean metallicity $\langle Z_{av} \rangle$
of the stellar population, weighted by the light fraction, are also presented. These
last two quantities are defined through Equations~\ref{idademedia} and 
and~\ref{metalicidademedia}:

\begin{equation}
\centering
(\mathrm{log}\mathit{t}_{a\textrm{v}})_{L}=\sum_{j=1}^{N_{*}}x_{j}\mathrm{log}\mathit{t}_{j}
\label{idademedia}
,\end{equation}

\begin{equation}
(\mathrm{Z}_{a\textrm{v}})_{L}=\sum_{j=1}^{N_{*}}x_{j}\mathrm{Z}_{j},
\label{metalicidademedia}
\end{equation}

where N$^\ast$ is the number of SSPs used in the base. 

The quality of the fits is measured by the reduced $\chi^2$ and the 
$\mathit{adev}$ parameter, which
represents the deviation average percentage over all adjusted pixels, determined 
by: 

\begin{equation}
adev=\frac{|\mathit{O}_\lambda - \mathit{M}_\lambda |}{\mathit{O}_\lambda},
\label{adev}
\end{equation}

where
$\mathit {O}_\lambda$ is the observed spectrum and $\mathit{M}_\lambda$ the model fitted to this spectrum.
From the values of $\chi^2$ and $\mathit{adev}$ in Tables~\ref{synthesisNS} and ~\ref{synthesisEW}
it can be seen that the outermost apertures, which have very low S/N, have results
with very large errors that cannot be trusted.

\begin{table*}
\centering
\small
\caption {Results of the stellar population fitting for the HII regions of the N-S slit and NGC~1232B.}
\begin{tabular}{ccccccccc}
\hline 
\multicolumn{9}{c}{BC03} \\
\hline
\textbf{Aperture} & \textbf{x$_Y$(\%)} & \textbf{x$_I$(\%)} & \textbf{x$_O$(\%)} & \textbf{A$_V$} & \textbf{log(tav)} & \textbf{Zav} & \textbf{$\chi^2$} & \textbf{adev} \\
\hline
    0001 &    6.979 &   51.668 &   41.352 &    1.74 &    9.21 &  0.0324 &    0.18 &    2.66\\
    0002 &    6.691 &   70.897 &   22.412 &    1.60 &    9.06 &  0.0391 &    0.15 &    2.93\\
    0003 &   14.050 &   72.720 &   13.230 &    1.73 &    8.70 &  0.0419 &    0.25 &    5.81\\
    0004 &   23.688 &   67.208 &    9.103 &    1.67 &    8.49 &  0.0408 &    0.33 &    8.57\\
    0005 &   18.235 &   43.437 &   38.328 &    1.63 &    8.88 &  0.0305 &    0.20 &    2.90\\
    0006 &   25.019 &   54.409 &   20.572 &    1.71 &    8.52 &  0.0359 &    0.33 &    4.94\\
    0007 &   29.848 &   47.436 &   22.716 &    1.78 &    8.39 &  0.0339 &    0.55 &    7.16\\
    0008 &   70.467 &    6.928 &   22.604 &    2.03 &    7.30 &  0.0210 &    0.78 &    8.07\\
    0009 &   72.915 &   18.955 &    8.129 &    2.00 &    7.58 &  0.0173 &    1.04 &   11.36\\
    0011 &   91.950 &    2.569 &    5.481 &    1.56 &    7.20 &  0.0140 &    1.09 &    7.60\\
    0012 &   70.524 &   28.181 &    1.295 &    1.56 &    7.43 &  0.0204 &    1.26 &   10.91\\
    0013 &   74.709 &   14.136 &   11.156 &    1.65 &    7.75 &  0.0147 &    0.92 &    5.89\\
    0014 &   67.214 &    2.389 &   30.397 &    1.96 &    7.46 &  0.0163 &    1.52 &    8.86\\
 NGC~1232B&   67.239 &   31.027 &    1.733 &    2.01 &    7.75 &  0.0060 &    1.50 &    5.27\\
\hline \hline
\multicolumn{9}{c}{V15} \\
\hline
\textbf{Aperture} & \textbf{x$_Y$(\%)} & \textbf{x$_I$(\%)} & \textbf{x$_O$(\%)} & \textbf{A$_V$} & \textbf{log(tav)} & \textbf{Zav} & \textbf{$\chi^2$} & \textbf{adev} \\
\hline
    0001 &   16.268 &   50.272 &   33.460 &    1.84 &    9.15 &  0.0341 &    0.20 &    2.82\\
    0002 &   16.022 &   40.974 &   43.004 &    1.59 &    9.17 &  0.0266 &    0.18 &    3.12\\
    0003 &   31.100 &   38.642 &   30.258 &    1.64 &    8.86 &  0.0244 &    0.27 &    6.10\\
    0004 &   32.808 &   29.796 &   37.396 &    1.62 &    8.84 &  0.0218 &    0.35 &    8.69\\
    0005 &   30.380 &    0.002 &   69.618 &    1.64 &    9.01 &  0.0207 &    0.23 &    3.17\\
    0006 &   34.235 &    0.000 &   65.765 &    1.65 &    8.88 &  0.0145 &    0.36 &    5.21\\
    0007 &   44.259 &    0.003 &   55.738 &    1.74 &    8.73 &  0.0151 &    0.59 &    7.50\\
    0008 &   90.042 &    0.000 &    9.958 &    2.00 &    7.74 &  0.0075 &    0.87 &    8.48\\
    0009 &   89.496 &    0.000 &   10.504 &    1.92 &    7.74 &  0.0078 &    1.07 &   11.70\\
    0011 &   97.755 &    0.000 &    2.245 &    1.37 &    7.54 &  0.0209 &    1.17 &    7.89\\
    0012 &   95.860 &    0.000 &    4.140 &    1.41 &    7.59 &  0.0312 &    1.36 &   11.26\\
    0013 &   90.039 &    0.000 &    9.961 &    1.60 &    7.74 &  0.0112 &    0.97 &    6.10\\
    0014 &   89.415 &    0.000 &   10.585 &    2.03 &    7.75 &  0.0087 &    1.61 &   11.08\\
NGC~1232B&   85.914 &   14.086 &    0.000 &    1.69 &    7.62 &  0.0091 &    1.43 &    5.13\\
\hline

\end{tabular}
\label{synthesisNS}
\end{table*}

\begin{table*}
\centering
\small
\caption {Results of the stellar population fitting for the HII regions of the E-W slit.}
\begin{tabular}{ccccccccc}
\hline 
\multicolumn{9}{c}{BC03} \\
\hline
\textbf{Aperture} & \textbf{x$_Y$(\%)} & \textbf{x$_I$(\%)} & \textbf{x$_O$(\%)} & \textbf{A$_V$} & \textbf{log(tav)} & \textbf{Zav} & \textbf{$\chi^2$} & \textbf{adev} \\
\hline
    0001 &   69.793 &    0.000 &   30.207 &    0.98 &    7.28 &  0.0259 &    0.42 &    4.99\\
    0002 &   71.499 &    0.000 &   28.501 &    0.66 &    7.16 &  0.0228 &    0.33 &    4.95\\
    0003 &   73.765 &    0.000 &   26.235 &    0.63 &    7.06 &  0.0253 &    0.37 &    5.63\\
    0004 &   75.094 &    0.000 &   24.906 &    0.66 &    6.99 &  0.0191 &    0.47 &    6.95\\
    0005 &   85.823 &    0.000 &   14.178 &    1.18 &    6.58 &  0.0127 &    0.58 &   12.97\\
    0006 &   86.709 &    0.000 &   13.291 &    0.82 &    6.55 &  0.0101 &    0.69 &   10.71\\
    0014 &   72.521 &    0.000 &   27.479 &    1.04 &    7.09 &  0.0213 &    0.57 &    4.98\\
    0015 &   73.407 &    0.000 &   26.593 &    0.93 &    7.15 &  0.0234 &    0.75 &    5.06\\
    0016 &   74.967 &    0.000 &   25.033 &    0.87 &    6.98 &  0.0233 &    0.95 &    5.36\\
    0017 &   86.911 &    0.000 &   13.089 &    1.20 &    6.96 &  0.0221 &    1.70 &    7.32\\
    0018 &   89.862 &    0.000 &   10.138 &    1.31 &    6.80 &  0.0160 &    1.59 &    8.75\\
\hline \hline
\multicolumn{9}{c}{V15} \\
\hline
\textbf{Aperture} & \textbf{x$_Y$(\%)} & \textbf{x$_I$(\%)} & \textbf{x$_O$(\%)} & \textbf{A$_V$} & \textbf{log(tav)} & \textbf{Zav} & \textbf{$\chi^2$} & \textbf{adev} \\
\hline
    0001 &   76.921 &    0.000 &   23.079 &    1.23 &    8.08 &  0.0276 &    0.65 &    6.05\\
    0002 &   79.503 &    0.000 &   20.497 &    0.89 &    8.02 &  0.0255 &    0.56 &    6.22\\
    0003 &   81.042 &    0.000 &   18.958 &    0.83 &    7.98 &  0.0280 &    0.52 &    6.70\\
    0004 &   82.456 &    0.000 &   17.544 &    0.79 &    7.94 &  0.0287 &    0.62 &    7.97\\
    0005 &   91.429 &    0.000 &    8.571 &    1.16 &    7.70 &  0.0349 &    0.68 &   13.61\\
    0006 &   94.218 &    0.000 &    5.782 &    0.88 &    7.63 &  0.0339 &    0.86 &   12.60\\
    0014 &   80.047 &    0.000 &   19.953 &    1.25 &    8.00 &  0.0290 &    0.79 &    6.04\\
    0015 &   81.530 &    0.000 &   18.470 &    1.16 &    7.96 &  0.0294 &    0.95 &    5.86\\
    0016 &   81.681 &    0.000 &   18.319 &    1.03 &    7.96 &  0.0319 &    1.19 &    6.21\\
    0017 &   87.895 &    0.000 &   12.105 &    1.29 &    7.80 &  0.0400 &    2.04 &    8.36\\
    0018 &   90.091 &    0.000 &    9.909 &    1.31 &    7.74 &  0.0400 &    1.74 &    9.56\\
\hline
\end{tabular} 
\label{synthesisEW}
\end{table*}

\begin{figure*}[!h]
\centering
\subfloat[BC03]{
\includegraphics [scale=0.43]{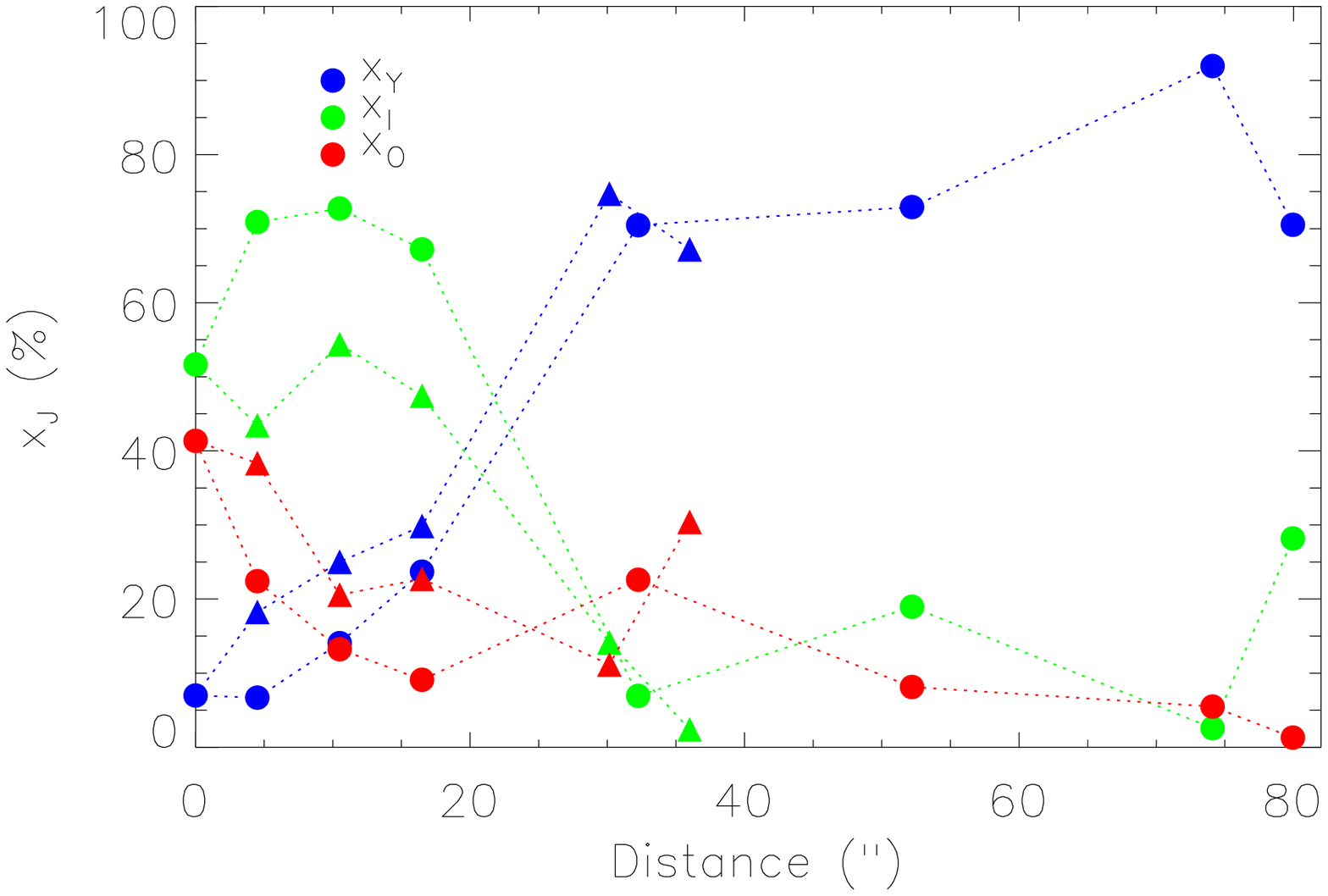}
}
\subfloat[V15]{
\includegraphics [scale=0.43]{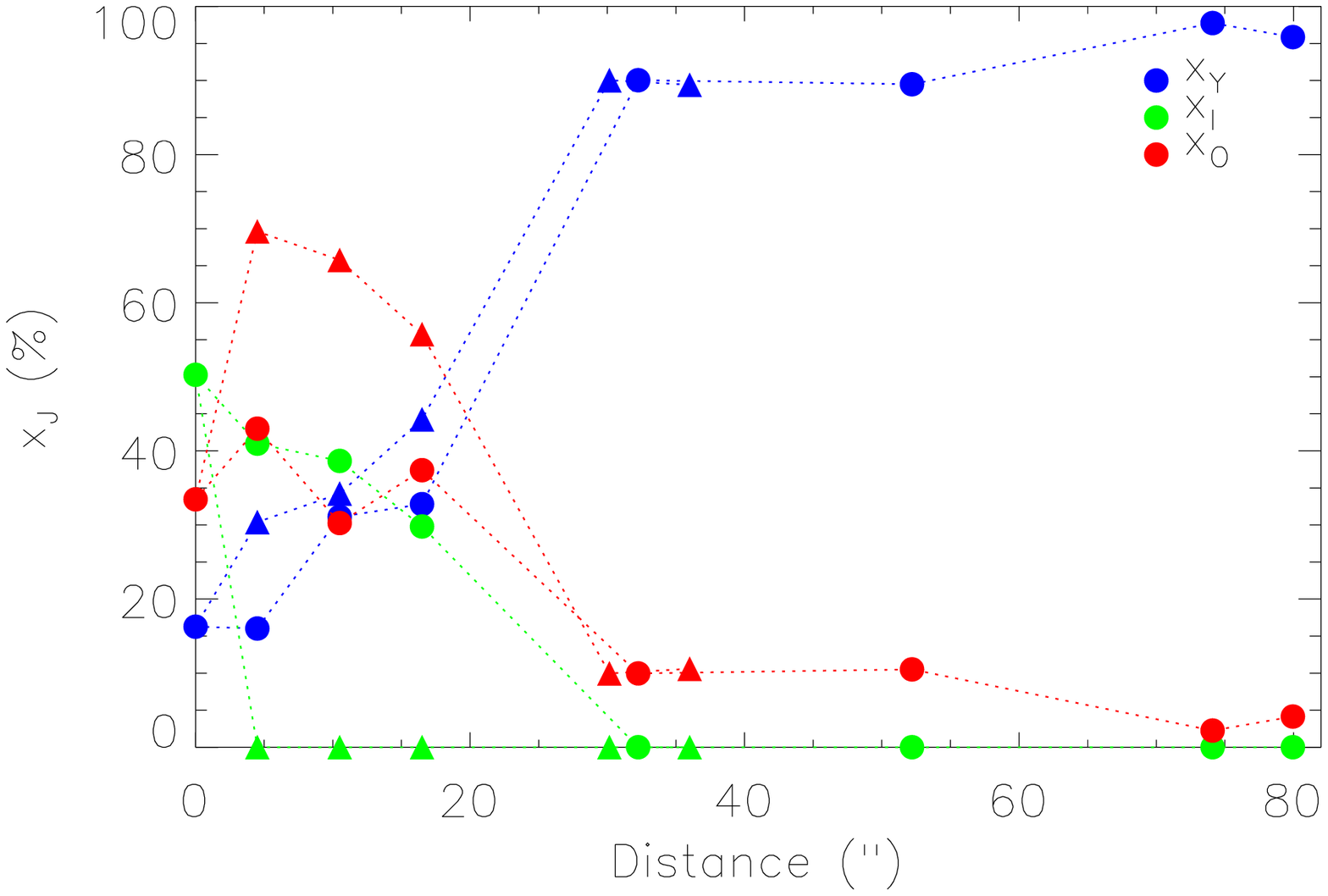}
}
\caption {Stellar population gradient for the N-S slit of NGC~1232, obtained with BC03
models (a) and V15 models (b). Triangles represent the southern region and circles the northern region.} 
\label{popgradNS}
\end{figure*}

\begin{figure*}[!h]
\centering
\subfloat[BC03]{
\includegraphics [scale=0.43]{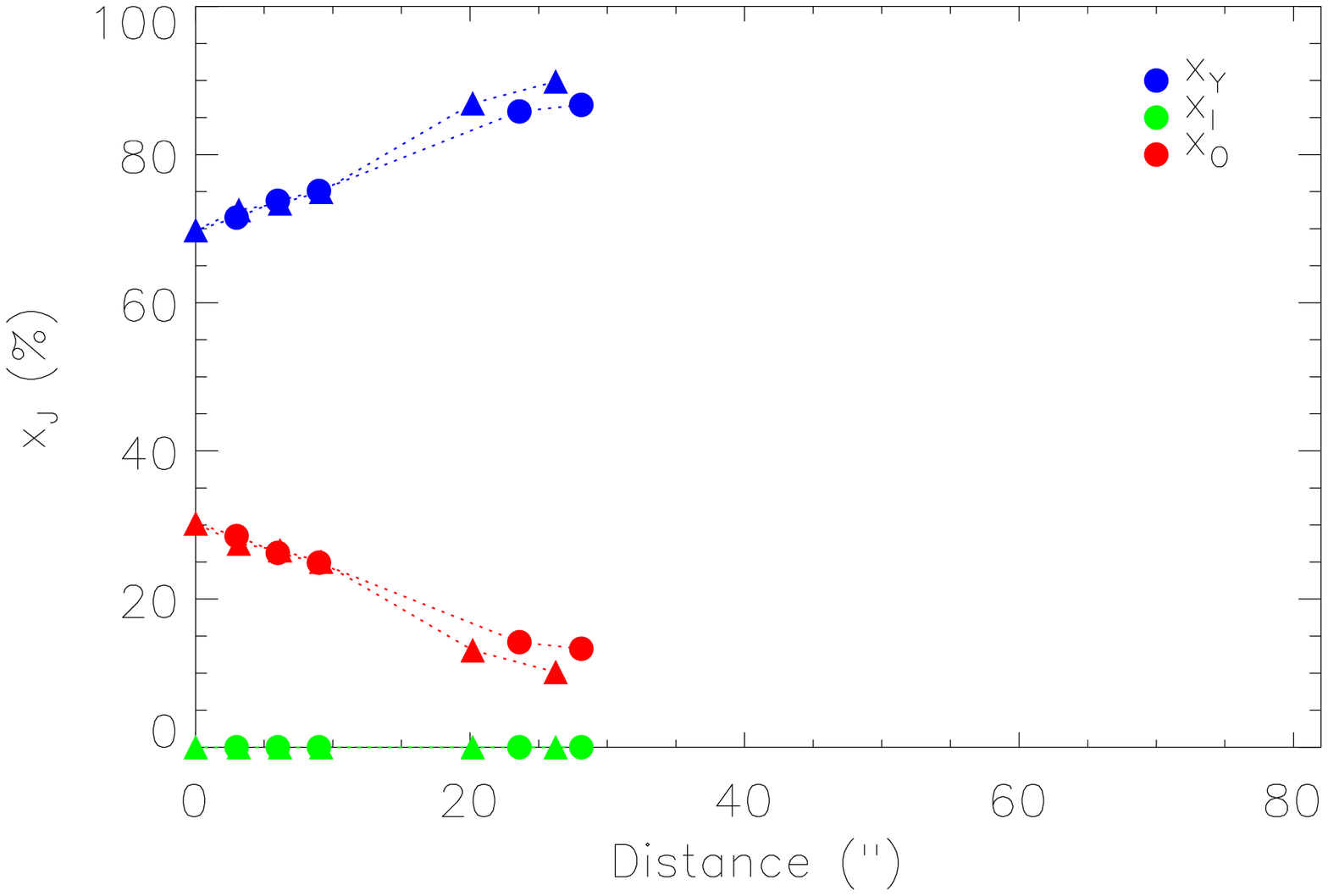} 
}
\subfloat[V15]{
\includegraphics [scale=0.43]{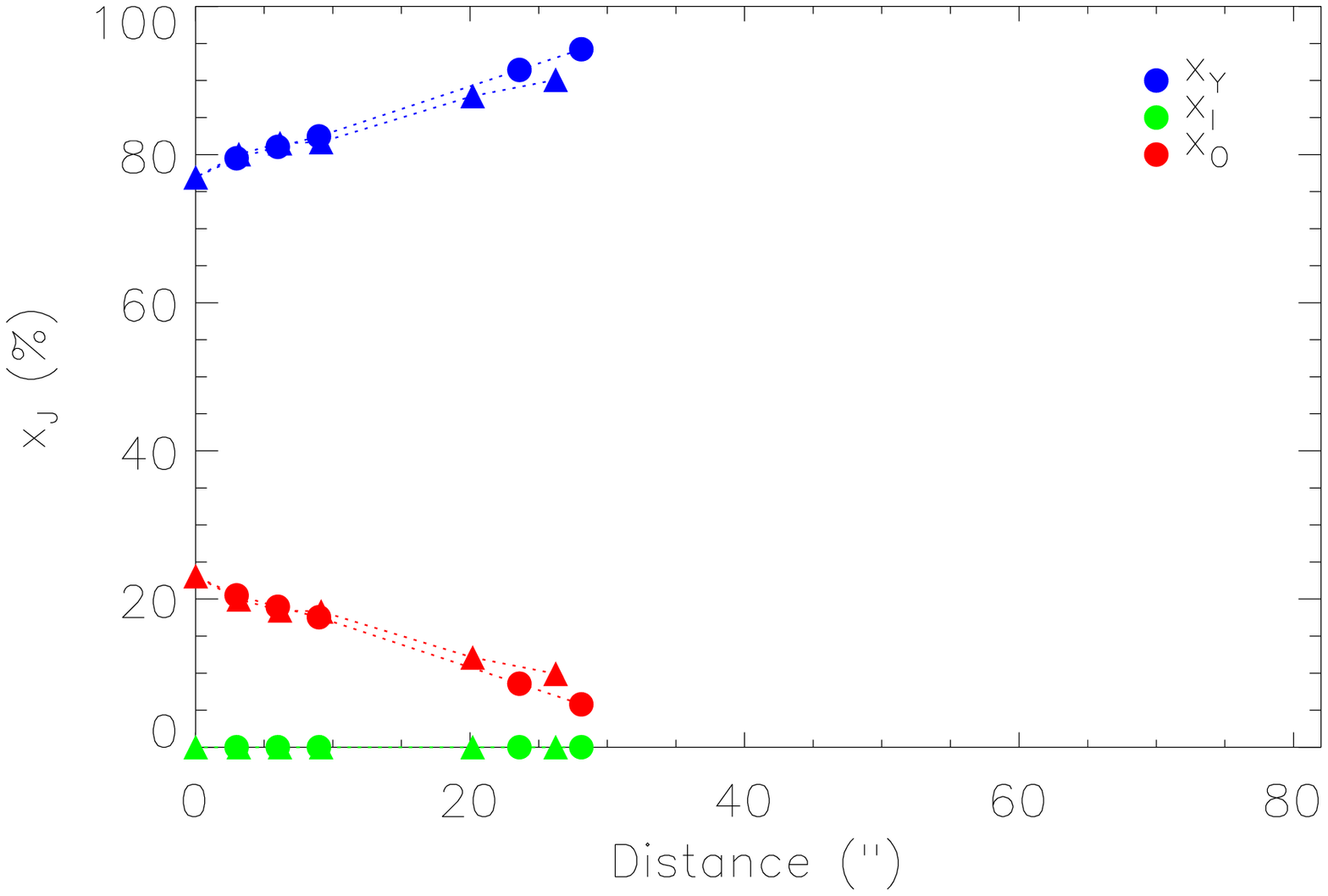} 
}

\caption {Stellar population gradient for the E-W slit of NGC~1232, obtained with BC03
models (a) and V15 models (b). Triangles represent the western region and circles the eastern region.
} 
\label{popgradEW}
\end{figure*}

Figures~\ref{popgradNS} and ~\ref{popgradEW} show the variation of the age vectors as a function
of the distance to the nucleus, for the N-S  and E-W  directions, respectively. 
Results from BC03 and V15 are in agreement and basically present the same trends. This gives us more
confidence in the results. These figures show an important difference between the
N-S  and E-W  directions: in the N-S direction, the central region presents an older population, 
becoming younger away from the nucleus. For the E-W  slit, the 
stellar population of the central HII regions
is much younger, and although there is still a trend of becoming even younger to the external regions,
the variations are much smaller. 
To be sure this was not a spurious result due to the lower S/N of the spectra, specially
in the blue, we tested the technique restricting the wavelength to 4300 to 7200 \AA (without
the blue part). The results were virtually the same (changes of less than 6\% in the young
population vector). 

We believe the difference between the N-S  and E-W  slits is related to the position of the slit:
the N-S slit was positioned slightly off the central luminosity peak of the galaxy, as can be
seen in Figure~\ref{slits}. This was done on 
purpose, so that the galaxy NGC~1232B would fall in the slit. The E-W slit, however, is placed
exactly through the central luminosity peak. Besides that, the brightest part of NGC~1232 nucleus
seems to be more extended in the
E-W  direction than in the N-S  direction, perhaps due to the presence of a small bar in this direction.  
This behavior (older population in the nucleus and a gradient towards younger population outwards)
is typical of disc galaxies \citep{sanchez-blazquez+14}.

\section{Emission Lines}

\subsection{Flux measurements}

After the subtraction of the underlying stellar population, the emission lines were measured.
The measurements were made with a multi-component fit, 
using single Gaussians per emission line, plus a linear 
function for the continuum.   
We chose to subtract the synthetic spectra obtained using BC03, 
because the S/N after subtraction is higher than that 
of V15. 
Emission lines are detected in all apertures, although 
in the nucleus and circumnuclear regions they are particularly
weak since the spectra are dominated by the stellar light. 
For some apertures, only H$\alpha$ and [NII]
were identified. 
Table~\ref{tabmedidas} presents the measured fluxes for both slits.

\begin{sidewaystable*}
\caption{Emission$^1$ line fluxes for the N-S and E-W apertures of NGC~1232 and for NGC~1232B.}
\centering
\footnotesize
\begin{tabular}{cccccccccccc}
\hline \hline
Aperture (N-S )   & H$\delta$     & H$\gamma$     & (H$\beta$)     & {[}OIII{]}$\lambda$ 4959 & {[}OIII{]}$\lambda$5007 & {[}HeI{]}$\lambda$ 5876 & {[}NII{]}$\lambda$6548 & (H$\alpha$)     & {[}NII{]}$\lambda$6583 & {[}SII{]}$\lambda$6716 & {[}SII{]}$\lambda$6731 \\
\hline
1         &.   .   .            & 6.13$\pm$0.92 & 5.85$\pm$0.80  & .   .   .                 & 5.27$\pm$0.73     & .   .   .             & 12.50$\pm$1.24         &17.84$\pm$1.72 & 25.77$\pm$1.08        & 4.67$\pm$0.56         & 5.19$\pm$0.71          \\
2         &.   .   .            &.   .   .      & .   .   .      & .   .   .                 & 2.81$\pm$0.59     & .   .   .             & .   .   .               & .   .   .            & .   .   .             & .   .   .              & .   .   .             \\   
3         &.   .   .            &.   .   .      & 0.88$\pm$0.41  & .   .   .                 & .   .   .         & .   .   .             & .   .   .               & 2.83$\pm$0.37        & 4.73$\pm$0.31         & 1.52$\pm$0.43         & 1.42$\pm$0.31          \\
4         &.   .   .            &.   .   .      & .   .   .      & .   .   .                 & .   .   .         & .   .   .             & .   .   .               & 1.98$\pm$0.29        & 1.80$\pm$0.25         & .   .   .              & .   .   .             \\
5         &.   .   .            &.   .   .      & .   .   .      & .   .   .                 & .   .   .         & .   .   .             & 1.15$\pm$0.27          & 1.40$\pm$0.15 & 1.32$\pm$0.14         & .   .   .             & .   .   .               \\
6         &.   .   .            &.   .   .      & 2.45$\pm$0.46  & .   .   .                 & .   .   .         & .   .   .             & 0.89$\pm$0.20          & 10.20$\pm$0.58        & 2.90$\pm$0.24         & 0.53$\pm$0.11         & 1.24$\pm$0.37          \\
7         &.   .   .            &.   .   .      & 2.40$\pm$0.28  & .   .   .                 & .   .   .         & .   .   .             & 0.84$\pm$0.14          & 9.81$\pm$0.81        & 2.77$\pm$0.26          & 0.85$\pm$0.13         & 0.65$\pm$0.07          \\
8         & .   .   .           & 2.05$\pm$0.57 & 3.95$\pm$1.05  & .   .   .                & .   .   .         & .   .   .             & .   .   .              & 33.78$\pm$1.73         & 9.51$\pm$0.51         & 1.93$\pm$0.38         & 1.89$\pm$0.27          \\
9         & .   .   .           & 4.05$\pm$0.82 & 7.72$\pm$0.87  & .   .   .                & .   .   .         & .   .   .             & 6.84$\pm$0.60          & 63.86$\pm$2.03       & 20.04$\pm$1.04        & 6.34$\pm$0.53         & 4.77$\pm$0.61          \\
10        & .   .   .           & .   .   .     & 9.76$\pm$0.97  & .   .   .                & .   .   .         & .   .   .             & 9.39$\pm$0.98          & 86.65$\pm$2.59       & 28.27$\pm$1.30        & 11.34$\pm$0.64        & 8.38$\pm$0.65          \\
11        & 3.02$\pm$0.87       & 6.91$\pm$1.04 & 17.16$\pm$0.93 & .   .   .                & 7.67$\pm$0.57     & .   .   .             & 14.70$\pm$1.02         & 100.69$\pm$2.60       & 41.11$\pm$12.24       & 20.21$\pm$0.71        & 15.80$\pm$1.09         \\
12        & 1.83$\pm$0.51       & 2.14$\pm$0.64 & 6.11$\pm$0.47  & 0.96$\pm$0.24            & 5.22$\pm$0.37     & .   .   .             & 4.42$\pm$0.44          & 34.53$\pm$1.26         & 11.53$\pm$0.72        & 7.87$\pm$0.45         & 5.43$\pm$0.37          \\
13        & 4.15$\pm$0.63       & .   .   .     & 21.14$\pm$0.78 & .   .   .                & 3.02$\pm$0.79     & .   .   .             & 15.94$\pm$0.39         & 94.26$\pm$3.34        & 36.40$\pm$0.88        & 7.61$\pm$0.32         & 7.01$\pm$0.57          \\
14        & .   .   .           & 1.90$\pm$0.61 & 6.72$\pm$0.57  & .   .   .                & .   .   .         & .   .   .             & 5.01$\pm$0.77          & 38.48$\pm$1.76       & 10.07$\pm$0.93        & 2.54$\pm$0.39         & 2.24$\pm$0.16          \\
15        & .   .   .           & .   .   .     & 4.11$\pm$0.84  & .   .   .                & .   .   .         & .   .   .             & 5.38$\pm$0.51          & 34.19$\pm$1.94       & 14.29$\pm$0.80        & 3.73$\pm$0.42         & 3.98$\pm$0.33          \\
16        & .   .   .           & 2.88$\pm$0.70 & 9.89$\pm$0.80  & .   .   .                & .   .   .         & .   .   .             & 4.11$\pm$0.60          & 60.27$\pm$2.49       & 13.13$\pm$0.56        & 3.73$\pm$0.42         & 2.75$\pm$0.30          \\
17        & .   .   .           & .   .   .     & 1.21$\pm$0.33  & .   .   .                & .   .   .         & .   .   .             & 1.59$\pm$0.39          & 18.92$\pm$1.08       & 5.40$\pm$0.53         & 1.72$\pm$0.26         & 2.24$\pm$0.32          \\
18        & .   .   .           & .   .   .     & 12.74$\pm$0.71 & 3.50$\pm$0.35            & 11.03$\pm$0.73    & 3.47$\pm$0.52         & 7.69$\pm$0.56          & 69.22$\pm$3.38         & 23.38$\pm$0.90        & 11.34$\pm$0.43        & 7.25$\pm$0.69          \\
NGC 1232B & .   .   .           & 5.27$\pm$0.77 & 12.84$\pm$0.74 & 3.56$\pm$0.36            & 11.08$\pm$0.85    & 3.42$\pm$0.54         & 8.25$\pm$0.66          & 69.42$\pm$2.92         & 23.55$\pm$0.93        & 11.11$\pm$0.48        & 7.20$\pm$0.65          \\
\hline
\hline \hline
Aperture(E-W ) & H$\delta$  & H$\gamma$ & H$\beta$  & {[}OIII{]}$\lambda$4959 & {[}OIII{]}$\lambda$5007 & {[}HeI{]}$\lambda$ 5876 & {[}NII{]}$\lambda$6548 & H$\alpha$  & {[}NII{]}$\lambda$6583 & {[}SII{]}$\lambda$6716  & {[}SII{]}$\lambda$6731  \\
\hline
1         & . . .          & . . .           & 3.15 $\pm$ 0.43   & . . .           & . . .           & . . .           & 6.48 $\pm$ 0.34    & 8.93 $\pm$ 0.46    & 21.17 $\pm$0.93   & 6.60 $\pm$ 0.75    & 5.12 $\pm$ 0.62    \\
2         & . . .          & . . .           & 1.30 $\pm$ 0.33   & . . .           & . . .           &5.12 $\pm$ 0.82  & 2.40 $\pm$ 0.18    & 5.51 $\pm$ 0.44    & 9.91 $\pm$0.55    & 2.90 $\pm$ 0.61    & 2.53 $\pm$ 0.49    \\
3         & . . .          & . . .           & 1.09 $\pm$ 0.30   & . . .           & . . .           & . . .           & 1.38 $\pm$ 0.16    & 3.68 $\pm$ 0.53    & 5.16 $\pm$0.37    & 1.35 $\pm$ 0.41    & 1.25 $\pm$ 0.30    \\
4         & . . .          & . . .           & . . .             & . . .           & . . .           & . . .           & 0.39 $\pm$ 0.09    & 2.17 $\pm$ 0.28    & 1.89 $\pm$0.24    & . . .              & . . .             \\
5         & . . .          & . . .           & 7.28 $\pm$ 0.75   & . . .           & . . .           & . . .           & 1.85 $\pm$ 0.31    & 28.20 $\pm$ 1.72   & 7.05 $\pm$0.54    & 2.17 $\pm$ 0.32    & 2.01 $\pm$ 0.23    \\
6         & . . .          & . . .           & 2.41 $\pm$ 0.48   & . . .           & . . .           & . . .           & 0.84 $\pm$ 0.20    & 10.16 $\pm$ 0.61   & 2.88 $\pm$0.30    & . . .              & 1.03 $\pm$ 0.29    \\
7         & . . .          & . . .           & 2.31 $\pm$ 0.27   & . . .           & . . .           & . . .           & 0.89 $\pm$ 0.18    & 9.83 $\pm$ 0.83    & 2.77 $\pm$0.29    & 0.85 $\pm$ 0.12    & 0.65 $\pm$ 0.14    \\
8         & . . .          & . . .           & 2.03 $\pm$ 0.23   & . . .           & . . .           & . . .           & 0.94 $\pm$ 0.09    & 8.90 $\pm$ 0.67    & 3.35 $\pm$0.29    & 0.87 $\pm$ 0.15    & 0.66 $\pm$ 0.08    \\
9         & . . .          & . . .           & 1.92 $\pm$ 0.25   & . . .           & . . .           & . . .           & 0.93 $\pm$ 0.19    & 6.61 $\pm$ 0.44    & 2.69 $\pm$0.20    & 0.68 $\pm $0.13    & 0.67 $\pm$ 0.12    \\
10        & . . .          & 1.91$\pm$0.39   & 2.66 $\pm$ 0.29   & . . .           & . . .           & . . .           & 0.93 $\pm$ 0.18    & 9.87 $\pm$ 0.62    & 3.57 $\pm$0.33    & 1.59 $\pm$ 0.17    & 1.33 $\pm$ 0.14    \\
11        &5.53 $\pm $1.01 & 9.17 $\pm $0.64 &15.85 $\pm$ 1.02   &2.06 $\pm$ 0.28  &5.75 $\pm$ 0.42  &1.71 $\pm$ 0.14  & 8.62 $\pm$ 0.56    & 71.81 $\pm$ 3.30   &25.92 $\pm$2.11    & 9.61 $\pm$ 0.72    & 7.16 $\pm$ 0.61    \\
12        & . . .          & . . .           & 2.09 $\pm$ 0.21   & . . .           & . . .           & . . .           & 0.99 $\pm$ 0.13    & 8.37 $\pm$ 0.45    & 2.99 $\pm$0.16    & 1.76 $\pm$ 0.07    & 1.36 $\pm$ 0.13    \\
13        & . . .          & . . .           & 1.55 $\pm$ 0.10   & . . .           & . . .           & . . .           & 0.87 $\pm$ 0.12    & 7.70 $\pm$ 0.46    & 2.67 $\pm$0.29    & 1.56 $\pm$ 0.17    & 0.86 $\pm$ 0.17    \\
14        & . . .          & . . .           & 1.21 $\pm $0.40   & . . .           &4.46 $\pm$ 0.57  & . . .           & 3.83 $\pm$ 0.36    & 6.23 $\pm$ 0.38    &13.06 $\pm$0.45    & 4.68 $\pm$ 0.69    & 3.20 $\pm$ 0.38    \\
15        & . . .          & . . .           & . . .             & . . .           & . . .           & . . .           & 1.90 $\pm$ 0.20    & 3.65 $\pm$ 0.32    & 6.11 $\pm$0.43    & 1.56 $\pm$ 0.23    & 1.14 $\pm$ 0.32    \\
16        & . . .          & . . .           & . . .             & . . .           & . . .           & . . .           & 0.91 $\pm$ 0.29    & 1.86 $\pm$ 0.37    & 3.34 $\pm$0.37    & 0.95 $\pm$ 0.24    & . . .             \\
17        & . . .          & . . .           & 1.85 $\pm$ 0.39   & . . .           & . . .           & . . .           & 0.41 $\pm$ 0.15    & 7.35 $\pm$ 0.43    & 2.72 $\pm$0.19    & . . .              & . . .             \\
18        & . . .          & . . .           & 3.12 $\pm$ 0.29   & . . .           & . . .           & . . .           & 0.96 $\pm$ 0.22    & 13.78 $\pm$ 0.78   & 3.94 $\pm$0.27    & 1.53 $\pm$ 0.23    & 0.77 $\pm$ 0.18    \\
19        & . . .          & . . .           & . . .             & . . .           & . . .           & . . .           & 1.36 $\pm$ 0.25    & 16.65 $\pm$ 0.85   & 4.62 $\pm$0.33    & 1.52 $\pm$ 0.18    & 0.73 $\pm$ 0.19    \\
20        & . . .          & 2.32$\pm$0.57   & 2.87 $\pm$ 0.27   & . . .           & . . .           & . . .           & . . .              & 10.46 $\pm$ 0.79   & 2.61 $\pm$0.24    & . . .              & . . .            \\
21        & . . .          & . . .           & 1.30 $\pm$ 0.19   & . . .           & . . .           & . . .           & 0.34 $\pm$ 0.10    & 6.41 $\pm$ 0.45    & 1.52 $\pm$0.13    & . . .              & . . .            \\
22        & . . .          & . . .           & 1.08 $\pm$ 0.17   & . . .           & . . .           & . . .           & 0.77 $\pm$ 0.09    & 4.53 $\pm$ 0.36    & 1.45 $\pm$0.14    & 0.90 $\pm$ 0.08    & 0.58 $\pm$ 0.10    \\
\hline 
\end{tabular}
\label{tabmedidas}
\\
\footnotemark[1]{Ergs units $^-$$^1$cm$^-$$^2$($\times$10$^{-15}$).} 
\end{sidewaystable*}

\subsection{Extinction}

The interstellar extinction was calculated using the Balmer decrement measured by 
H$\alpha$ and H$\beta$, and the reddening law of
\citet{cardelli+89}, assuming a total-to-selective extinction
ratio R$_V$ = A$_V$/E$_{B-V}$$=$3.1, and case B theoretical ratios at 10,000K
\citep{osterbrock+06}.
We calculated the extinction coefficient \emph{c} for each HII region using the ratio H$\alpha$/H$\beta$,
as defined by Equation~\ref{ext1}, when both lines were measured:\

\begin{equation}
\centering
c=3.01 log \left\{\frac{H\alpha/H\beta }{2.86}\right \}.
\label{ext1}
\end{equation}

 The values
obtained for each HII region are presented in 
Table~\ref{fluxocor}. The errors were estimated through a Monte Carlo simulation,
varying the fluxes within the errors a thousand times.

\begin{figure*}[!h]
\centering
\subfloat[]{
\includegraphics[scale=0.57]{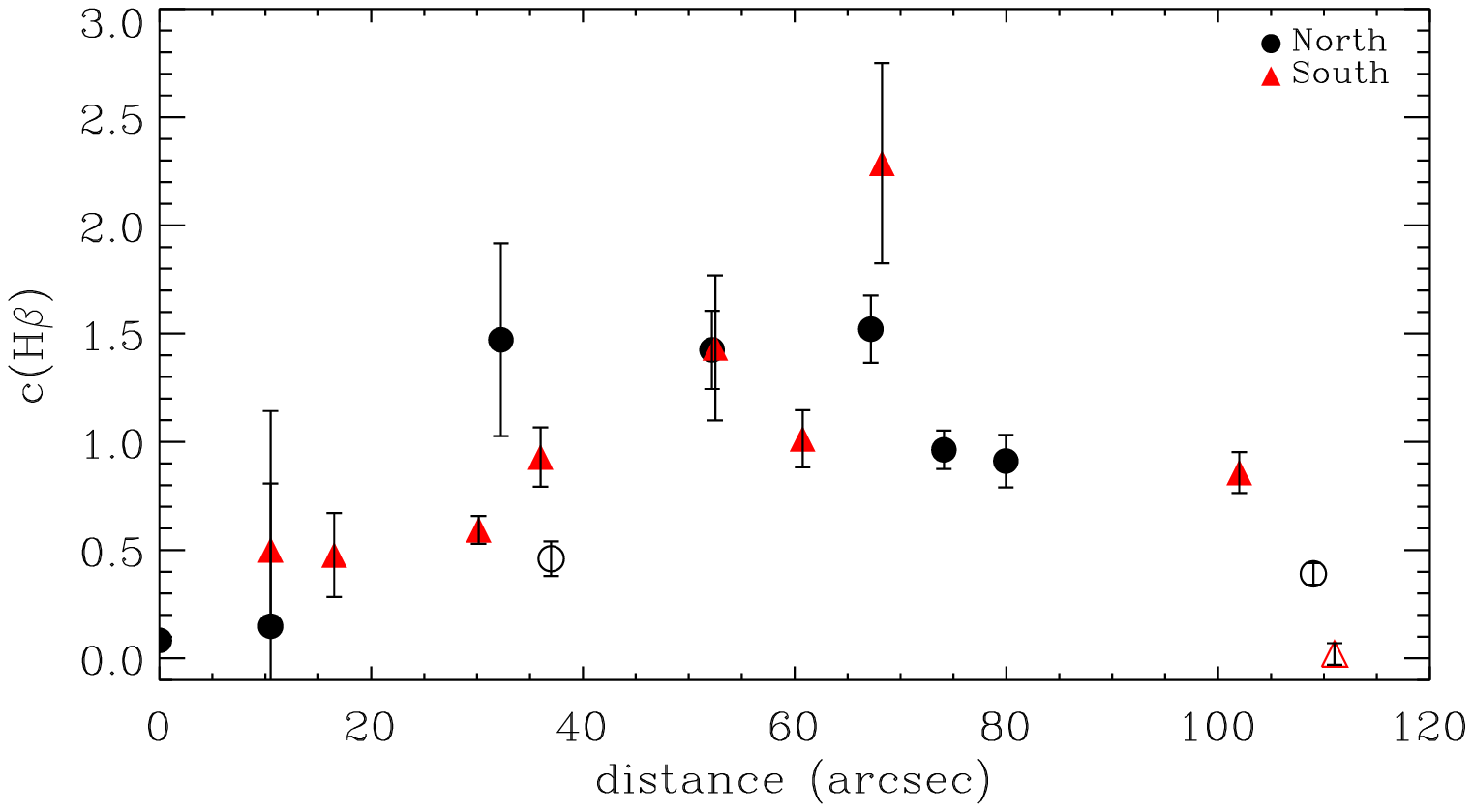}
}
\subfloat[]{
\includegraphics[scale=0.57]{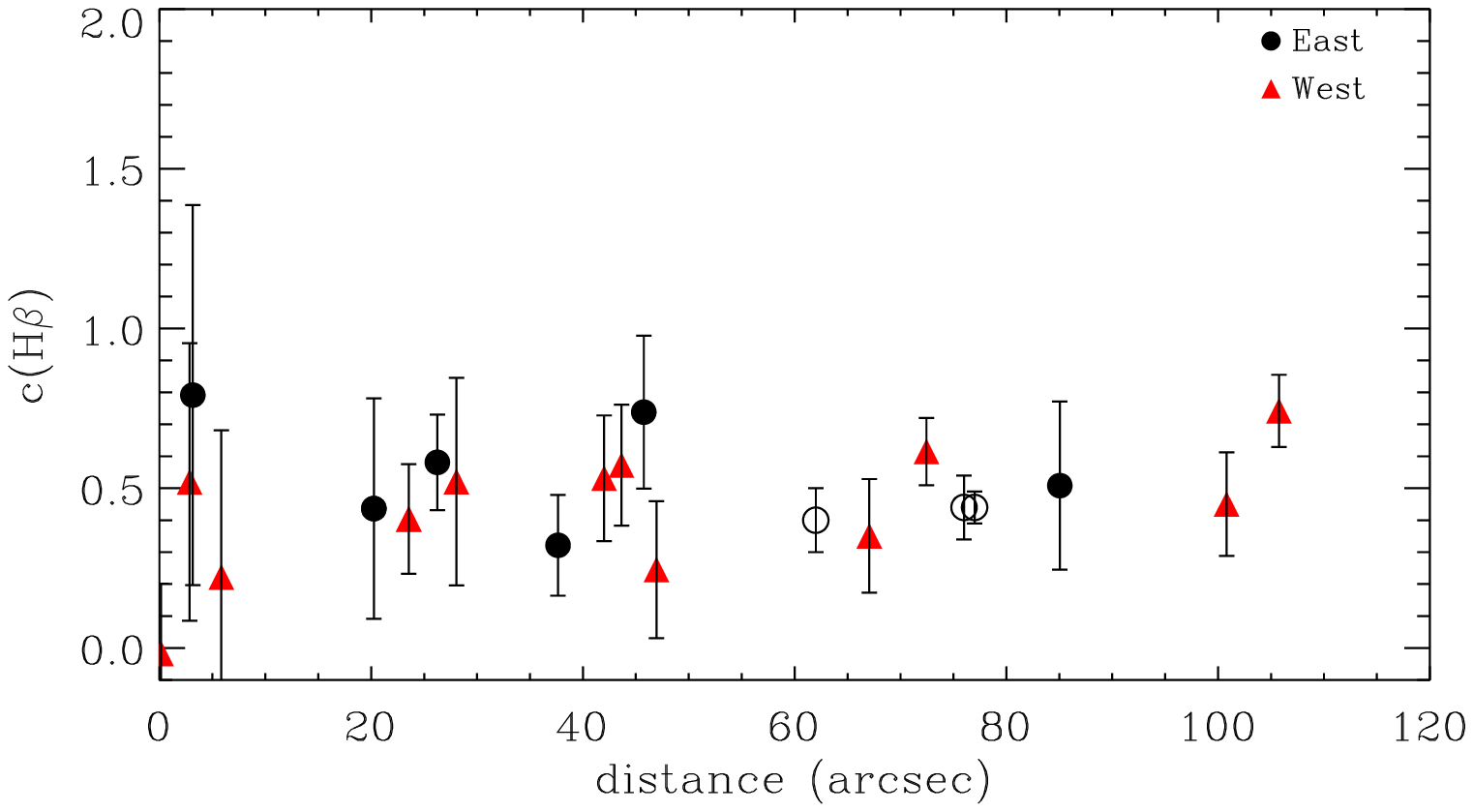}
}
\caption{Extinction coefficient as 
a function of distance to the nucleus of the galaxy NGC~1232, for the N-S  (a) and
E-W  (b) slit.
Open symbols in both figures are the HII regions from \citet{bresolin2005vlt}}.
\label{distconst}
\end{figure*}

Figure~\ref{distconst} shows the variation of the calculated extinction coefficient 
as a function of the distance to the galaxy center. 
We also include in this figure the extinction coefficients obtained by \citet{bresolin2005vlt}, who
measured line fluxes of 13 metal-rich HII regions in NGC~1232 (open symbols in this figure). From their
13 objects, we considered 2 to the north (objects 05 and 08 in their paper), 
1 to the south (object 14 in their paper), and 2 to the east (objects 10 and 11 in their paper), which
were the objects included in this figure. The other 8 objects were
located at different regions in the galaxy and were not included in this plot.

Trends of decreasing extinction
towards the outskirts of spiral galaxies are typically reported \citep[e.g.][]{vanzee+98}.
However, our data shows that both the N-S  and E-W  directions 
show no extinction gradient. The absence of a extinction gradient for NGC~1232 was also
found in \citet{bresolin2005vlt}. 
However, we note that the E-W regions seem to have, on average,
smaller extinction than the ones from N-S , at least outside the nucleus.

With the calculated extinction coefficient, it was possible to correct the 
emission line fluxes by the extinction effect
using the extinction law from \citet{cardelli+89}. Dereddened fluxes 
for the HII regions where extinction could be derived
are also presented in Table~\ref{fluxocor}.
All the further analysis in this paper is done using the extinction-corrected
line fluxes.

\begin{sidewaystable*}
\caption{Reddening corrected $^2$ emission line fluxes (relative to H$\beta$) for the N-S and E-W slits.}
\centering
\footnotesize
{\renewcommand{\arraystretch}{}
\begin{tabular}{ccccccccccccc}
\hline \hline
Aperture(N-S ) & c(H$\beta$) & H$\delta$             & H$\gamma$                &{[}OIII{]}$\lambda$ 4959  & {[}OIII{]}$\lambda$5007  & {[}HeI{]}$\lambda$ 5876 & {[}NII{]}$\lambda$6548 & (H$\alpha$)     & {[}NII{]}$\lambda$6583 & {[}SII{]}$\lambda$6716 & {[}SII{]}$\lambda$6731 \\
\hline
 1        &  0.08 $\pm$ 0.53 &       .  .  .         &   108.0 $\pm$15.7        &            .  .  .       &     89.6 $\pm$12.5       &   .  .  .               &  200.8 $\pm$21.2   &   286.5 $\pm$29.5 &   413.6 $\pm$18.4 &    74.7 $\pm$ 9.5 &    83.0 $\pm$12.2 \\
 3        &  0.15 $\pm$ 1.33 &       .  .  .         &            .  .  .       &            .  .  .       &            .  .  .       &   .  .  .               &  94.2 $\pm$60.2    &   286.0 $\pm$42.1 &   477.5 $\pm$35.2 &   152.7 $\pm$48.9 &   141.8 $\pm$34.5 \\
 6        &  0.50 $\pm$ 0.92 &       .  .  .         &            .  .  .       &            .  .  .       &            .  .  .       &   .  .  .               &  24.7 $\pm$ 8.0    &   283.7 $\pm$23.5 &    80.7 $\pm$ 9.8 &    14.3 $\pm$ 4.3 &    33.6 $\pm$15.2 \\
 7        &  0.48 $\pm$ 0.73 &       .  .  .         &            .  .  .       &            .  .  .       &            .  .  .       &   .  .  .               &  24.3 $\pm$ 6.0    &   283.9 $\pm$33.7 &    79.8 $\pm$10.8 &    23.9 $\pm$ 5.4 &    18.3 $\pm$ 2.8 \\
 8        &  1.47 $\pm$ 2.34 & 277.4 $\pm$43.8       &    77.3 $\pm$12.9        &            .  .  .       &    14.7 $\pm$ 9.7        &   .  .  .               &  31.9 $\pm$38.7    &   277.4 $\pm$43.8 &    77.3 $\pm$12.9 &    14.7 $\pm$ 9.7 &    14.3 $\pm$ 6.8 \\
 9        &  1.43 $\pm$ 0.93 & 40.9 $\pm$ 9.9        &    87.5 $\pm$10.6        &            .  .  .       &     9.8 $\pm$ 5.0        &   .  .  .               &  30.0 $\pm$ 7.8    &   277.7 $\pm$26.2 &    86.3 $\pm$13.5 &    25.7 $\pm$ 6.9 &    19.2 $\pm$ 7.9 \\
 10       &  1.52 $\pm$ 0.94 &         .  .  .       &            .  .  .       &            .  .  .       &            .  .  .       &   .  .  .               &  30.3 $\pm$10.1    &   277.1 $\pm$26.5 &    89.5 $\pm$13.3 &    33.6 $\pm$ 6.5 &    24.7 $\pm$ 6.7 \\
 11       &  0.96 $\pm$ 0.35 & 26.6 $\pm$ 5.1        &    56.9 $\pm$ 6.0        &            .  .  .       &    41.3 $\pm$ 3.3        &   .  .  .               &  41.2 $\pm$ 5.9    &   280.7 $\pm$15.1 &   113.9 $\pm$71.3 &    53.7 $\pm$ 4.2 &    41.8 $\pm$ 6.3 \\     
 12       &  0.91 $\pm$ 0.72 & 44.4 $\pm$ 8.3        &    48.6 $\pm$10.5        &    15.0 $\pm$ 3.9        &    79.2 $\pm$ 6.1        &   .  .  .               &  36.1 $\pm$ 7.2    &   281.0 $\pm$20.7 &    93.3 $\pm$11.8 &    61.2 $\pm$ 7.4 &    42.0 $\pm$ 6.1 \\ 
 13       &  0.59 $\pm$ 0.22 & 25.4 $\pm$ 3.0        &            .  .  .       &            .  .  .       &    13.6 $\pm$ 3.7        &   .  .  .               &  48.0 $\pm$ 1.8    &   283.1 $\pm$15.8 &   108.9 $\pm$ 4.1 &    22.2 $\pm$ 1.5 &    20.4 $\pm$ 2.7 \\
 14       &  0.93 $\pm$ 0.52 &         .  .  .       &    39.6 $\pm$ 9.1        &            .  .  .       &            .  .  .       &   .  .  .               &  36.7 $\pm$11.5    &   280.9 $\pm$26.2 &    73.1 $\pm$13.8 &    17.7 $\pm$ 5.8 &    15.5 $\pm$ 2.4 \\         
 15       &  1.43 $\pm$ 1.77 &         .  .  .       &    50.2 $\pm$13.6        &            .  .  .       &            .  .  .       &   .  .  .               &  44.0 $\pm$12.4    &   277.7 $\pm$47.2 &   114.9 $\pm$19.4 &    28.2 $\pm$10.2 &    29.9 $\pm$ 8.1 \\          
 16       &  1.01 $\pm$ 0.57 &         .  .  .       &    41.9 $\pm$ 7.1        &            .  .  .       &            .  .  .       &   .  .  .               &  19.2 $\pm$ 6.0    &   280.4 $\pm$25.2 &    60.7 $\pm$ 5.7 &    16.5 $\pm$ 4.2 &    12.1 $\pm$ 3.0 \\    
 17       &  2.29 $\pm$ 5.13 &         .  .  .       &            .  .  .       &            .  .  .       &            .  .  .       &   .  .  .               &  23.1 $\pm$31.9    &   272.2 $\pm$89.7 &    76.5 $\pm$44.0 &    22.1 $\pm$21.3 &    28.4 $\pm$26.7 \\        
 18       &  0.97 $\pm$ 0.47 &         .  .  .       &    61.3 $\pm$63.6        &    28.3 $\pm$ 3.0        &    86.8 $\pm$ 6.2        &  18.0 $\pm$ 4.4         &  31.3 $\pm$ 4.8    &   280.7 $\pm$28.8 &    94.2 $\pm$ 7.7 &    51.5 $\pm$ 3.6 &    27.9 $\pm$ 5.9 \\         
 NGC1232B  & 0.85 $\pm$ 0.39 &         .  .  .       &    55.7 $\pm$ 6.0        &    26.5 $\pm$ 2.8        &    80.5 $\pm$ 6.6        &  17.3 $\pm$ 4.2         &  33.6 $\pm$ 5.1    &   281.4 $\pm$22.7 &    94.9 $\pm$ 7.3 &    43.1 $\pm$ 3.7 &    27.9 $\pm$ 5.1 \\      
\hline
\hline \hline
Aperture(E-W )  & c(H$\beta$)& H$\delta$ & H$\gamma$  &  {[}OIII{]}$\lambda$4959 & {[}OIII{]}$\lambda$5007 & {[}HeI{]}$\lambda$ 5876 & {[}NII{]}$\lambda$6548 & H$\alpha$  & {[}NII{]}$\lambda$6583 & {[}SII{]}$\lambda$6716  & {[}SII{]}$\lambda$6731  \\
\hline
 1  &  0.00 $\pm$ 0.34 &           . . . &             . . . &             . . . &             . . . &             . . . &   208.5 $\pm$10.7 &   287.1 $\pm$ 5.5 &   680.9 $\pm$29.6 &   212.3 $\pm$24.0 &   164.7 $\pm$19.6 \\
 2  &  0.52 $\pm$ 1.26 &           . . . &             . . . &             . . . &             . . . &   301.3 $\pm$63.2 &   123.8 $\pm$13.7 &   283.6 $\pm$33.8 &   508.5 $\pm$42.4 &   145.4 $\pm$46.6 &   127.0 $\pm$37.5 \\
 3  &  0.22 $\pm$ 1.32 &           . . . &             . . . &             . . . &             . . . &             . . . &   107.3 $\pm$14.7 &   285.5 $\pm$48.4 &   399.7 $\pm$34.5 &    87.1 $\pm$37.3 &    95.9 $\pm$28.0 \\
 5  &  0.40 $\pm$ 0.46 &           . . . &             . . . &             . . . &             . . . &             . . . &    18.7 $\pm$ 4.3 &   284.3 $\pm$23.6 &    70.9 $\pm$ 7.5 &    21.4 $\pm$ 4.3 &    19.8 $\pm$ 3.2 \\
 6  &  0.52 $\pm$ 0.97 &           . . . &             . . . &             . . . &             . . . &             . . . &    23.6 $\pm$ 8.1 &   283.6 $\pm$25.5 &    80.3 $\pm$12.7 &    13.9 $\pm$10.6 &    27.8 $\pm$11.9 \\
 7  &  0.53 $\pm$ 0.78 &           . . . &             . . . &             . . . &             . . . &             . . . &   255.0 $\pm$ 7.6 &   283.5 $\pm$36.1 &    79.6 $\pm$12.7 &    23.9 $\pm$ 5.2 &    18.1 $\pm$ 6.2 \\
 8  &  0.57 $\pm$ 0.32 &           . . . &             . . . &             . . . &             . . . &             . . . &    29.9 $\pm$ 4.6 &   283.2 $\pm$33.2 &   106.1 $\pm$14.5 &    26.8 $\pm$ 7.7 &    20.3 $\pm$ 4.1 \\
 9  &  0.25 $\pm$ 0.20 &           . . . &             . . . &             . . . &             . . . &             . . . &    40.0 $\pm$ 8.9 &   285.4 $\pm$22.9 &   116.1 $\pm$10.3 &    28.4 $\pm$ 6.6 &    28.7 $\pm$ 6.4 \\
10  &  0.35 $\pm$ 0.58 &           . . . &    81.7 $\pm$14.7 &             . . . &             . . . &             . . . &    26.8 $\pm$ 6.7 &   284.7 $\pm$23.5 &   102.5 $\pm$12.3 &    45.1 $\pm$ 9.7 &    37.5 $\pm$ 5.2 \\
11  &  0.61 $\pm$ 0.35 &  45.5 $\pm$ 6.4 &    72.2 $\pm$ 4.1 &    12.5 $\pm$ 1.8 &    34.5 $\pm$ 2.6 &     7.9 $\pm$ 0.9 &    34.1 $\pm$ 3.5 &   283.0 $\pm$20.8 &   101.7 $\pm$13.3 &    36.7 $\pm$ 4.5 &    27.3 $\pm$ 3.9 \\
12  &  0.45 $\pm$ 0.18 &           . . . &             . . . &             . . . &             . . . &             . . . &    33.8 $\pm$ 6.3 &   284.0 $\pm$21.8 &   101.4 $\pm$ 7.6 &    58.6 $\pm$ 3.3 &    45.2 $\pm$ 6.3 \\
13  &  0.74 $\pm$ 0.34 &           . . . &             . . . &             . . . &             . . . &             . . . &    32.0 $\pm$ 8.0 &   282.1 $\pm$29.9 &    97.5 $\pm$19.1 &    55.2 $\pm$10.7 &    30.2 $\pm$11.3 \\
14  &  0.79 $\pm$ 1.69 &           . . . &             . . . &             . . . &   347.1 $\pm$47.5 &             . . . &   174.1 $\pm$29.6 &   281.8 $\pm$31.9 &   588.0 $\pm$37.0 &   203.8 $\pm$57.0 &   138.6 $\pm$31.8 \\
17  &  0.44 $\pm$ 0.84 &           . . . &             . . . &             . . . &             . . . &             . . . &    15.8 $\pm$ 8.2 &   284.1 $\pm$23.4 &   105.0 $\pm$10.6 &             . . . &             . . . \\
18  &  0.58 $\pm$ 0.59 &           . . . &             . . . &             . . . &             . . . &             . . . &    19.7 $\pm$ 7.2 &   283.2 $\pm$25.0 &    80.7 $\pm$ 8.8 &    30.4 $\pm$ 7.4 &    15.4 $\pm$ 6.0 \\
20  &  0.32 $\pm$ 0.56 &           . . . &    90.9 $\pm$20.0 &             . . . &             . . . &             . . . &     6.9 $\pm$ 7.2 &   284.9 $\pm$27.7 &    71.1 $\pm$ 8.4 &             . . . &             . . . \\
21  &  0.74 $\pm$ 0.29 &           . . . &             . . . &             . . . &             . . . &             . . . &    15.3 $\pm$ 8.0 &   282.2 $\pm$34.5 &    66.5 $\pm$10.3 &             . . . &             . . . \\
22  &  0.51 $\pm$ 0.35 &           . . . &             . . . &             . . . &             . . . &             . . . &    48.1 $\pm$ 8.4 &   283.7 $\pm$33.5 &    90.7 $\pm$12.7 &    54.9 $\pm$ 7.3 &    35.5 $\pm$ 9.4 \\
\hline 
\\
\end{tabular}}
\label{fluxocor}
\\
\footnotemark[2]{Values for H$\beta$=100.}
\end{sidewaystable*}

\subsection{Comparison with the literature}

\begin{figure}[]
\centering
\includegraphics[width=\hsize]{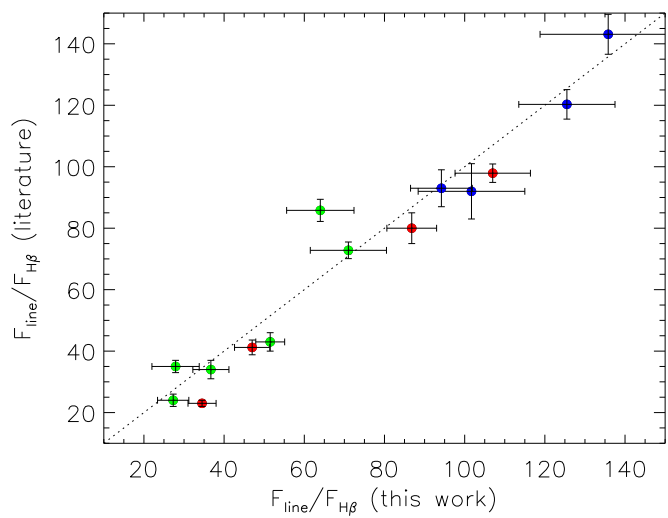}
\caption{Comparison of reddening-corrected line intensities (in units
of H$\beta$$=$100), of the most common lines, measured in the current 
work (horizontal axis) and values from the literature (vertical axis). Red points refer to
[OIII]$\lambda$5007, blue points to [NII]$\lambda$6583 and green points to
[SII]$\lambda$6716 and [SII]$\lambda$6731.}
\label{comparisons}
\end{figure}

Previous works have obtained spectra of HII regions of NGC~1232, but they were mostly targeted
at bright regions in the outskirts of the galaxy. We found two works that have HII regions
in common with our objects, \citet{vanzee+98} and \citet{bresolin2005vlt}, with two regions in each. 
We compare the fluxes of the brightest lines of these objects in Figure~\ref{comparisons}, where
[OIII]$\lambda$5007 correspond to the red points, 
[NII]$\lambda$6583 the blue points, and [SII]$\lambda$6716 and [SII]$\lambda$6731 to the green points. 
We note that the data from \citet{vanzee+98} come from low-resolution
spectroscopy and many of their measurements are actually a blend of two lines:  
[OIII]$\lambda$4959+5007, [NII]$\lambda$6548+6583 and [SII]$\lambda$6716+6731.
The resulting comparison displays the reddening-corrected line intensities
(in units of H$\beta$ $=$ 100) from this work (horizontal axis) with
the literature values (vertical axis). We can see from this figure that there 
is no evidence for systematic deviations from the dashed line, which represents
the location for equal values. Considering the effects of varying slit apertures,
orientation, and centering of the objects, the agreement is excellent.

\section{Empirical Diagrams}

\begin{figure}[!h]
\begin{center}
\includegraphics [scale=0.55]{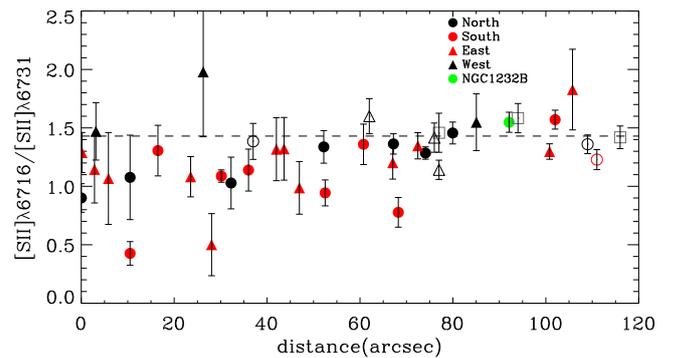} 
\end{center}
\caption {Density-sensitive diagnostic diagram using the line ratio [SII]$\lambda$6716/[SII]$\lambda$6731
as a function of the distance to the galaxy nucleus.
Open symbols are data from \citet{bresolin2005vlt}. Colors and symbols are the same as for our data, and
open gray squares are the HII regions that are in a different direction than the ones covered by our slits.} 
\label{diagdensity}
\end{figure}

Diagrams of a number of crucial line ratios can be used to assess general properties
of the extracted HII regions. For example, Figure~\ref{diagdensity}
shows the density-sensitive ratio [SII]$\lambda$6716/[SII]$\lambda$6731 as a function
of the distance to the galaxy center. The dashed line in this figure represents the 
`zero-density' limit reached by this line ratio, for T$_e$=10,000K, which is 1.43.
 Values of 9 of the 13 HII regions of \citet{bresolin2005vlt} were plot as open 
symbols for comparison. The other 4 objects have very large distances from the nucleus
(r $>$ 170"). 
What can be seen in this figure is that most of the objects lie close to the ratio limit, 
meaning densities of tens to a few hundred particles per cm$^3$.

To determine the electron density for each HII region we used the recipe from \citet{2014A&A...561A..10P}. 
They recalculated some well-known diagnostic diagrams using the
 photoionisation code CLOUDY \citep{2013RMxAA..49..137F} 
and developed a simpler analytical expression to determine the electron density than
 the traditional nonlinear equations \citep{mcc+85}. 
The equation for electron density is based on [SII] lines such that:

\begin{equation}
R_{S_2} = \frac{I_{\lambda 6716}}{I_{\lambda 6731}},
\label{eq4.5}
\end{equation}

\noindent and the electron density according to \citet{2014A&A...561A..10P} is given by:

\begin{eqnarray}
\log(n_e[cm^{-3}]) &=& 0.0543 \tan (-3.0553R_{S_2}+2.8506)      \nonumber \\
                 &+& 6.98 - 10.6905 R_{S_2}                     \nonumber \\
                 &+& 9.9186R^2_{S_2} - 3.5442R^3_{S_2}
\label{eq4.6}
.\end{eqnarray}
 
\begin{figure}[!h]
\begin{center}
\includegraphics [scale=0.55]{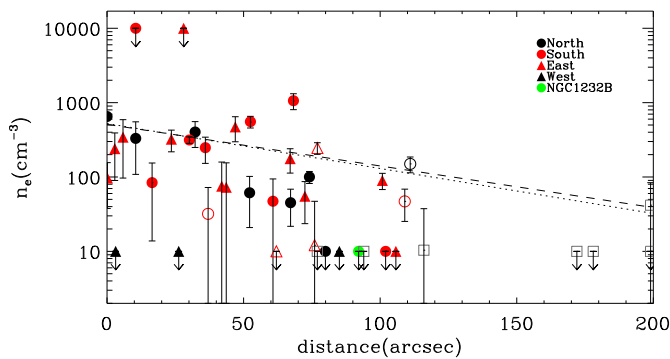} 
\end{center}
\caption {Density of the HII regions as a function of the distance to the galaxy nucleus. 
The dotted line represents the fit of Equation~\ref{eqne} to the data, excluding the upper limits. 
Symbols are the same as in Figure~\ref{diagdensity}.} 
\label{nenucleo}
\end{figure}

The values obtained for the electron density in each HII region, as well as for NGC~1232B
are presented in Table~\ref{abundances}. 
Figure~\ref{nenucleo} shows the density as a function of the distance to the nucleus of the galaxy.
Again here we plot the HII regions from \citet{bresolin2005vlt} for comparison.
For the objects where the R$_{S_2}$ was larger than the theoretical limit of 1.43, we assumed a lower limit 
of n$_e$ $=$ 10~cm$^{-3}$, and for two objects, where the ratio was too small and Equation~\ref{eq4.6} was not
valid anymore, we assumed an upper limit of n$_e$ $=$ 10000~cm$^{-3}$.

\citet{Guti_rrez_2010} found that the density of the HII regions of the galaxies M51 and NGC~4449 
tended to decrease with 
galactocentric distance, with scale lengths similar to the column densities of HI. 
In its first approximation, 
it would indicate that the HII regions can be considered in pressure equilibrium with the surrounding 
environment. Figure~\ref{nenucleo} shows the same tendency for the HII regions of this work if the upper limits are ignored. The possible exception 
is region 17 of the N-S  slit (region to the south,
which is at a distance of 68" from the nucleus and has a density of about 1000~cm$^{-3}$). 
These authors found that 
for the two galaxies studied the electron density variation with galactocentric distance can be very
well described by an exponential function in the form of:

\begin{equation}
\langle\mathrm{ne}\rangle=\langle\mathrm{ne}{\rangle}_0 e^{-r/h},
\label{eqne}
\end{equation}

where $\langle\mathrm{ne}{\rangle}_0$ is a central value of the electron density and h is a 
scale-length. We have a much smaller number of HII regions than \citet{Guti_rrez_2010}, but fitting this 
equation to our data (not considering the upper and lower density limits and taking the error bars into
consideration) returns values of 
$\langle\mathrm{n_e}{\rangle}_0$ = 524$\pm$53~cm$^{-3}$ and h = 7.97$\pm$0.99~kpc. This fit is represented by 
the dotted line in 
Fig.~\ref{nenucleo}. 
 We can also include the data from \citet{bresolin2005vlt} in the fit, which 
then returns the values $\langle\mathrm{n_e}{\rangle}_0$ = 510$\pm$55~cm$^{-3}$ and 
h = 8.62$\pm$1.16~kpc. The two fits are very similar.
This is represented by the dashed line in this figure.
 \citet{Guti_rrez_2010} found a scale-length of 9~kpc for M51 and 11~kpc for 
NGC~4449. For M51, the density scale-length is the same as the HI scale-length
measured by \citet{tilanus+91}. They did not have the HI scale-length of
NGC~4449, but they argue that since the HI galaxy radius of M51 is 15~kpc and of 
NGC~4449 is 18~kpc, by comparison they could also claim that the electron densities in the HII
regions follow the column density of the HI gas for this galaxy too. 
NGC~1232 is much larger than these
galaxies, and its estimated HI radius is about 37~kpc \citep{vanZee+99}. If we follow
the same argument, we can say that in the case of NGC~1232, the electron
densities of these regions are not correlated with the column density of the HI gas.
On the other hand, as mentioned in the introduction, NGC~1232 has an extended envelope,
which might be the result of a collision with a dwarf galaxy. In this case, 
the current radius of the HI might not be representative of the HI scale
length. So this conclusion has to be taken with a grain of salt.

\begin{figure}[!h]
\centering
\subfloat[{[NII]}]{
\includegraphics [scale=0.41]{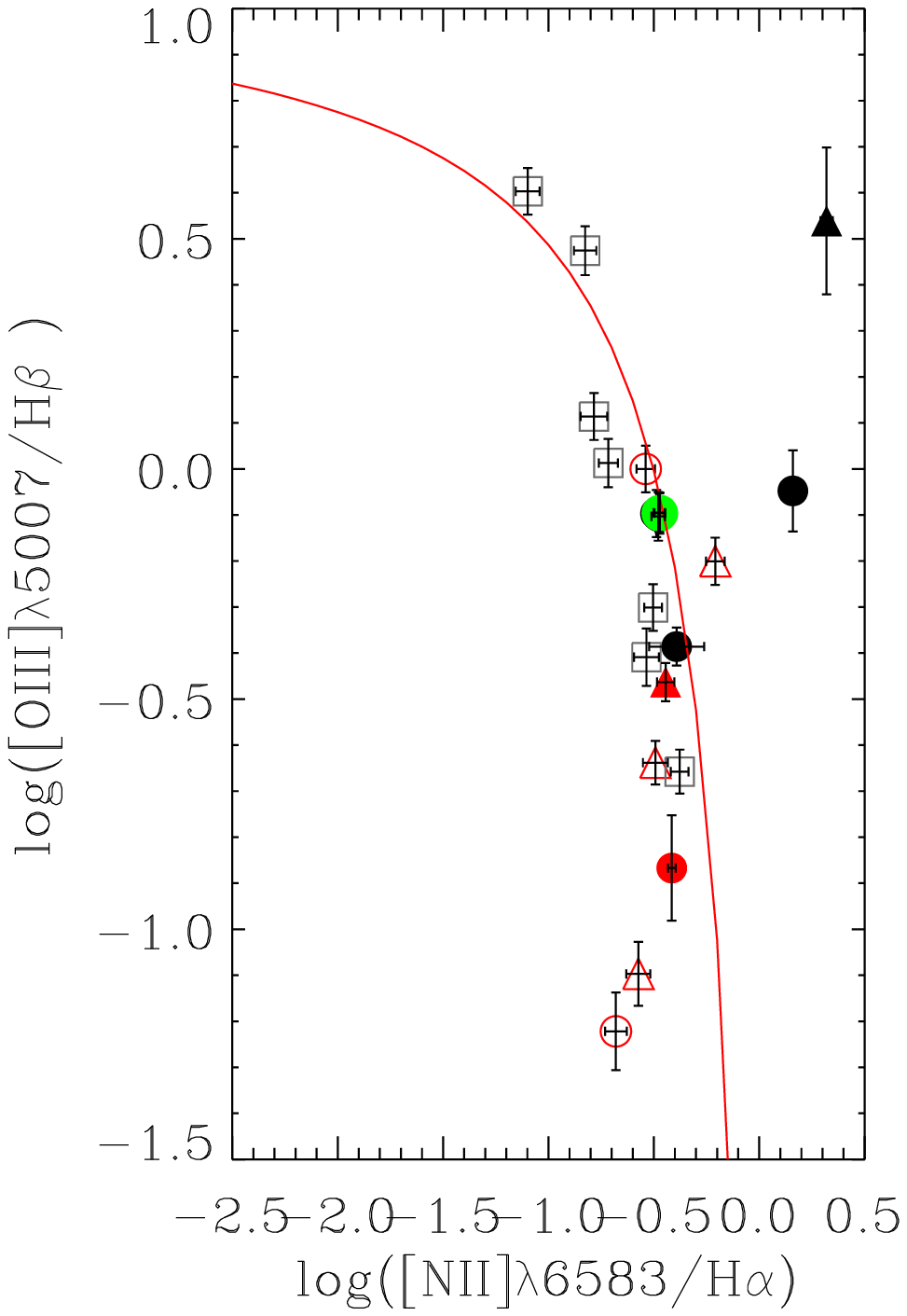}
}
\subfloat[{[SII]}]{
\includegraphics [scale=0.41]{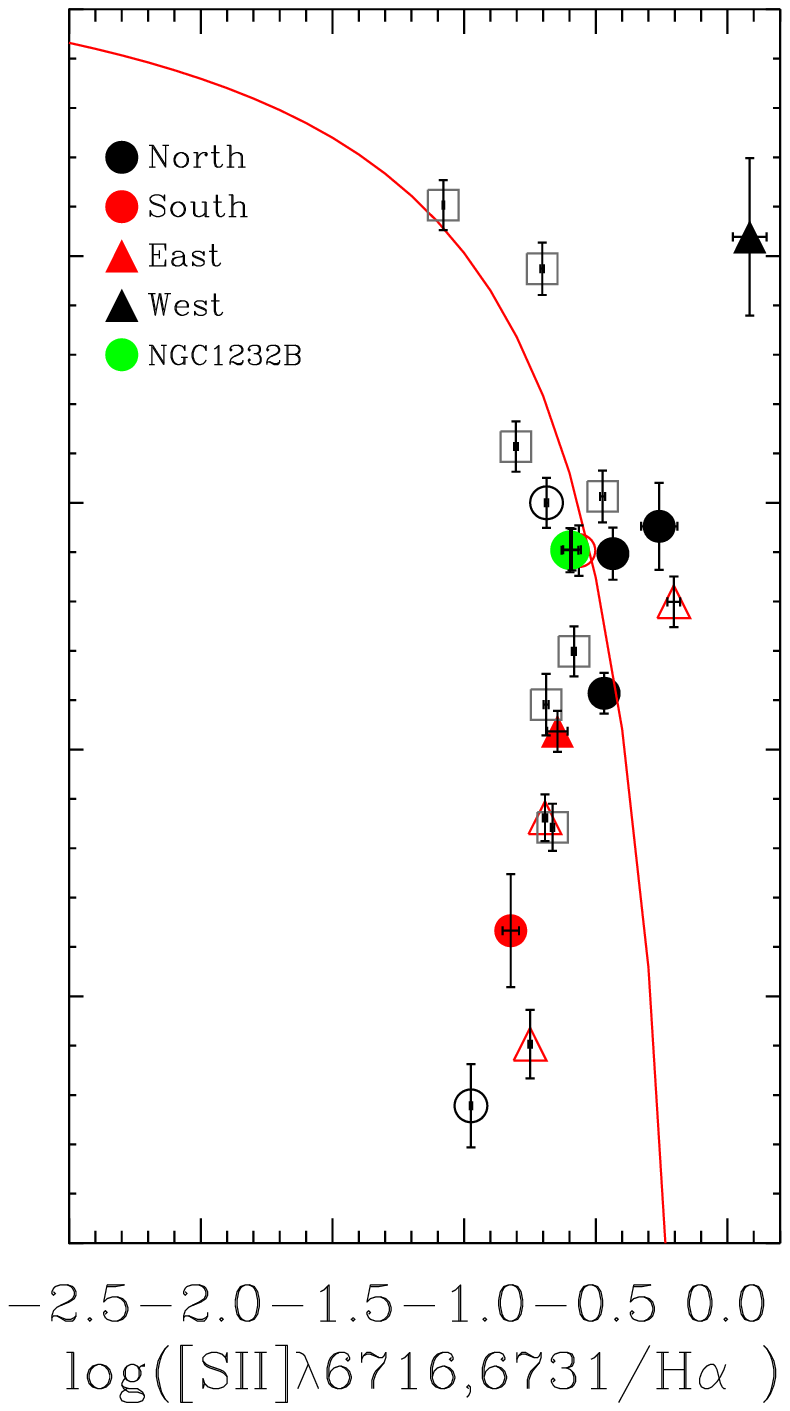}
}
\caption {Diagnostic diagram to show the excitation properties of the detected HII regions. 
The red solid line
represents the theoretical upper boundaries for HII regions from \citet{dopita2000theoretical}.
Symbols are the same as in Figure~\ref{diagdensity}.}
  
\label{diagNIISII}
\end{figure}

Diagnostic diagrams like the ones presented in Figure~\ref{diagNIISII}  
can be used to study the excitation properties of the HII regions. These
diagrams relate the low excitation lines
[NII]$\lambda$6583/H$\alpha$ and [SII]$\lambda$6716,6731/H$\alpha$ with
a high excitation one, namely, [OIII]$\lambda$5007/H$\beta$. 
Again, we added the data from \citet{bresolin2005vlt} to the plot for comparison, with open symbols.
Although for all our regions with emission lines, the [NII]$\lambda$6583/H$\alpha$ was measured, and
for most of them, the [SII]$\lambda$6716,6731/H$\alpha$ was also measured, 
it was only possible to measure the [OIII]$\lambda$5007/H$\beta$ line for a few of them. 
This already hints that these are low-excitation
objects. For the few where the [OIII]$\lambda$5007/H$\beta$ line was measured, this is confirmed. 
These objects lie in the low excitation part of the diagram, where the theoretical boundaries from 
\citet{dopita2000theoretical} are plotted (full red lines). High-excitation objects would
be found in the upper part of this diagram, where the line turns sharply to the left. 

There are two points outside the region associated with star-forming objects:
region 1 of the N-S  slit and region 14 of the E-W  slit. These are two apertures very near the center
of the galaxy, where the stellar population dominates the spectra. There might be
two reasons for the location of these points in this figure: the H$\alpha$ and H$\beta$ lines
are seen in absorption before the stellar population subtraction. It is very possible that 
the model fitted to the underlying stellar population is underestimating these lines,
artificially making the line ratios larger. If, however, we believe the models are correct, 
it would mean the the line excitation in this regions have contribution of other mechanisms 
besides hot stars, such as shocks, for example (potentially induced by the recent merger with a dwarf). 
 We discard the possibility of an AGN contribution to the emission lines because neither 
the evidence found here nor that already reported in the literature (in other wavelength intervals) suggest 
the presence of an active nucleus in the center of this object.

\subsection{NGC~1232B}

\begin{figure}[!h]
\begin{center}
\includegraphics[scale=0.5]{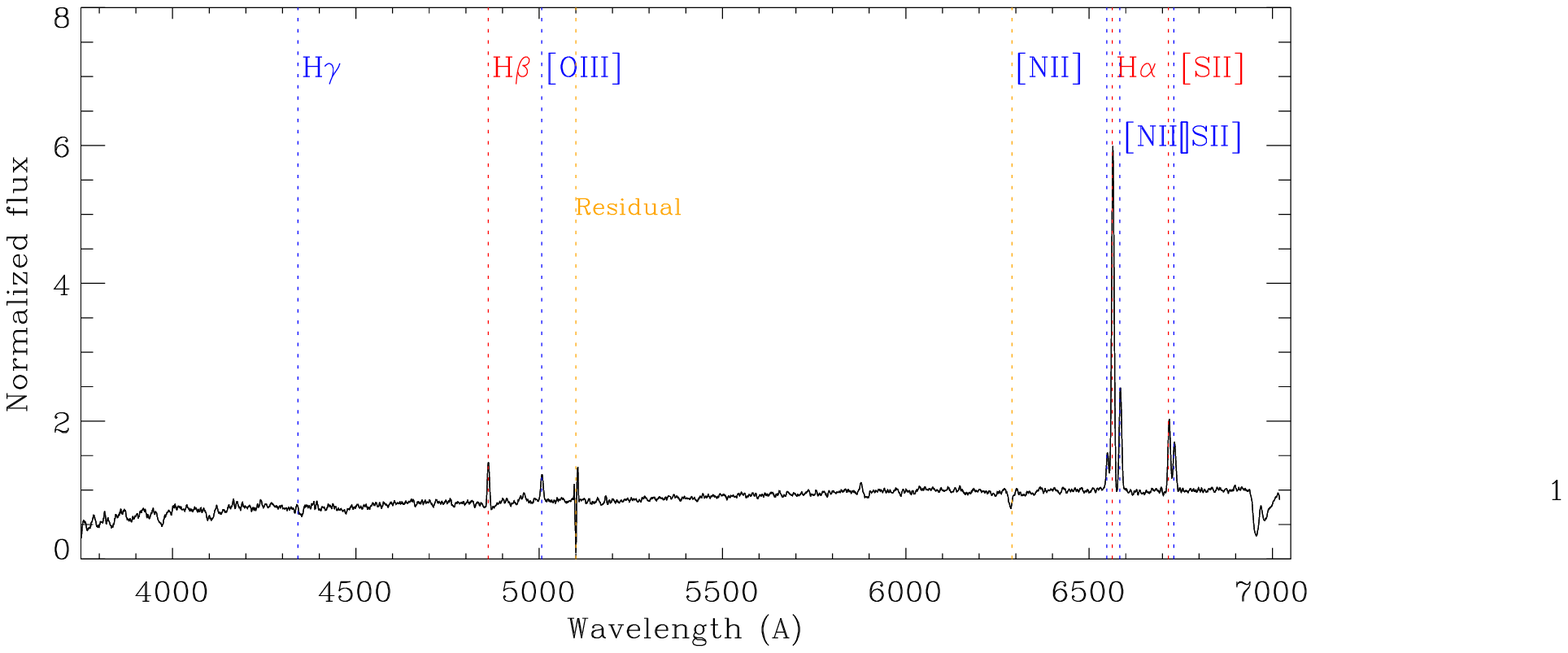} 
\end{center}
\caption {NGC~1232B galaxy spectrum. Red and blue dotted lines mark the main emission lines detected. Orange dotted lines
mark artificial effects due to instrumental problems.} 
\label{espectroNGC1232B}
\end{figure}

The galaxy NGC~1232B was first cited by \citet {1982ApJ...263...54A} with no 
further information in the literature beyond the image in this paper.
The positioning of the slit in the N-S  direction allowed us to include this object 
and thus obtain its spectrum, as shown in Figure~\ref{espectroNGC1232B}. As can be 
seen in this figure, the spectrum shows strong emission lines. 
Using these lines. we measured a redshift of z=0.09261. To the
best of our knowledge, this is the first report of the measurement of z for this object.

In order to measure the emission line fluxes, we followed the same
procedure for the HII regions in NGC~1232. That is, fitting the
stellar continuum with STARLIGHT, subtraction of the stellar population,
and Gaussian fitting of the line profiles. 
Table~\ref{fluxocor} lists the line emission fluxes.

As for the HII regions, we obtained the extinction coefficient through the H$\alpha$ 
and H$\beta$ line ratio. The galaxy 
does not seem to be very dusty, with
c(H$\beta$)$\approx$0.85, close to that found for the HII regions. 

\begin{figure*}[!h]
\centering
\subfloat[]{
\includegraphics[scale=0.38]{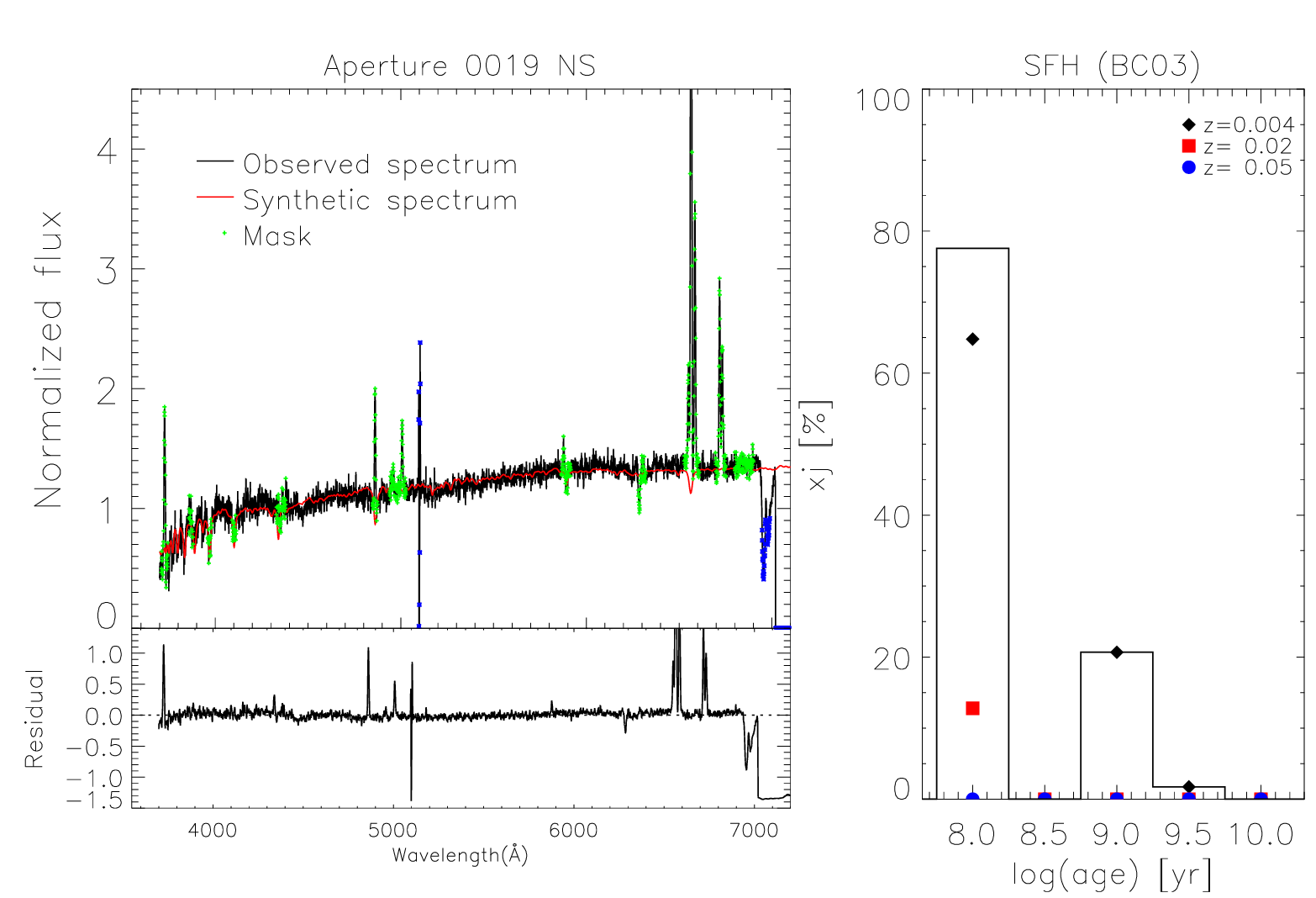}    
}
\subfloat[]{
\includegraphics[scale=0.38]{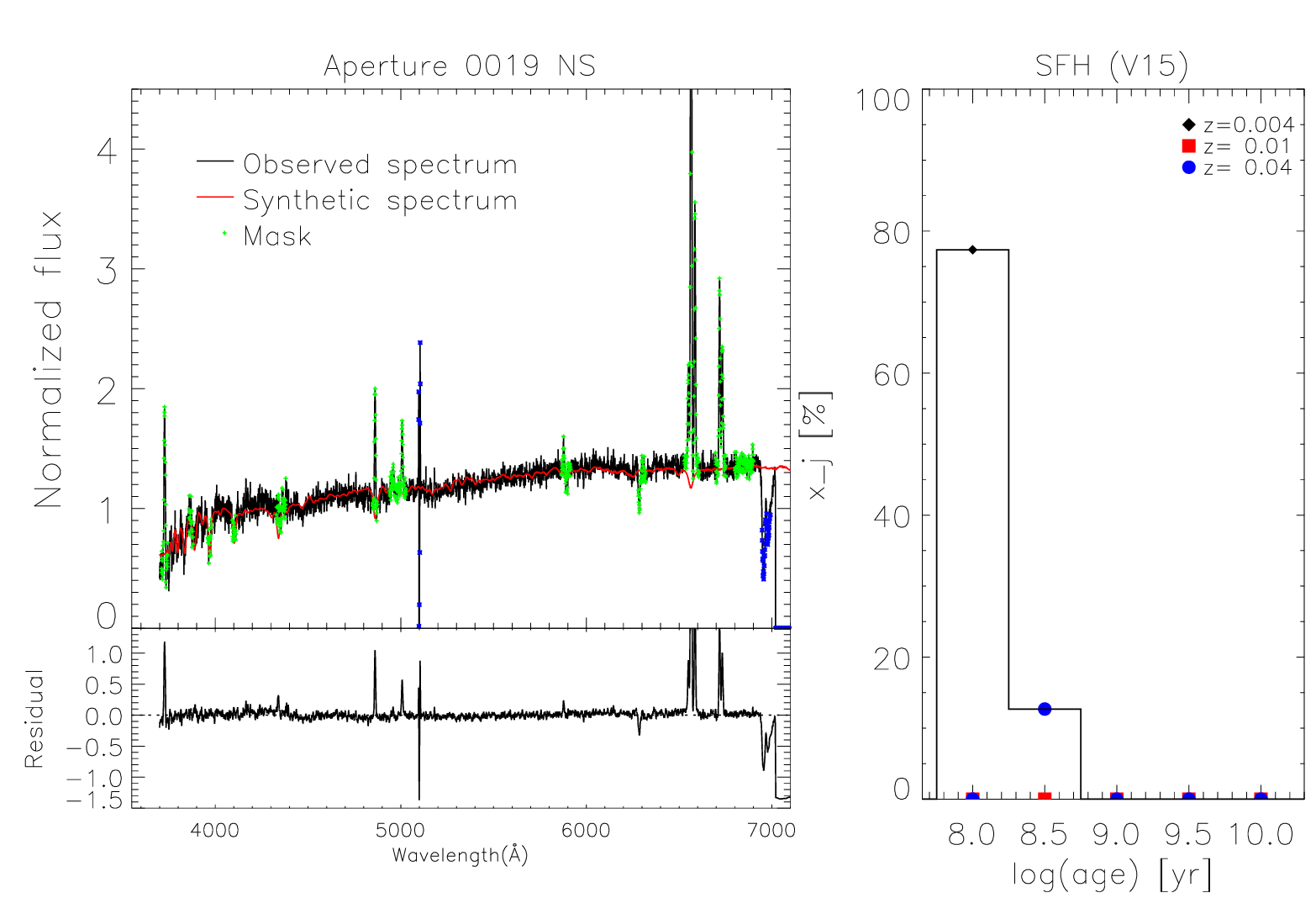}
}
\caption {Stellar population spectral fitting of the galaxy NGC~1232B. 
Figures (a) and (b) correspond to the spectrum fitted 
with BC03 and V15 models, respectively. 
In each figure, the top-left panel shows the observed spectrum in black and the
fitted synthetic spectrum in red. The bottom-left panel shows the residual of the
difference between the two. The right panel shows the SFH, which indicates the contribution
of each population age and metallicity to the final synthetic spectrum. 
} 
\label{popstellar02}
\end{figure*}

The stellar population fitting of the spectrum reveals it is dominated by
a young stellar population, as can be seen in Figure~\ref{popstellar02}. 

We also estimated the galaxy's abundance using the bright line method
values from
$N_2$ and $O_3N_2$
 (see next section for definitions), which are presented in Table~\ref{abundances}, as well as the
abundances derived from these. The results indicate that the abundance of the galaxy is close to  the solar value. We can also use the emission lines measured for NGC~1232B for ionisation source diagnostics.
The green dot in Figure~\ref{diagNIISII} represents NGC~1232B.
Its place in the diagram indicates the ionisation due to stars, 
which is also supported by the stellar population fitting. 
To classify this galaxy as a starburst, we adopt the criteria
 from \citet{bergvall+16}, 
based on the H$\alpha$ emission line equivalent width (E-W ), its FWHM (full width at half maximum)
and a relation between the [OIII] and [NII] emission lines, namely:

\begin{itemize}

\item EW$_{H\alpha}$ $\geq$ 60 \AA,
\item FWHM$_{H\alpha}$ $\leq$ 540 km s$^{-1}$,
\item log([OIII]$\lambda$5007/H$\beta$) $<$ 0.71/log([NII]$\lambda$6584/H$\alpha$–0.25)+1.25.

\end{itemize}

From the measured H$\alpha$ emission line of NGC~1232B, we obtained EW=803$\AA$
and a FWHM of 320 km s$^{-1}$. From the measured [OIII] and [NII] lines,
we obtained log([OIII]$\lambda$5007/H$\beta$) = 0.09 and
0.71/log([NII]$\lambda$6584/H$\alpha$–0.25)+1.25 = 0.58. This means that
NGC~1232B satisfies all the necessary criteria for it to be considered a starburst galaxy.




\section{Abundance from statistical methods}

Studying the chemical composition of spiral galaxies is an important tool in 
improving our knowledge of
the evolution of these complex systems.
Optical emission lines can be used to determine ionic abundances of many key elements, but
their line emissivities strongly depend on the electron density and temperature of the gas. The
electron temperature can be obtained through line ratios involving auroral lines
(like [OIII]$\lambda$4363 or [NII]$\lambda$5755), which
is a major obstacle for extragalactic  HII regions. These lines are intrinsically weak, and because
the cooling efficiency of the gas increases with the gas abundance, they become too faint 
to be observed
with the largest telescopes even at modest metallicities. 
In these cases, it is still possible to obtain estimates of nebular abundances through 
an alternative method using only the intensity of the brightest emission lines. This method is called an
"empirical" or "semi-empirical" approach, or, more descriptively, the "bright lines" method
\citep{dinerstein1990abundances}. This method has many applications besides extragalactic HII 
regions and it has been used in chemical abundance studies of a variety objects, such as
low surface-brightness galaxies \citep{deNaray+04} and star-forming galaxies  at
intermediate and high redshifts \citep{Kobulnicky+04, shapley+04,sanders+20}. 

The lines ([OII]3727\AA + [OIII]4959,5007\AA) were first calibrated and used as 
an abundance indicator by \citet{1979MNRAS.189...95P} and have since been recalibrated many times. 
Variations of these lines and combinations with other lines of [NII], [SII], and [SIII] 
were also adopted \citep{mcc+85, 1985BAAS...17Q.596M}.

The most common abundance indicators using the bright-line method in the literature are:

\begin{equation}
R_{23}=\frac {\mathrm{[OII}] \lambda 3727+\mathrm{[OIII]} \lambda 4949,5007}{\mathrm{H\beta}}, 
\label {eq4.3}
\end{equation} 

\begin{equation}
S_{23}=\frac{\mathrm{[SII}] \lambda 6716, 6731 +\mathrm{[SIII]} \lambda 9069, 9532}{\mathrm{H\beta}},
\label {eq4.4}
\end{equation}

\begin{equation}
N_2=log\frac { \mathrm{[NII]} \lambda 6583}{\mathrm{H\alpha}},
\label {eq4.5} 
\end{equation}

\begin{equation}
O_3N_2=log \frac {(\mathrm{[OIII]}\lambda 5007 / \mathrm{H\beta})}{(\mathrm{[NII]} \lambda 6583/\mathrm{H \alpha)}},
\label {eq4.6}
\end{equation}

which were defined by \citet[][Equation~\ref{eq4.3}]{1979MNRAS.189...95P}, 
\citet[][Equation~\ref{eq4.4}]{2000MNRAS.312..130D}, 
\citet[][Equation~\ref{eq4.5}]{2002MNRAS.330...69D},
\citet{1979A&A....78..200A} and \citet[][Equation~\ref{eq4.6}]{2004MNRAS.348L..59P}.

In this work, it was not possible to calculate R$_{23}$ because of the weak oxygen 
line at 3727$\AA$ that we were not able to measure. Also, the
lines necessary for the S$_{23}$ are out of range of our spectra. 
We then use the N$_2$ and O$_3$N$_2$ indexes to estimate the abundances.
It's important to be aware, however, that these indexes might also be sensitive to the ionisation
parameter and not only to the metallicity \citep{poetrodjojo+18}. 
These values are presented in Table~\ref{abundances}. We used these indexes to obtain the
oxygen abundances, adopting the calibration from \citet{marino+13}. These authors recalibrated
these indexes using T$_e$ based abundances from 603 HII regions extracted from the literature,
together with new measurements from the CALIFA survey \citep{Sanchez+12}. 
They obtained:

\begin{equation}
12 + log (O/H)= 8.743 + 0.462 N_2
\label {eqO/HN2}
\end{equation}
 
and

\begin{equation}
12 + log (O/H)= 8.533 - 0.214 O_3N_2. 
\label {eqO/HO3N2}
\end{equation}

The oxygen abundance values obtained for each HII region from these indexes
are presented in Table~\ref{abundances}.
To derive a radial distribution of abundances that can be compared with values from
the literature, we used the galactocentric distances normalised by the disc effective radius
(r$_e$). This parameter was obtained from its relation with the disc scale-length (r$_d$)
as suggested by \citet{sanchez+14}: r$_e$~$=$~1.67835~r$_d$. We derived the 
disc scale-length from the surface brightness profile of NGC~1232 using its
g-band image from \citet{2018AJ....155..234A}. This profile was fitted with the classical exponential decline
from \citet{freeman70}: 

\begin{equation}
I(r) = I_0^D e^{-r/r_d},
\label {isophot}
\end{equation}

where I$_0^D$ is the extrapolated central surface brightness and r$_d$ the scale-length of the disc.
For NGC~1232, we found r$_d$ = 28.6", which gives r$_e$ = 48.0" (or 5.3~kpc).

\begin{figure}[!h]
\begin{center}
\includegraphics [scale=0.55]{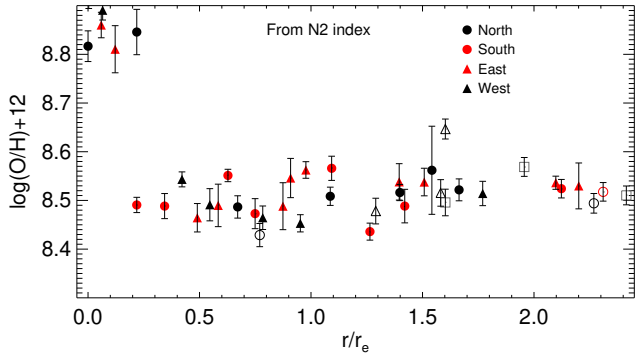} 
\end{center}
\caption{Oxygen radial abundance gradient obtained from the $N_2$ index.
Distances are normalised to the disc effective radius.
 Symbols are the same as in Figure~\ref{diagdensity}.} 
\label{N2}
\end{figure}

\begin{figure}[!h]
\begin{center}
\includegraphics [scale=0.55]{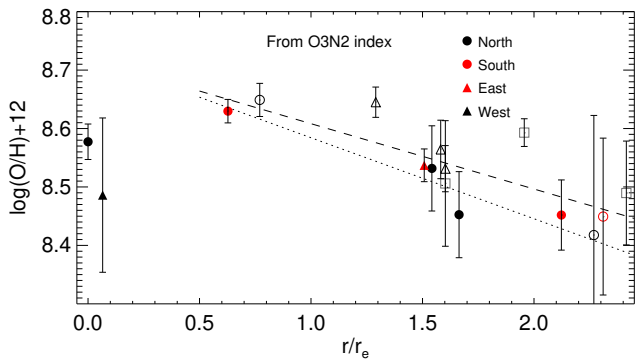} 
\end{center}
\caption{Oxygen radial abundance gradient obtained from the $O_3N_2$ 
index. Distances are normalised to the disc effective radius. The dotted line
represents the fit to the distribution with r/r$_e$~$>$ 0.5.
The dashed line is the same fit, including the data from
\citet{bresolin2005vlt}. Symbols are the same as in Figure~\ref{diagdensity}.} 
\label{O3N2}
\end{figure}

Figures~\ref{N2} and \ref{O3N2} show the oxygen abundance of the HII regions
as a function of the distance to the galaxy center (normalised to the disc effective radius). 
HII regions from \citet{bresolin2005vlt} are also included in the plot,
with open symbols.
As mentioned by other authors \citep[e.g.][]{bresolin2005vlt, sanchez+18}, the 
$N_2$ index does not seem to be very sensitive for abundances higher than solar, 
leveling off at this metallicity. The exceptions are the very central apertures, where we already mentioned that
the ratio between [NII] and H$\alpha$ might be overestimated.

While we have fewer objects
measured, the abundances obtained from the $O_3N_2$ index should better represent the abundance gradient of NGC~1232.
The small number of points make it more difficult to fully characterise the
abundance gradient of the galaxy, but Figure~\ref{O3N2} suggests the presence of 
an inner drop in abundance.
This is a relatively common features in spiral galaxies. 
\citet{sanchez+18} found about 36\% of spiral galaxies present a broken gradient
profile due to a inner flattening or drop of the curve. Because of the
limited number of points, it is not possible to determine where the profile 
changes, or even if it is, in fact, a real effect. The two points close to the nucleus
are the points in the diagnostic diagram of Fig.~\ref{diagNIISII} which are outside the region expected for
HII regions and might have an overestimated [NII]$\lambda$6583/H$_{\alpha}$ ratio (although,
if that is true, the real oxygen abundance from this index would be even lower). 
If the reason for the high ratio comes from, on the other hand,  other mechanisms that contribute
to the emission lines, the abundance measurement cannot be trusted. 
Unfortunately, \citet{bresolin2005vlt} did not observe any HII region closer to the nucleus,
so their data do not help us to verify this trend.
For 
distances larger then R/R$_e$ $=$ 0.5, a linear fit gives a gradient of -0.16 dex/r$_e$,
and if the \citet{bresolin2005vlt} data are included, we obtain 
a gradient of -0.11 dex/r$_e$.
These values are very close to the values published 
by \citet{bresolin2005vlt} and recently by \citet{poetrodjojo+19},
who find a gradient of 
-0.10 dex/r$_e$.

\begin{table*}[]
\centering
\footnotesize
\caption{Abundances of the HII regions of galaxy NGC~1232 and NGC~1232B.}
{\renewcommand{\arraystretch}{}}
\begin{tabular}{cccccc}
\hline
\textbf{Apertures} & &\multicolumn{4}{c}{\textbf{Abundances}} \\ \hline
                   &n$_e$ (cm$_{-3})$ & N$_2$    &O$_3$N$_2$     & log(O/H)+12 (N$_2$)  & log(O/H)+12 (O$_3$N$_2$)  \\ \hline \hline
1                  & 651  &  0.160   & -0.207   &  8.82        & 8.58         \\
3                  & 331  &  0.222   & -        &  8.85        & -        \\
6                  &10000** & -0.546   & -      &  8.49        & -        \\
7                  &  84  & -0.551   & -        &  8.49        & -        \\
8                  & 403  & -0.555   & -        &  8.49        & -         \\ 
9                  &  61  & -0.507   & -0.505   &  8.51        & 8.64      \\ 
10                 &  45  & -0.491   & -        &  8.52        & -         \\ 
11                 & 100  & -0.392   & 0.006    &  8.56        & 8.53      \\ 
12                 &  10*  & -0.480   & 0.376   &  8.52        & 8.45      \\ 
13                 & 315  & -0.416   & -0.452   &  8.55        & 8.63      \\ 
14                 & 248  & -0.585   & -        &  8.47        & -         \\ 
15                 & 554  & -0.383   & -        &  8.57        & -         \\ 
16                 &  47  & -0.665   & -        &  8.44        & -         \\ 
17                 & 1062 & -0.551   & -        &  8.49        & -         \\ 
18                 &   10* & -0.474   & 0.410   &  8.52        & 8.45      \\ 
NGC 1232B          &   10* & -0.472   & 0.376   &  8.52        & 8.45      \\ 
1                  &   97 & 0.375    & -        &  8.92        & -         \\ 
2                  &  241 & 0.253    & -        &  8.86        & -         \\ 
3                  &  633 & 0.147    & -        &  8.81        & -         \\ 
5                  &  323 & -0.603   & -        &  8.46        & -         \\ 
6                  & 10000**& -0.548   & -      &  8.49        & -         \\ 
7                  &   75 & -0.552   & -        &  8.49        & -         \\ 
8                  &   73 & -0.426   & -        &  8.55        & -         \\ 
9                  &  472 & -0.390   & -        &  8.56        & -         \\ 
10                 &  176 & -0.443   & -        &  8.54        & -         \\ 
11                 &   55 & -0.444   & -0.019   &  8.54        & 8.54      \\ 
12                 &   90 & -0.447   & -        &  8.54        & -         \\ 
13                 &   10* & -0.461   & -       &  8.53        & -         \\ 
14                 &   10* &  0.319   & 0.219   &  8.89        & -         \\ 
17                 &   -  & -0.432   & -        &  8.54        & -         \\ 
18                 &   10* & -0.545   & -       &  8.49        & -         \\ 
20                 &   -  & -0.603   & -        &  8.46        & -         \\ 
21                 &   -  & -0.628   & -        &  8.45        & -         \\ 
22                 &   10* & -0.495   & -       &  8.51        & -         \\ \hline
\end{tabular}

*[SII] lines ratio is larger than the theoretical limit and an
upper limit of 10 particles/cm$^3$ was assumed.\\
**[SII] lines ratio is smaller than the theoretical limit and 
an upper limit of 10000 particles/cm$^3$ was assumed.

\label{abundances}
\end{table*}


\section{SFR distribution}

The SFR is an important factor in the chemical evolution of a
galaxy. Its value gives the total amount of gas converted into
stars over a given time interval, which may depend on several
environmental properties.
The SFR is related to the H$\alpha$ luminosity, as defined by \citet{calzetti+07}: 

\begin{equation}
SFR(M_\odot~yr^{-1}) = 5.3 \times 10^{-42} L(H\alpha)~~(erg~ s^{-1}).
\end{equation}

The SFR for each HII region is also presented in Table~\ref{abundances}.
As found by \citet{2018AJ....155..234A}, the central region of the galaxy has a lower SFR
than the most remote regions. For the slit positions of this work, 
the N-S regions seem to have higher
star formation compared to the E-W direction for the outter regions. 
On average, the values obtained for SFR in this study are
in agreement with the values of \citet{2018AJ....155..234A}.

\begin{figure}[!h]
\begin{center}
\includegraphics[scale=0.57]{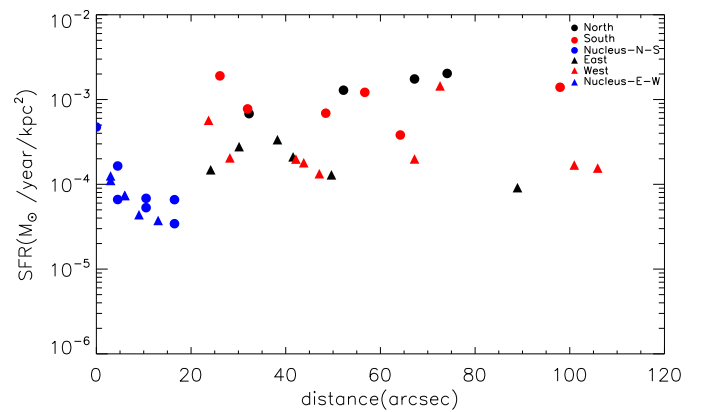} 
\end{center}
\caption{SFR as a function of the distance to the nucleus of NGC~1232.} 
\label{grafsrf}
\end{figure}


\section {Discussion and conclusions}

In this work, we study the galaxy NGC~1232 with long-slit spectroscopy at two different, perpendicular
positions. NGC~1232 is a nearby spiral galaxy, practically face-on, which makes it 
an ideal laboratory for the study of star formation. Recent X-ray studies have suggested that 
it recently collided with a dwarf galaxy \citep{garmire+13}, of which there is no apparent remnant. The effects
of such a collision might have been a localised increase in star formation in the galaxy,
as suggested by the results of \citet{2018AJ....155..234A}.

From the H$\alpha$ emission peaks along the slits positioned in the N-S  and E-W  directions, we extracted
the HII regions analysed in this work. Before measuring the emission lines, we fit the underlying stellar 
population of the galaxy using the spectral fitting technique. We found that both the N-S  and E-W 
 directions present a stellar population gradient typical of spiral galaxies, with the fraction of the 
young component increasing progressively when moving outwards from the center. Conversely, the 
old component becomes less important out of the galaxy nucleus.
The E-W  direction seems to have a younger population 
in the nucleus compared to the N-S  direction. We believe that this is both connected with the slit position
(the N-S slit is positioned slightly to the west of the nucleus) and the presence of a small bar
in the E-W  direction. 

After the subtraction of the stellar population, we measured the emission lines in each HII region.
For many of them, it was only possible to measure H$\alpha$ and [NII] lines but when possible, we
determined the extinction coefficient and corrected the lines for it. These lines were used to determine
the electron density profile of the galaxy and the oxygen abundance gradient.

The existence of a radially decreasing abundance gradient in spiral galaxies has been well-established by several
observations over the years
 \citep {searle1971evidence, martin+92, zaritsky+94,
vanzee+98, martins+00,
kennicutt2003sings, bresolin2012gas,
sanchez+14, sanchez2016evidence, kaplan+16, belfiore+17, poetrodjojo+19, kreckel+19}, which support the inside-out disk scenario.
However, deviations from a simple linear decrease have been reported in recent years, with a 
flattening or drop of the relation in the central regions and sometimes a flattening of the curve
in the outer parts \citep{belley+92, martin+95,vilchez+96, roy+97,
sanchez+18}. 
 Even the purely radial picture has been questioned of late \citep{kreckel+19} 
since spiral arms, bars, and stellar feedback drive mixing and gas flows,
affecting the gas-phase abundances azimuthally \citep{grand+16, sanchez-menguiano+16,ho+17,
yang+12,petit+15, deAvillez+02, krumholz+18}. This is a difficult effect 
to measure because a large number of objects per galaxy are needed to detect it.
Evidence has been 
found to support it \citep{kreckel+19} but it is, nonetheless, still inconclusive. 
Aside from that, there is the complication of defining the real boundary of the HII regions 
and, in many instances, the surrounding diffuse ionised gas (DIG) contaminates the observed
spectrum of these objects \citep{poetrodjojo+19}. This contamination might cause a flattening
of the metallicity gradient.

With our data, it is difficult to fully characterise the abundance
gradient of NGC~1232 due to the weaknesses of some important lines for the diagnostics
and the small number of objects detected. We were able to
determine two abundance indexes from our data, N$_2$ and O$_3$N$_2$. Despite having more measurements for
the N$_2$, this index is known to be less sensible to higher abundances, which hampers its efficacy to
characterise the galaxy's gradient. From the few determinations we have of O$_3$N$_2$, we were able
to measure an oxygen abundance gradient of -0.16 dex/r$-e$, which is within the expected value for 
a spiral galaxy. 
We also believe there might be an abundance drop towards the center, although it is not
possible to determine where the drop begins. Many different mechanisms have been proposed in the literature
to explain the deviation from the negative gradient in spiral galaxies, such as satellite accretion
\citep{10.1086/680484, bird2012radial}. It would be tempting to assume that the broken gradient
profile found for NGC~1232 could be an extra evidence of the collision seen in X-ray, but recent
observational works have found no dependence of the presence of the broken profile with either environment,
morphological type, or the presence of bars \citep{sanchez+18}. 
 
For all the characteristics studied in this paper (electron density profile, stellar populations,
SFR distribution, and abundance gradients), we found no deviation from a typical spiral galaxy
and only tenuous differences between some of the directions.
Therefore, through our long-slit analysis, we 
found no clear evidence of the collision seen in X-ray. We believe that if the collision caused a
disturbance in the galaxy, it will be a small disturbance not clearly observed in a small number of objects.
A final understanding of the
effect or not of this possible collision can only be achieved with a deep spectroscopic observation
covering the whole galaxy (maybe with an IFU observation).

The northernmost H$\alpha$ emission peak measured in the N-S slit was not an HII region of NGC~1232, but
a background galaxy, named by \citet{arp82} as NGC~1232B. 
To the best of our knowledge, we provide the first characterisation of that galaxy so far in the literature. 
We derived a redshift of z = 0.09261 using the emission lines detected on its spectrum. Moreover, by means of the 
stellar population analysis, Baldwin, Phillips \& Terlevich (BPT) diagrams, and emission line criteria by
\citet{bergvall+16}, we classified NGC~1232B as a starburst galaxy.


\begin{acknowledgements}
We thank the referee for the valuable comments that greatly improved the paper. 
Based on observations obtained at the Southern Astrophysical Research (SOAR) telescope, 
which is a joint project of the Minist\'{e}rio da Ci\^{e}ncia, Tecnologia, e Inova\c{c}\~{a}o (MCTI) da Rep\'{u}blica Federativa do Brasil, the U.S. National Optical Astronomy Observatory (NOAO), the University of North Carolina at Chapel Hill (UNC), and Michigan State University (MSU).
F.L.C. acknowledges CAPES for financial support.
L.M. thanks CNPQ for financial support through grant  306359/2018-9 and FAPESP through
grant 2015/14575-0. 
A.R.A. thanks CNPq for partial
support to this work.

\end{acknowledgements}


\bibliographystyle{aa} 
\bibliography{bib} 



\onecolumn
\begin{appendix} 
\label{appendix}
\section{Stellar population spectral fitting for all apertures of NGC~1232}

\begin{figure*}[!h]
\centering
\subfloat[]{
\includegraphics[scale=0.37]{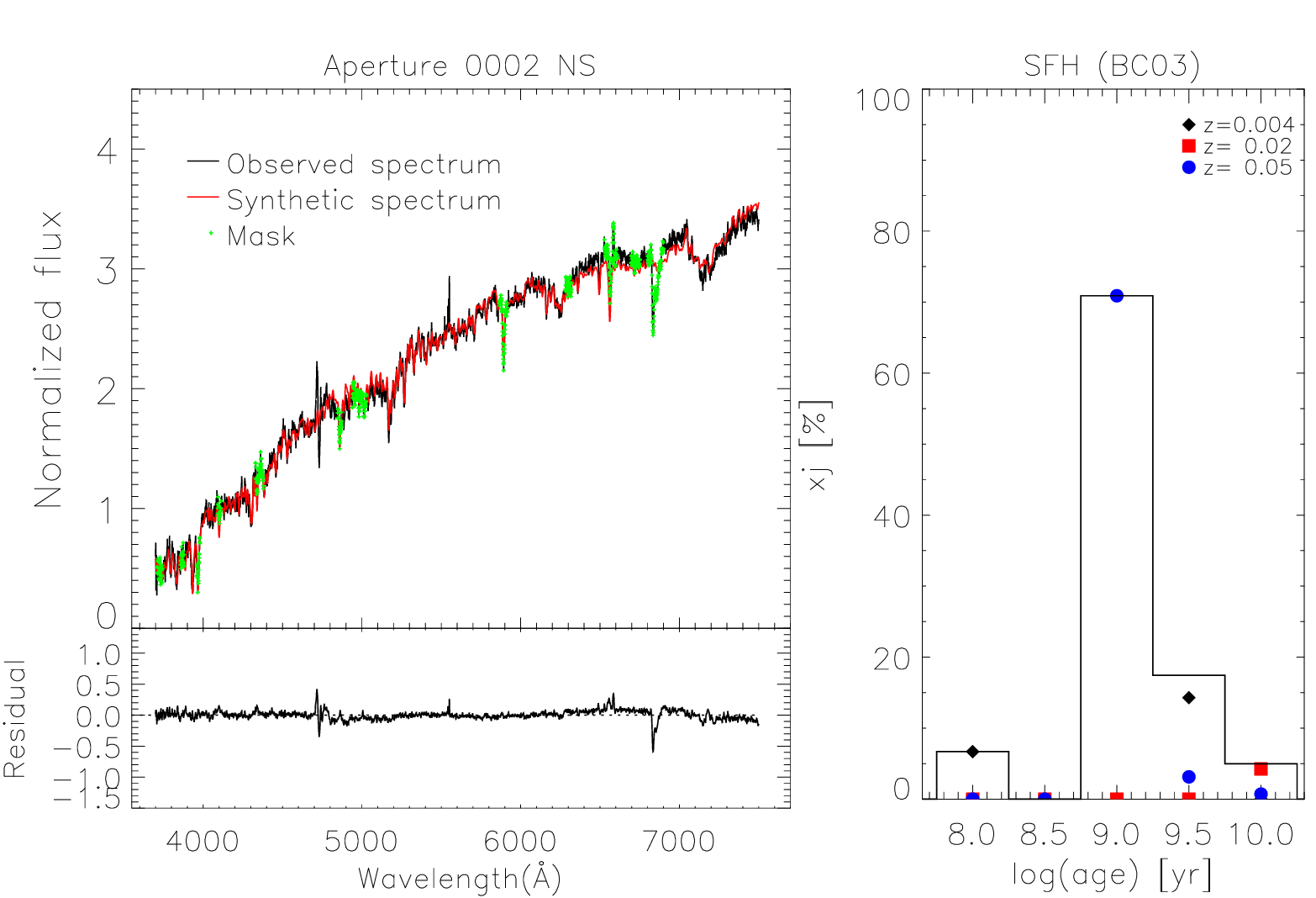}
}
\subfloat[]{
\includegraphics[scale=0.37]{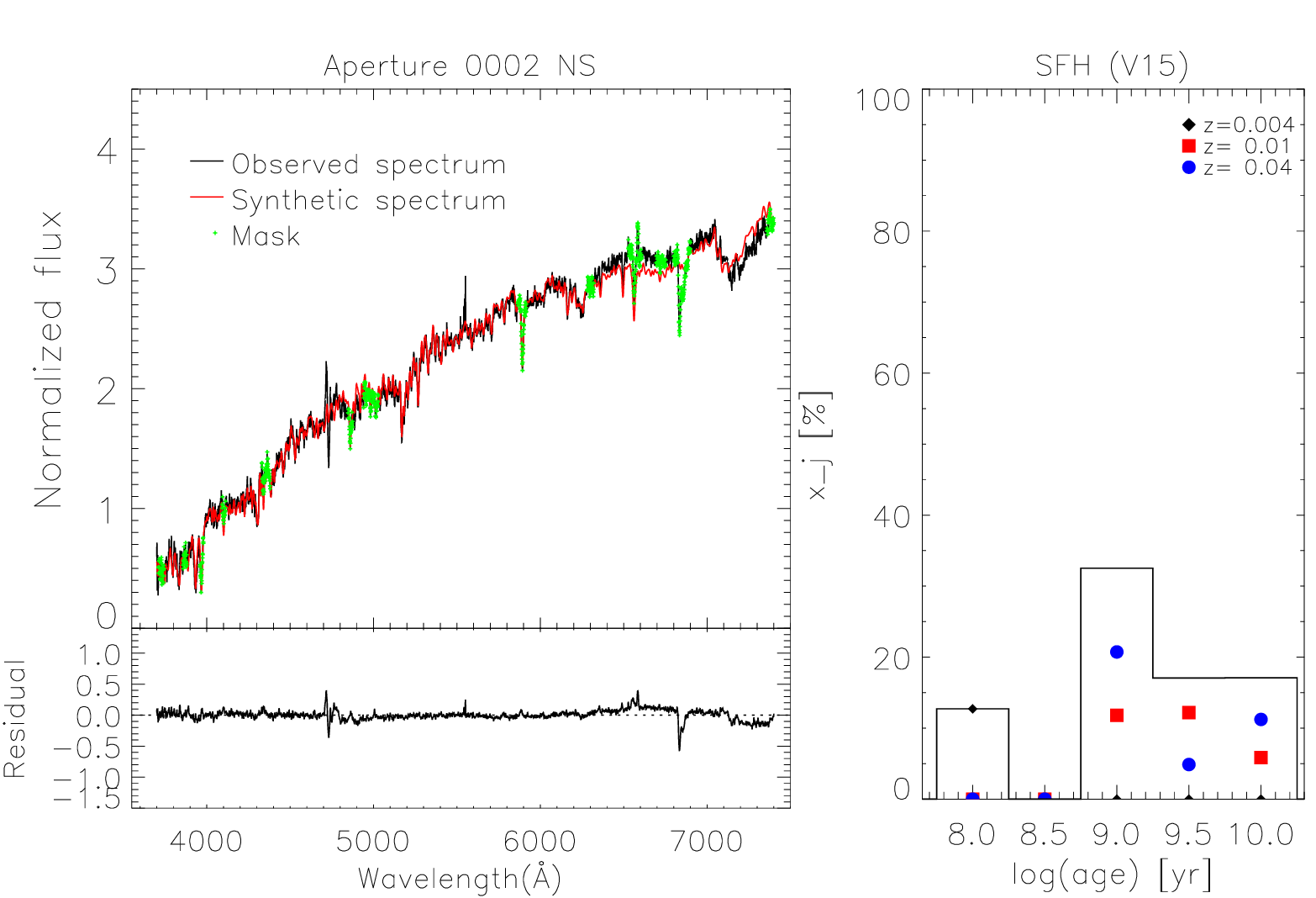}

}
\\
\subfloat[]{
\includegraphics[scale=0.37]{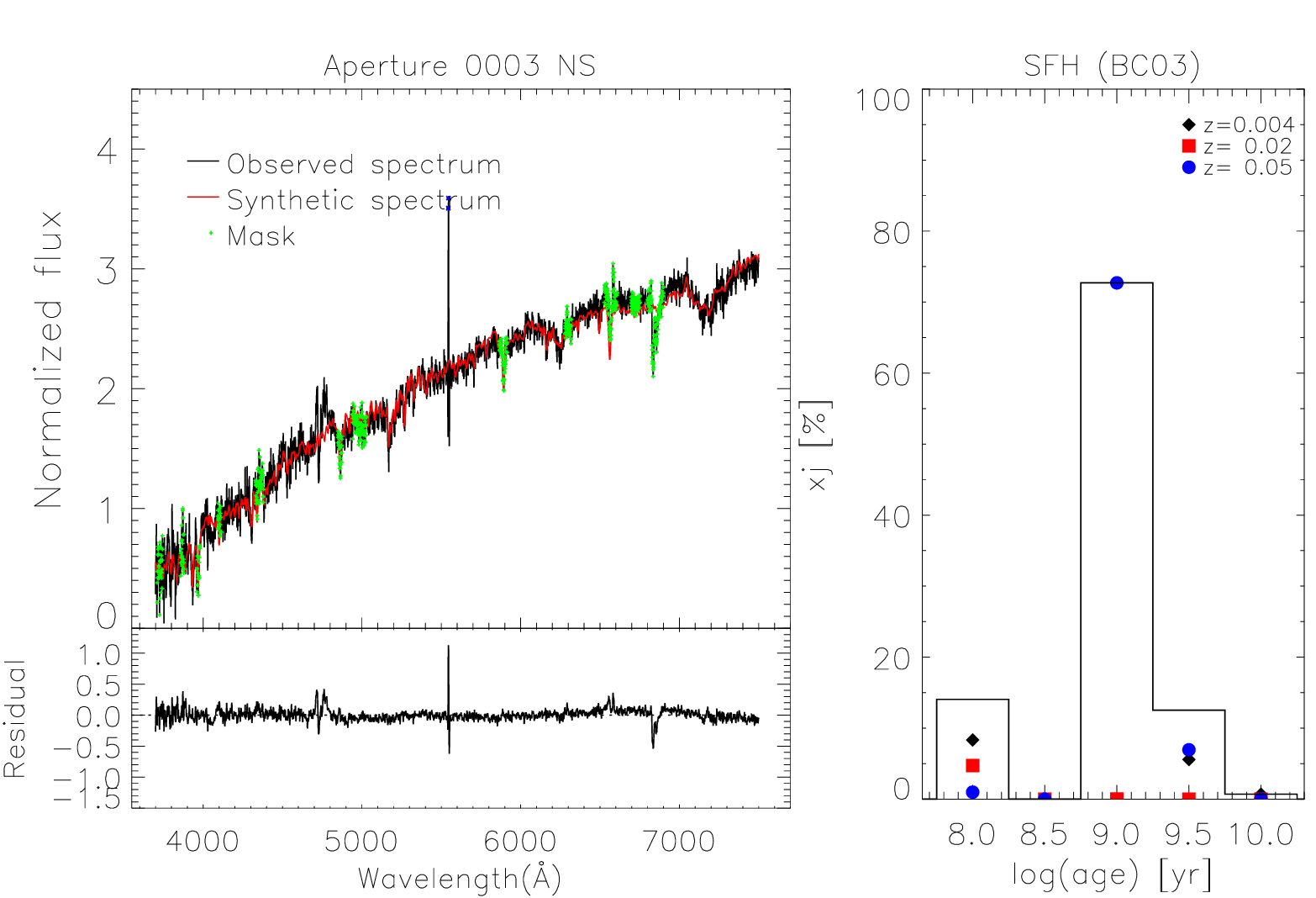}
}
\subfloat[]{
\includegraphics[scale=0.37]{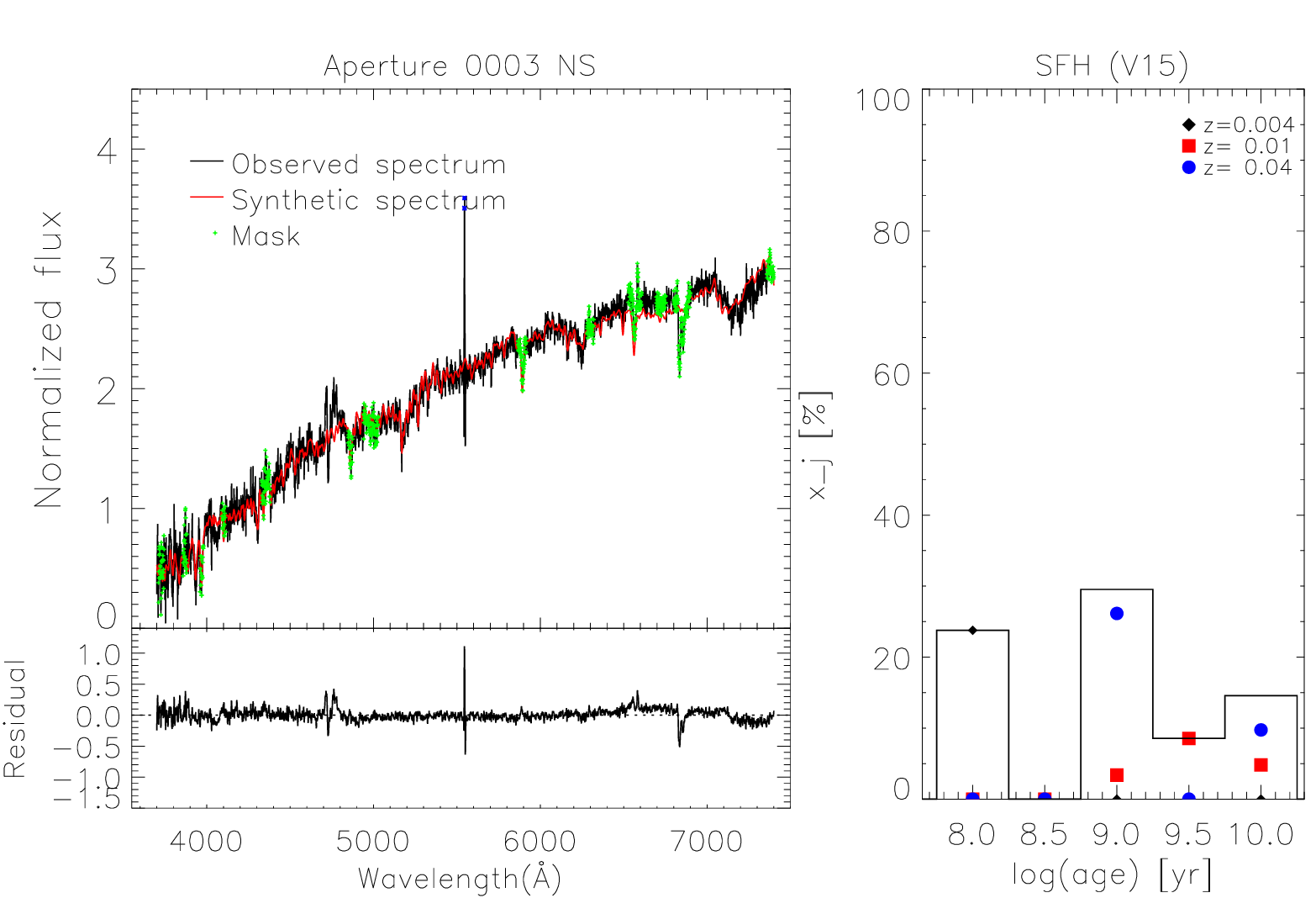}
}
\\
\subfloat[]{
\includegraphics[scale=0.37]{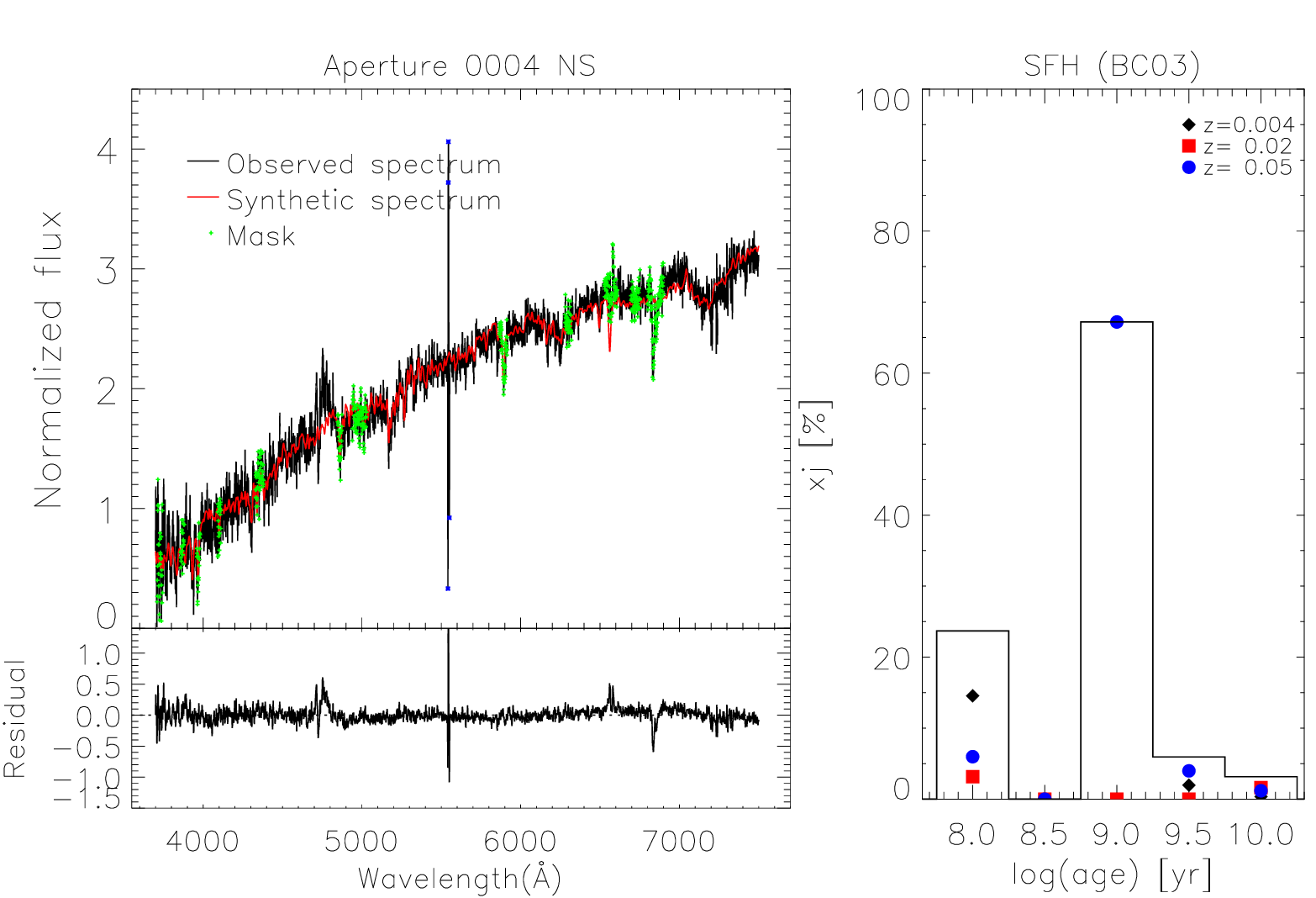}
}
\subfloat[]{
\includegraphics[scale=0.37]{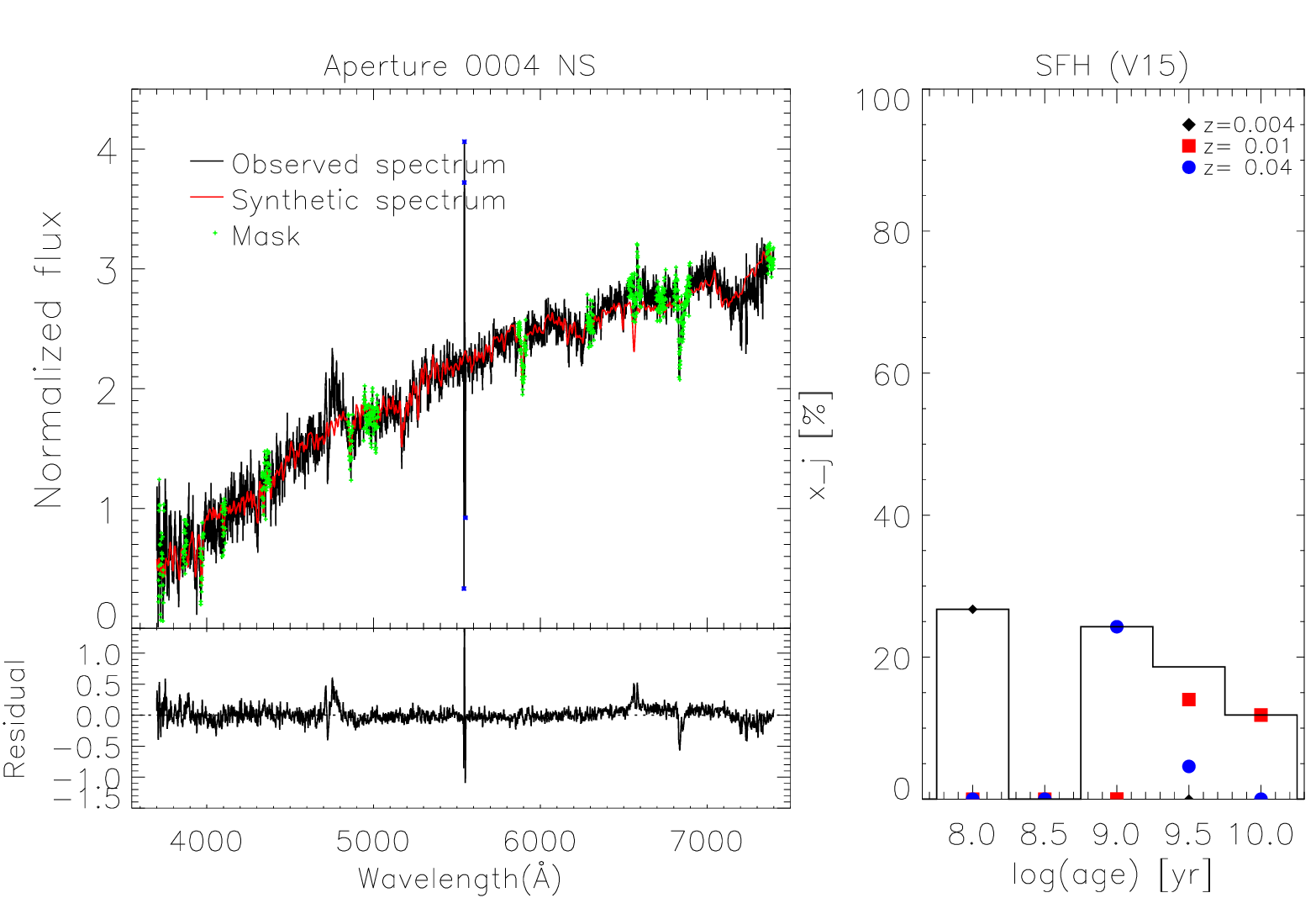}
}

\caption {Stellar-population spectral fitting of the HII regions of NGC~1232, a
apertures 02, 03, and 04 of the N-S slit. Left figures show
the spectra fitted 
with BC03 and right with V15 models.
In each figure, the top left panel shows the observed spectrum in black, and the
fitted synthetic spectrum in red. The bottom left panel shows the residual of the
difference between the two. The right panel shows the star formation history (SFH), 
which indicates the contribution
of each population age and metallicity to the final synthetic spectrum. 
} 
\label{popstellarA1}
\end{figure*}


\begin{figure*}[!h]
\centering
\subfloat[]{
\includegraphics[scale=0.38]{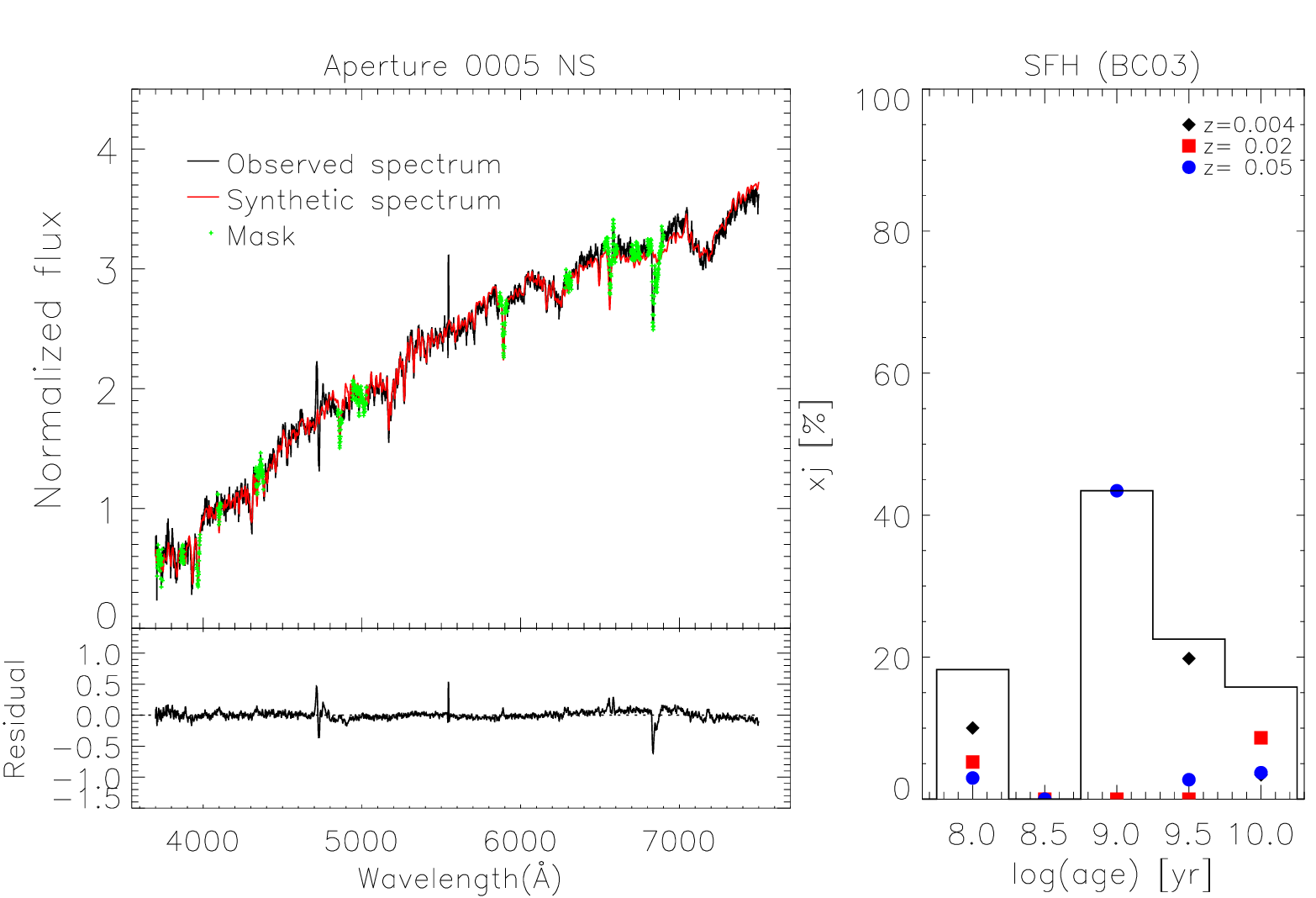}
}
\subfloat[]{
\includegraphics[scale=0.38]{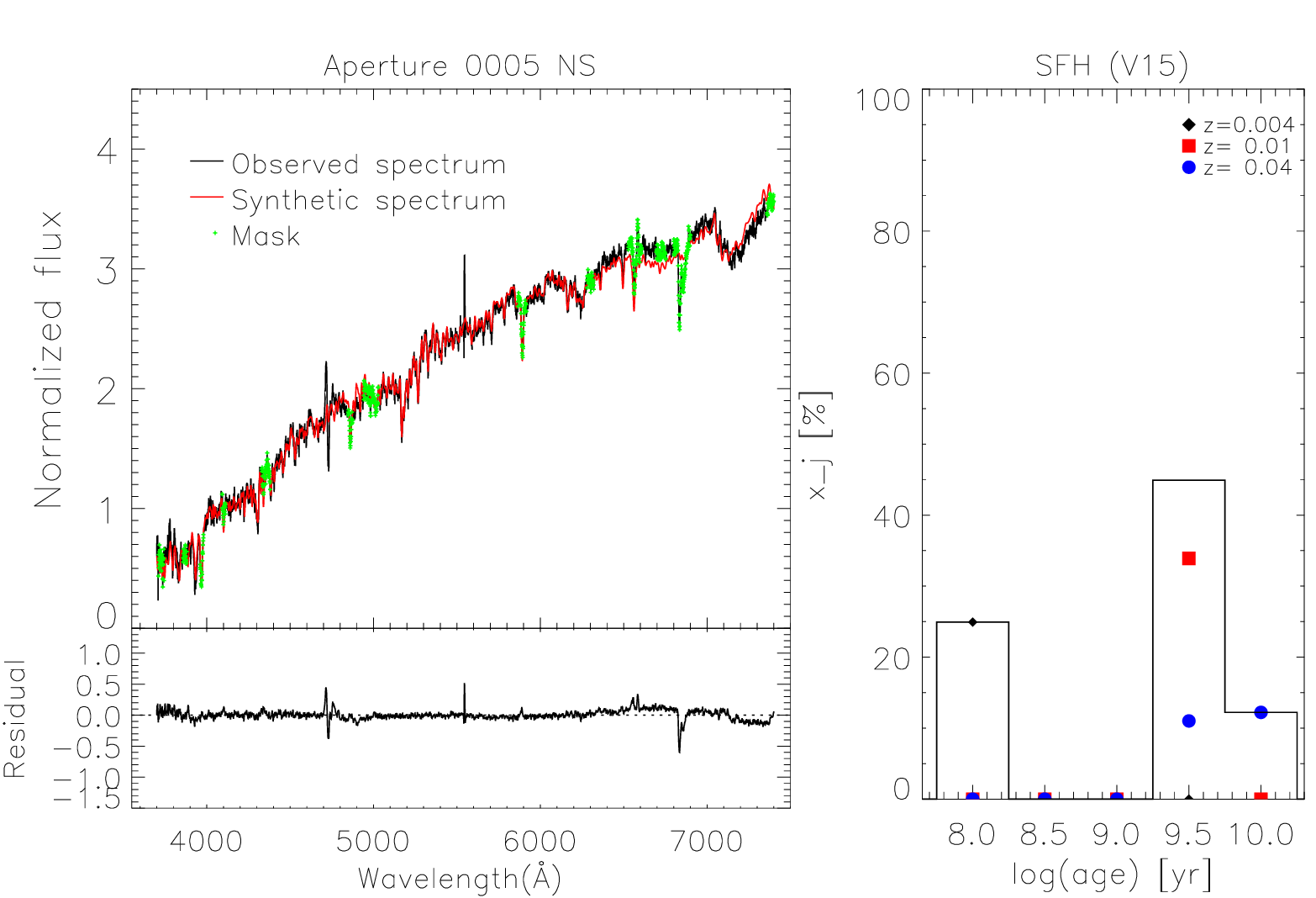}

}
\\
\subfloat[]{
\includegraphics[scale=0.38]{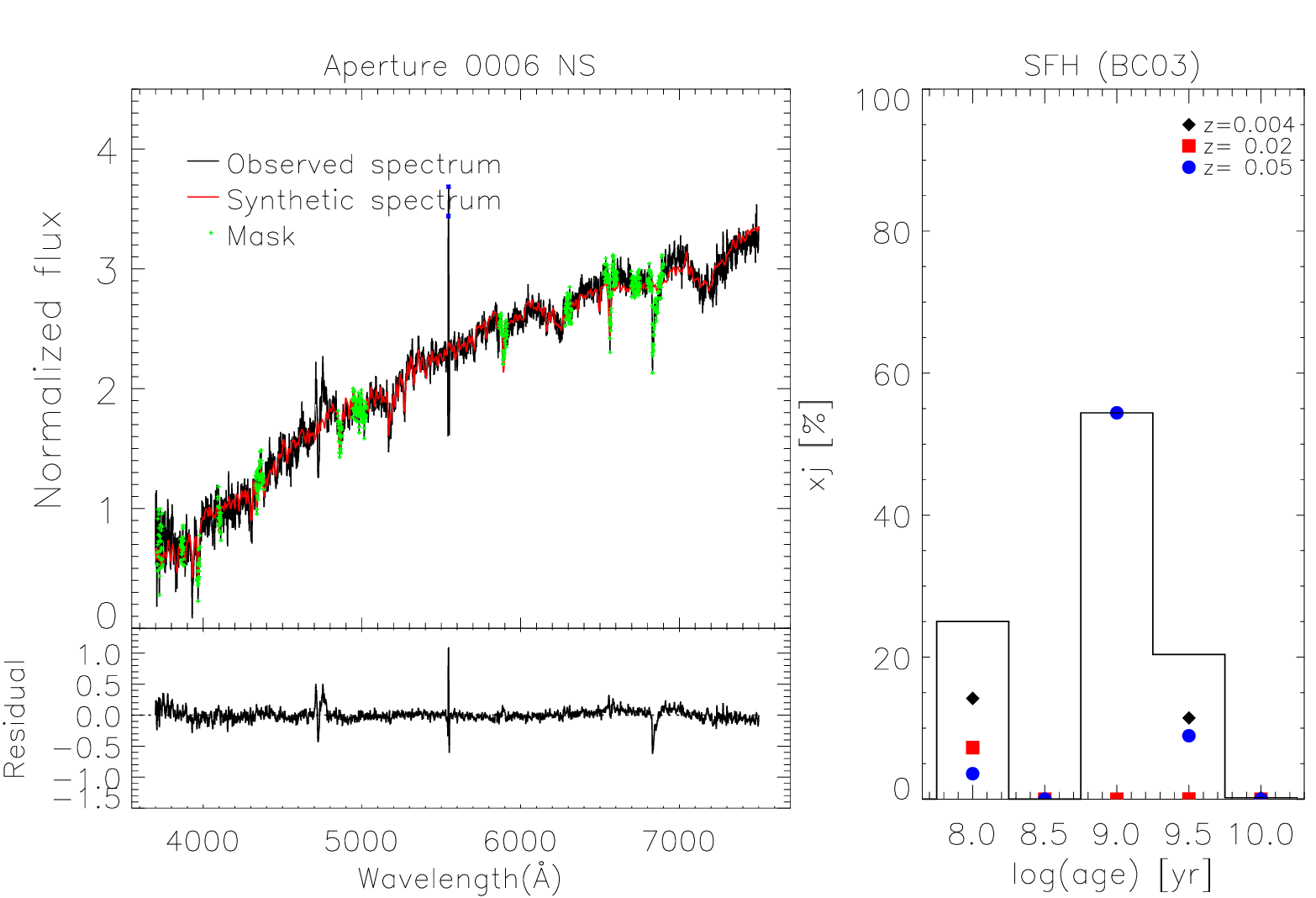}
}
\subfloat[]{
\includegraphics[scale=0.38]{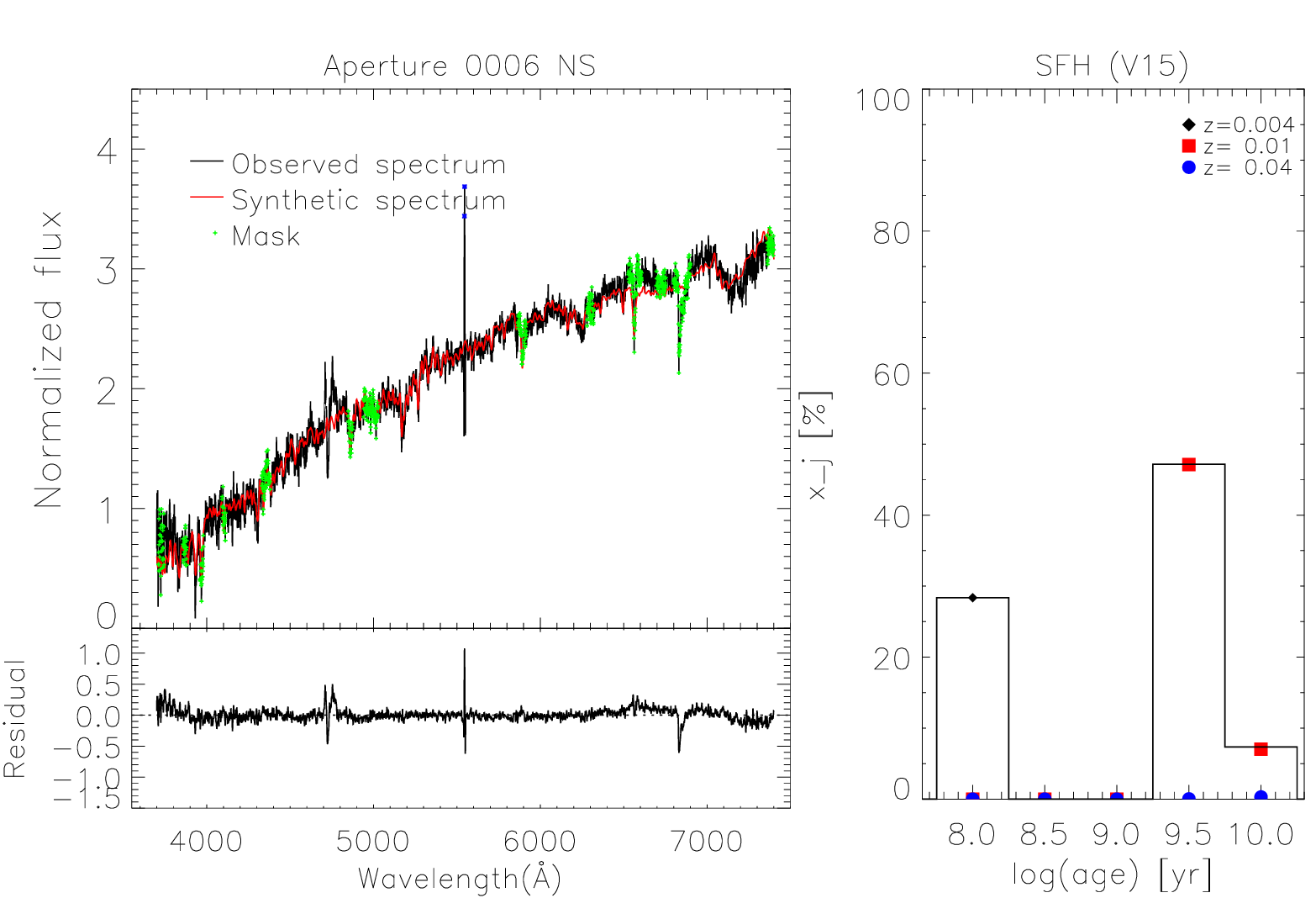}
}
\\
\subfloat[]{
\includegraphics[scale=0.38]{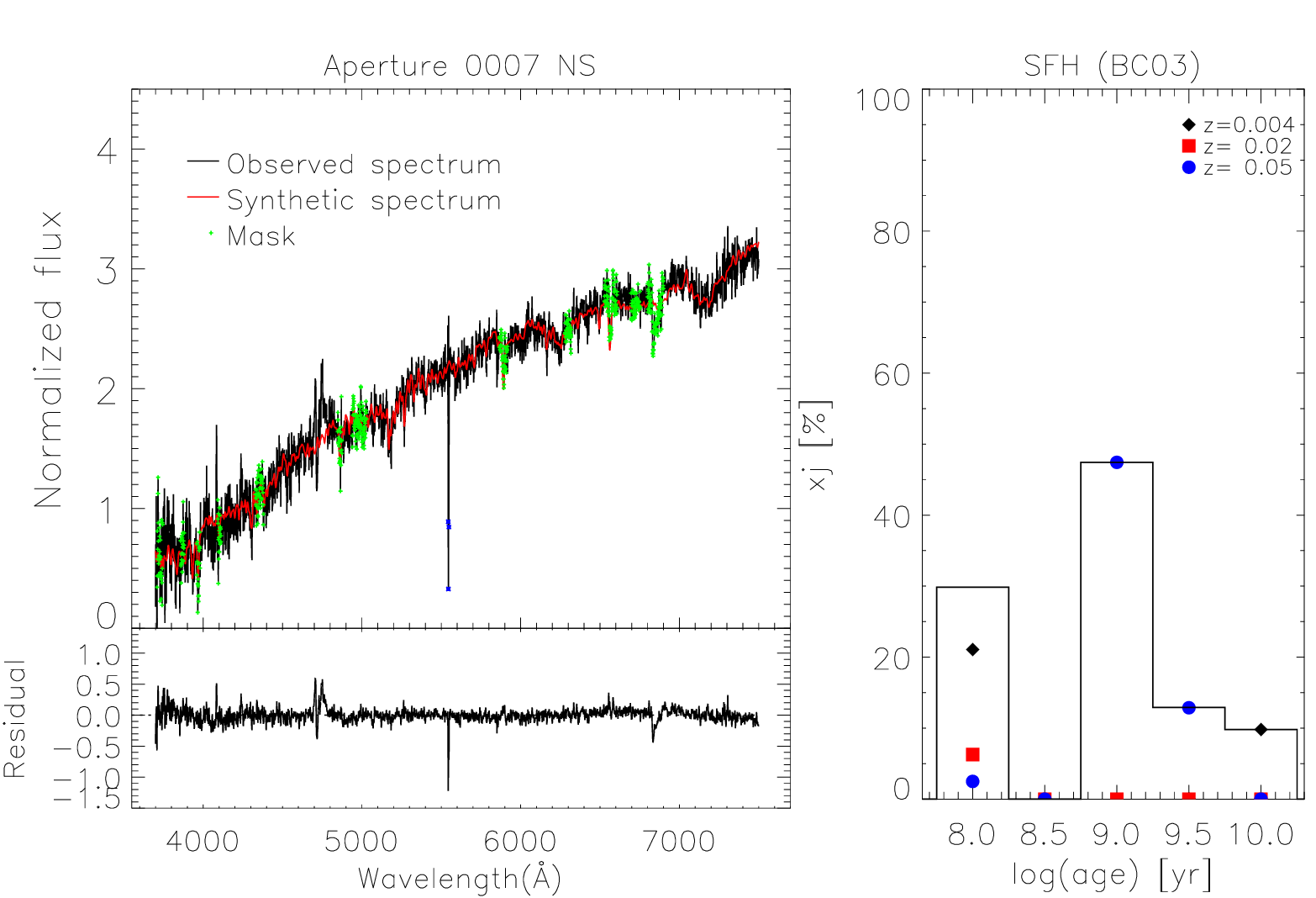}
}
\subfloat[]{
\includegraphics[scale=0.38]{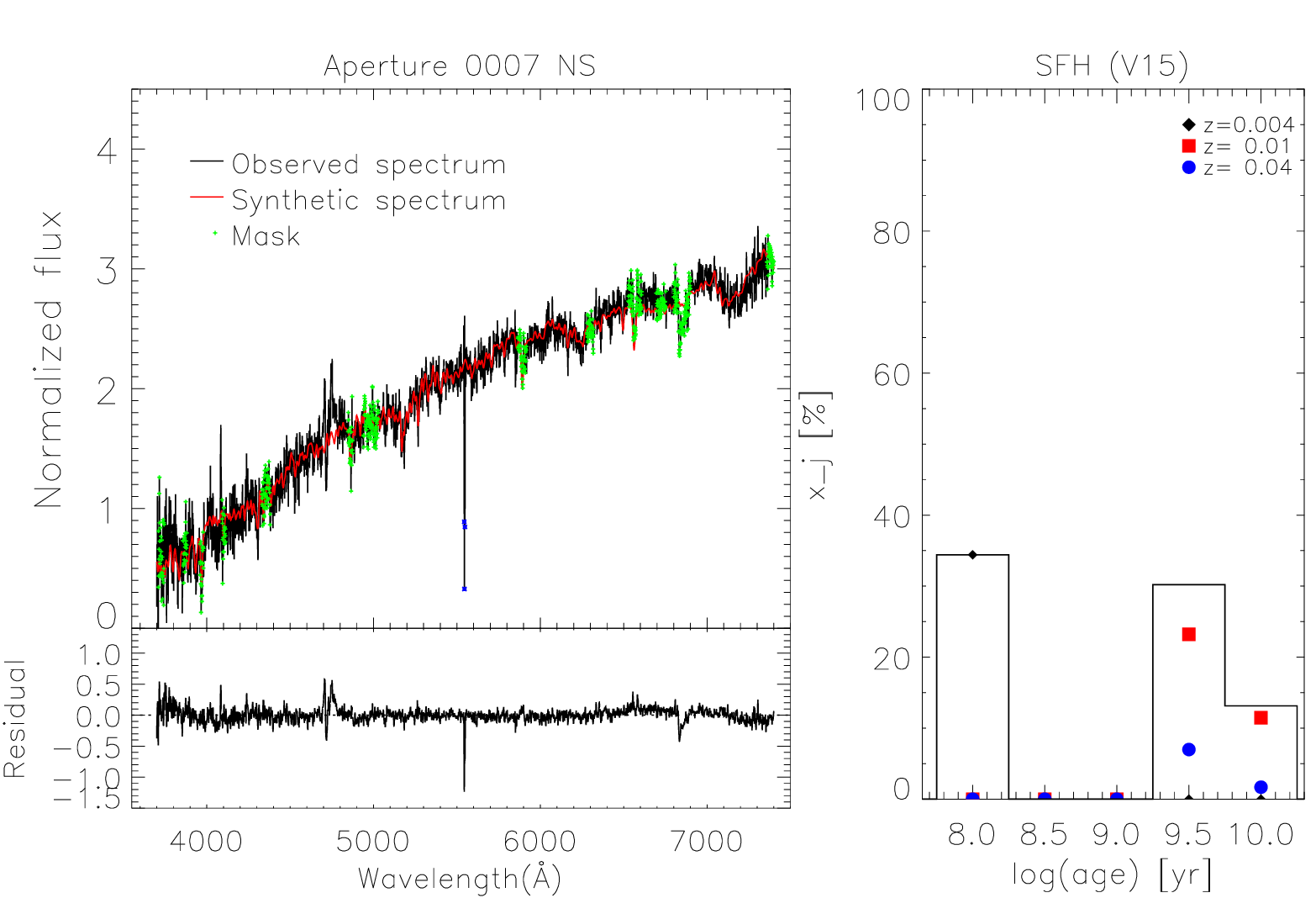}
}

\caption {Same as Figure~\ref{popstellarA1} but for apertures 05, 06, and 07
of the N-S  slit.
} 
\label{popstellarA2}
\end{figure*}


\begin{figure*}[!h]
\centering
\subfloat[]{
\includegraphics[scale=0.38]{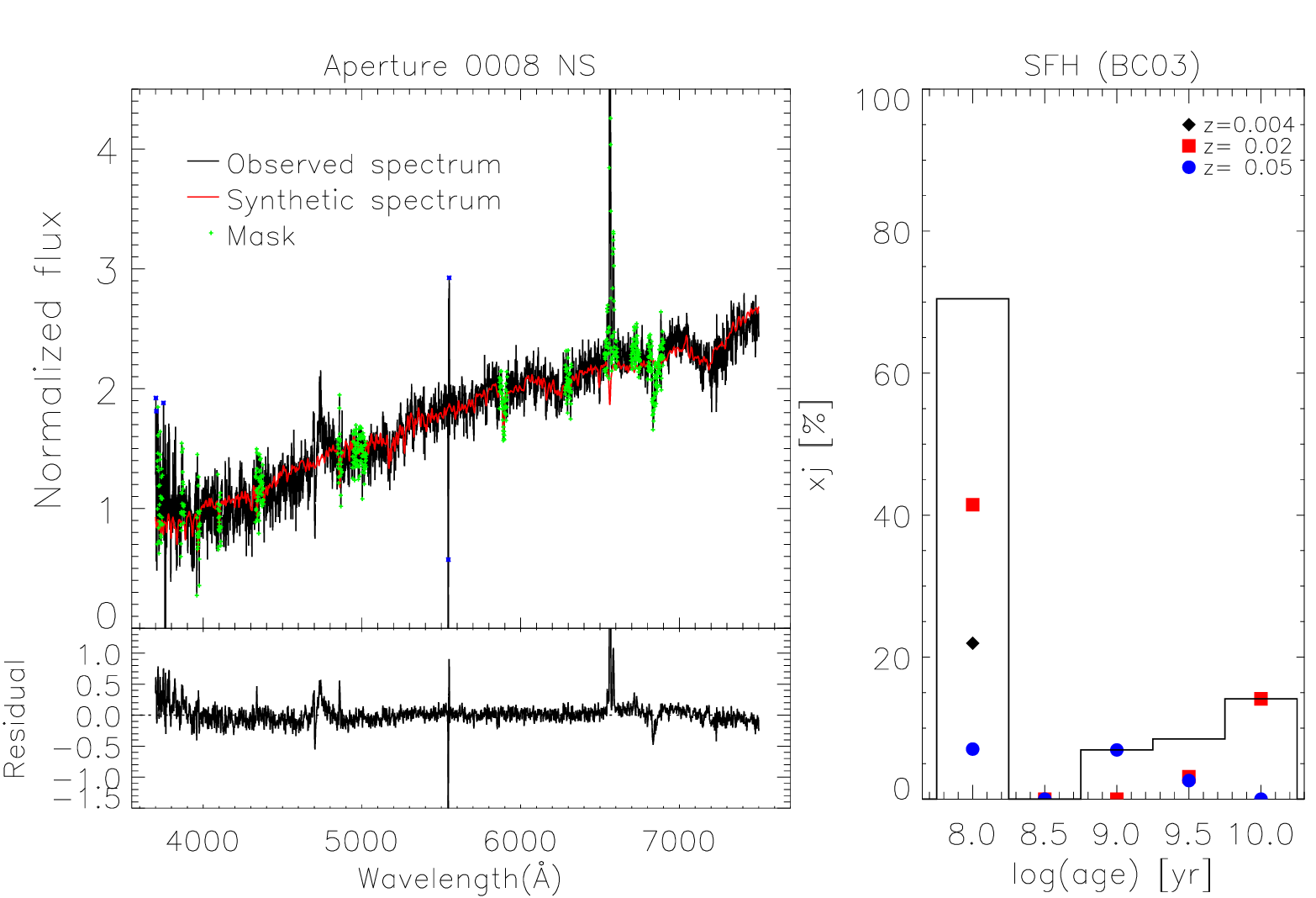}
}
\subfloat[]{
\includegraphics[scale=0.38]{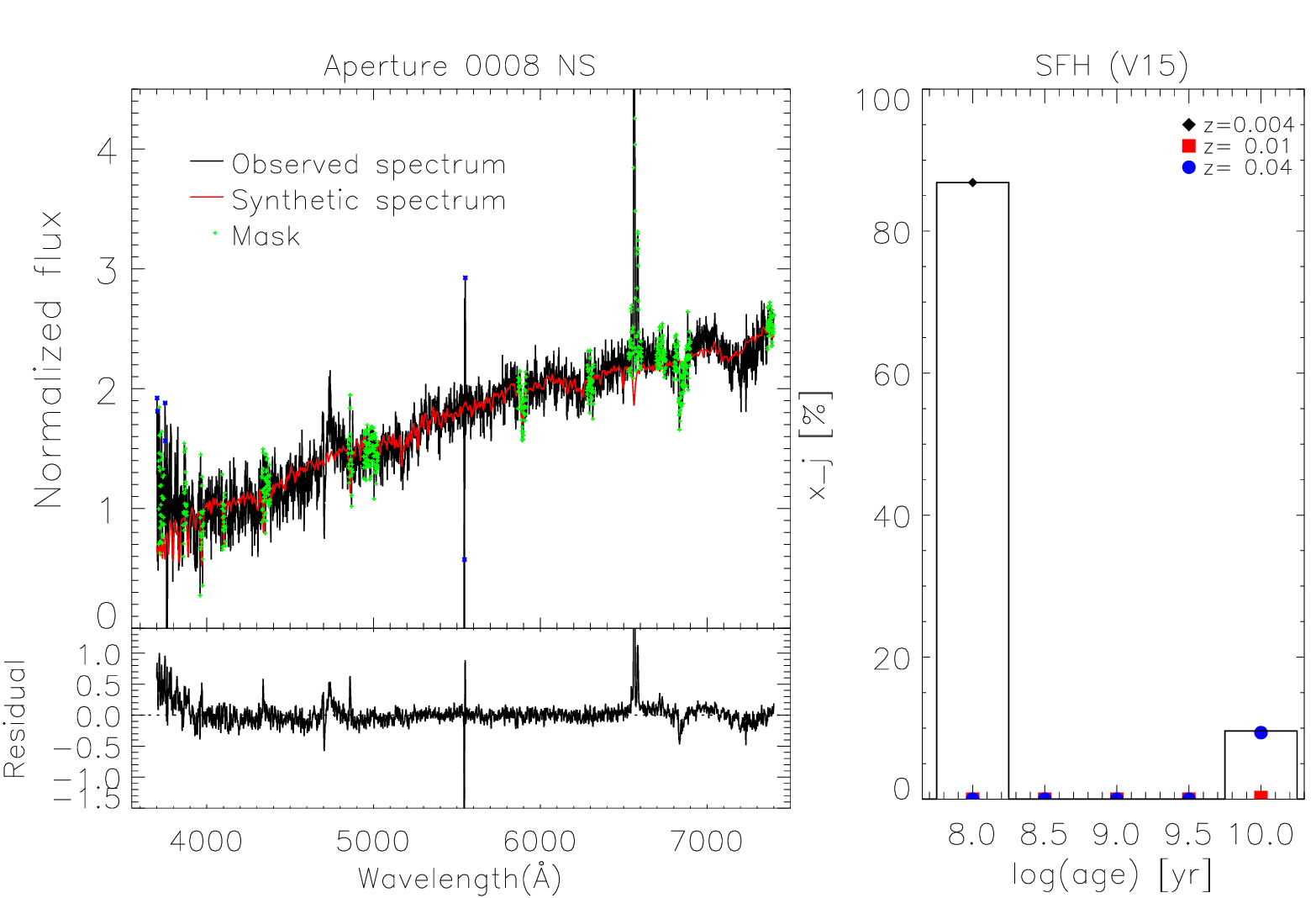}

}
\\
\subfloat[]{
\includegraphics[scale=0.38]{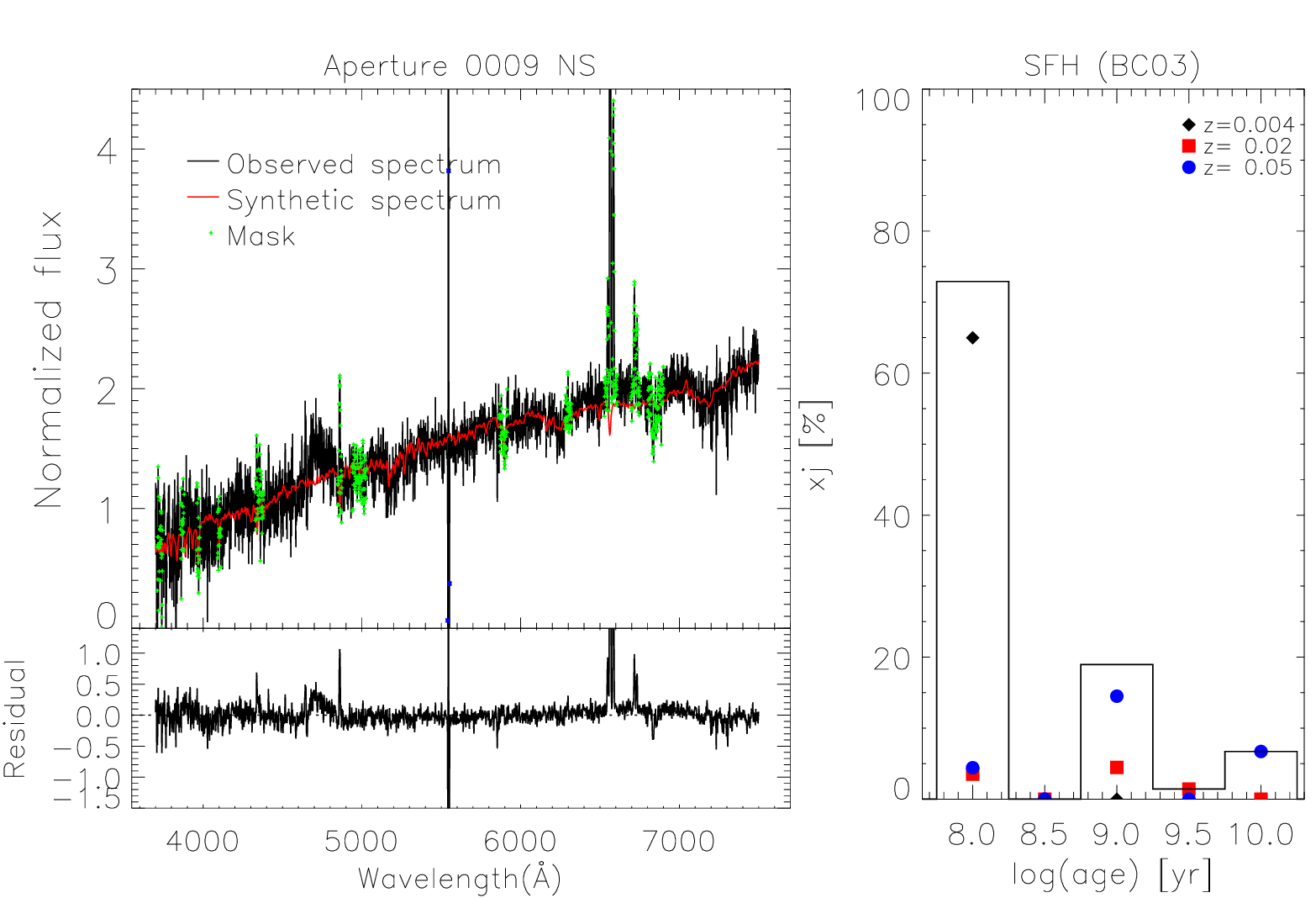}
}
\subfloat[]{
\includegraphics[scale=0.38]{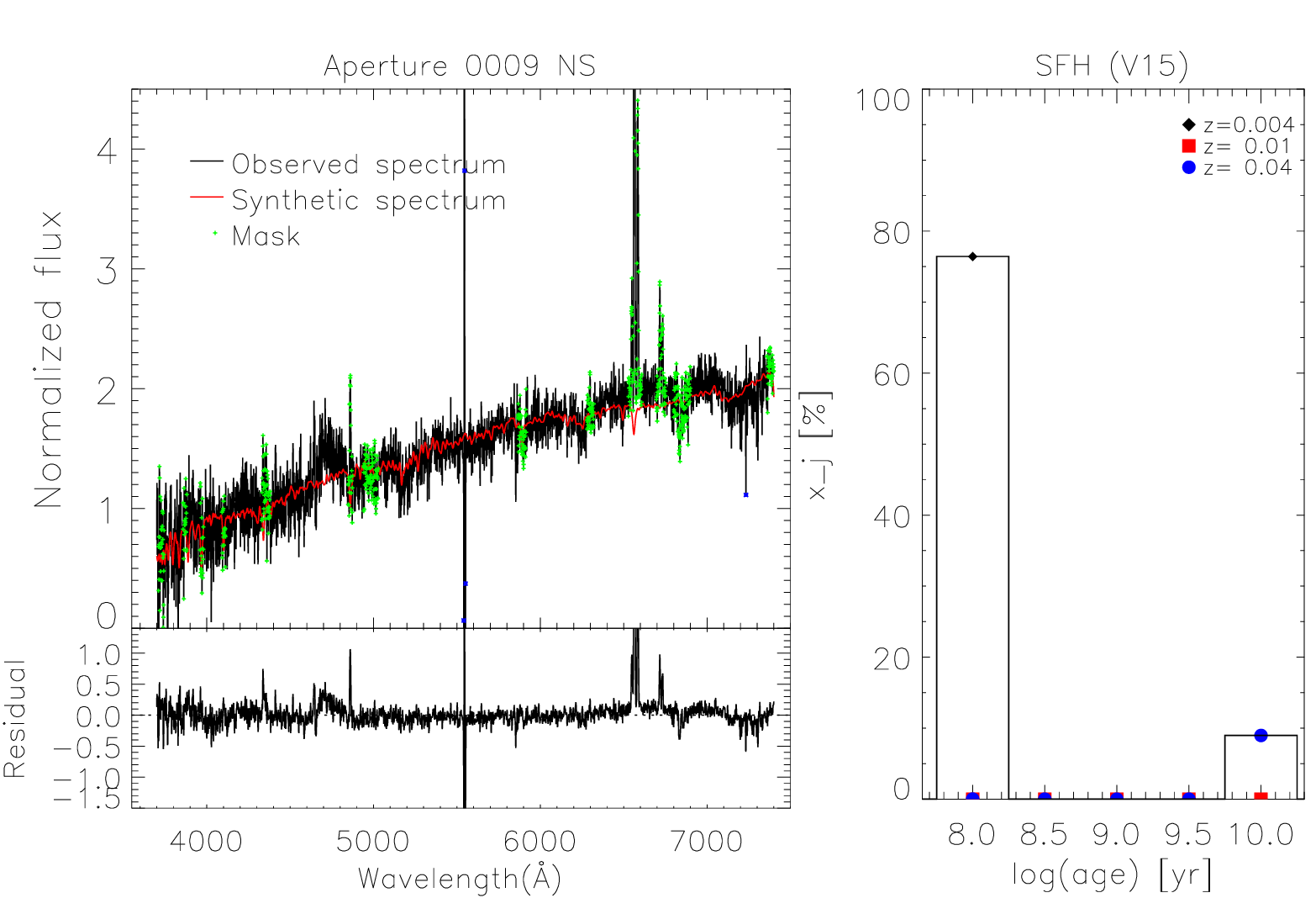}
}
\\
\subfloat[]{
\includegraphics[scale=0.38]{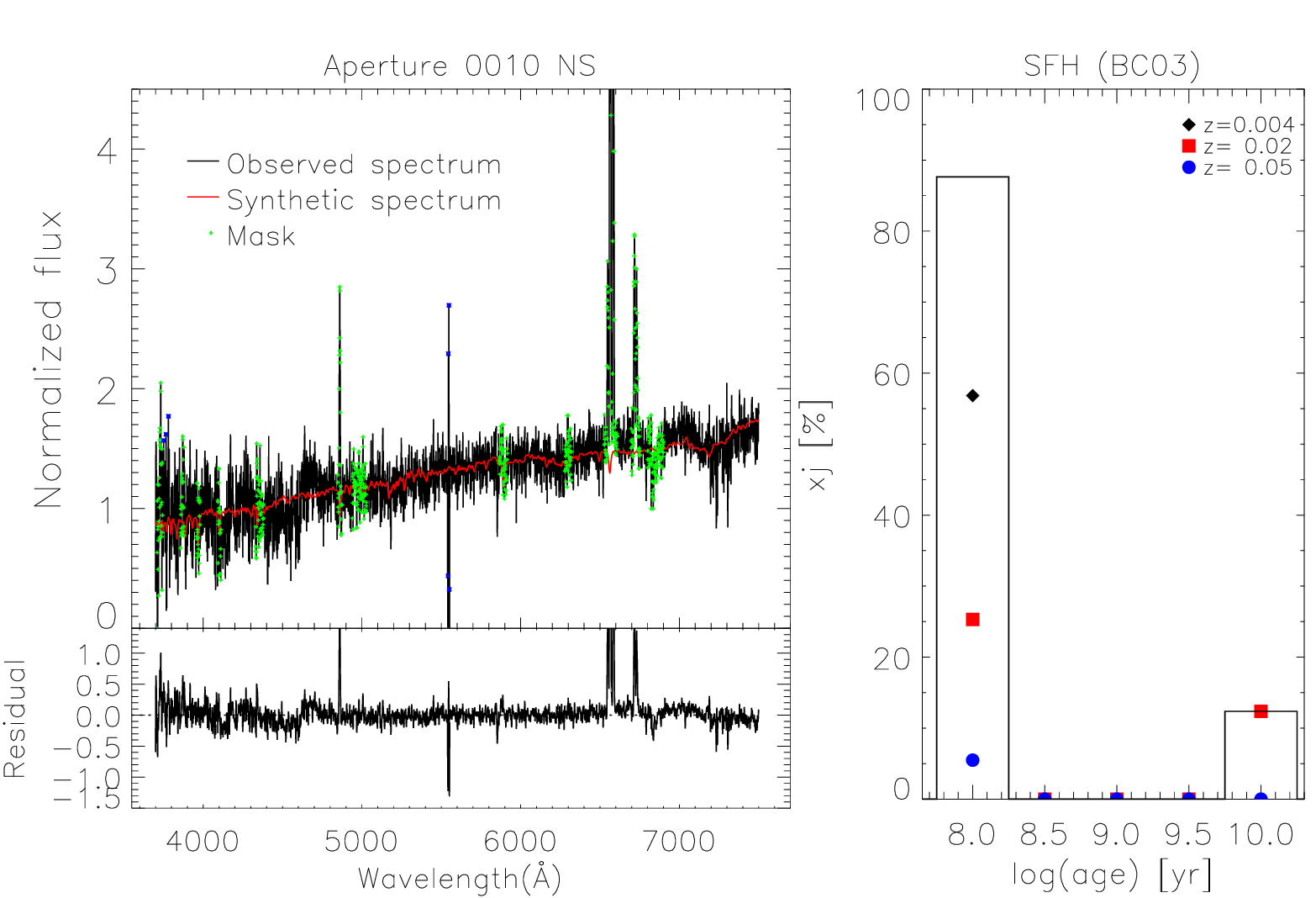}
}
\subfloat[]{
\includegraphics[scale=0.38]{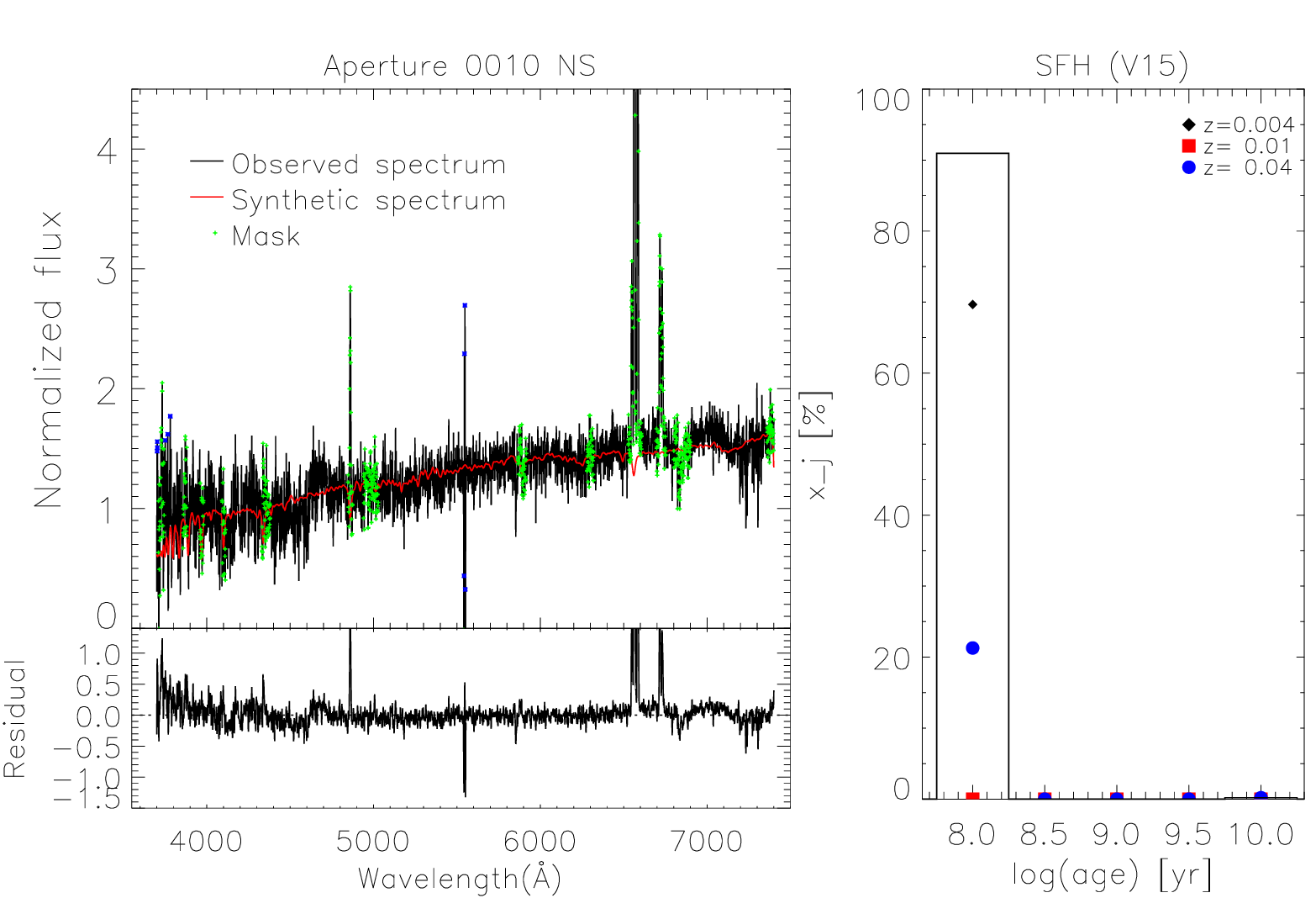}
}

\caption {Same as Figure~\ref{popstellarA1} but for apertures 08, 09, and 10
of the N-S  slit.
} 
\label{popstellarA3}
\end{figure*}


\begin{figure*}[!h]
\centering
\subfloat[]{
\includegraphics[scale=0.38]{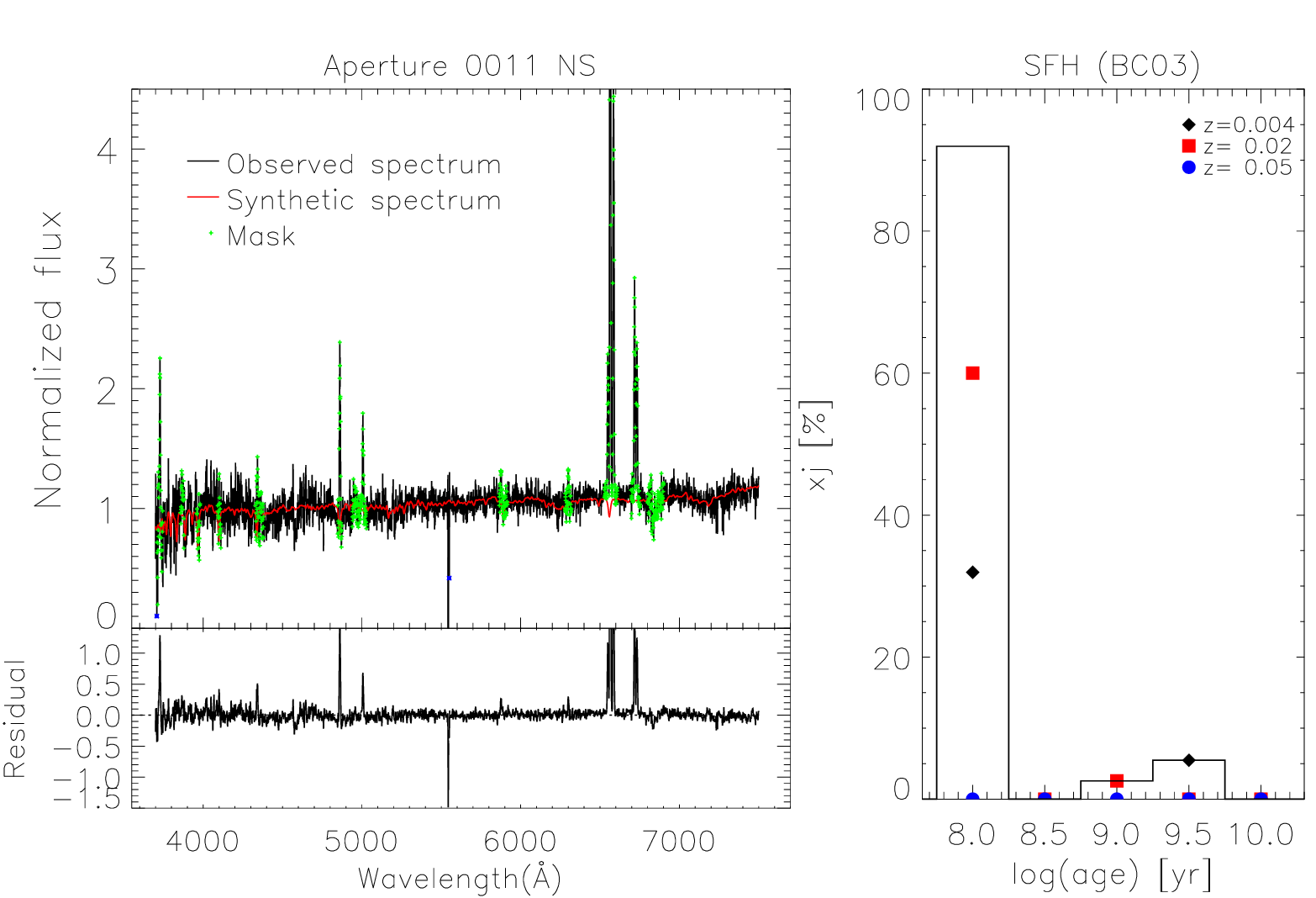}
}
\subfloat[]{
\includegraphics[scale=0.38]{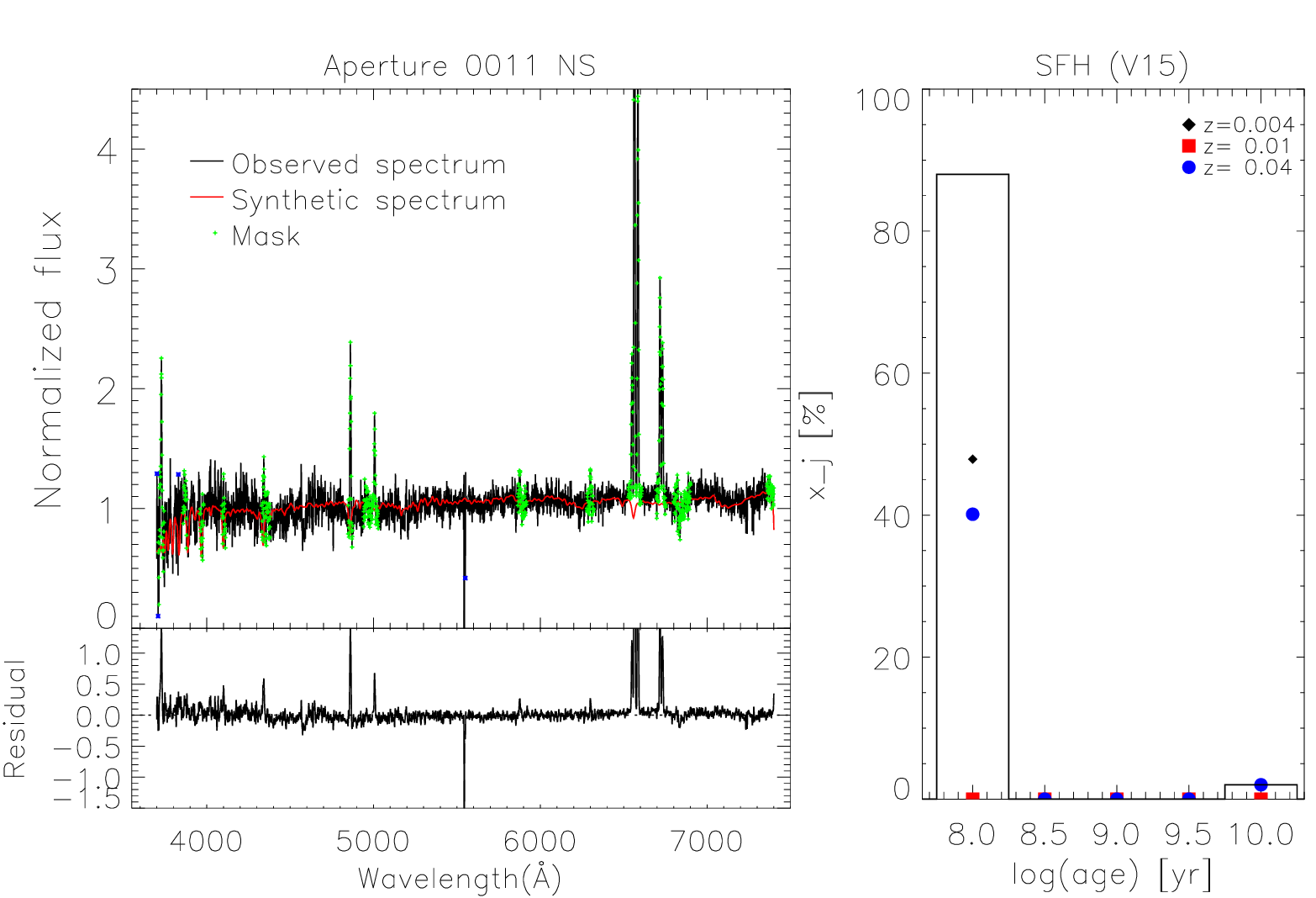}

}
\\
\subfloat[]{
\includegraphics[scale=0.38]{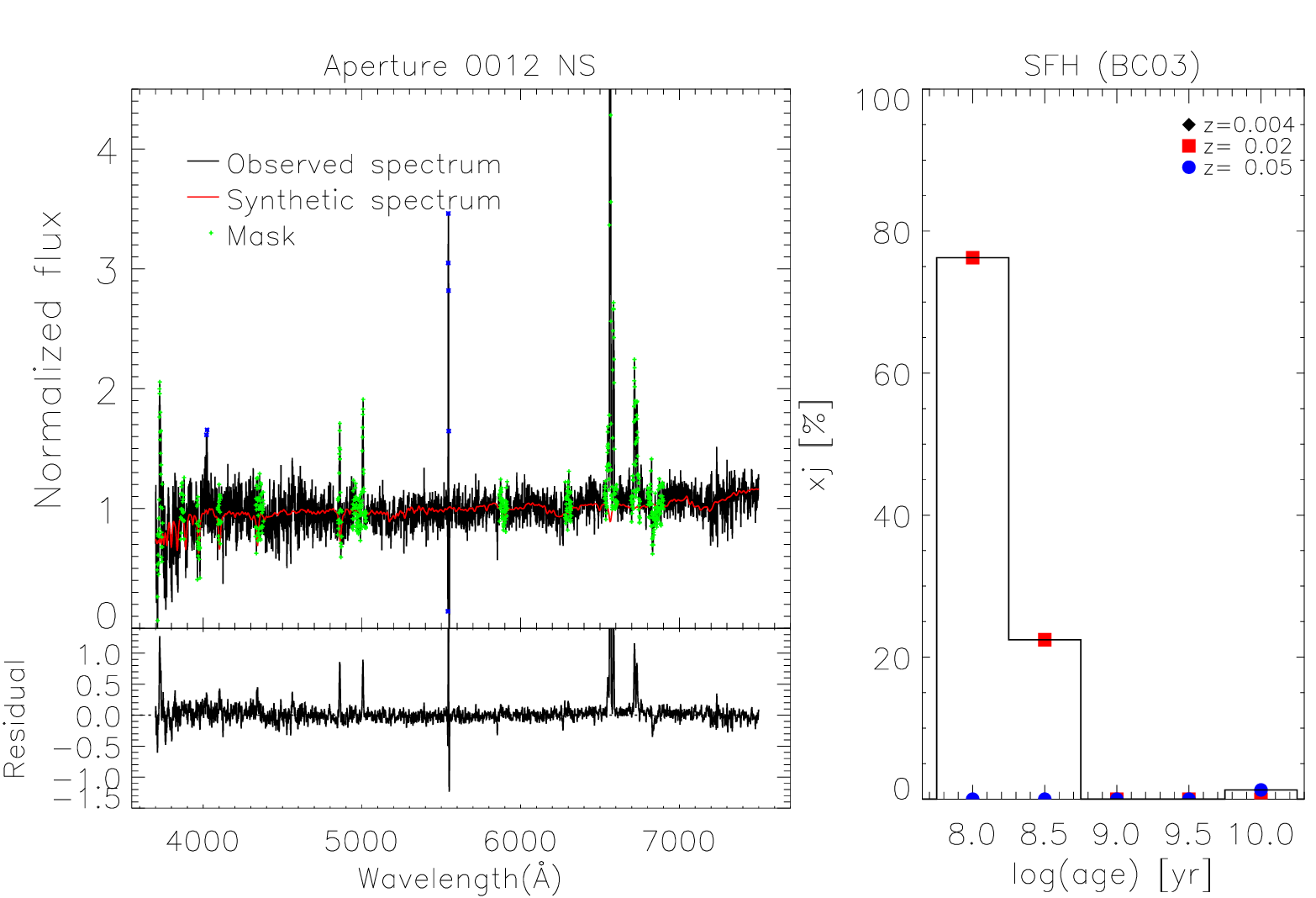}
}
\subfloat[]{
\includegraphics[scale=0.38]{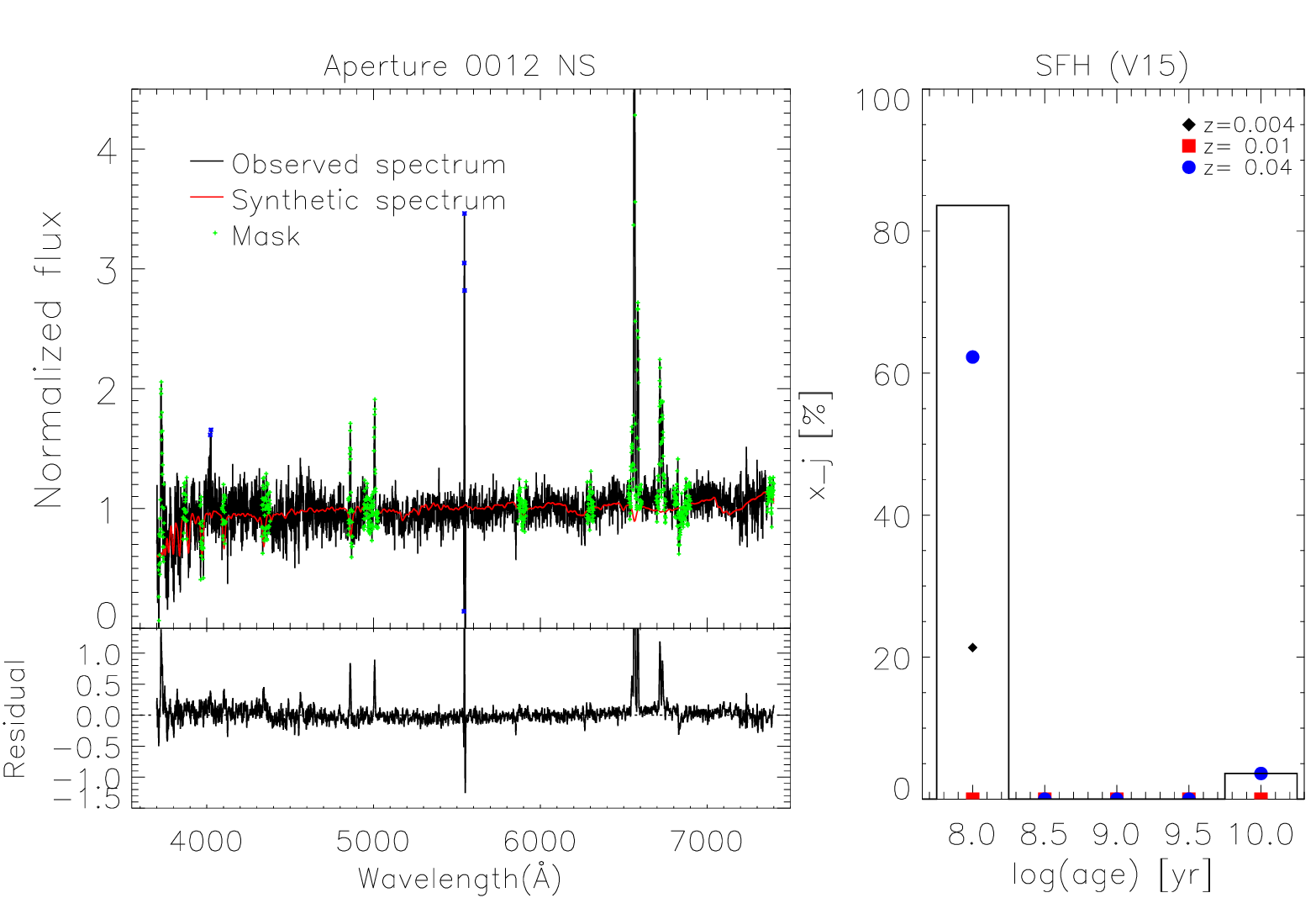}
}
\\
\subfloat[]{
\includegraphics[scale=0.38]{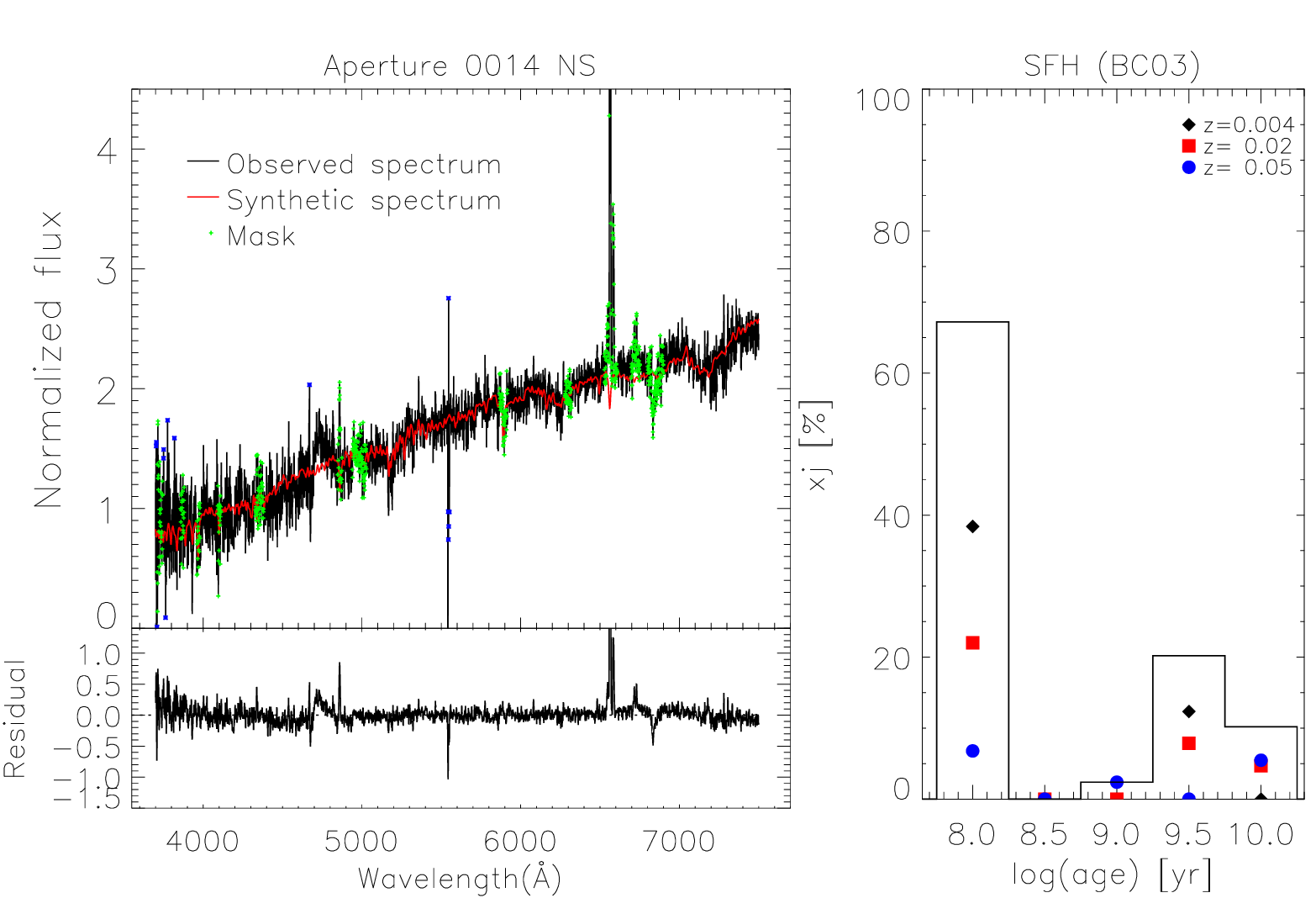}
}
\subfloat[]{
\includegraphics[scale=0.38]{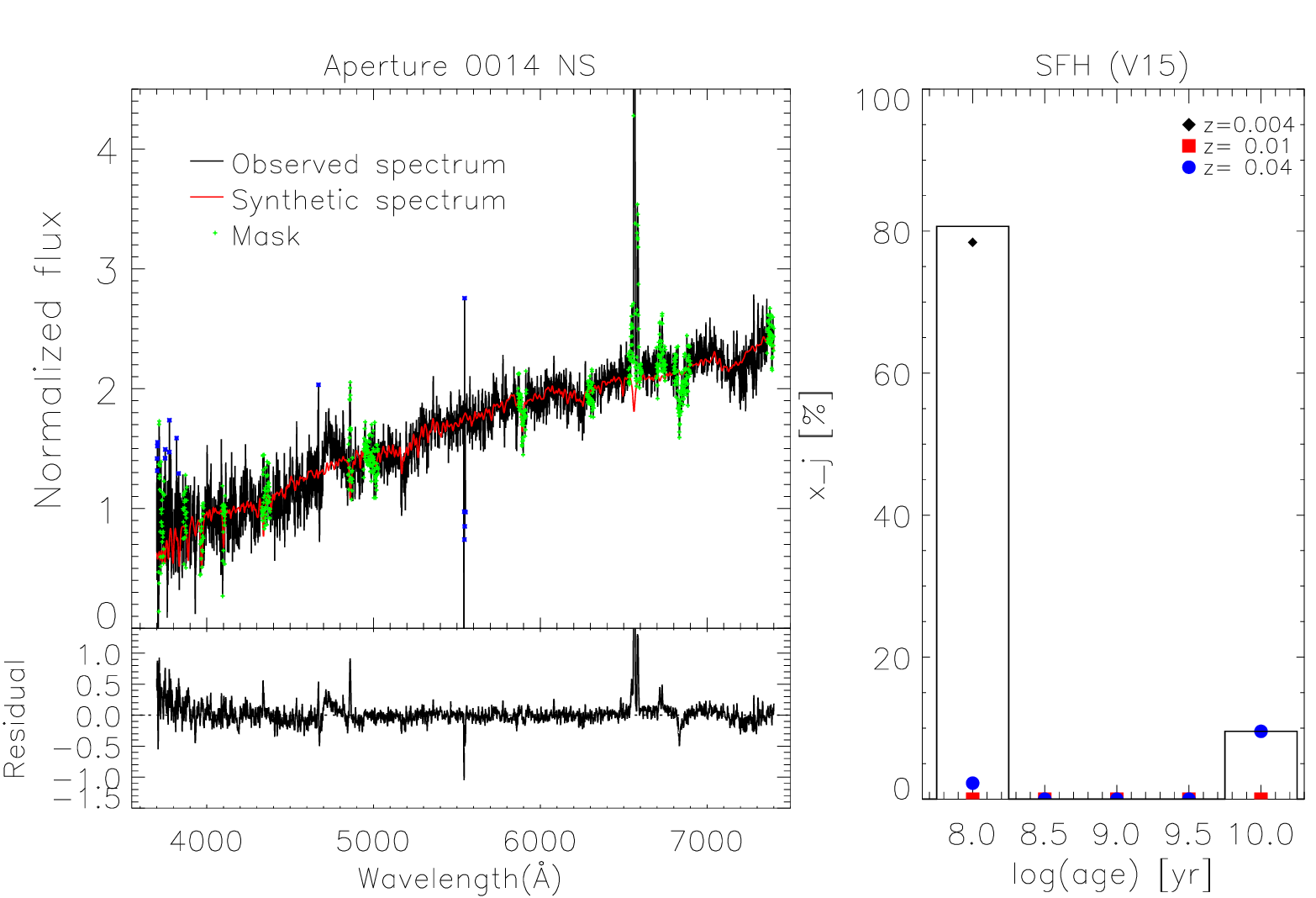}
}

\caption {Same as Figure~\ref{popstellarA1} but for apertures 11, 12, and 14
of the N-S  slit.
} 
\label{popstellarA4}
\end{figure*}


\begin{figure*}[!h]
\centering
\subfloat[]{
\includegraphics[scale=0.38]{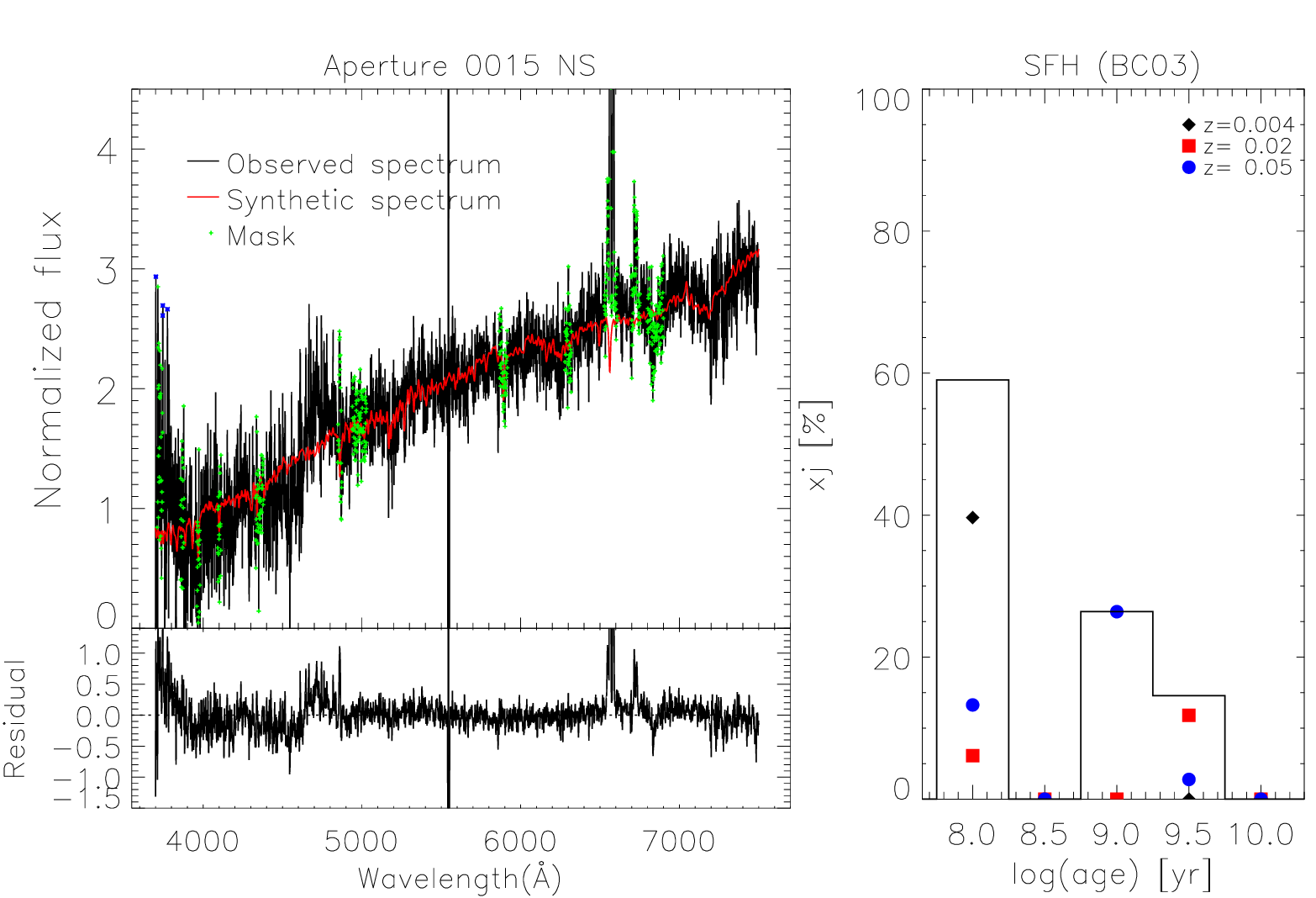}
}
\subfloat[]{
\includegraphics[scale=0.38]{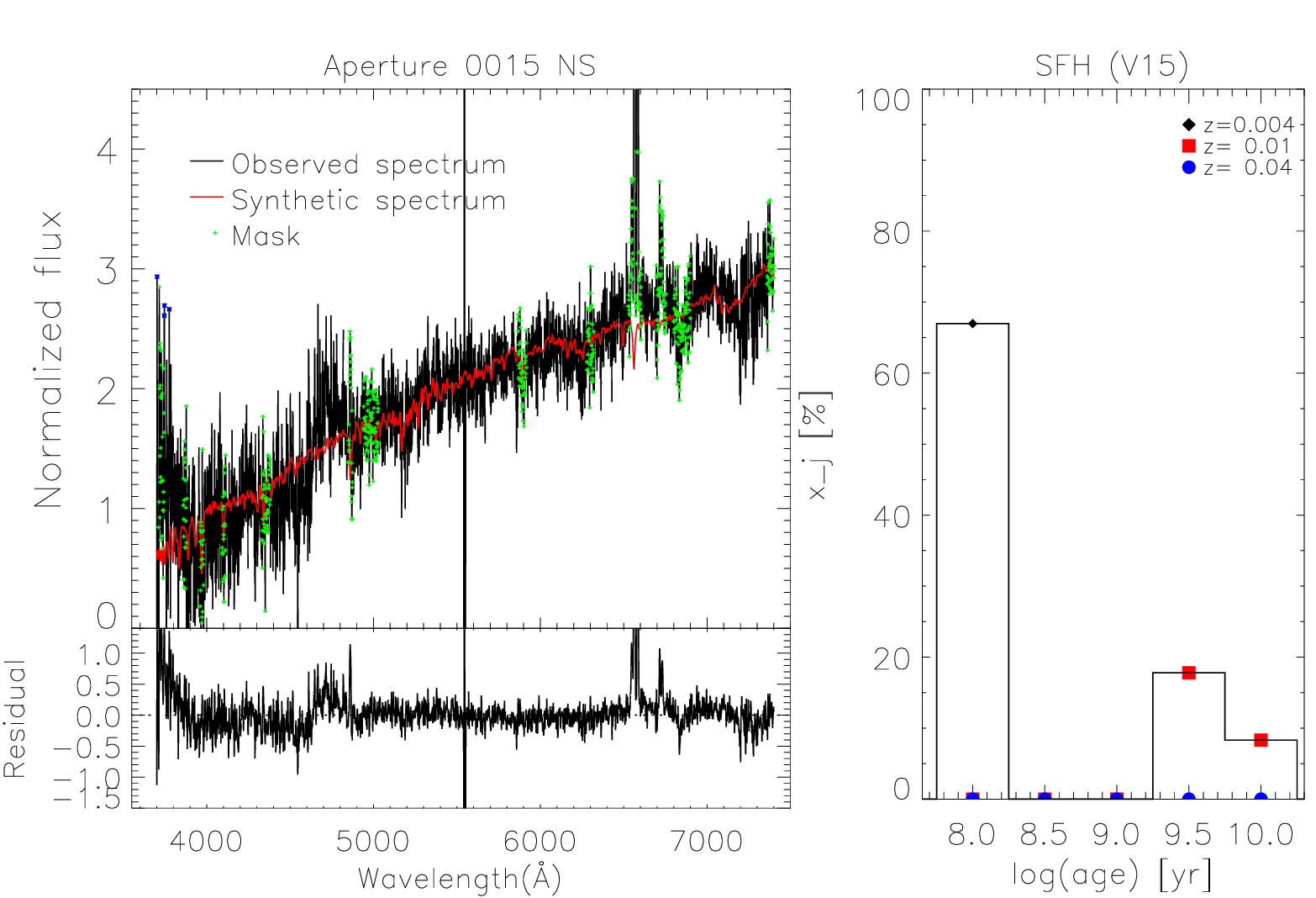}

}
\\
\subfloat[]{
\includegraphics[scale=0.38]{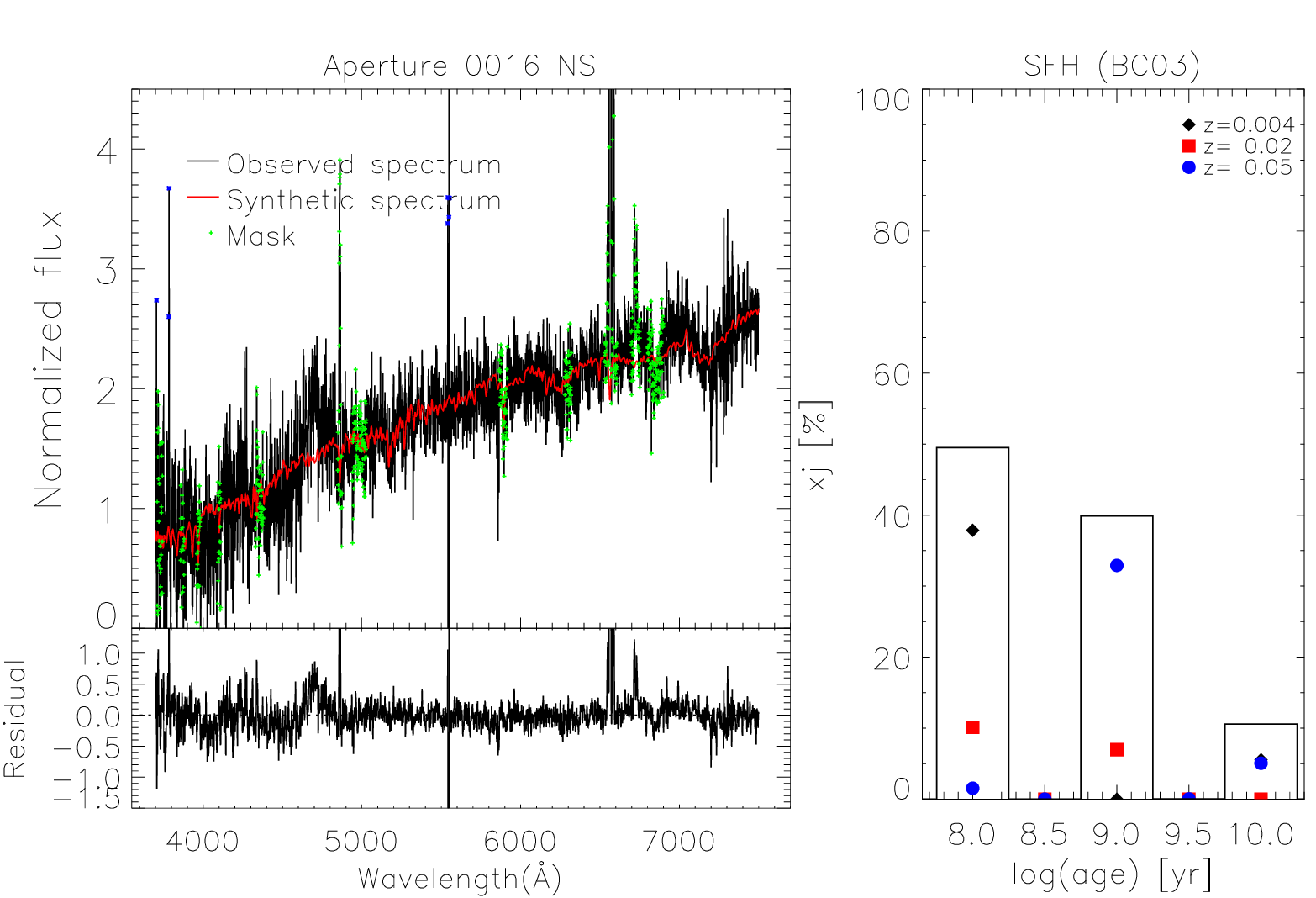}
}
\subfloat[]{
\includegraphics[scale=0.38]{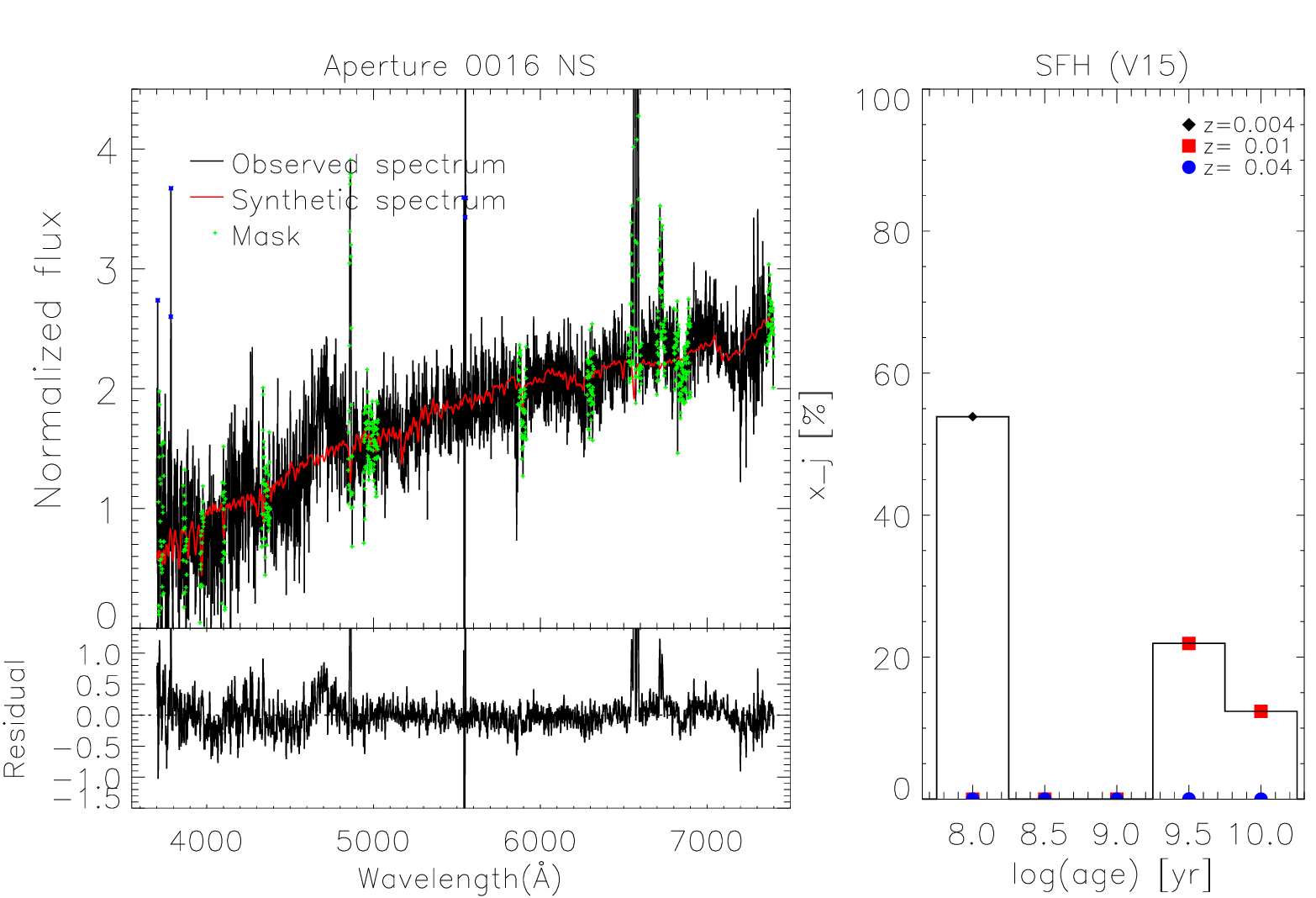}
}
\\
\subfloat[]{
\includegraphics[scale=0.38]{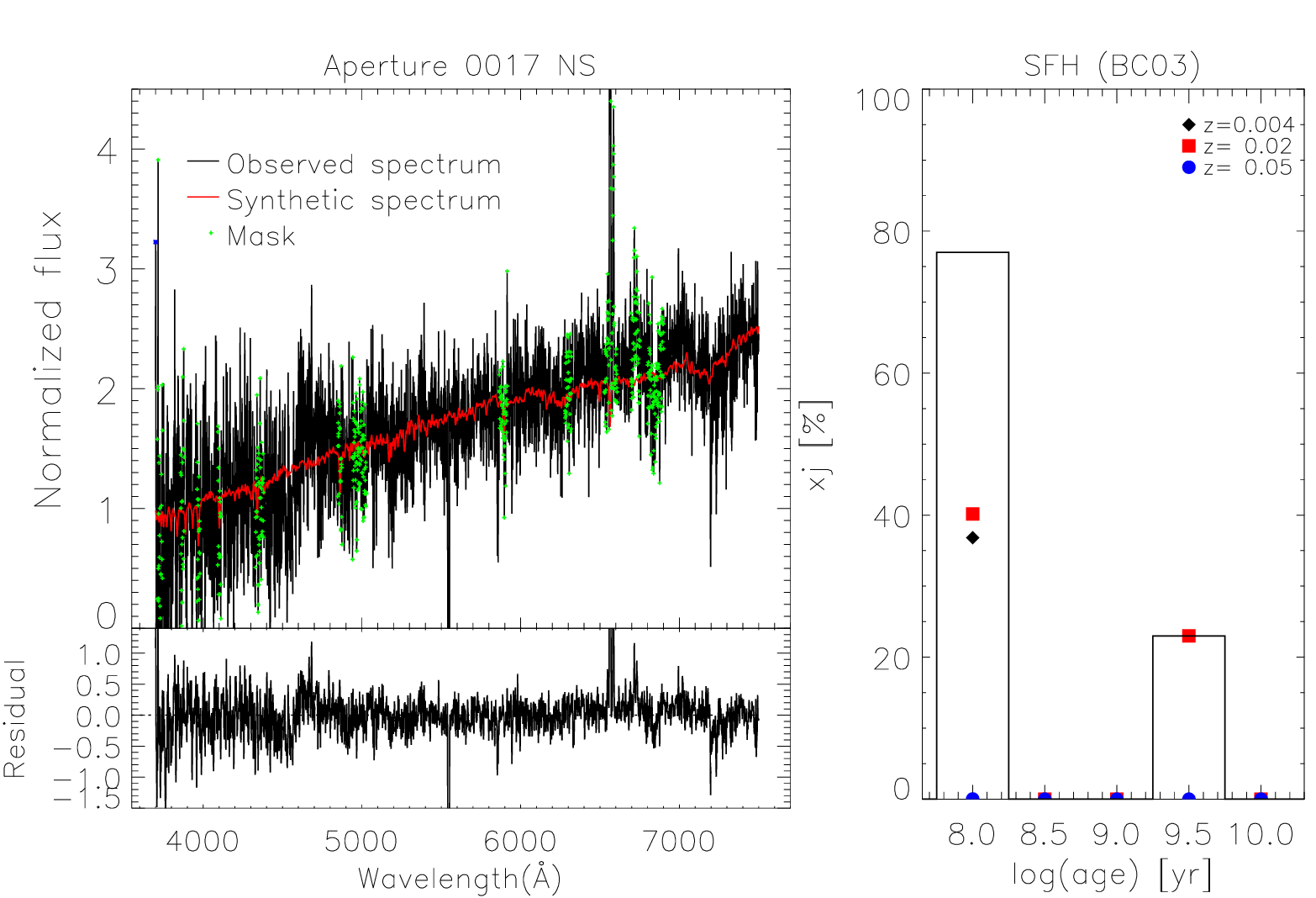}
}
\subfloat[]{
\includegraphics[scale=0.38]{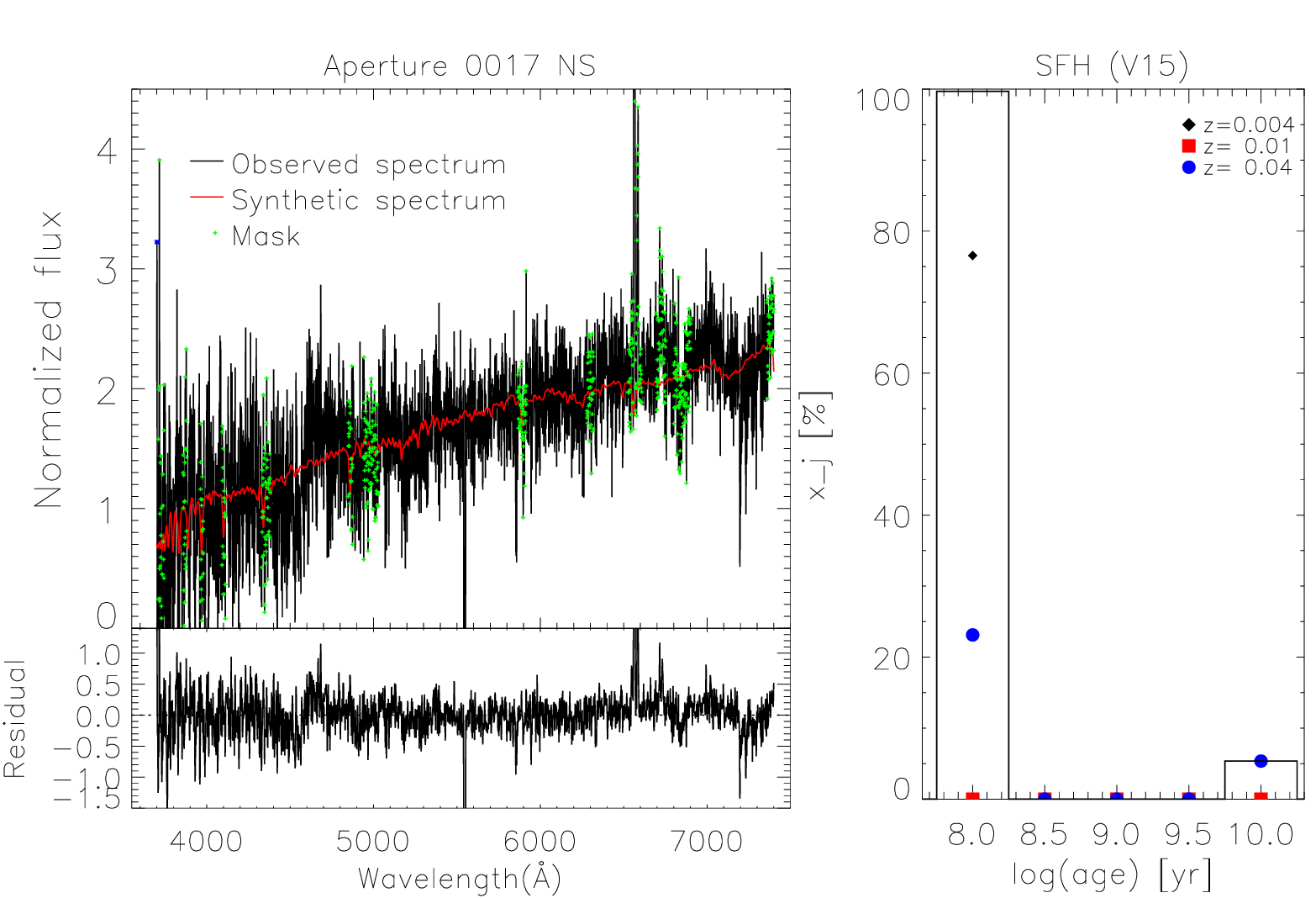}
}

\caption {Same as Figure~\ref{popstellarA1} but for apertures 15, 16, and 17
of the N-S  slit.
} 
\label{popstellarA5}
\end{figure*}


\begin{figure*}[!h]
\centering
\subfloat[]{
\includegraphics[scale=0.38]{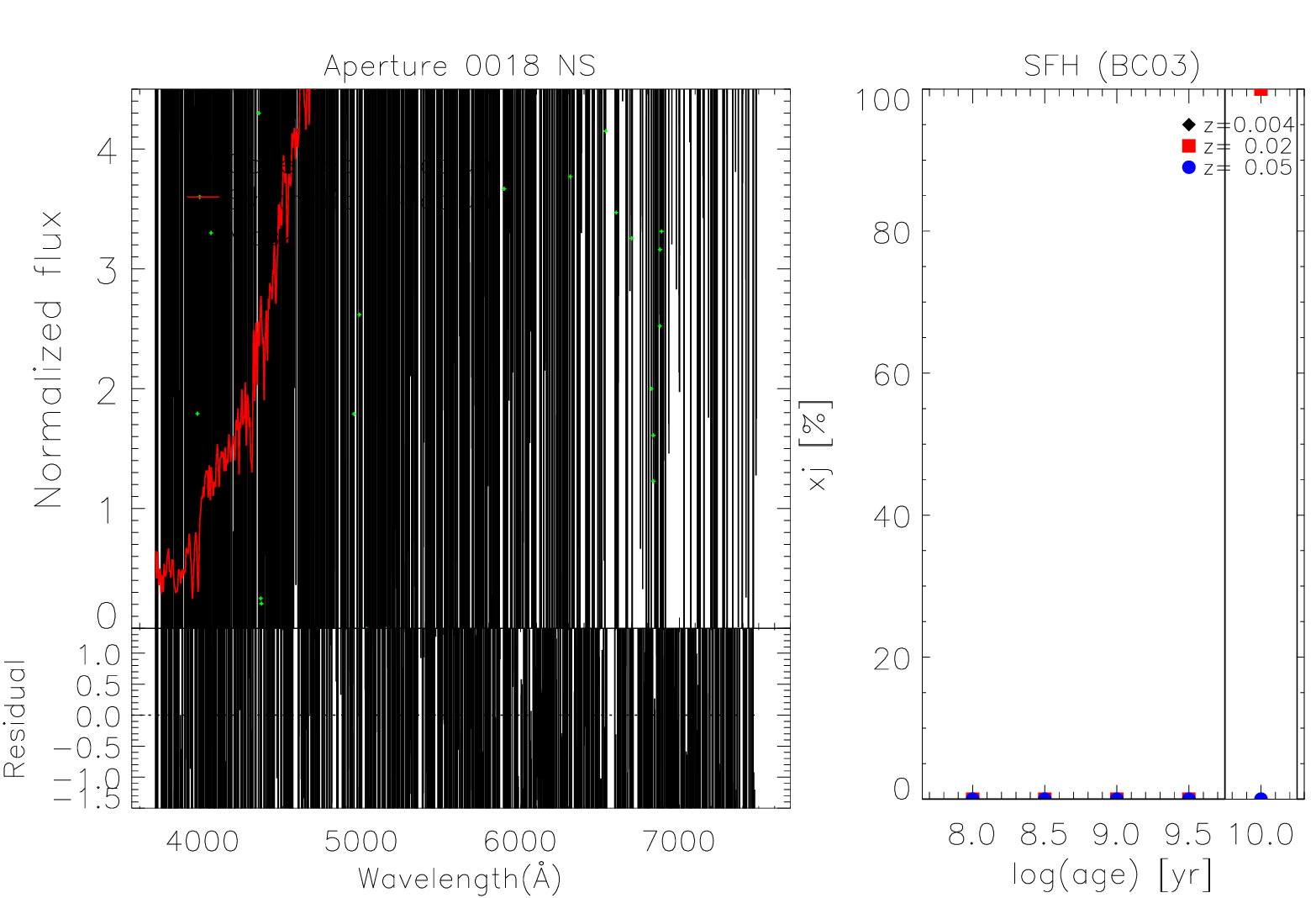}
}
\subfloat[]{
\includegraphics[scale=0.38]{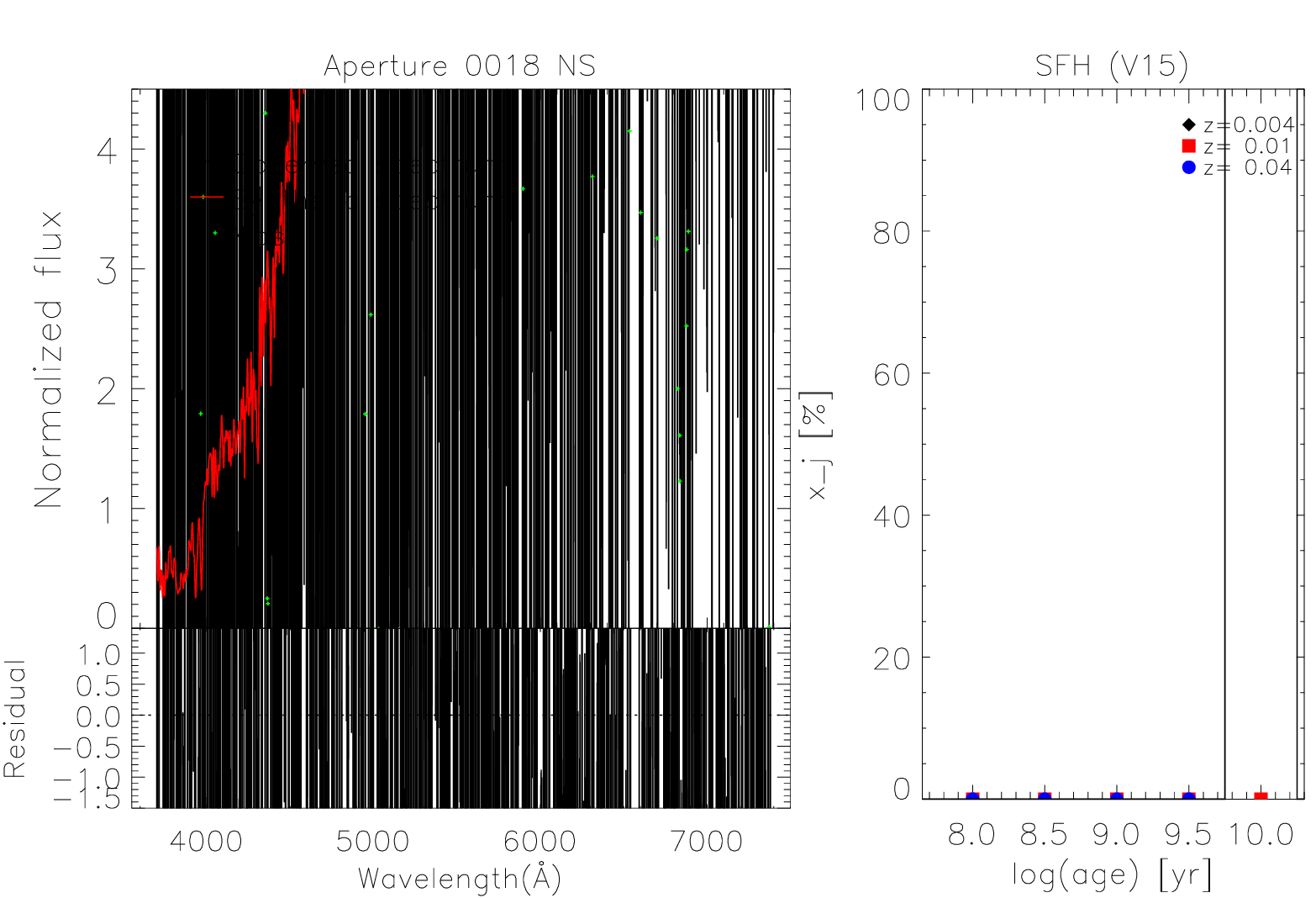}

}
\\
\subfloat[]{
\includegraphics[scale=0.38]{star_out_ngc1232_0019.eps}
}
\subfloat[]{
\includegraphics[scale=0.38]{star_out_n1232_0019_MIL.eps}
}
\\
\subfloat[]{
\includegraphics[scale=0.38]{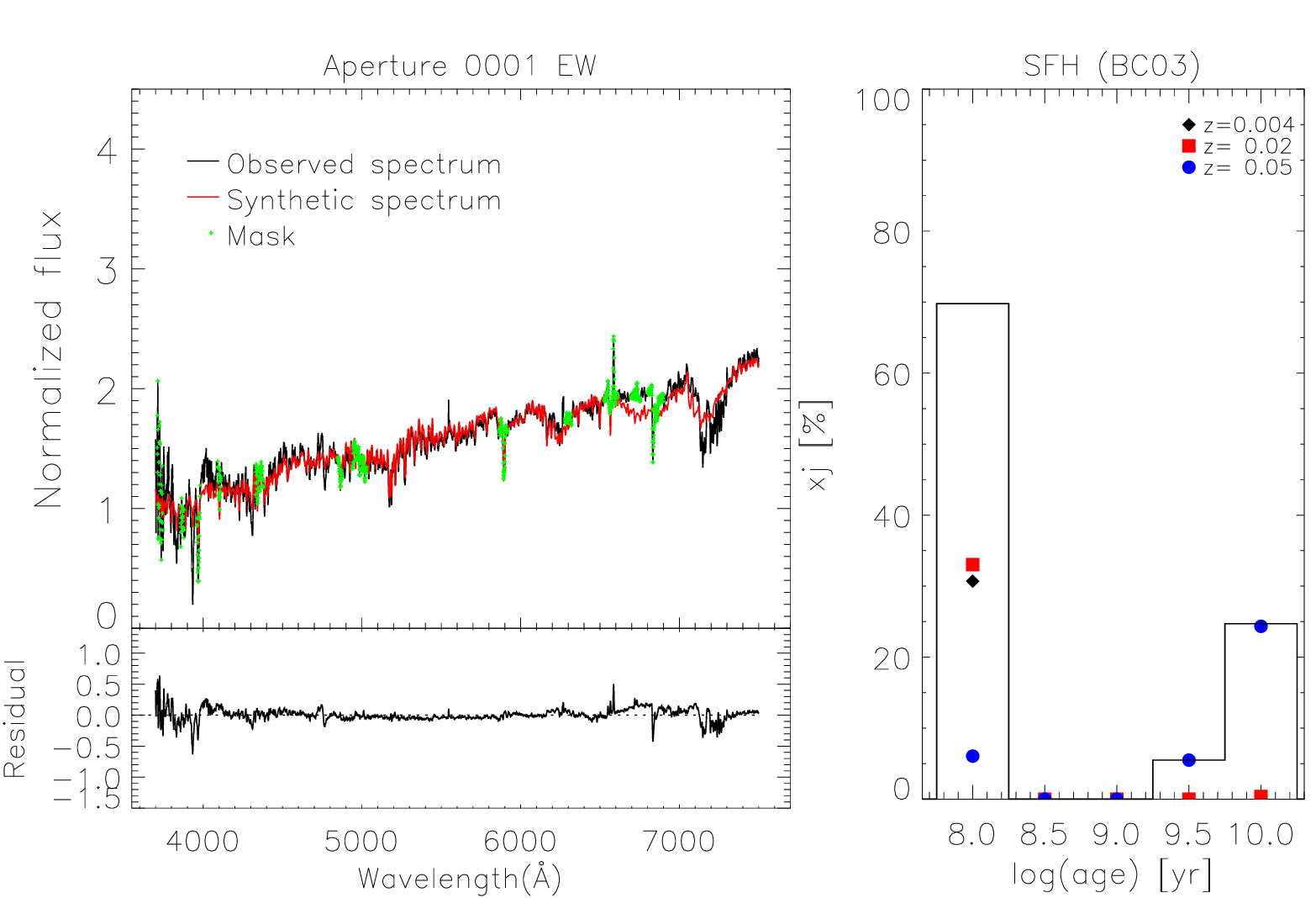}
}
\subfloat[]{
\includegraphics[scale=0.38]{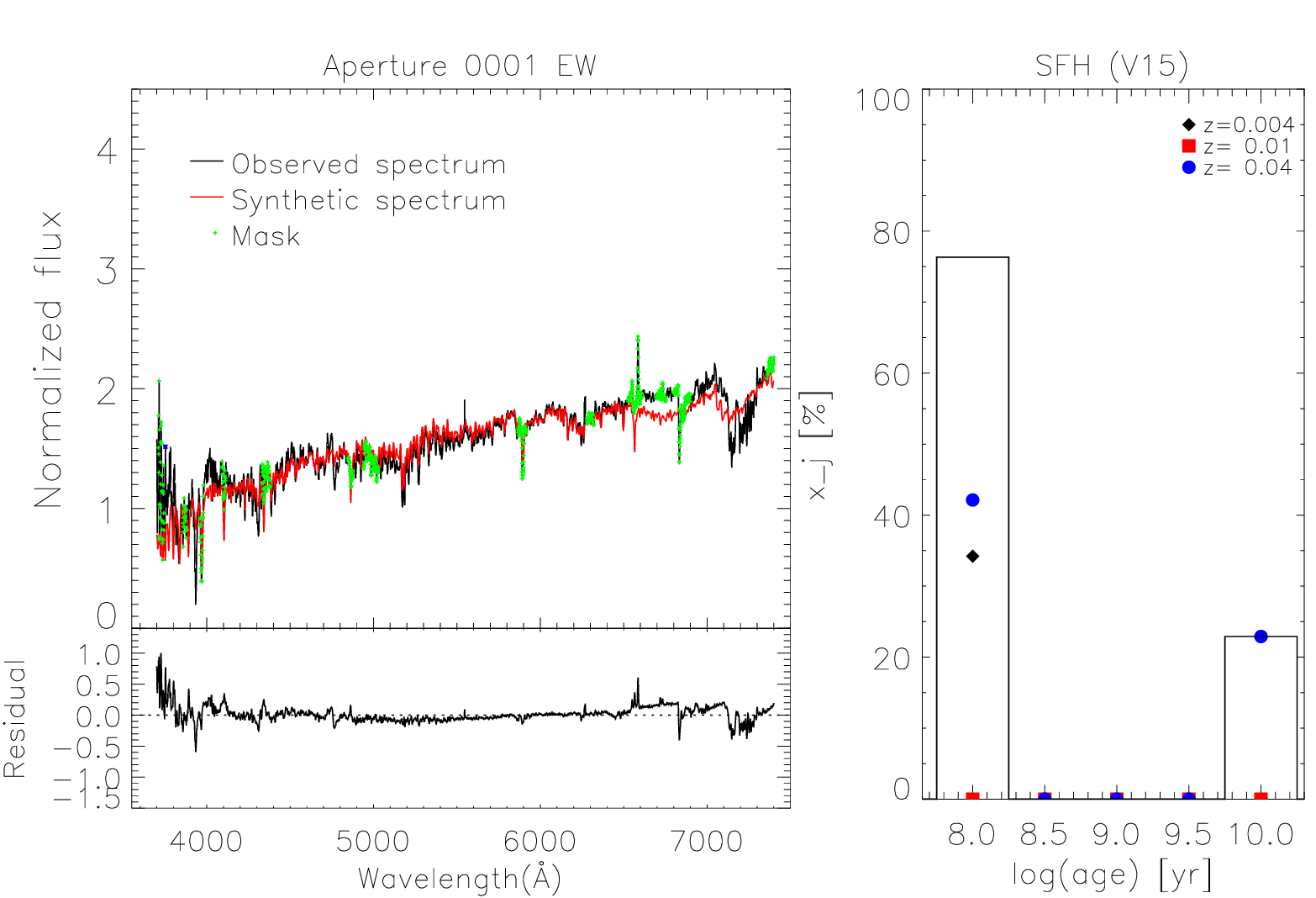}
}

\caption {Same as Figure~\ref{popstellarA1} but for apertures 18 and 19
of N-S  slit and aperture 01 of E-W  slit.
} 
\label{popstellarA6}
\end{figure*}


\begin{figure*}[!h]
\centering
\subfloat[]{
\includegraphics[scale=0.38]{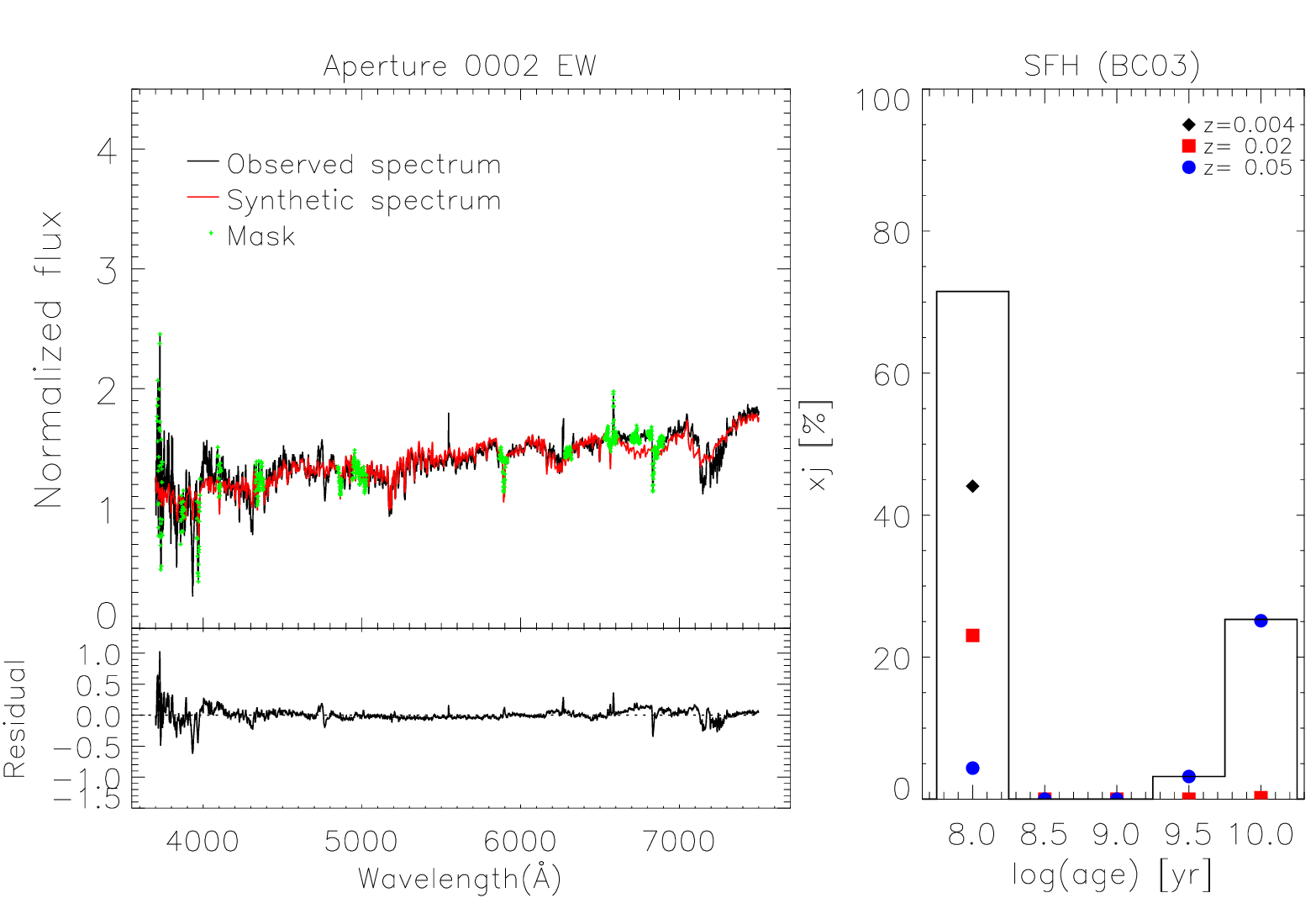}
}
\subfloat[]{
\includegraphics[scale=0.38]{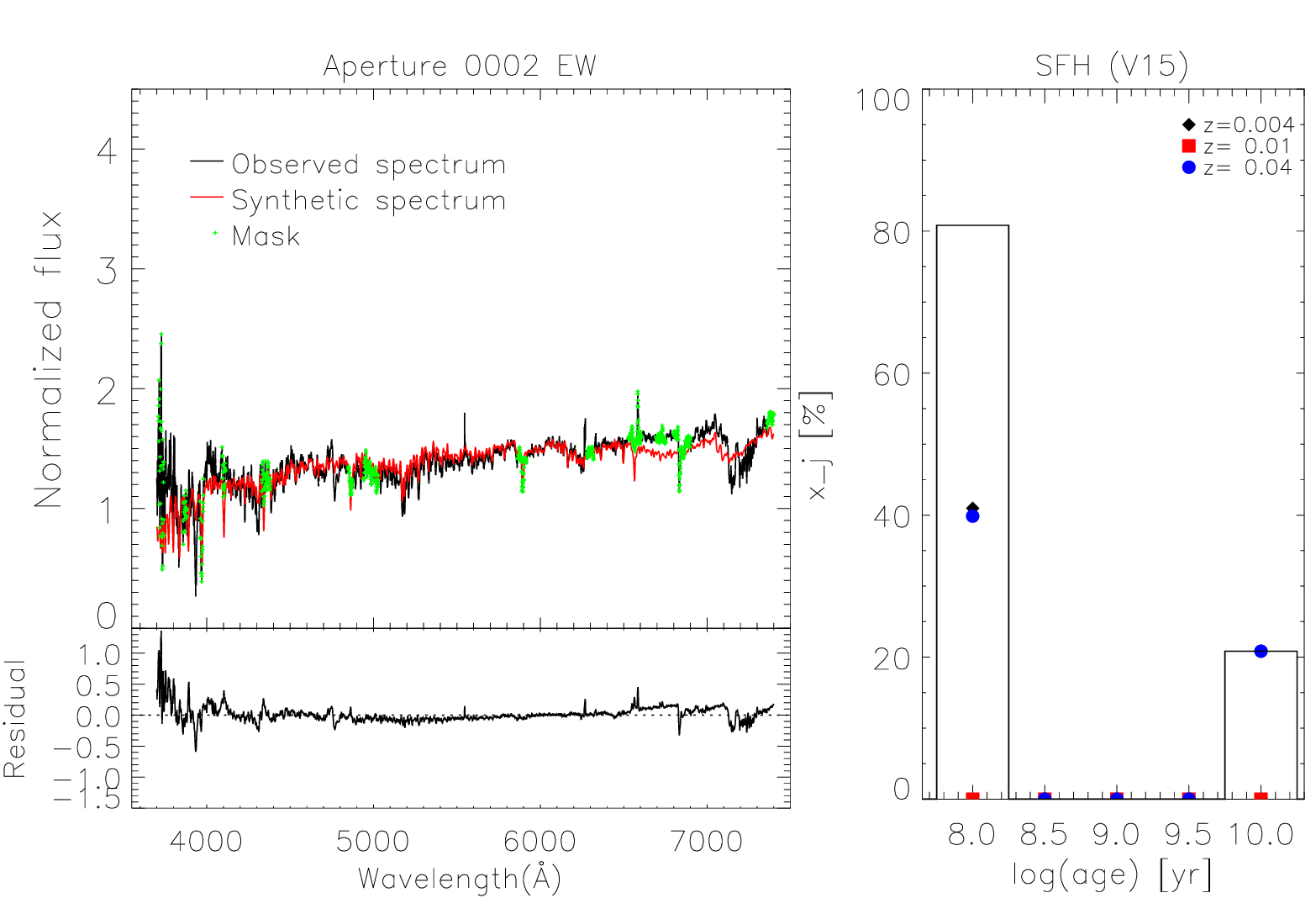}

}
\\
\subfloat[]{
\includegraphics[scale=0.38]{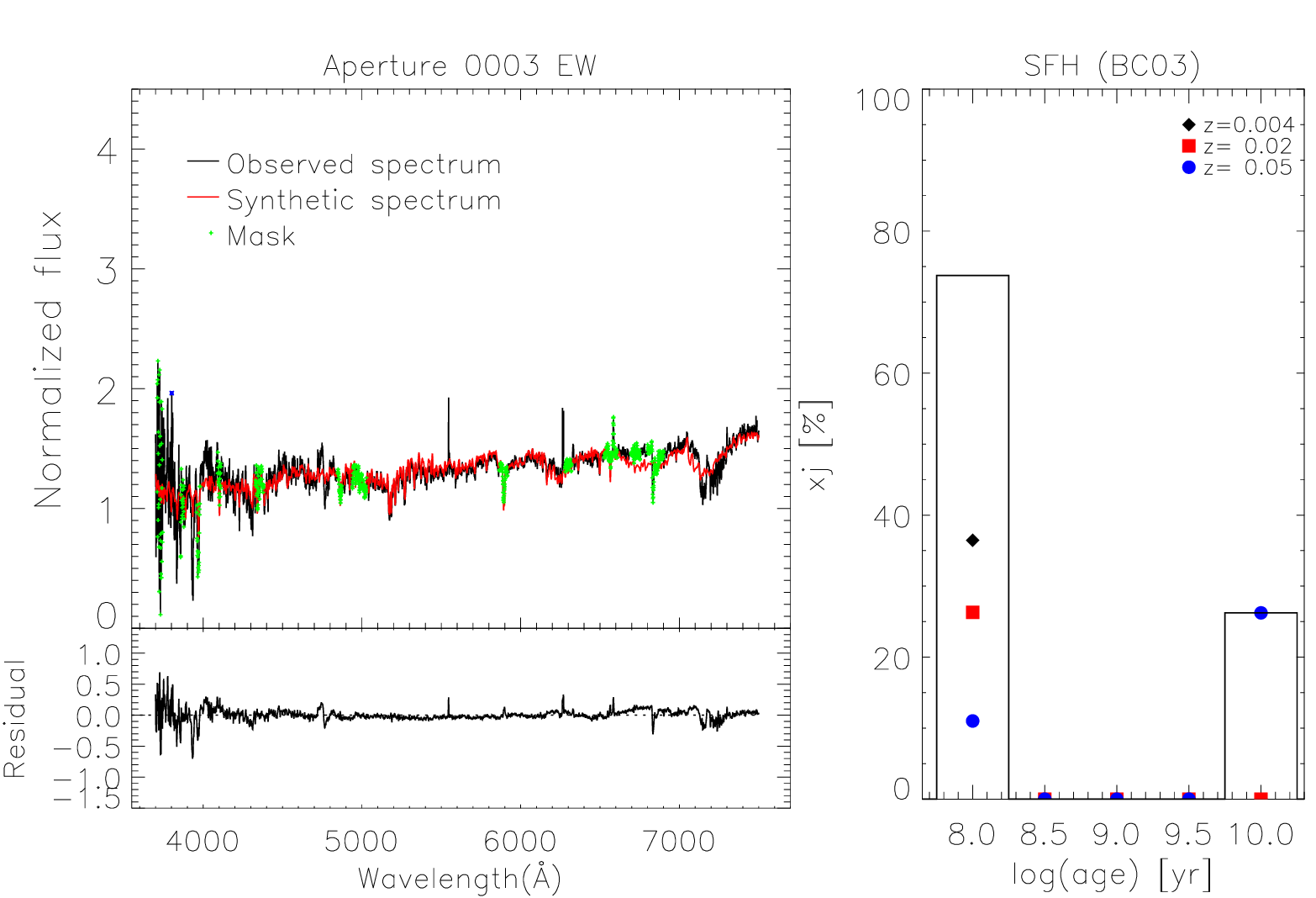}
}
\subfloat[]{
\includegraphics[scale=0.38]{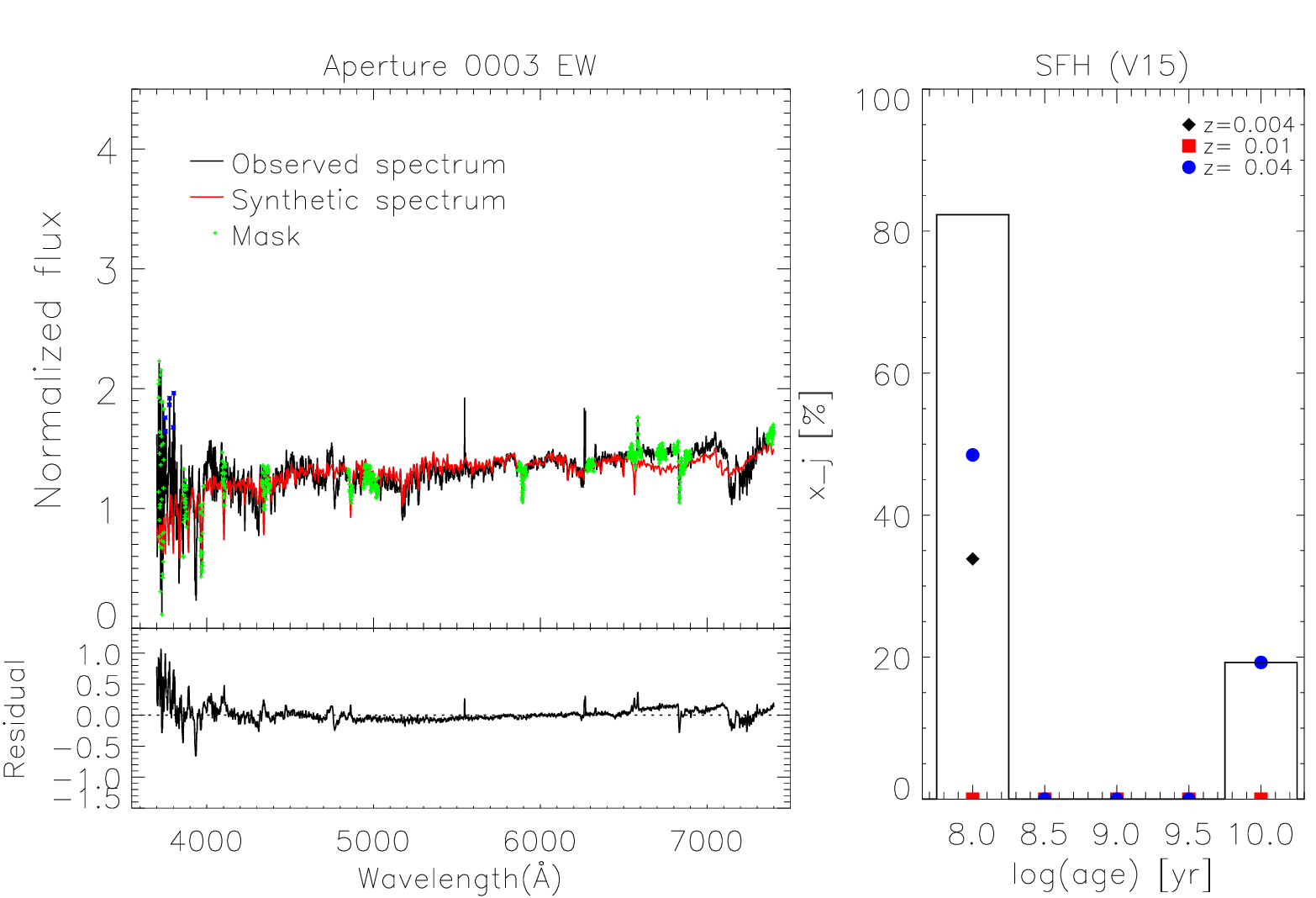}
}
\\
\subfloat[]{
\includegraphics[scale=0.38]{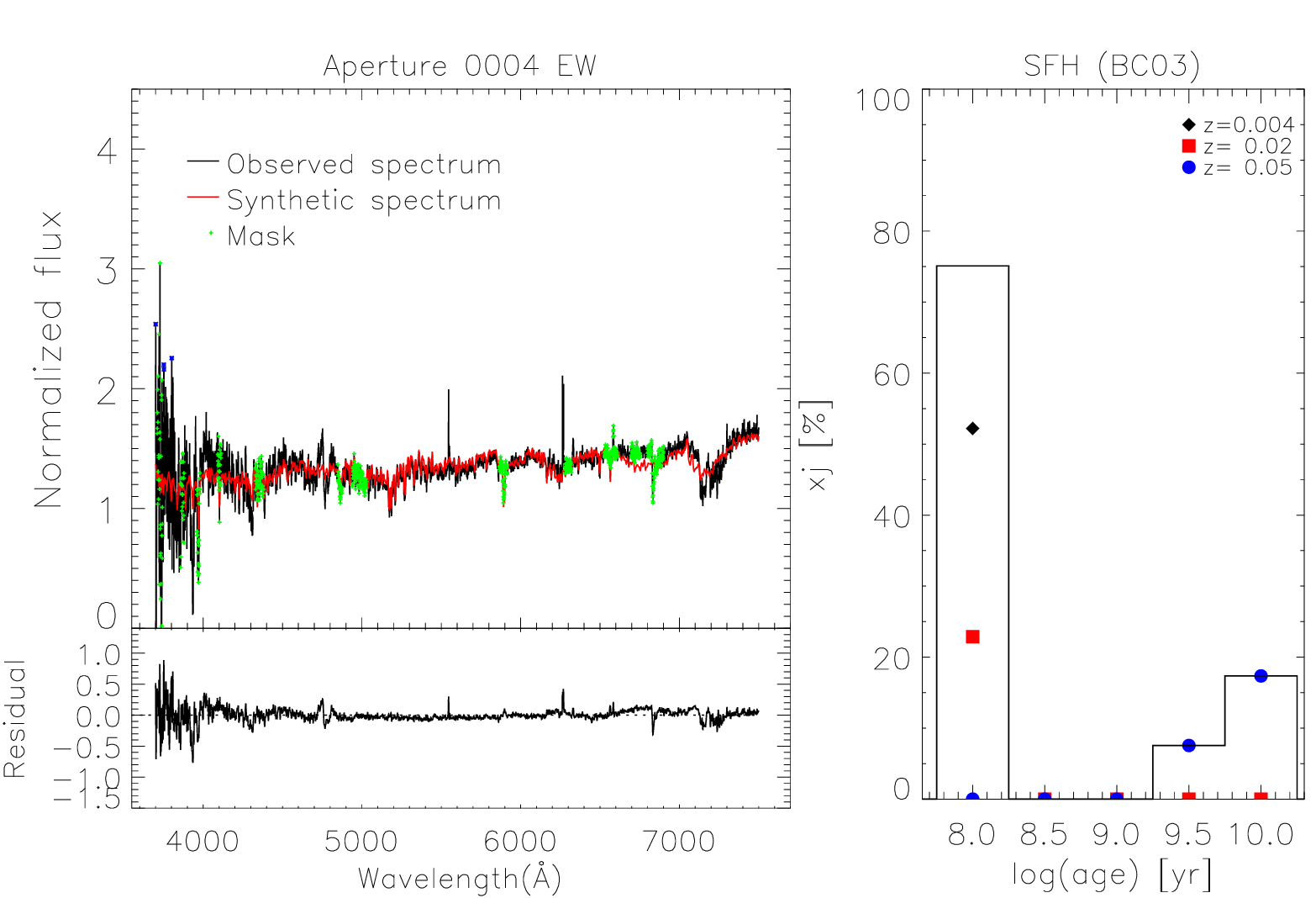}
}
\subfloat[]{
\includegraphics[scale=0.38]{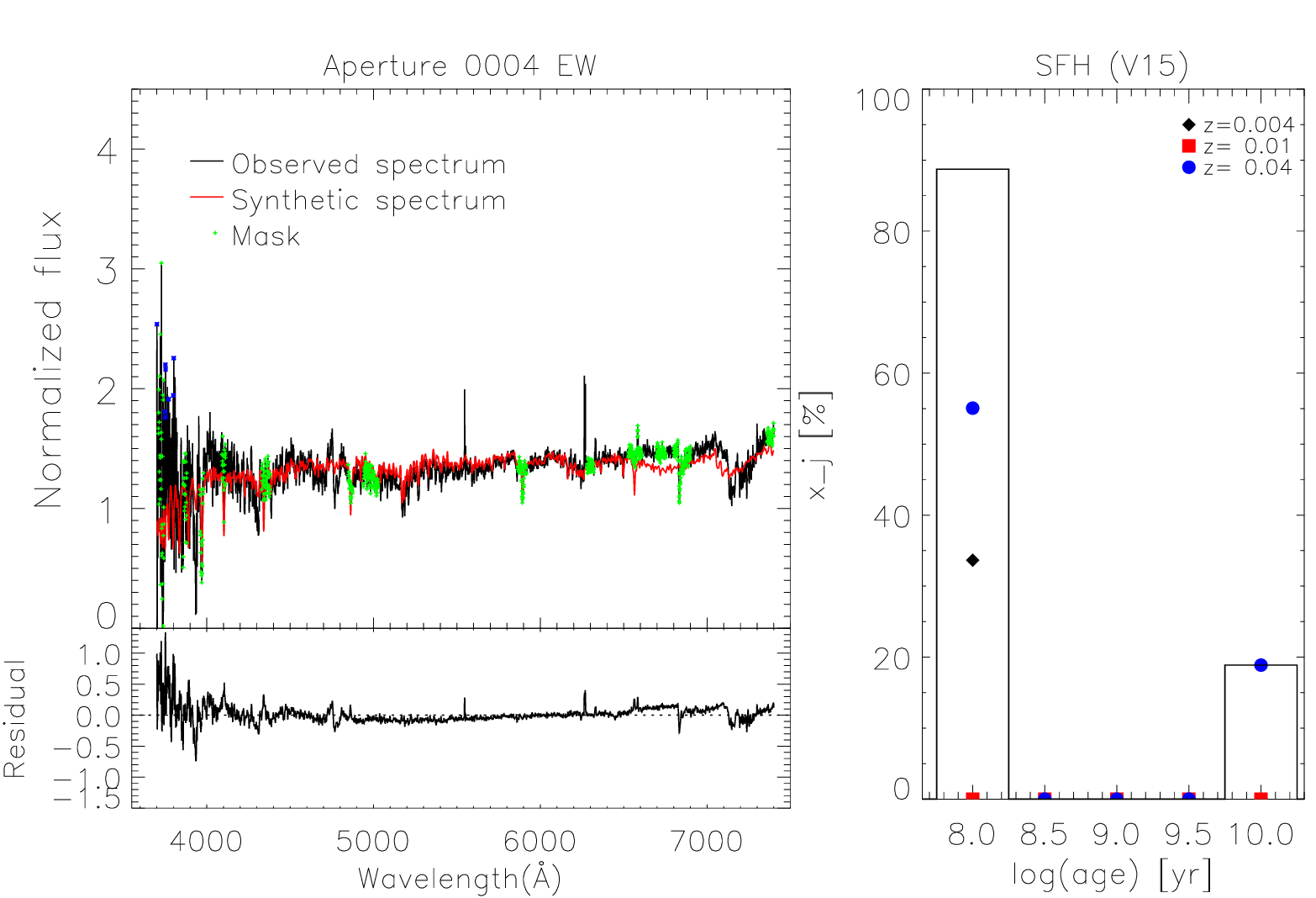}
}

\caption {Same as Figure~\ref{popstellarA1} but for apertures 02, 03,
 and 04 of E-W  slit.
} 
\label{popstellarA2}
\end{figure*}


\begin{figure*}[!h]
\centering
\subfloat[]{
\includegraphics[scale=0.38]{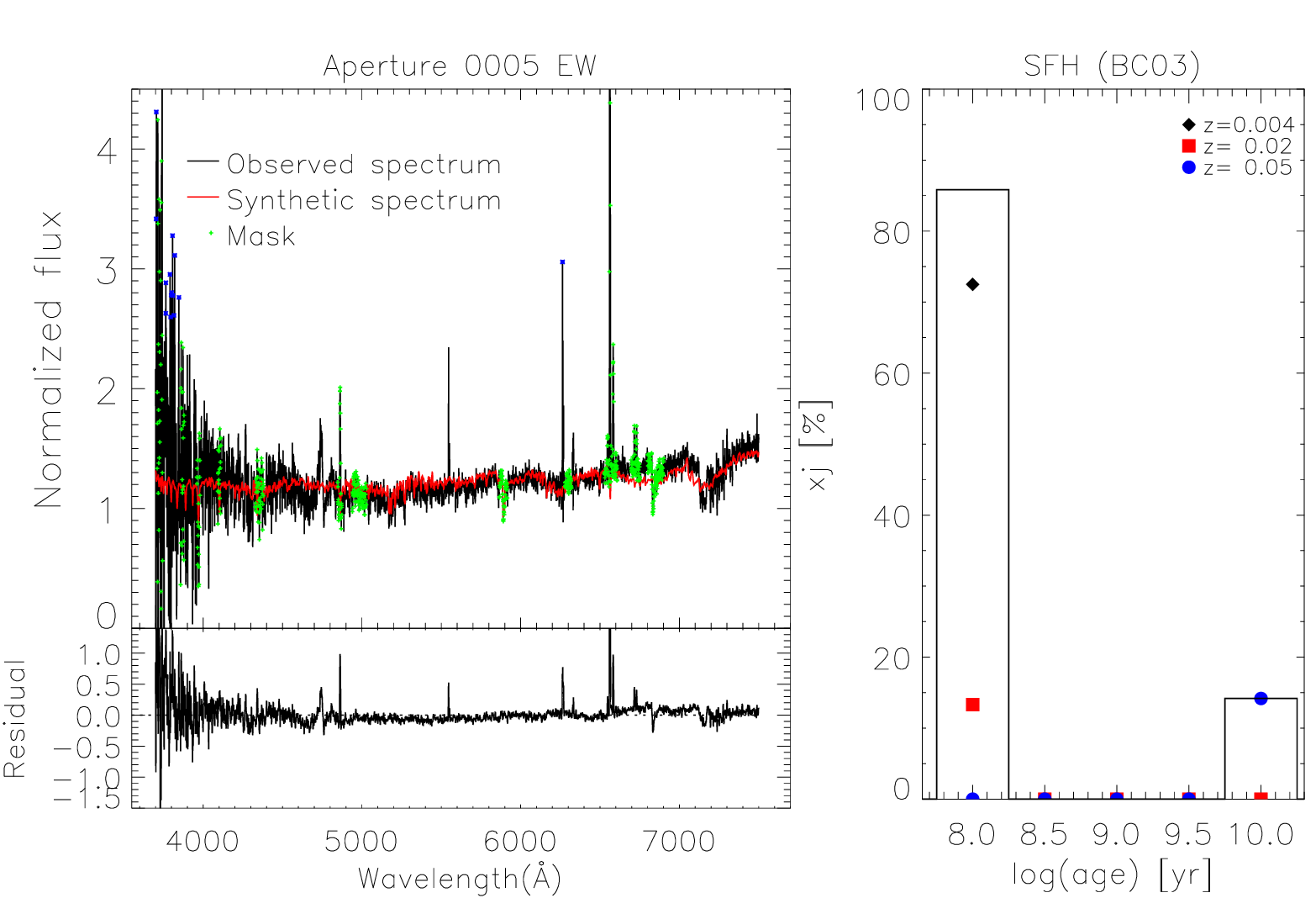}
}
\subfloat[]{
\includegraphics[scale=0.38]{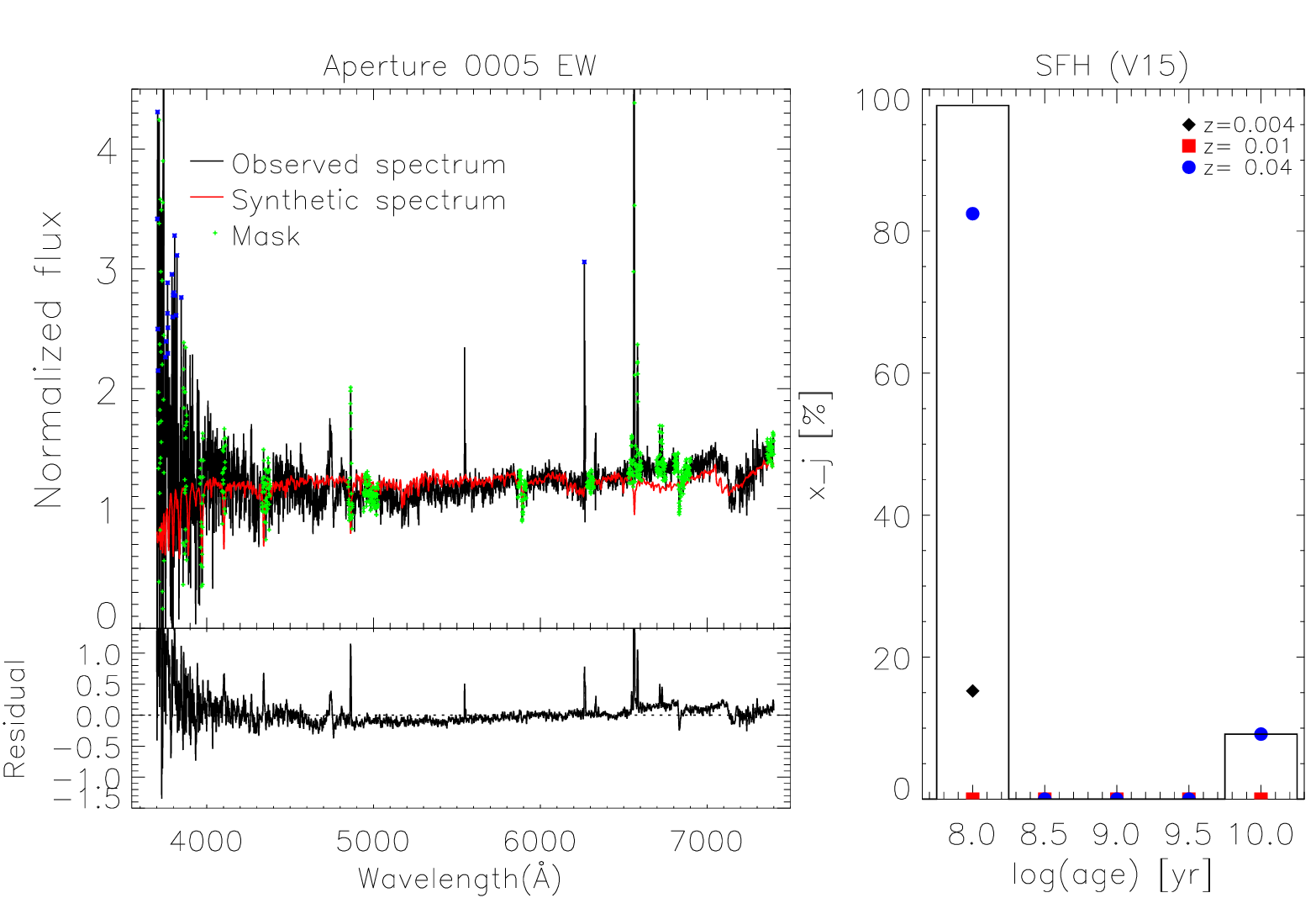}

}
\\
\subfloat[]{
\includegraphics[scale=0.38]{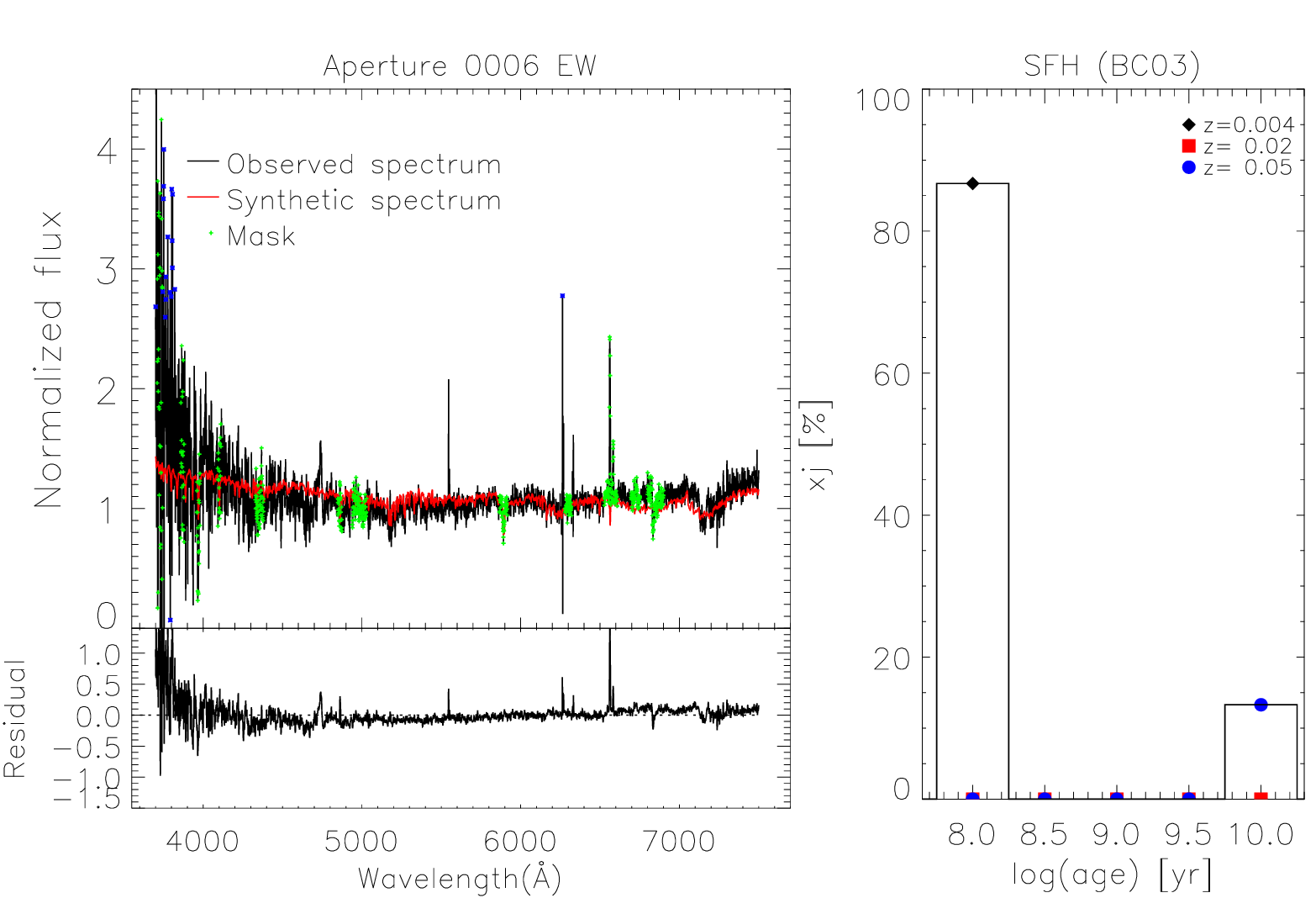}
}
\subfloat[]{
\includegraphics[scale=0.38]{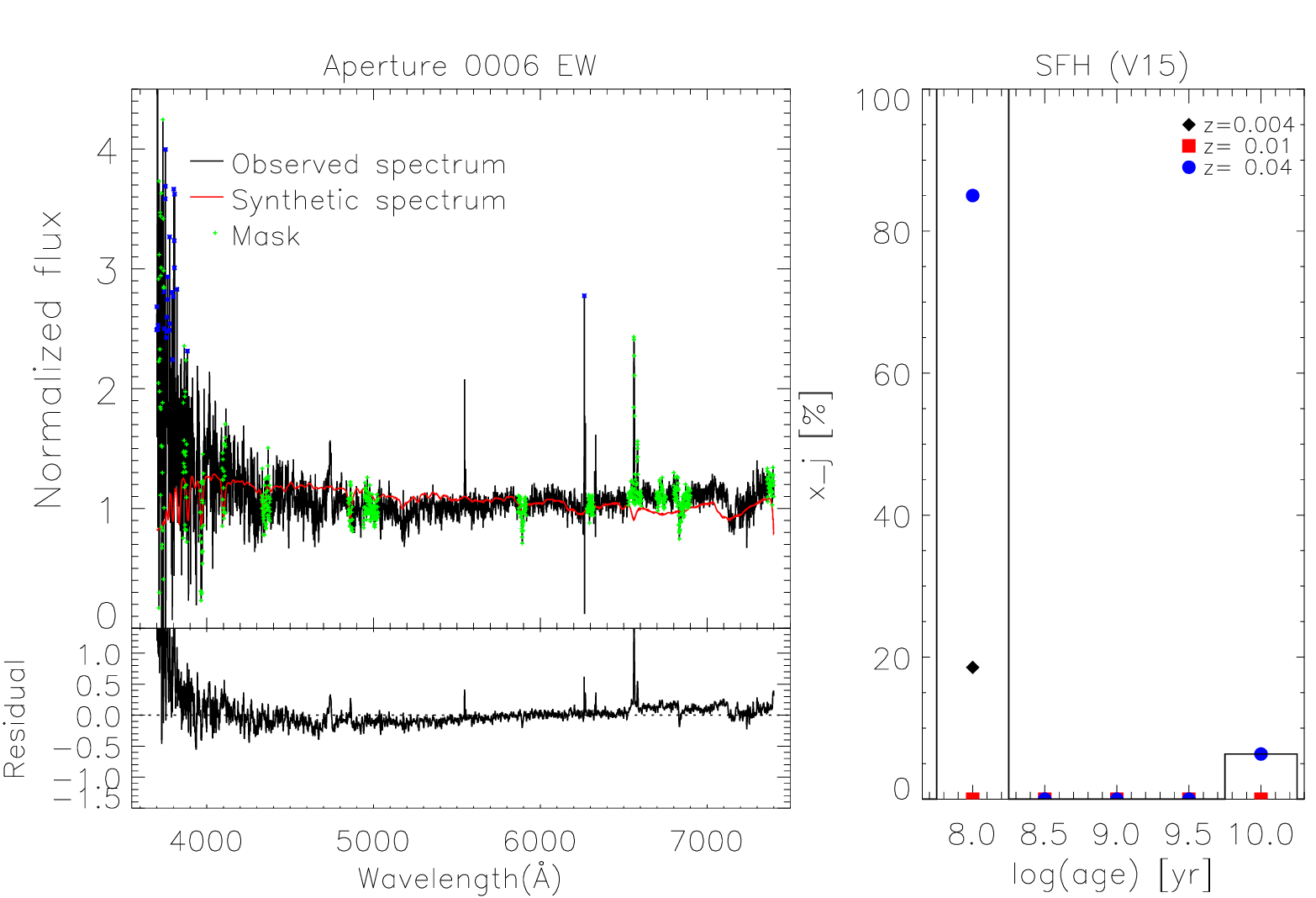}
}
\\
\subfloat[]{
\includegraphics[scale=0.38]{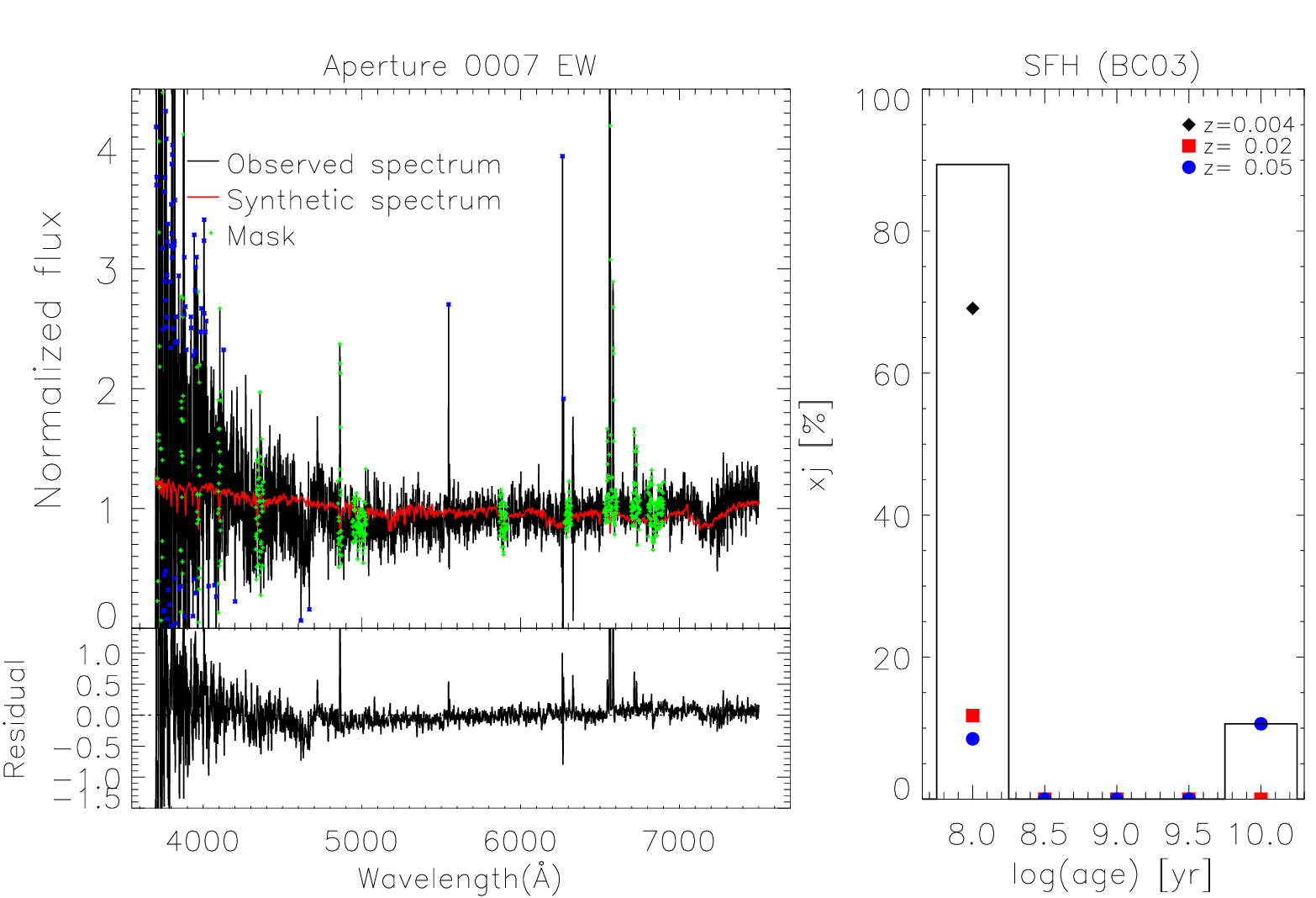}
}
\subfloat[]{
\includegraphics[scale=0.38]{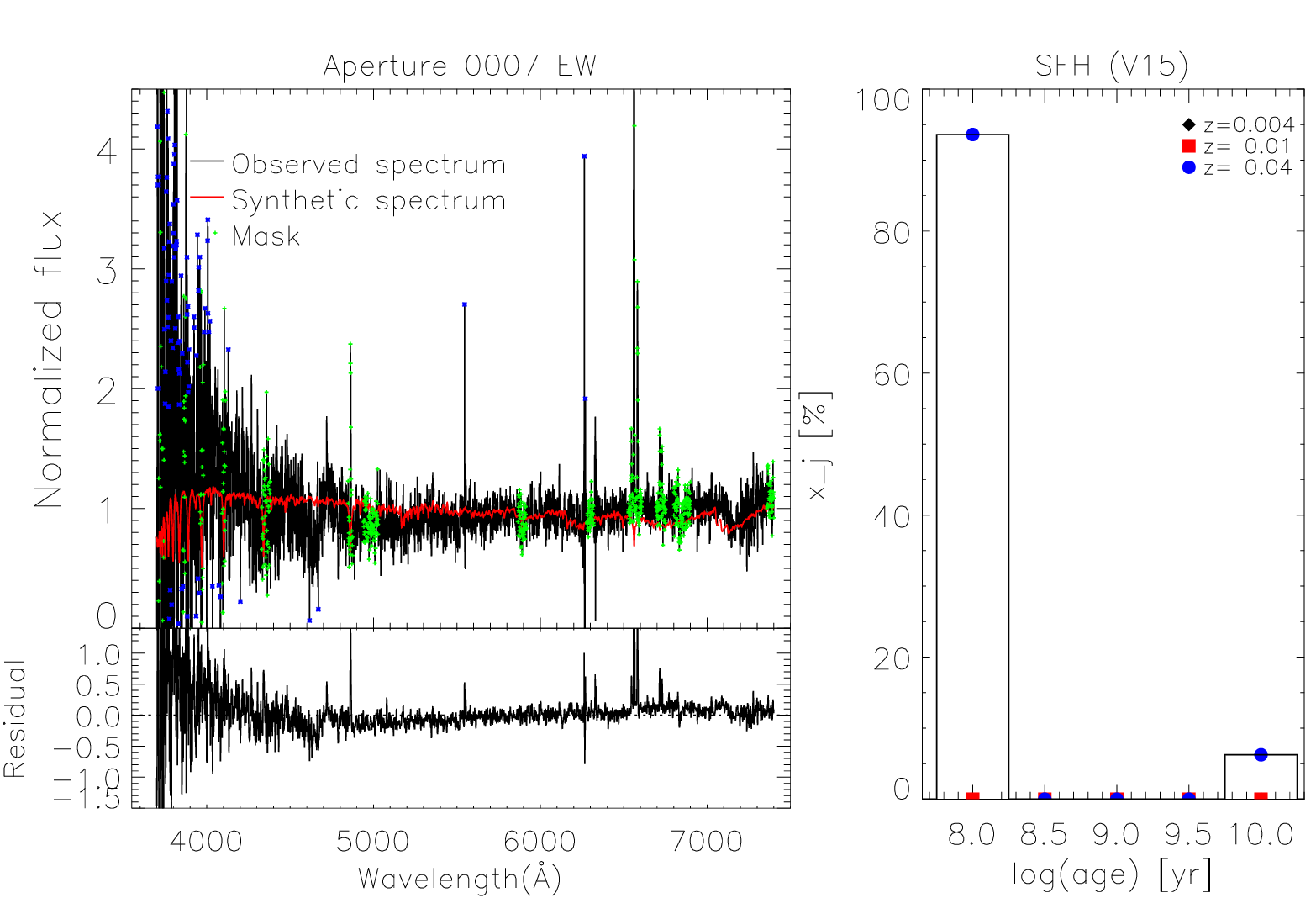}
}

\caption {Same as Figure~\ref{popstellarA1} but for apertures 05, 06,
 and 07 of the E-W  slit.
} 
\label{popstellarA7}
\end{figure*}


\begin{figure*}[!h]
\centering
\subfloat[]{
\includegraphics[scale=0.38]{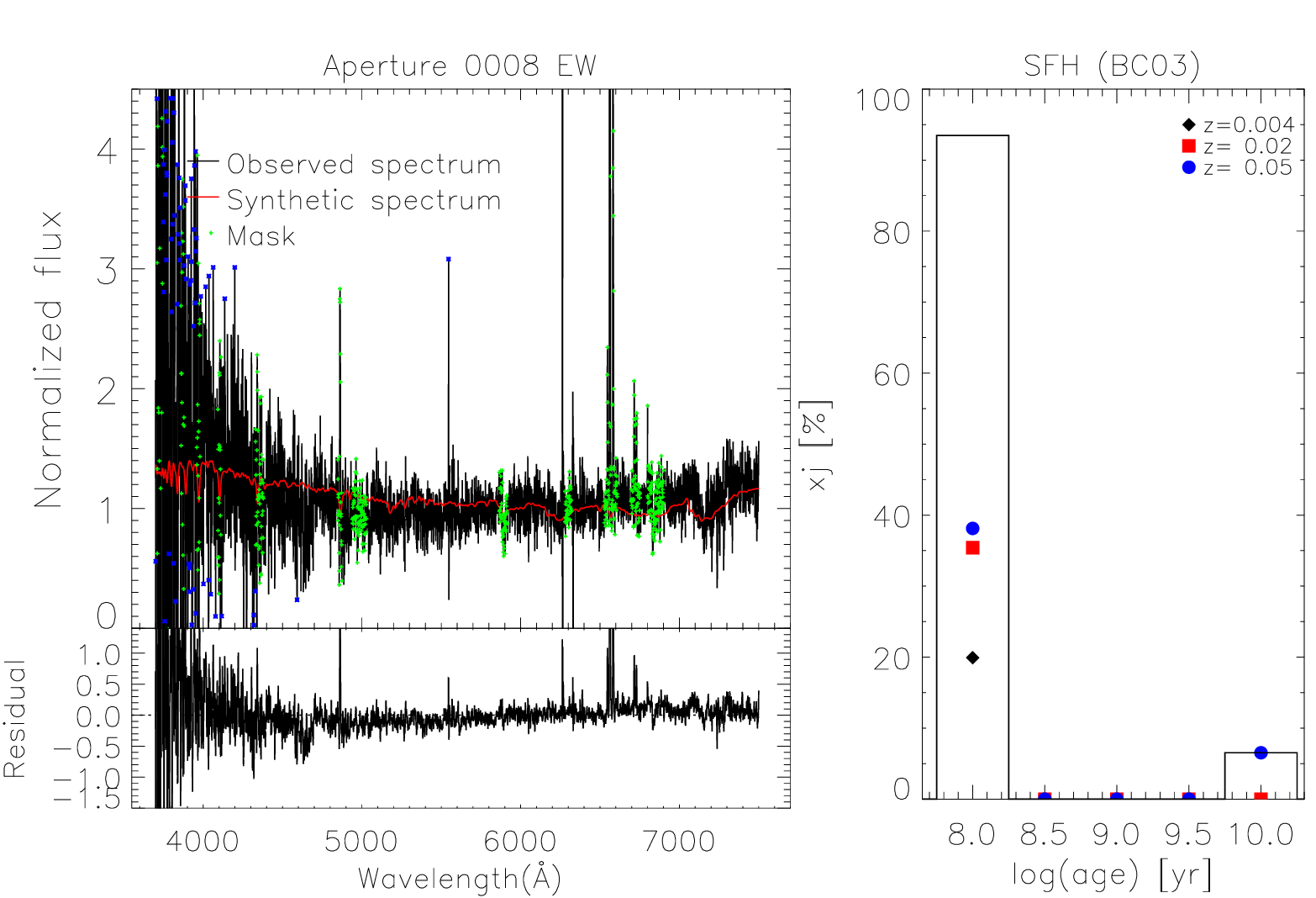}
}
\subfloat[]{
\includegraphics[scale=0.38]{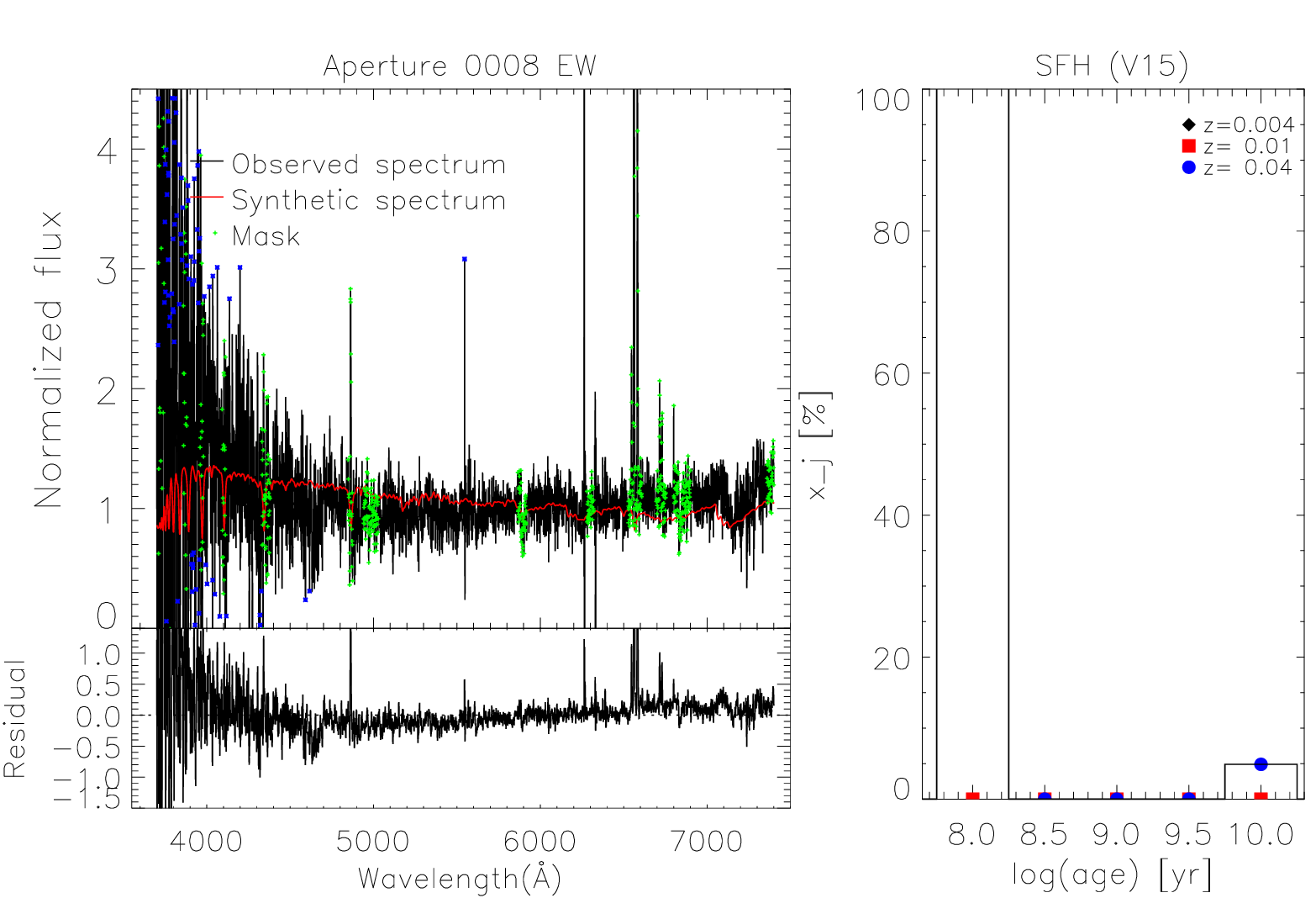}

}
\\
\subfloat[]{
\includegraphics[scale=0.38]{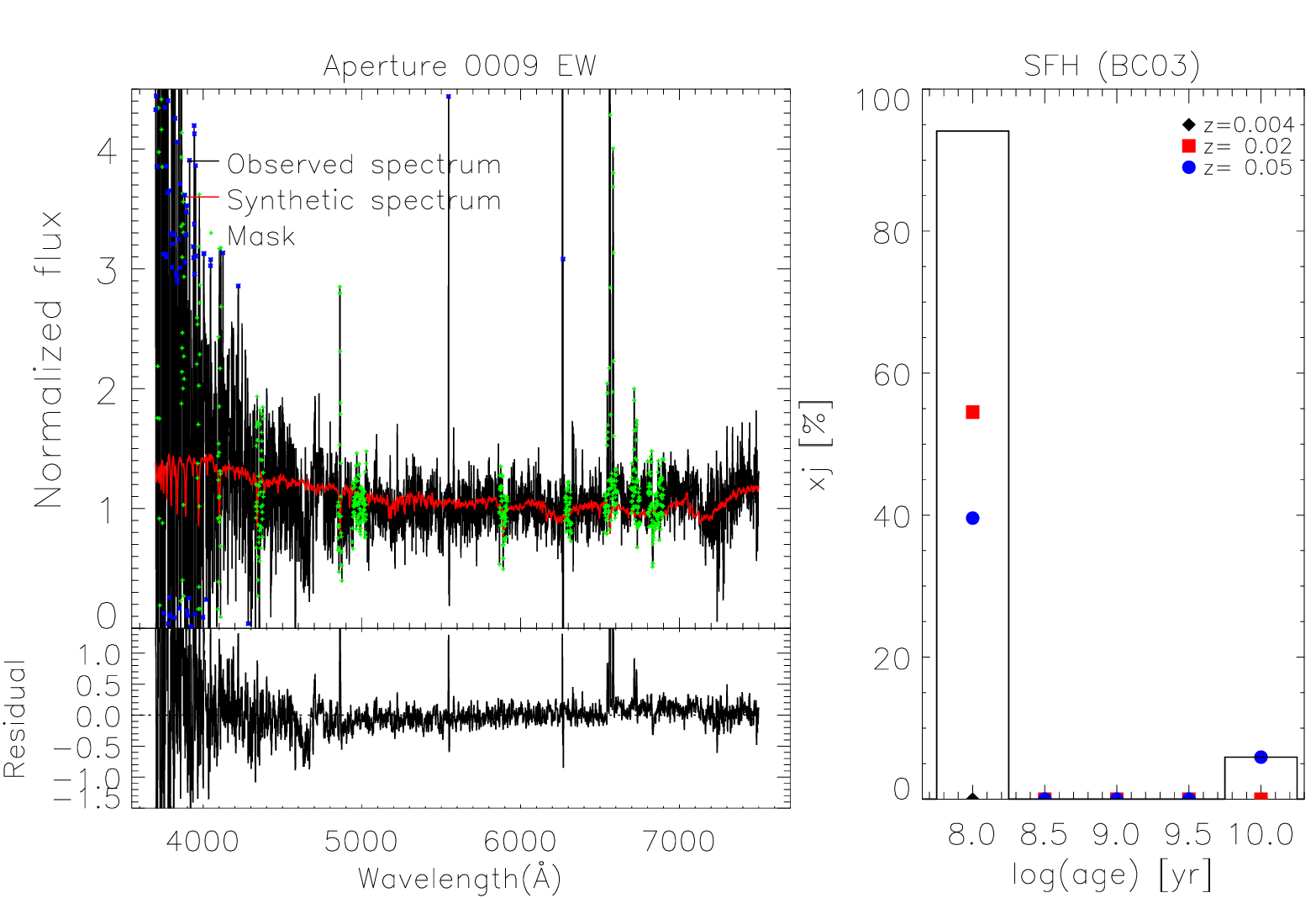}
}
\subfloat[]{
\includegraphics[scale=0.38]{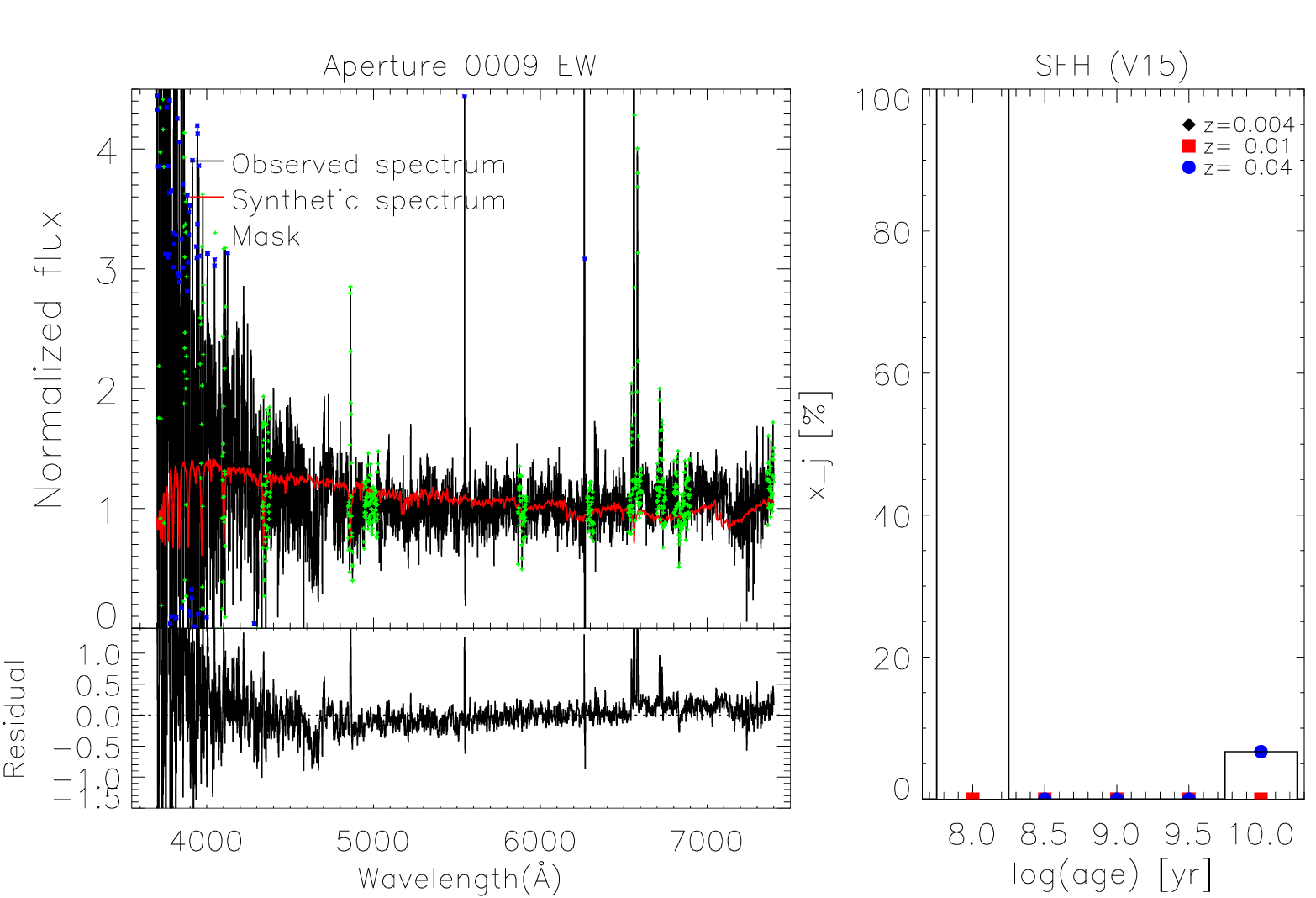}
}
\\
\subfloat[]{
\includegraphics[scale=0.38]{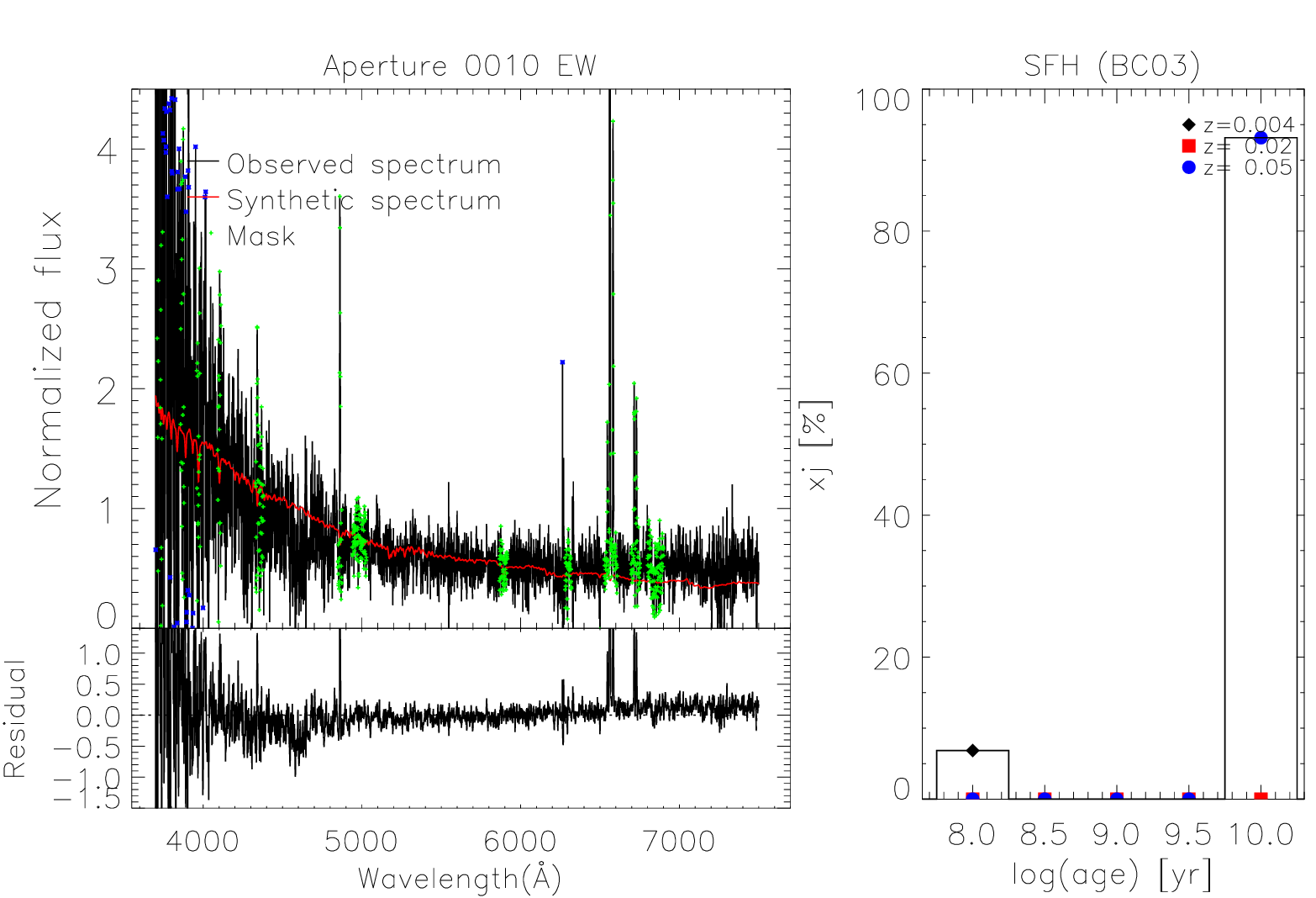}
}
\subfloat[]{
\includegraphics[scale=0.38]{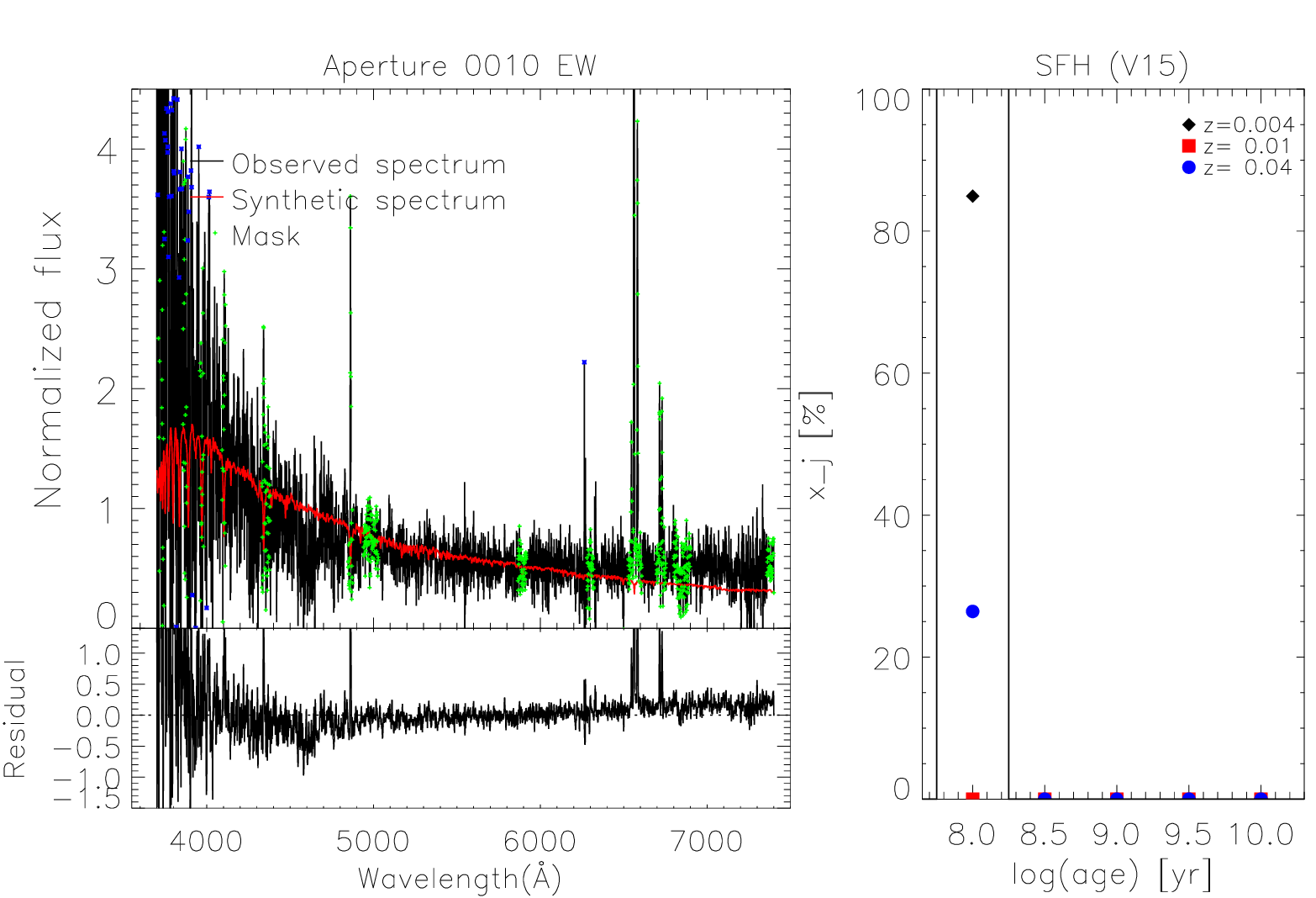}
}

\caption {Same as Figure~\ref{popstellarA1} but for apertures 08, 09,
 and 10 of the E-W  slit.
} 
\label{popstellarA8}
\end{figure*}


\begin{figure*}[!h]
\centering
\subfloat[]{
\includegraphics[scale=0.38]{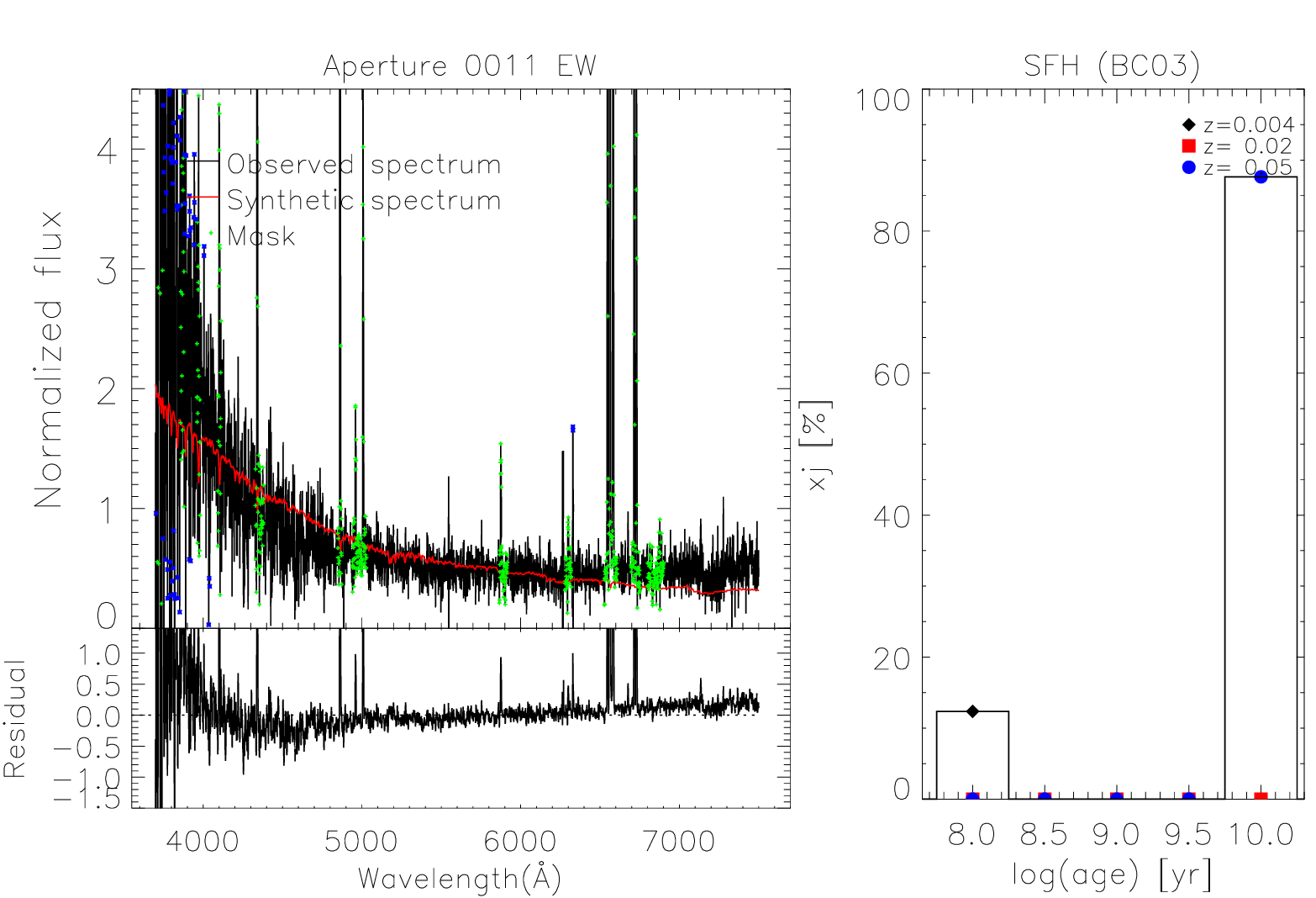}
}
\subfloat[]{
\includegraphics[scale=0.38]{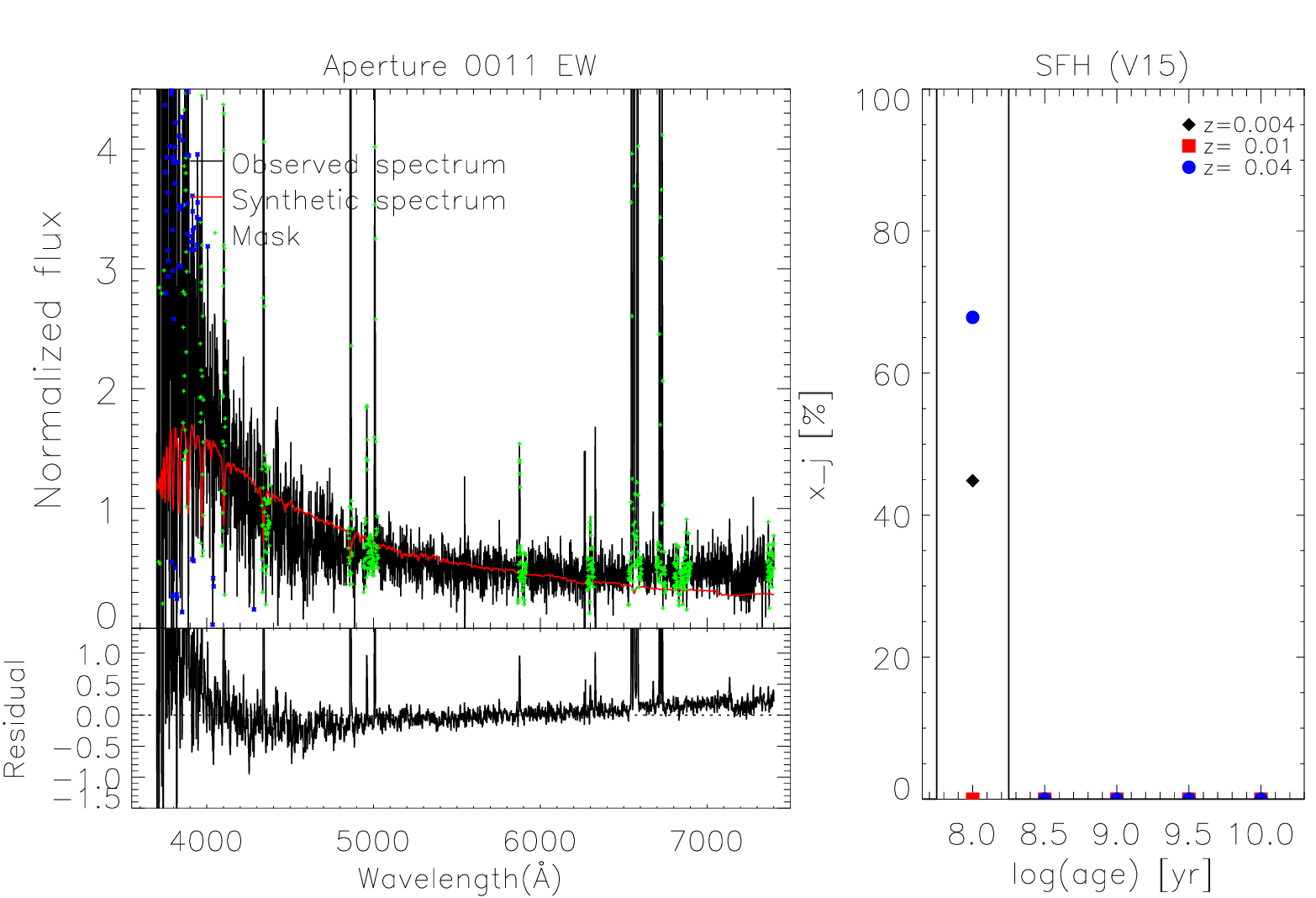}
}
\\
\subfloat[]{
\includegraphics[scale=0.38]{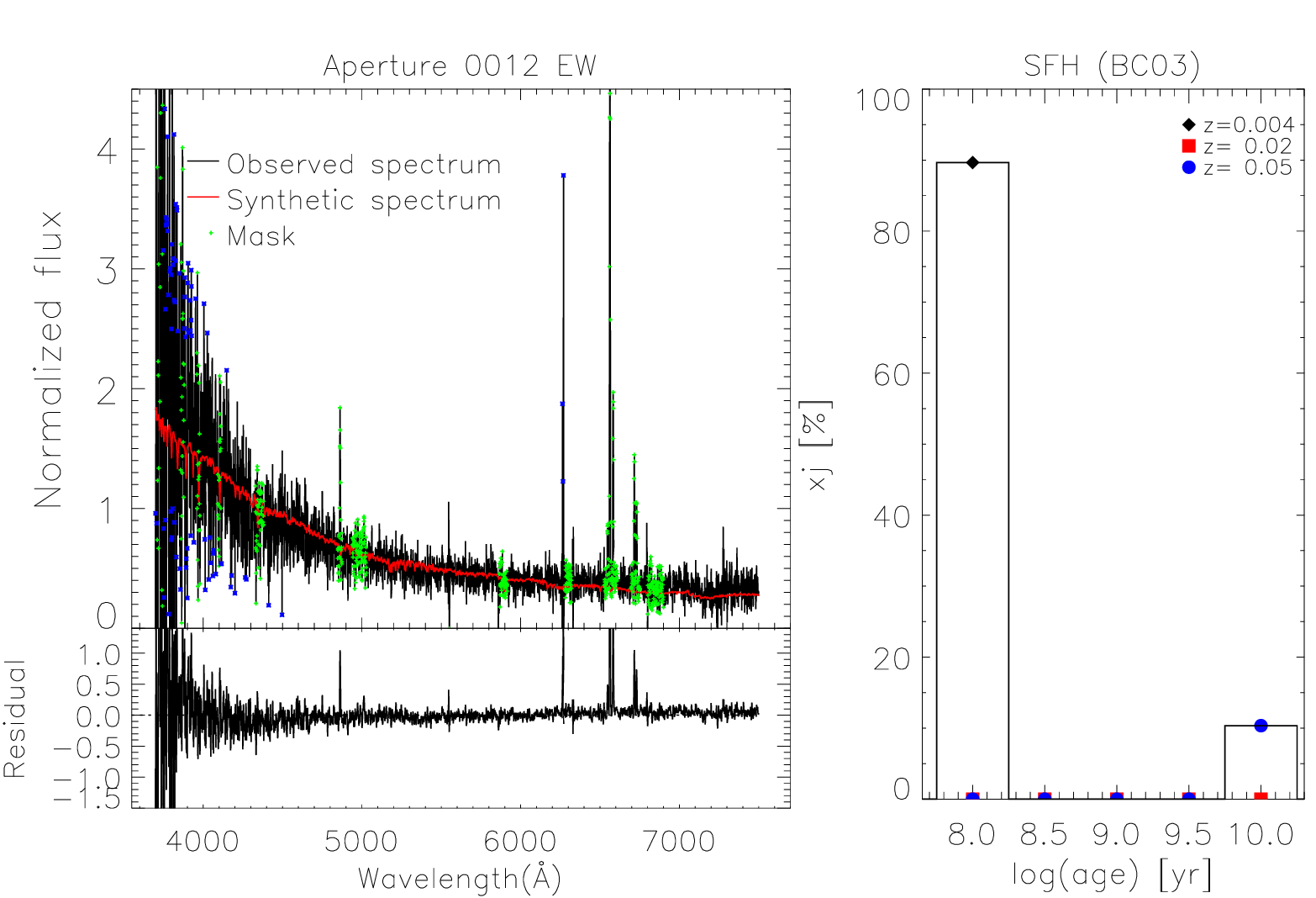}
}
\subfloat[]{
\includegraphics[scale=0.38]{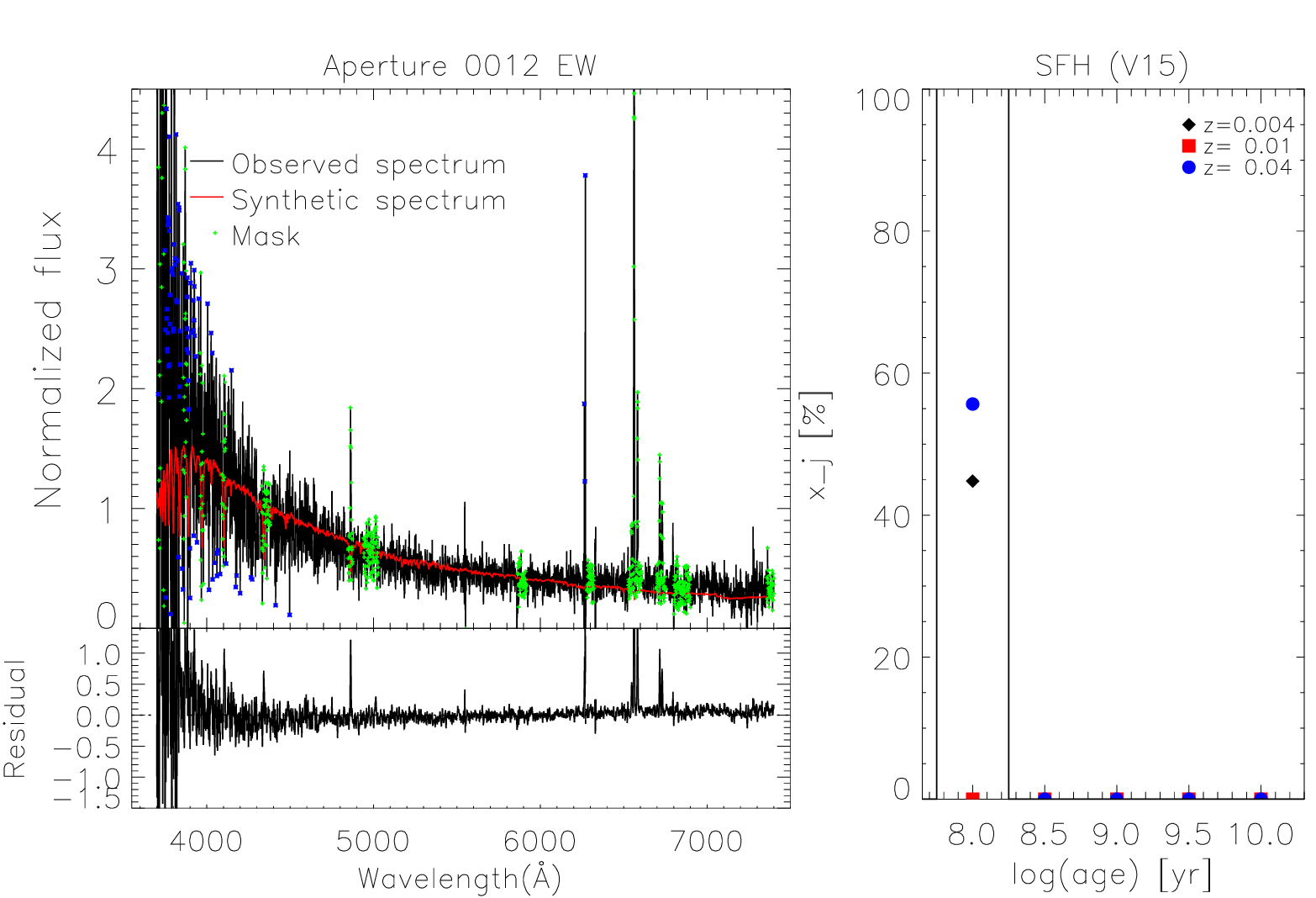}
}
\\
\subfloat[]{
\includegraphics[scale=0.38]{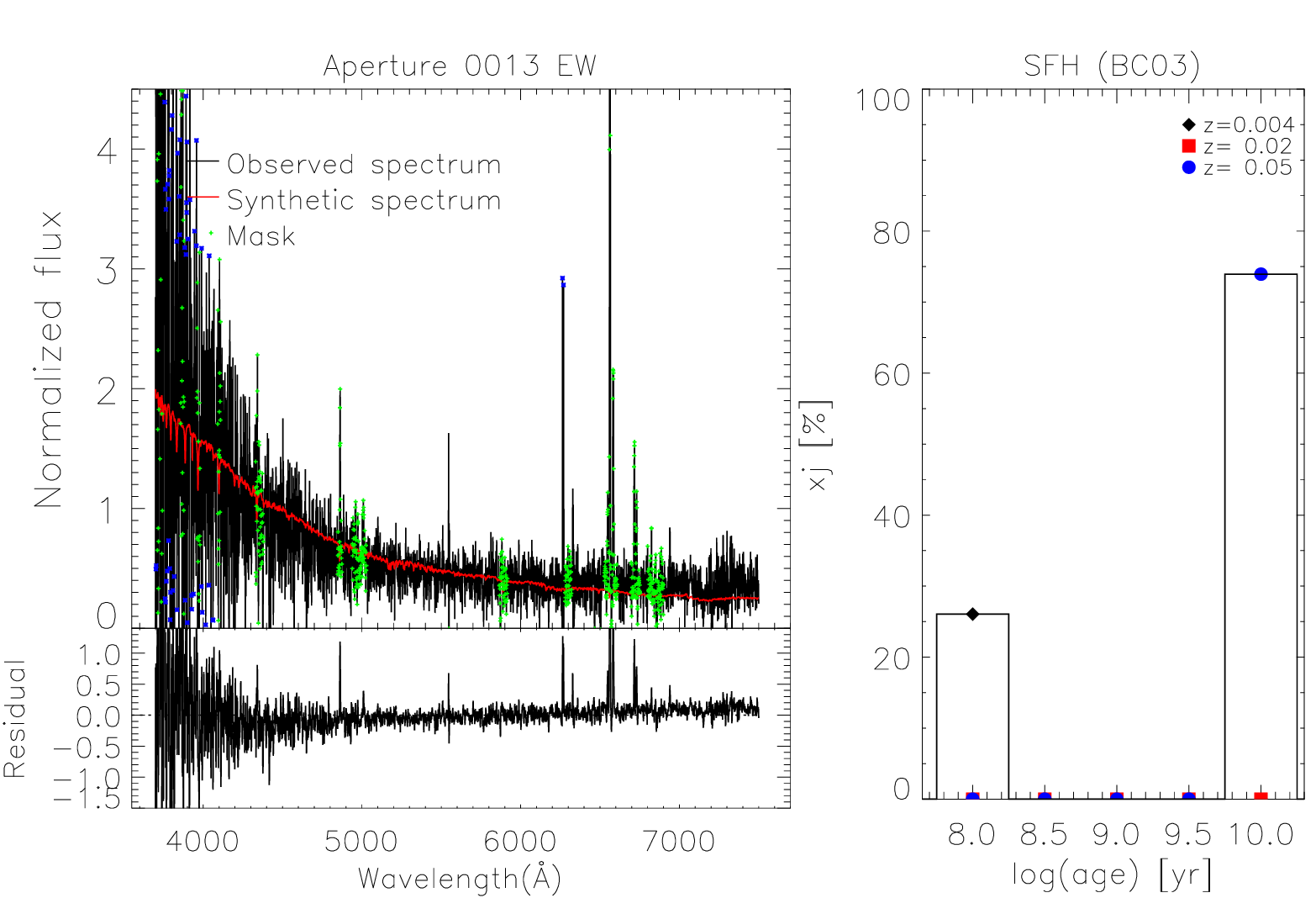}
}
\subfloat[]{
\includegraphics[scale=0.38]{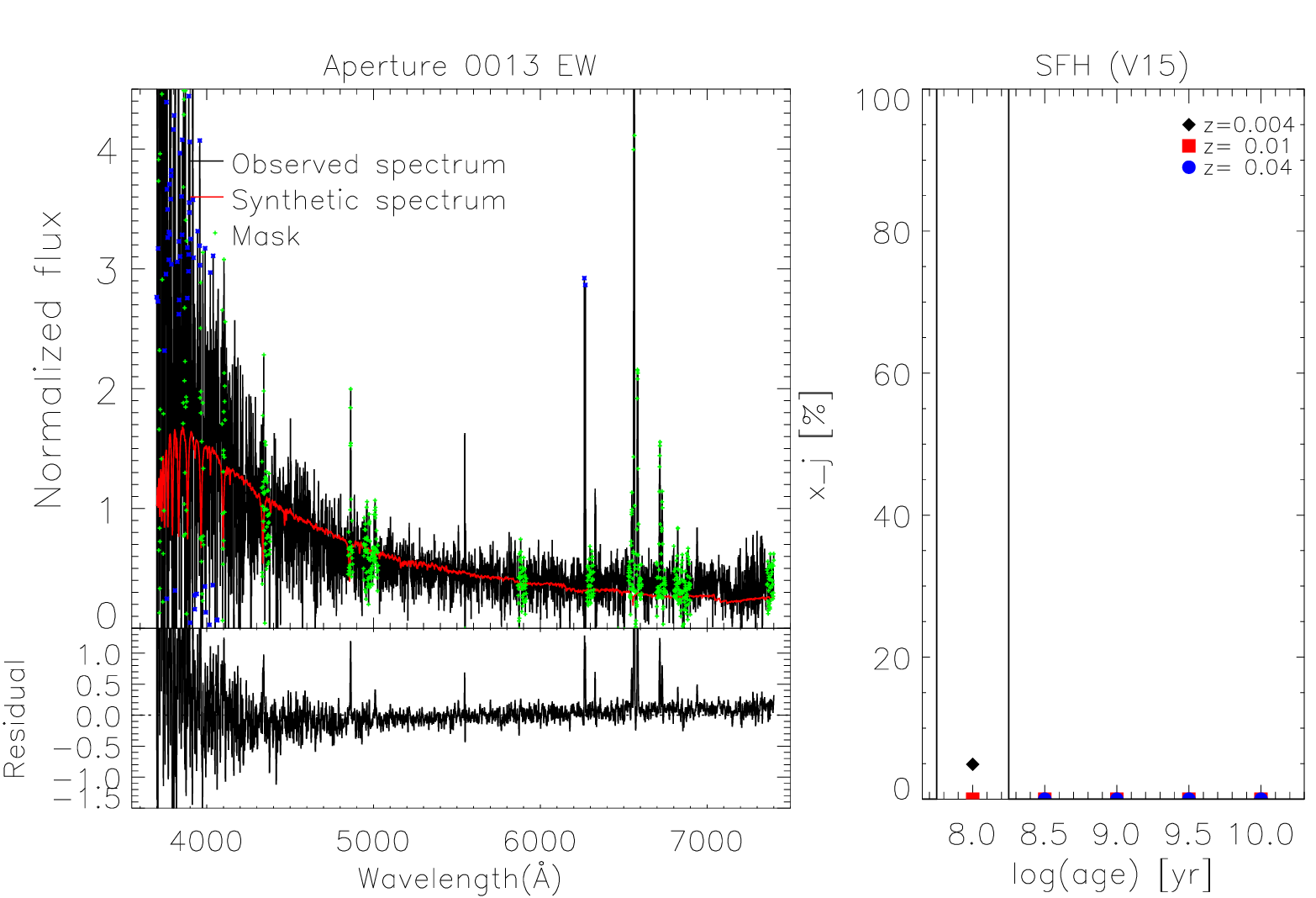}
}

\caption {Same as Figure~\ref{popstellarA1} but for apertures 11, 12
, and 13 of the E-W  slit.
} 
\label{popstellarA9}
\end{figure*}


\begin{figure*}[!h]
\centering
\subfloat[]{
\includegraphics[scale=0.38]{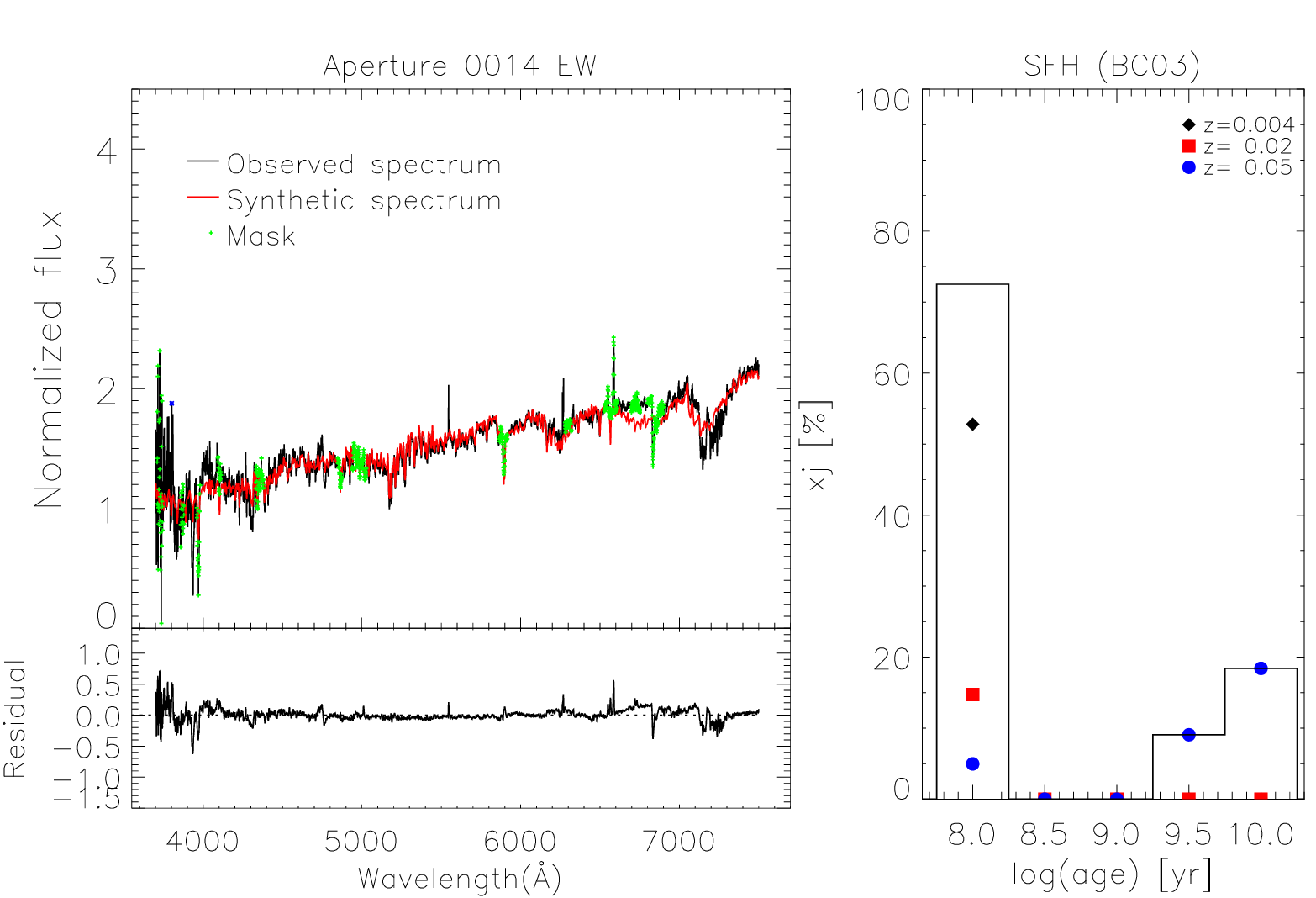}
}
\subfloat[]{
\includegraphics[scale=0.38]{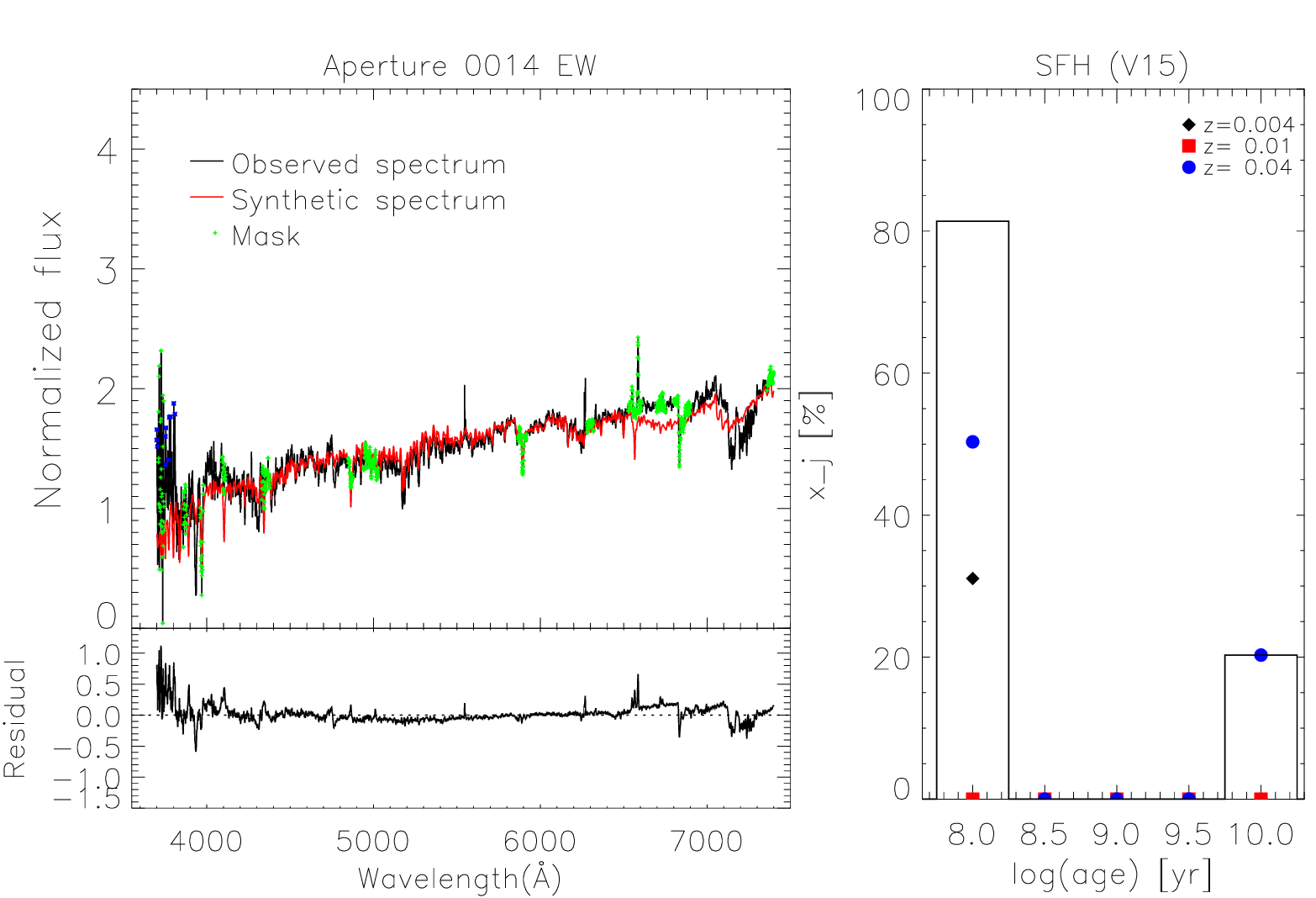}
}
\\
\subfloat[]{
\includegraphics[scale=0.38]{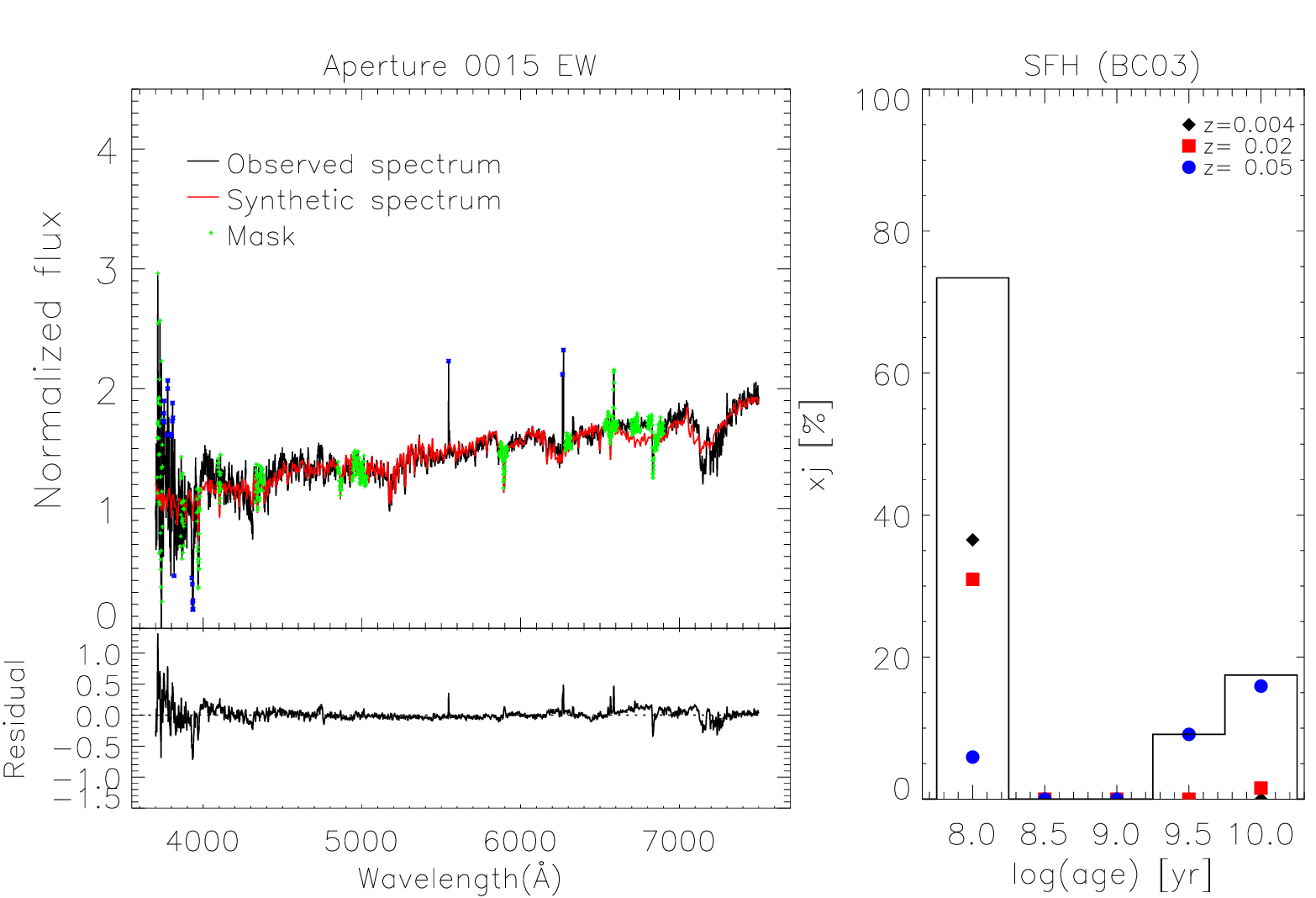}
}
\subfloat[]{
\includegraphics[scale=0.38]{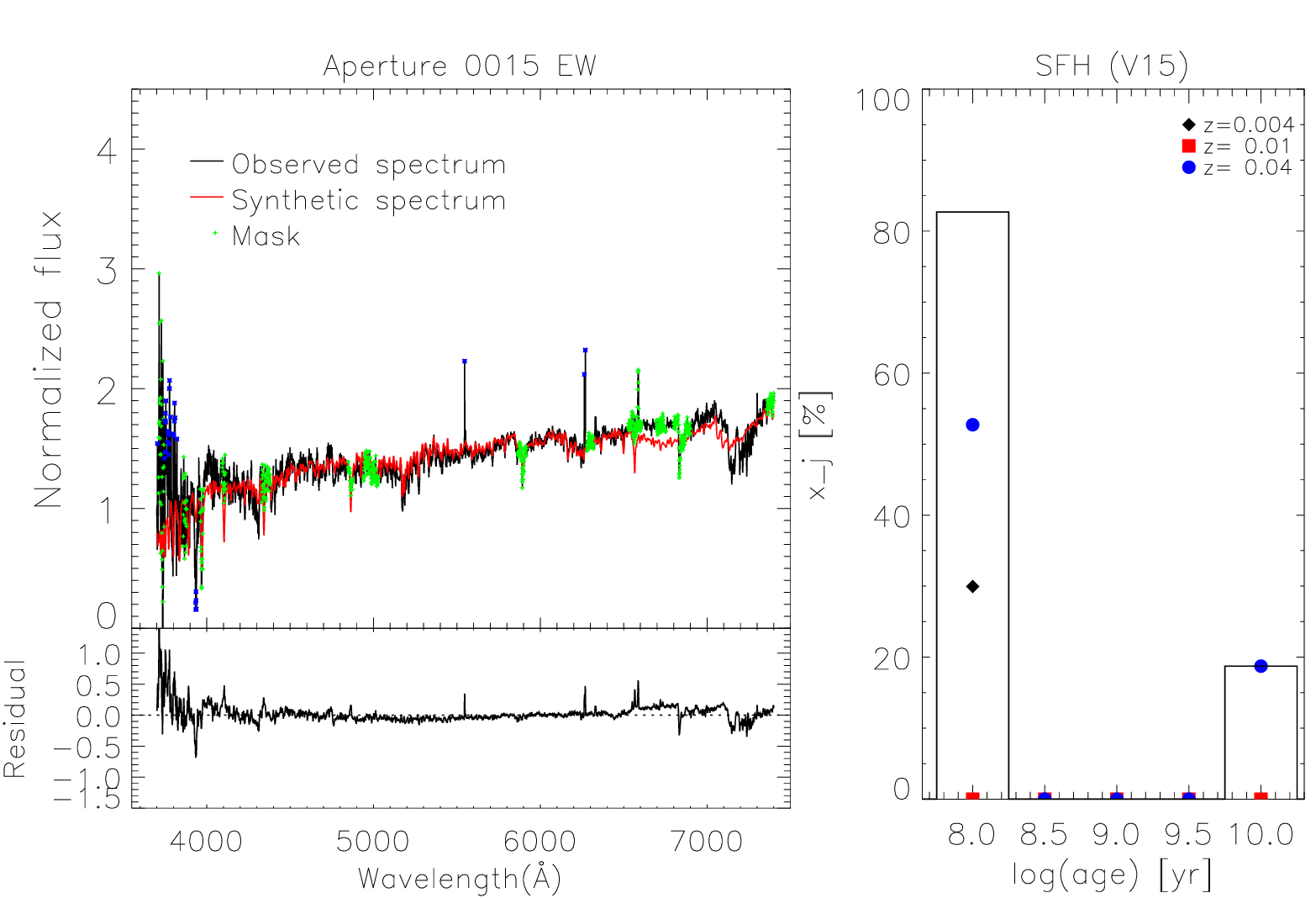}
}
\\
\subfloat[]{
\includegraphics[scale=0.38]{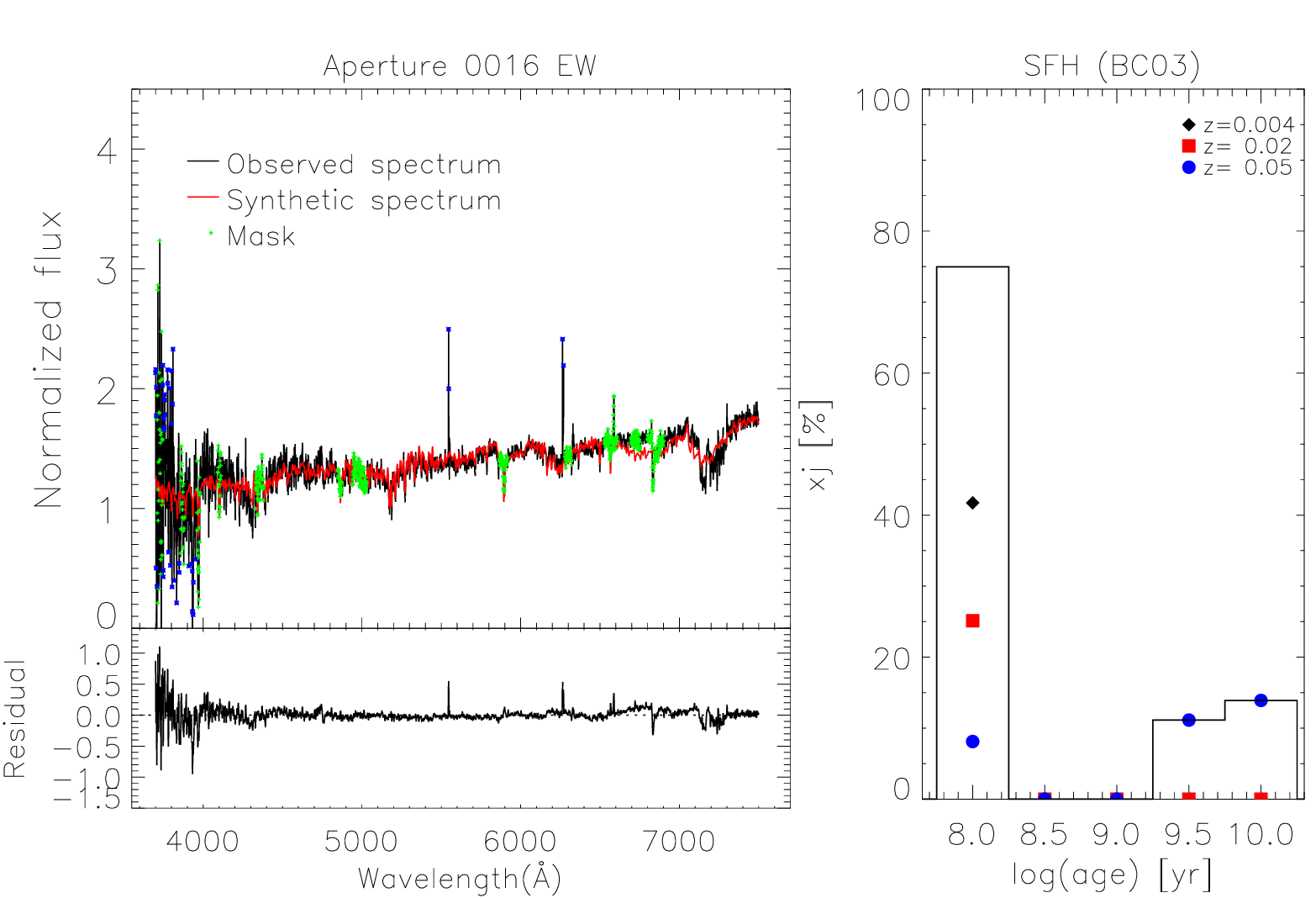}
}
\subfloat[]{
\includegraphics[scale=0.38]{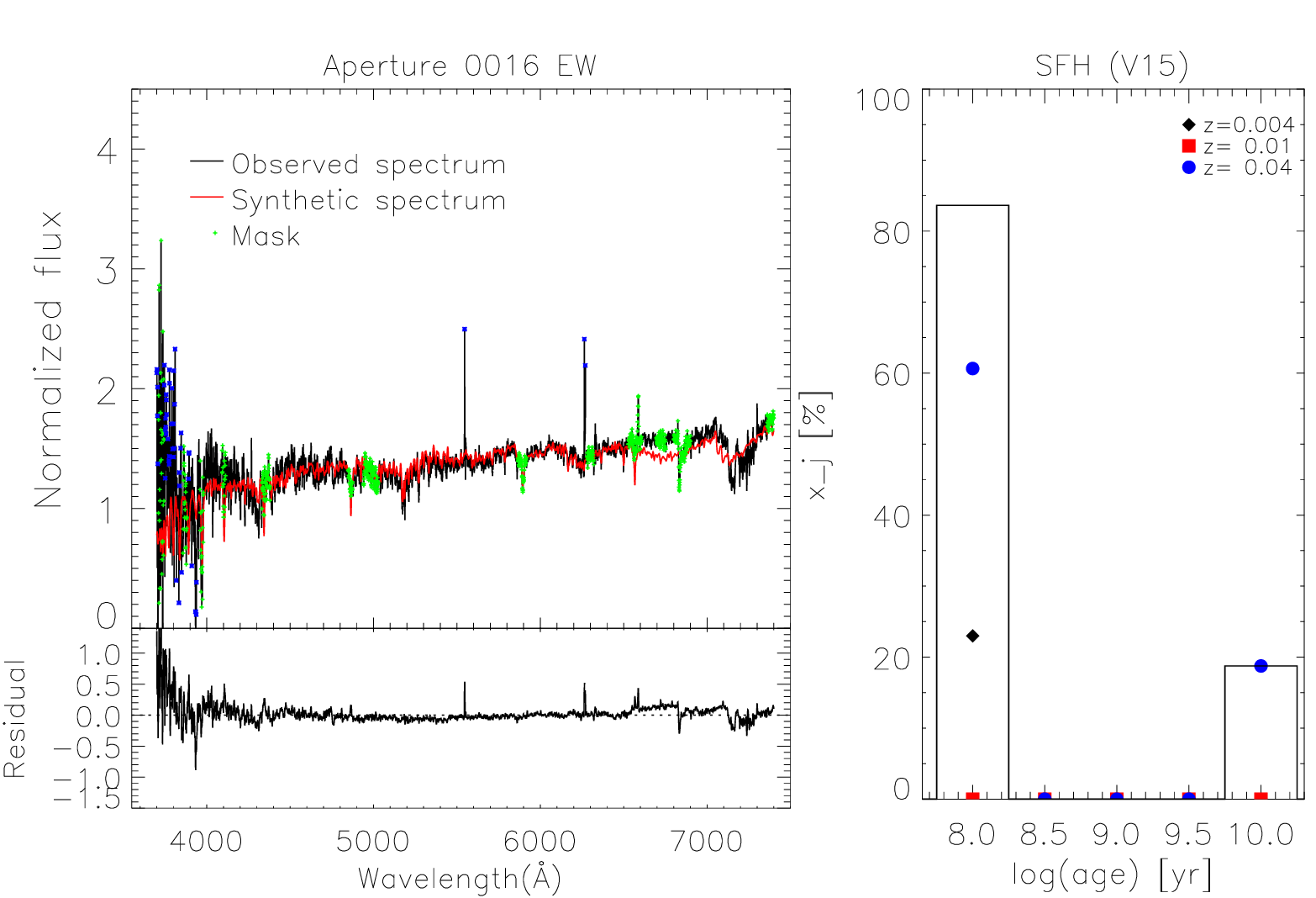}
}

\caption {Same as Figure~\ref{popstellarA1} but for apertures 14, 15, and 16 of the E-W  slit.
} 
\label{popstellarA10}
\end{figure*}


\begin{figure*}[!h]
\centering
\subfloat[]{
\includegraphics[scale=0.38]{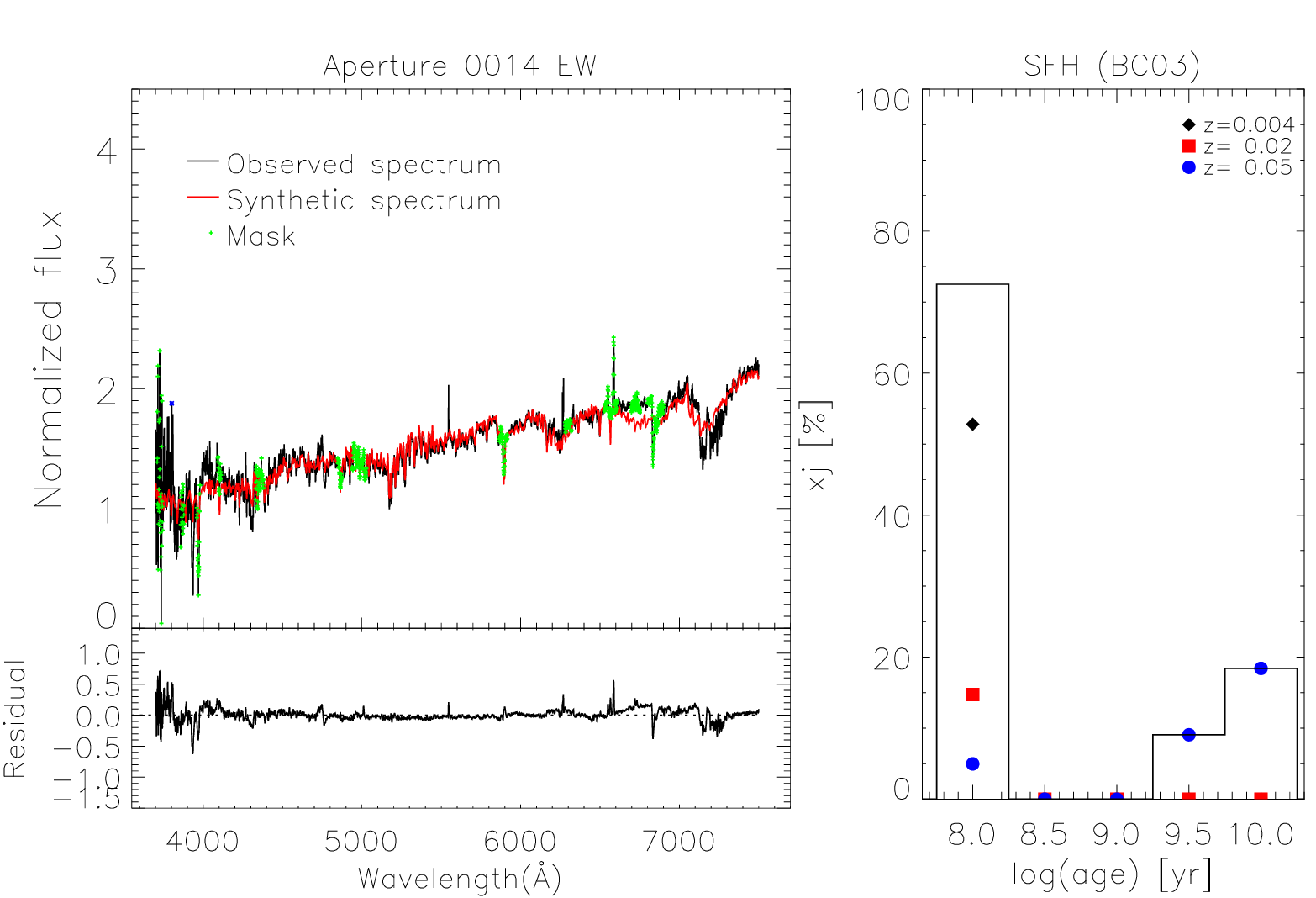}
}
\subfloat[]{
\includegraphics[scale=0.38]{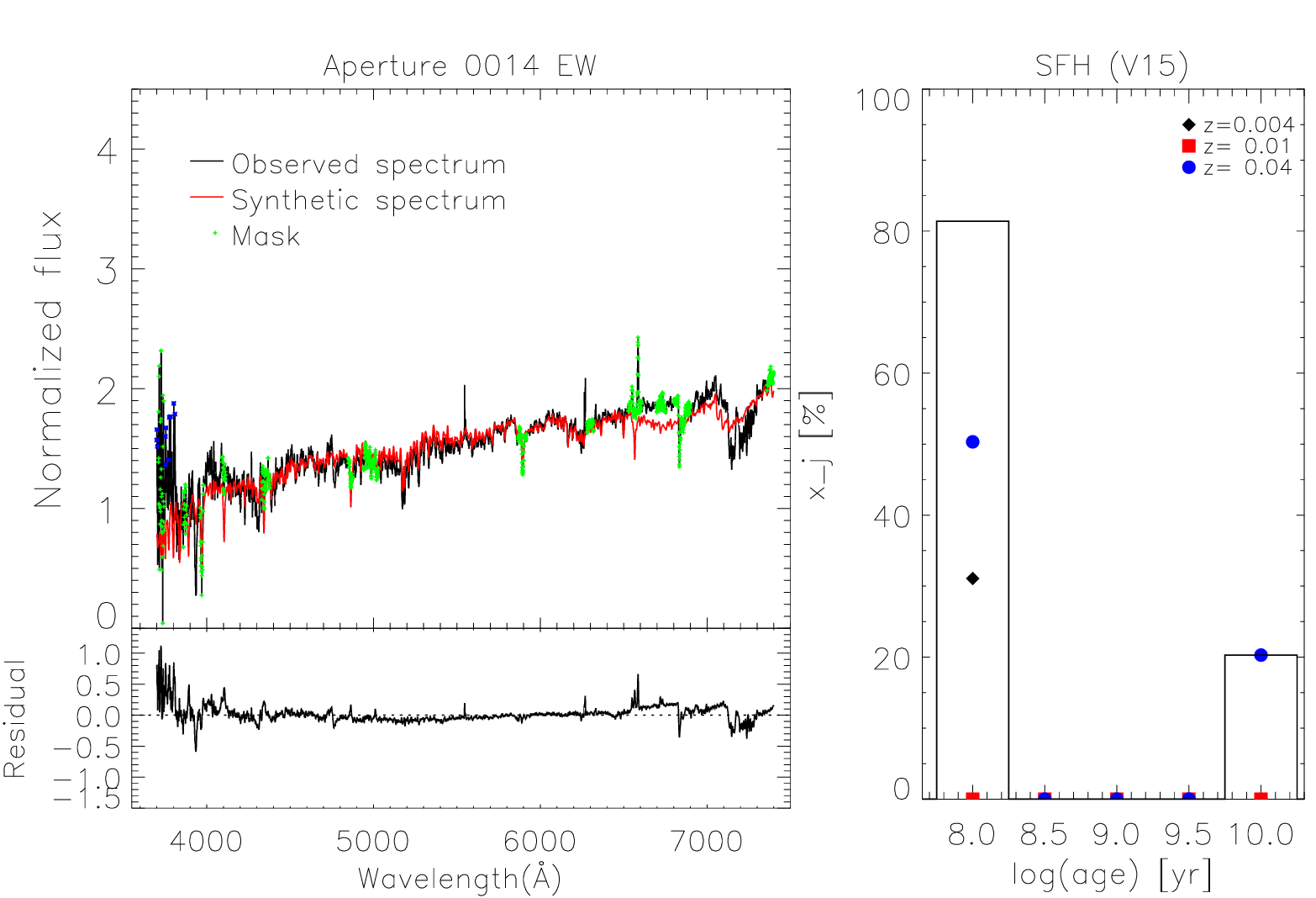}
}
\\
\subfloat[]{
\includegraphics[scale=0.38]{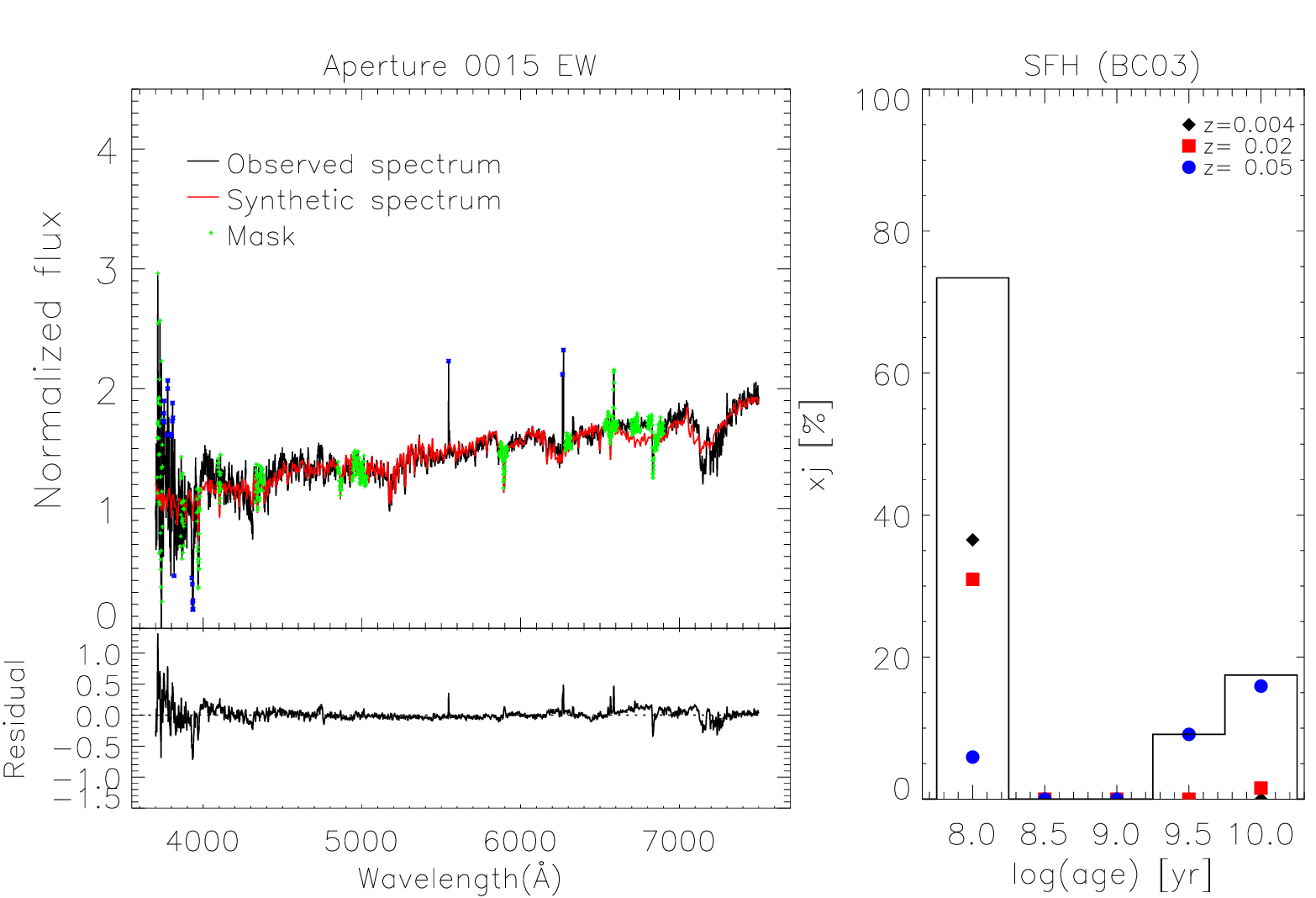}
}
\subfloat[]{
\includegraphics[scale=0.38]{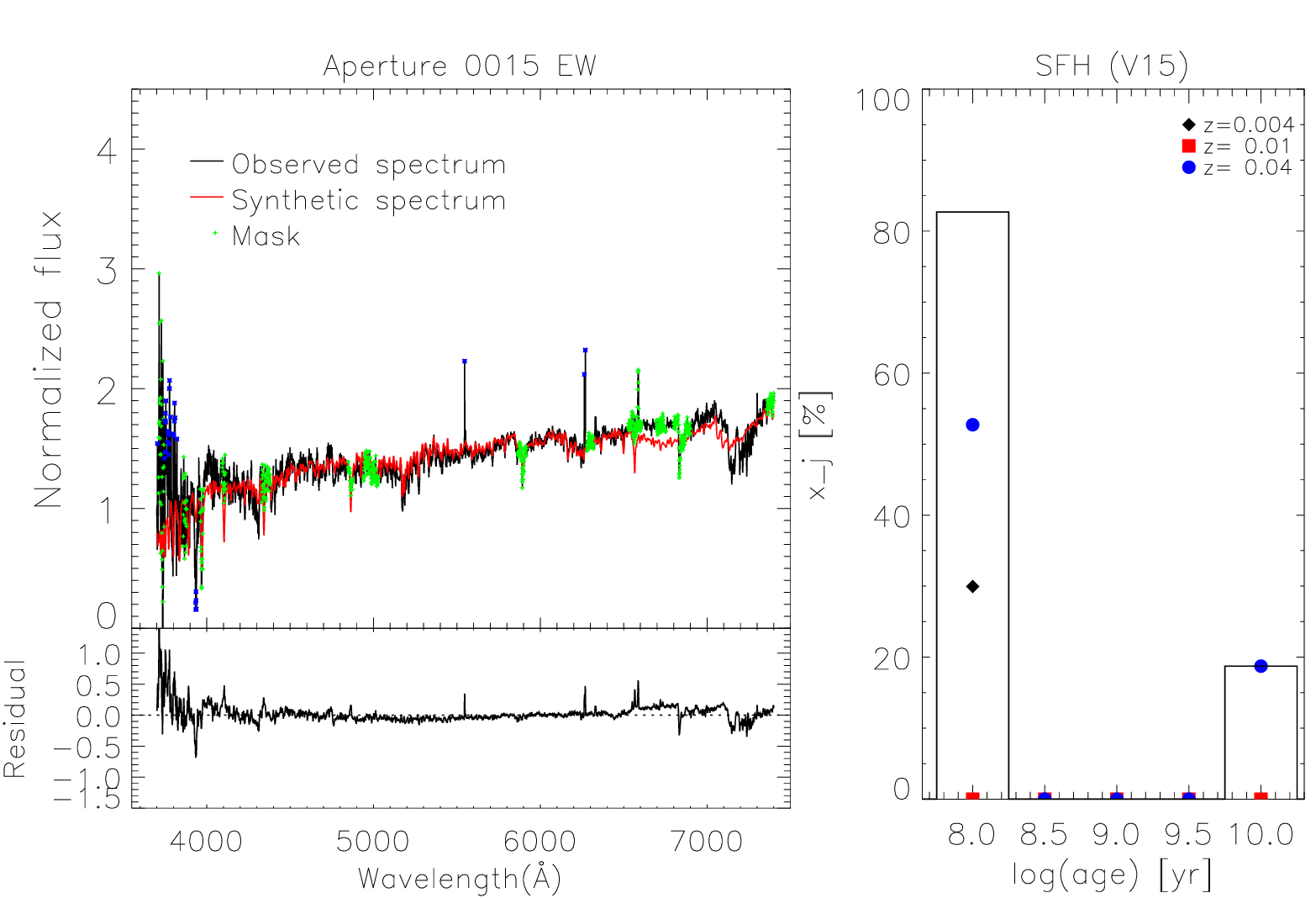}
}
\\
\subfloat[]{
\includegraphics[scale=0.38]{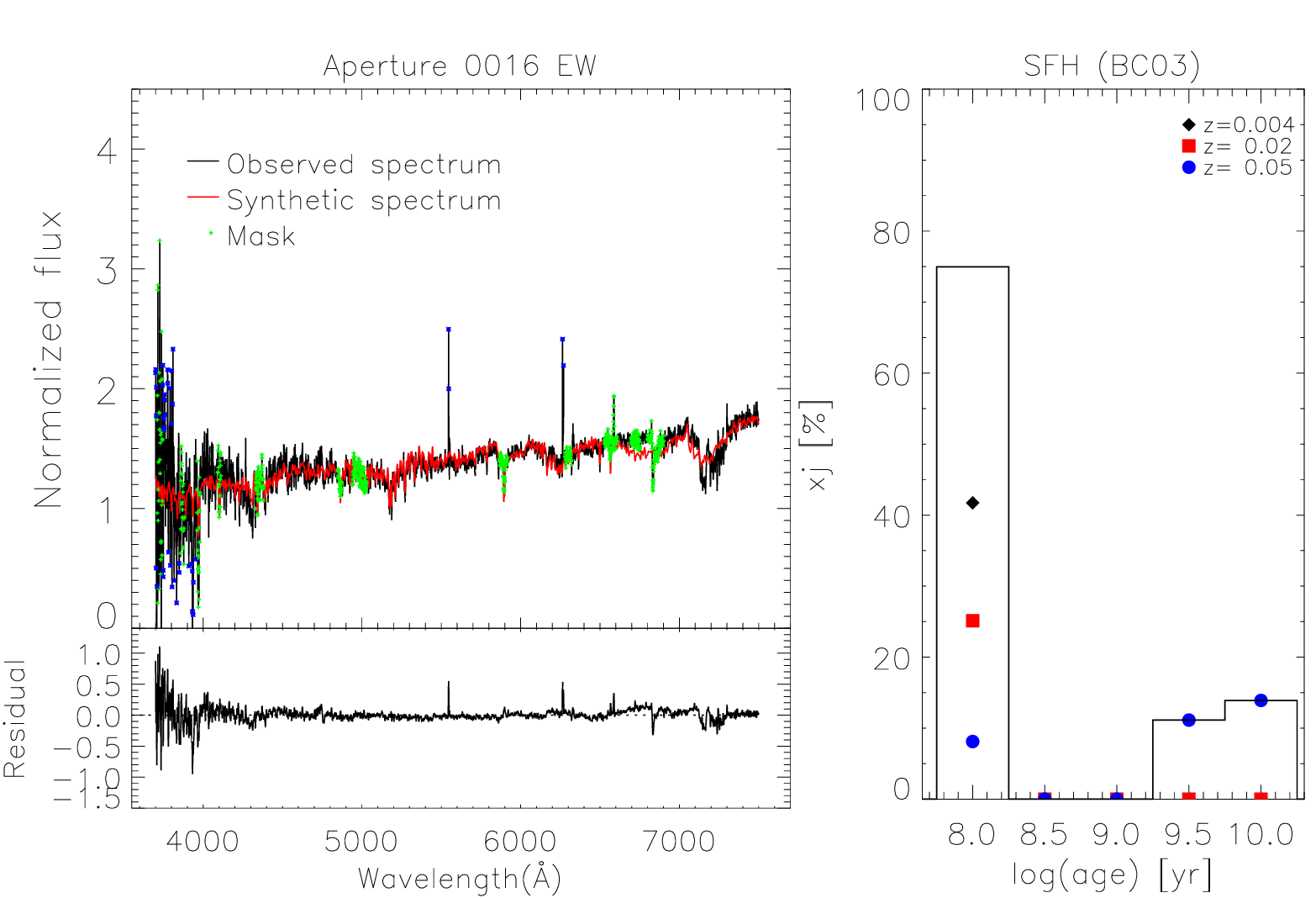}
}
\subfloat[]{
\includegraphics[scale=0.38]{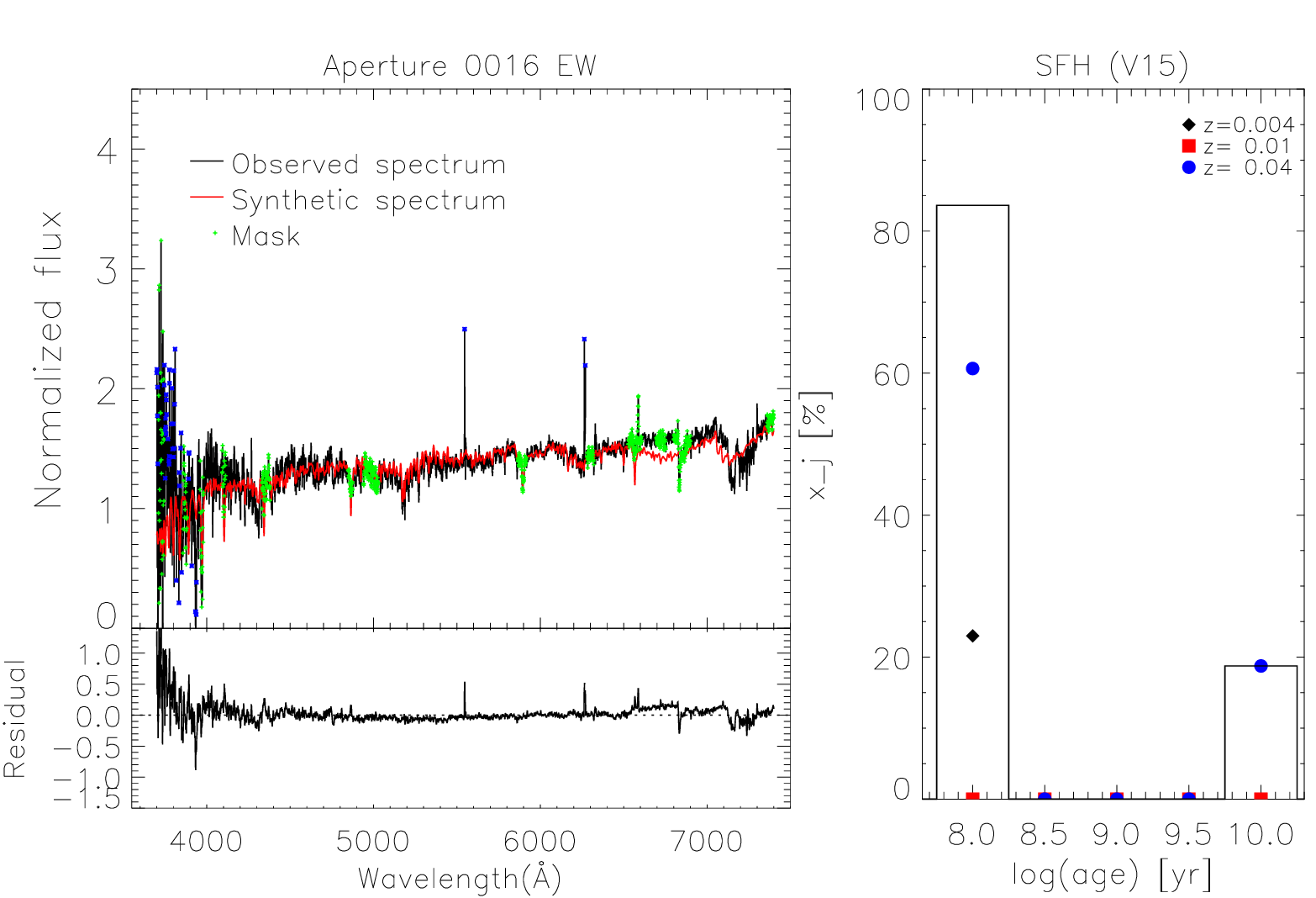}
}

\caption {Same as Figure~\ref{popstellarA1} but for apertures 14, 15,
 and 16 of the E-W  slit.
} 
\label{popstellarA11}
\end{figure*}


\begin{figure*}[!h]
\centering
\subfloat[]{
\includegraphics[scale=0.38]{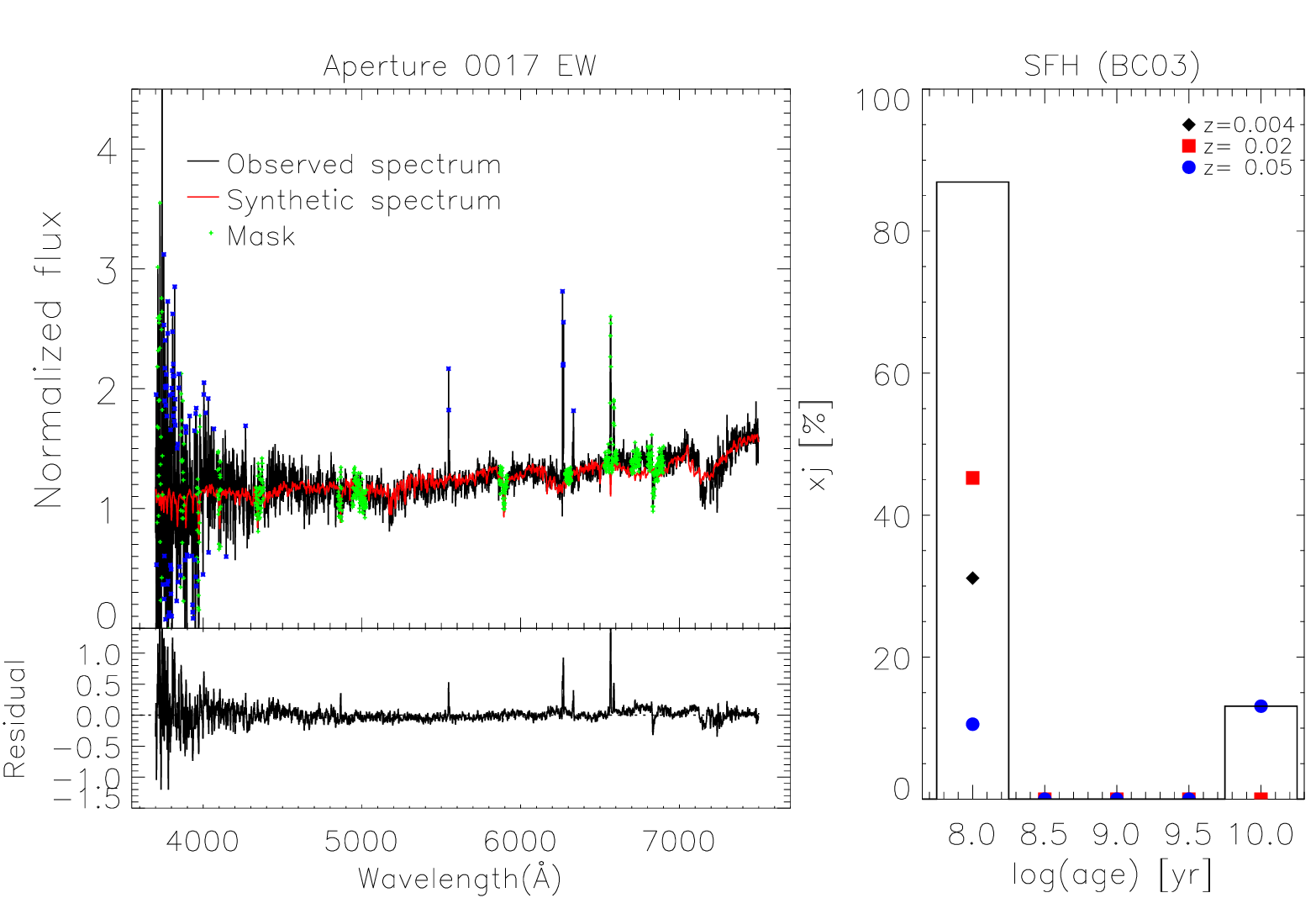}
}
\subfloat[]{
\includegraphics[scale=0.38]{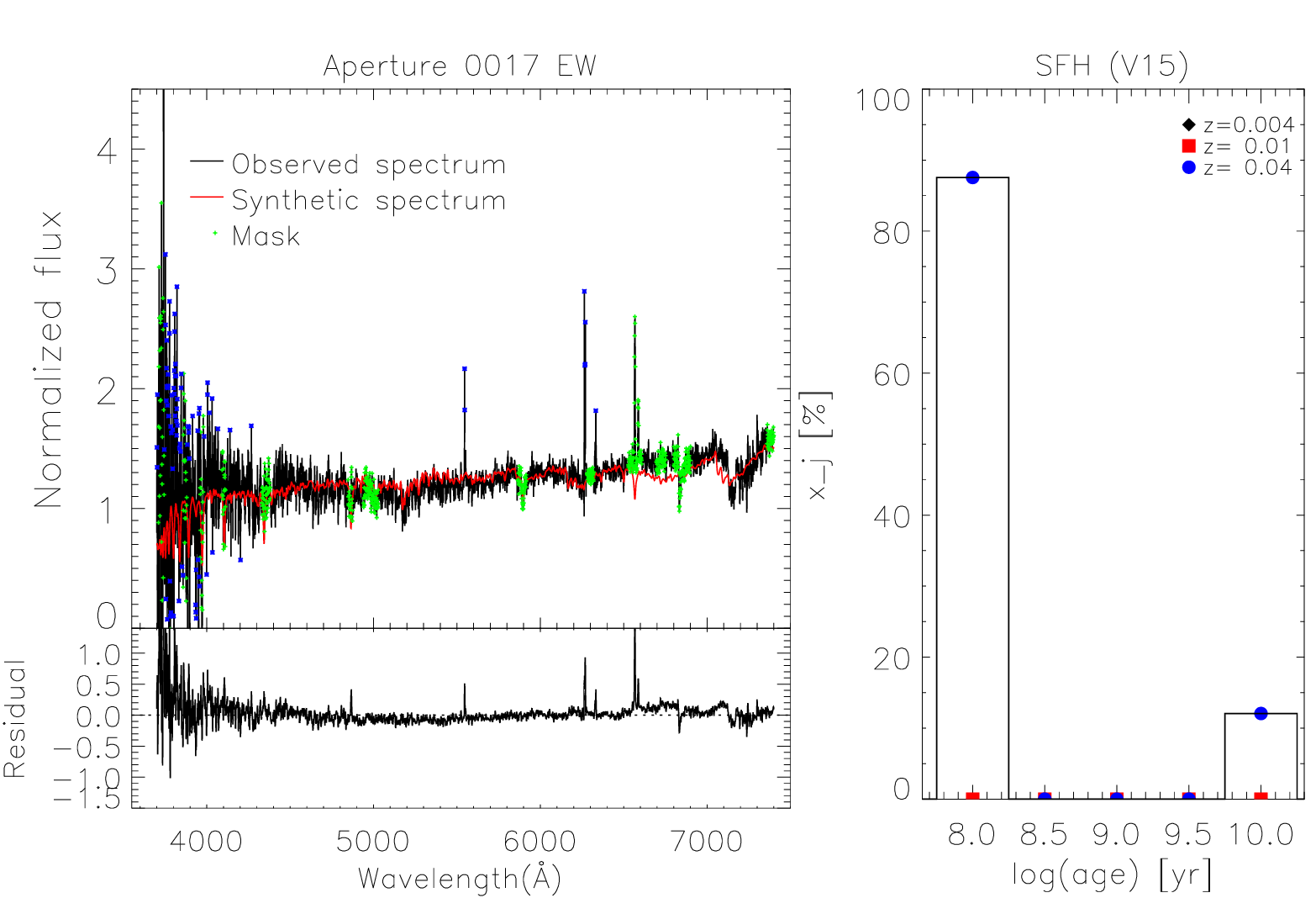}
}
\\
\subfloat[]{
\includegraphics[scale=0.38]{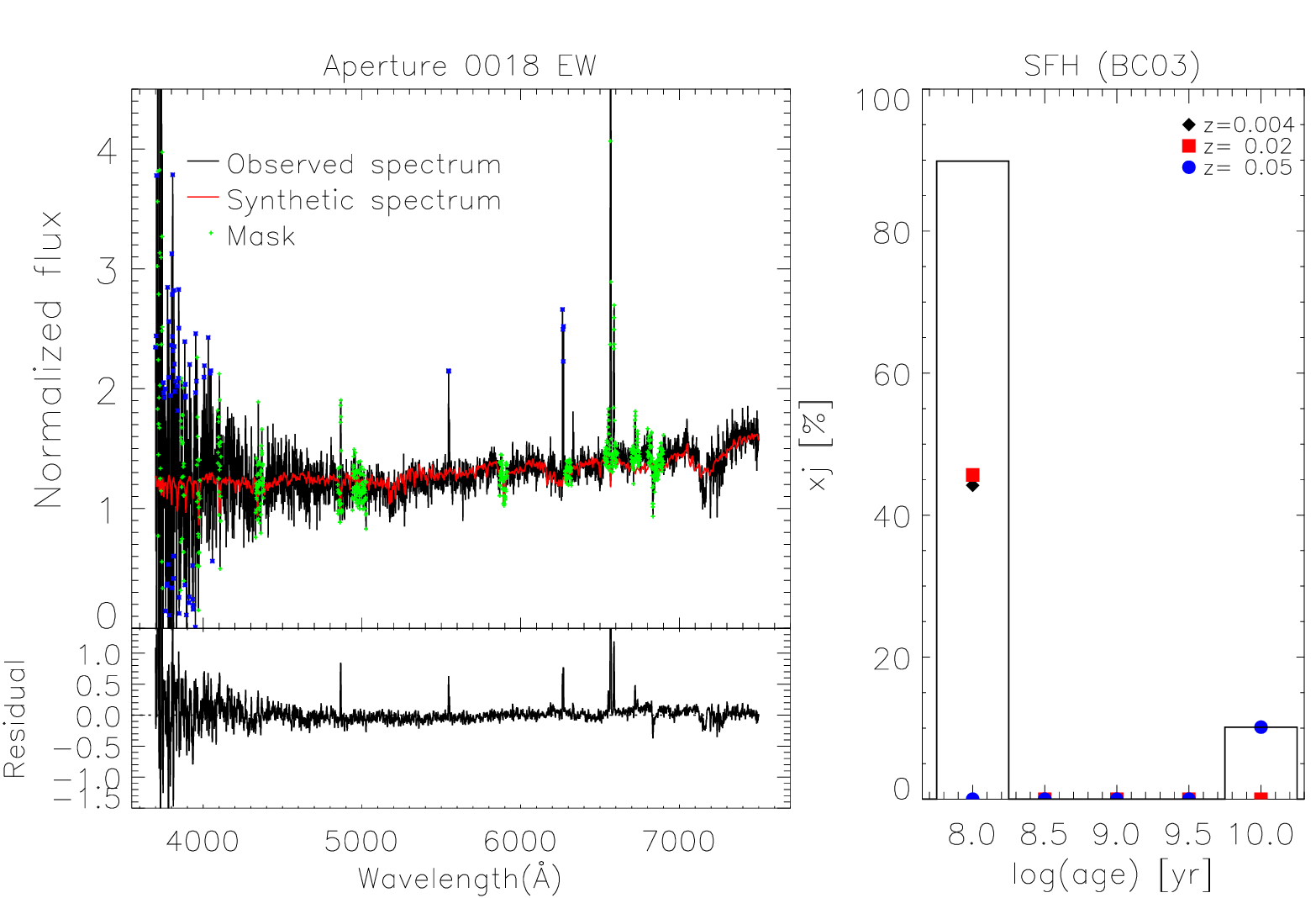}
}
\subfloat[]{
\includegraphics[scale=0.38]{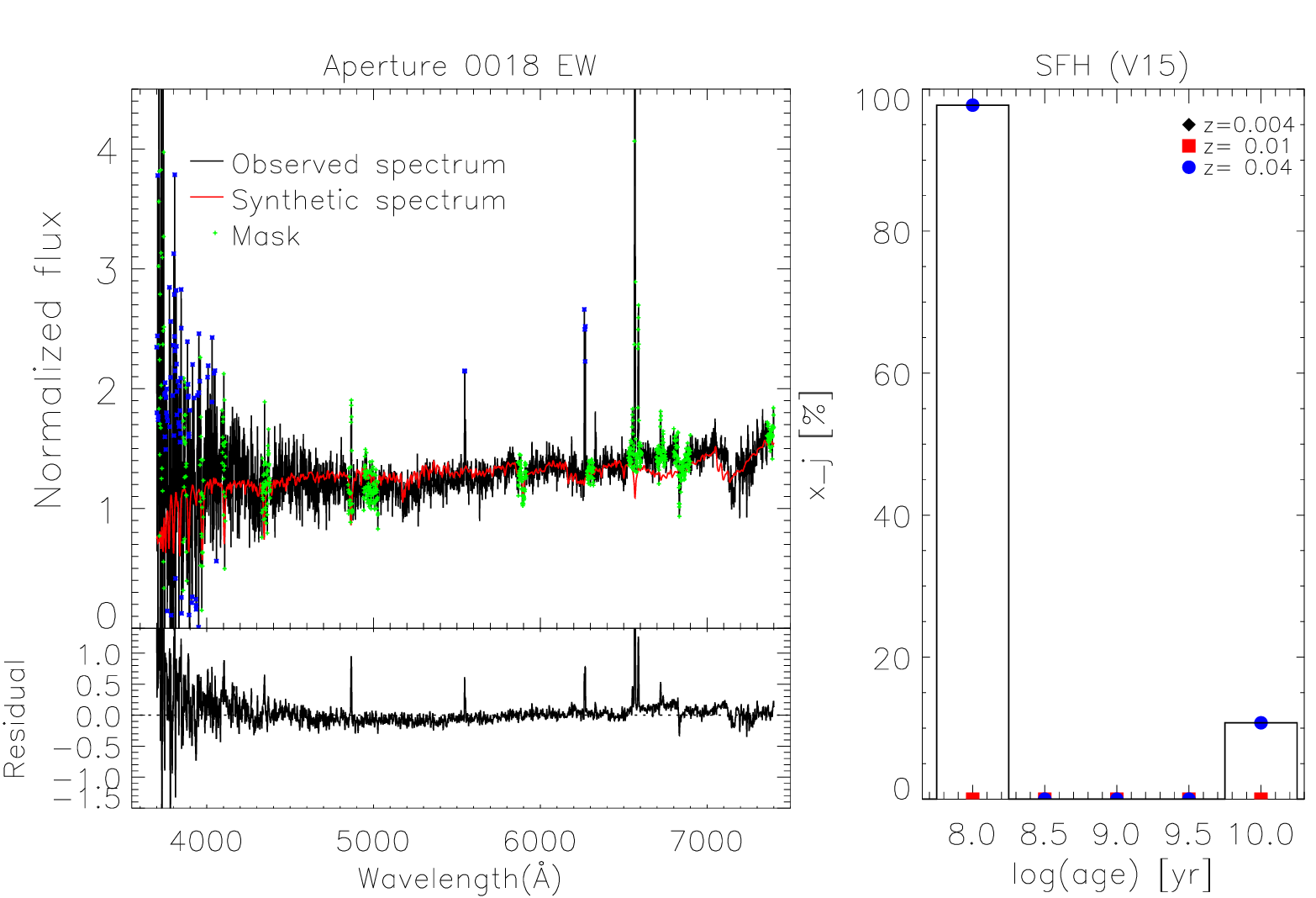}
}
\\
\subfloat[]{
\includegraphics[scale=0.38]{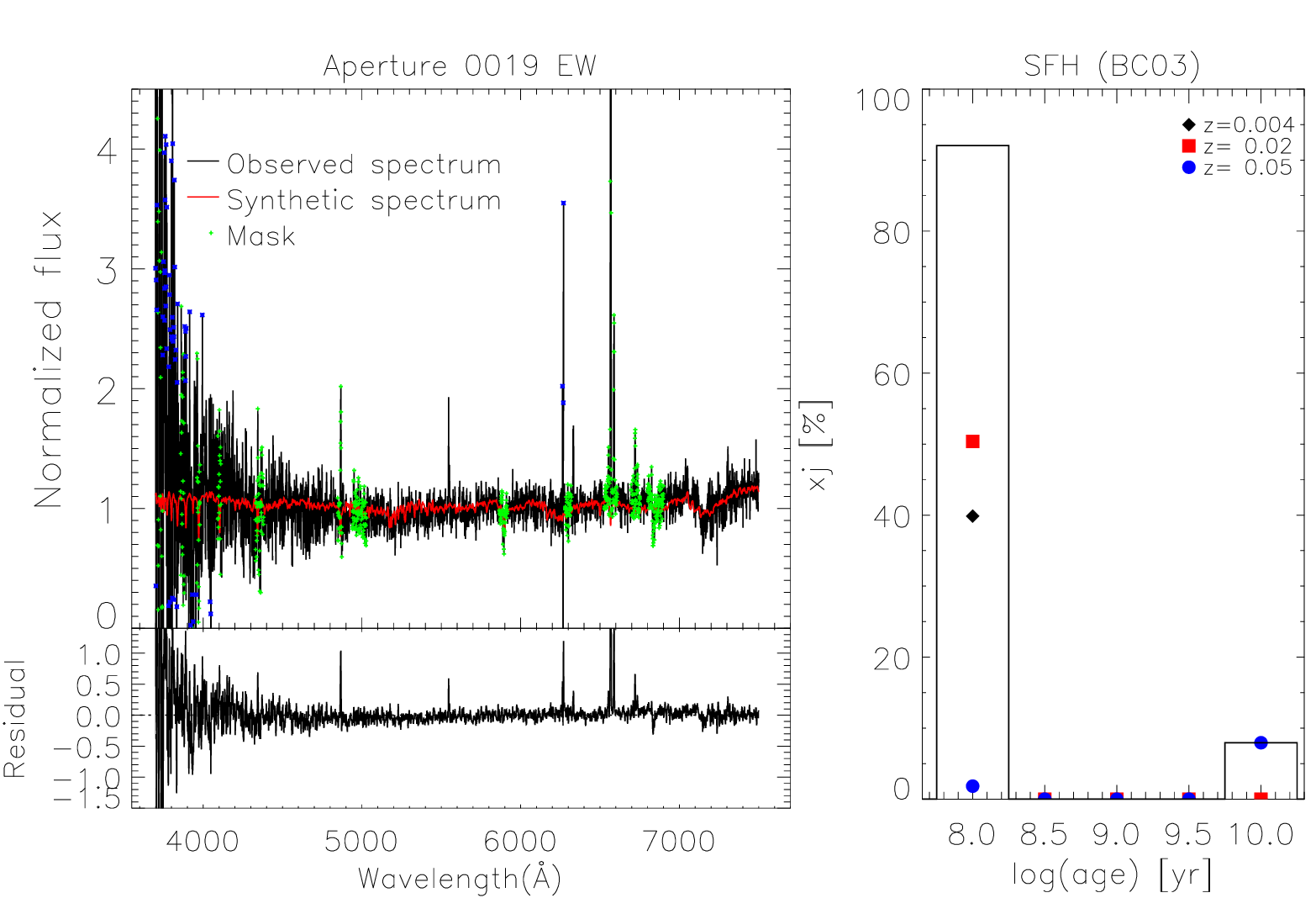}
}
\subfloat[]{
\includegraphics[scale=0.38]{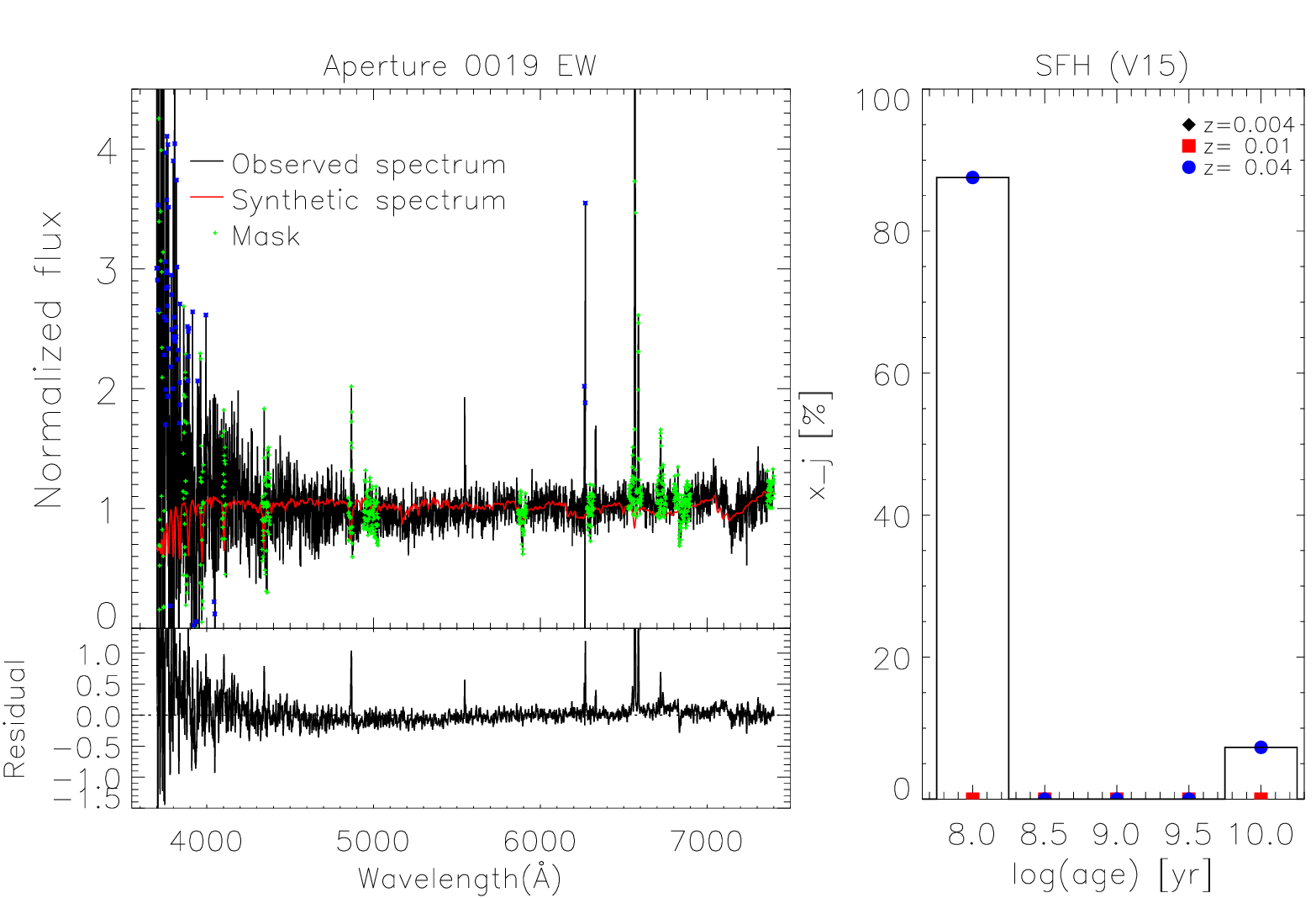}
}

\caption {Same as Figure~\ref{popstellarA1} but for apertures 17, 18,
 and 19 of the E-W  slit.
} 
\label{popstellarA12}
\end{figure*}


\begin{figure*}[!h]
\centering
\subfloat[]{
\includegraphics[scale=0.38]{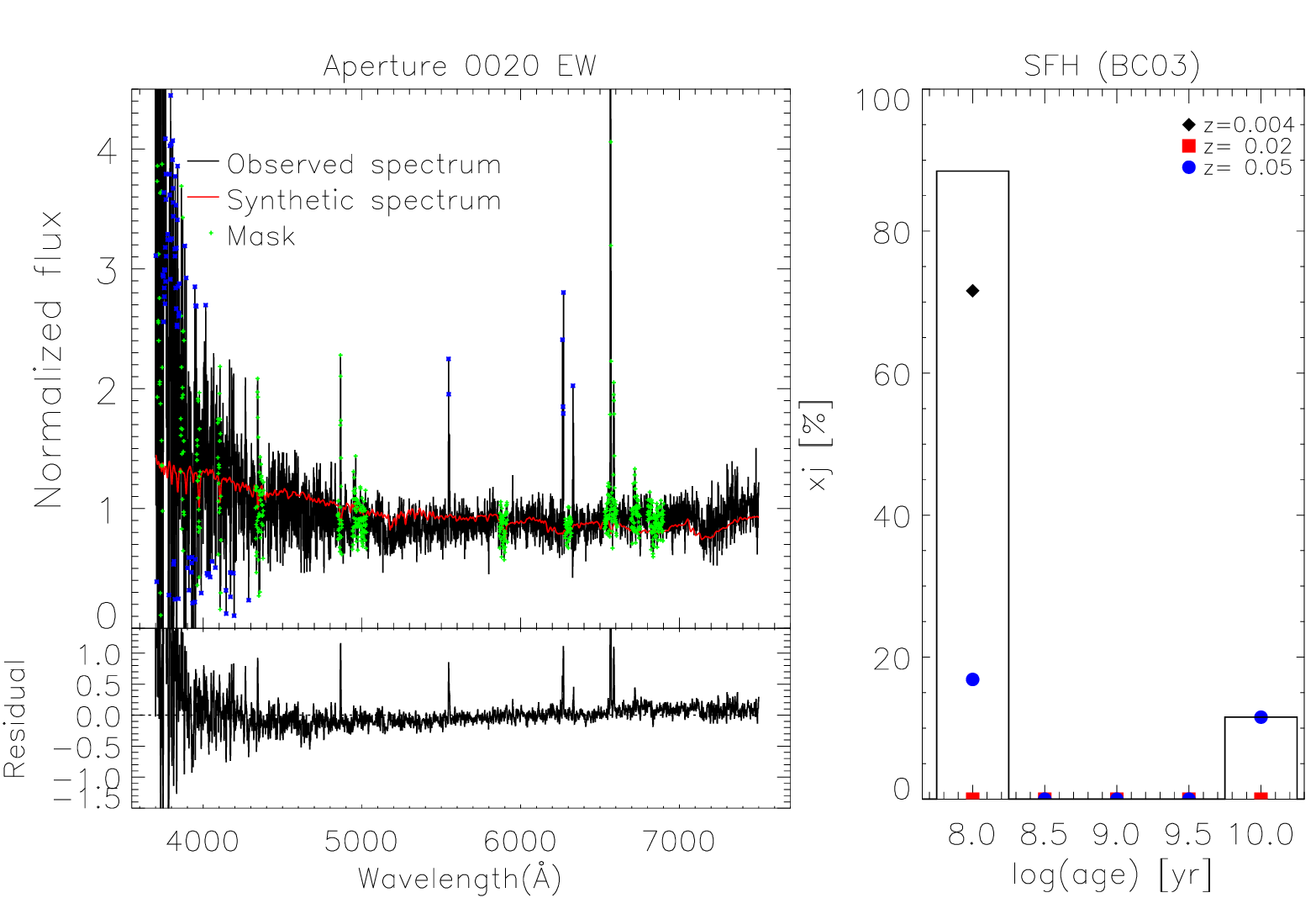}
}
\subfloat[]{
\includegraphics[scale=0.38]{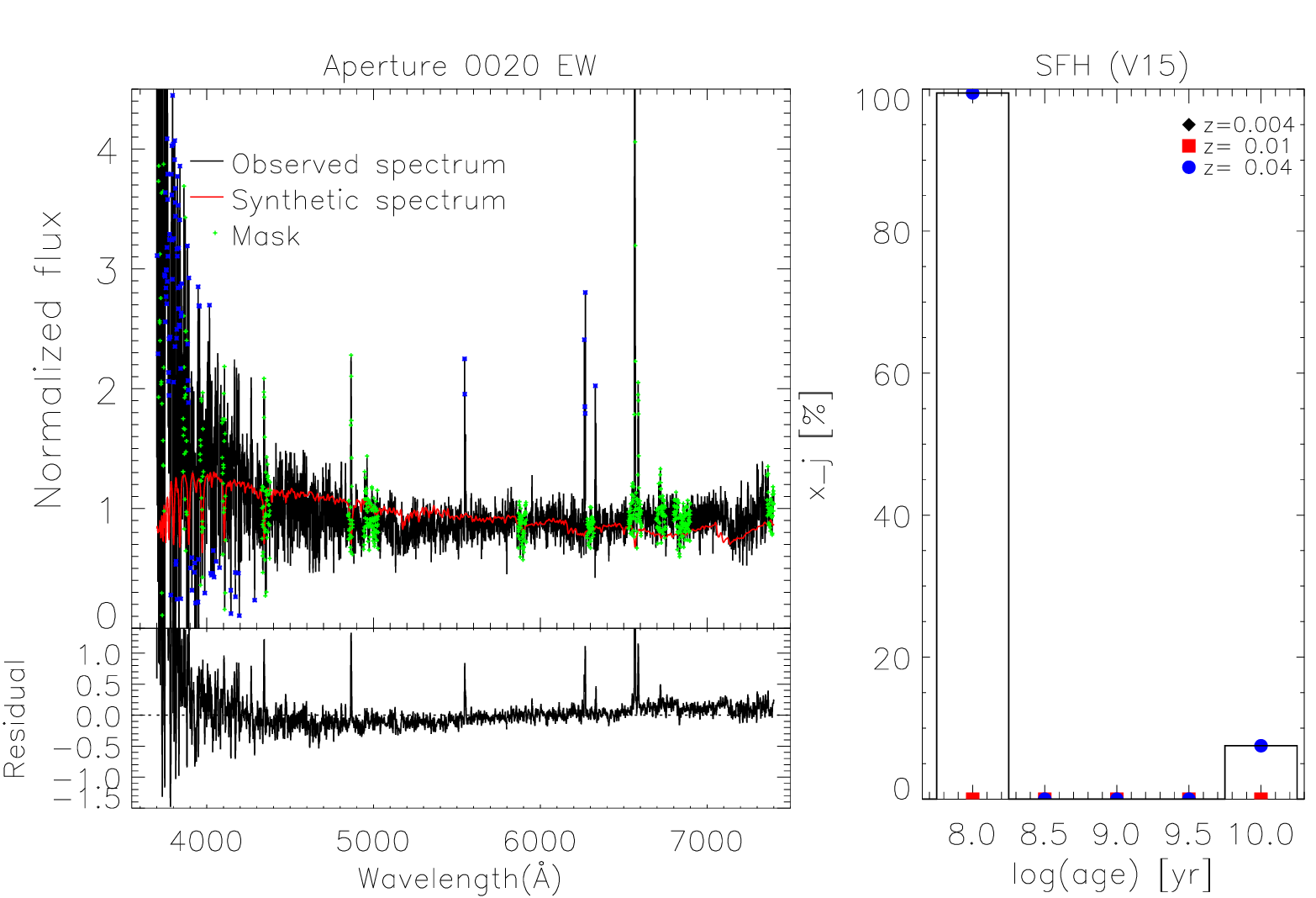}
}
\\
\subfloat[]{
\includegraphics[scale=0.38]{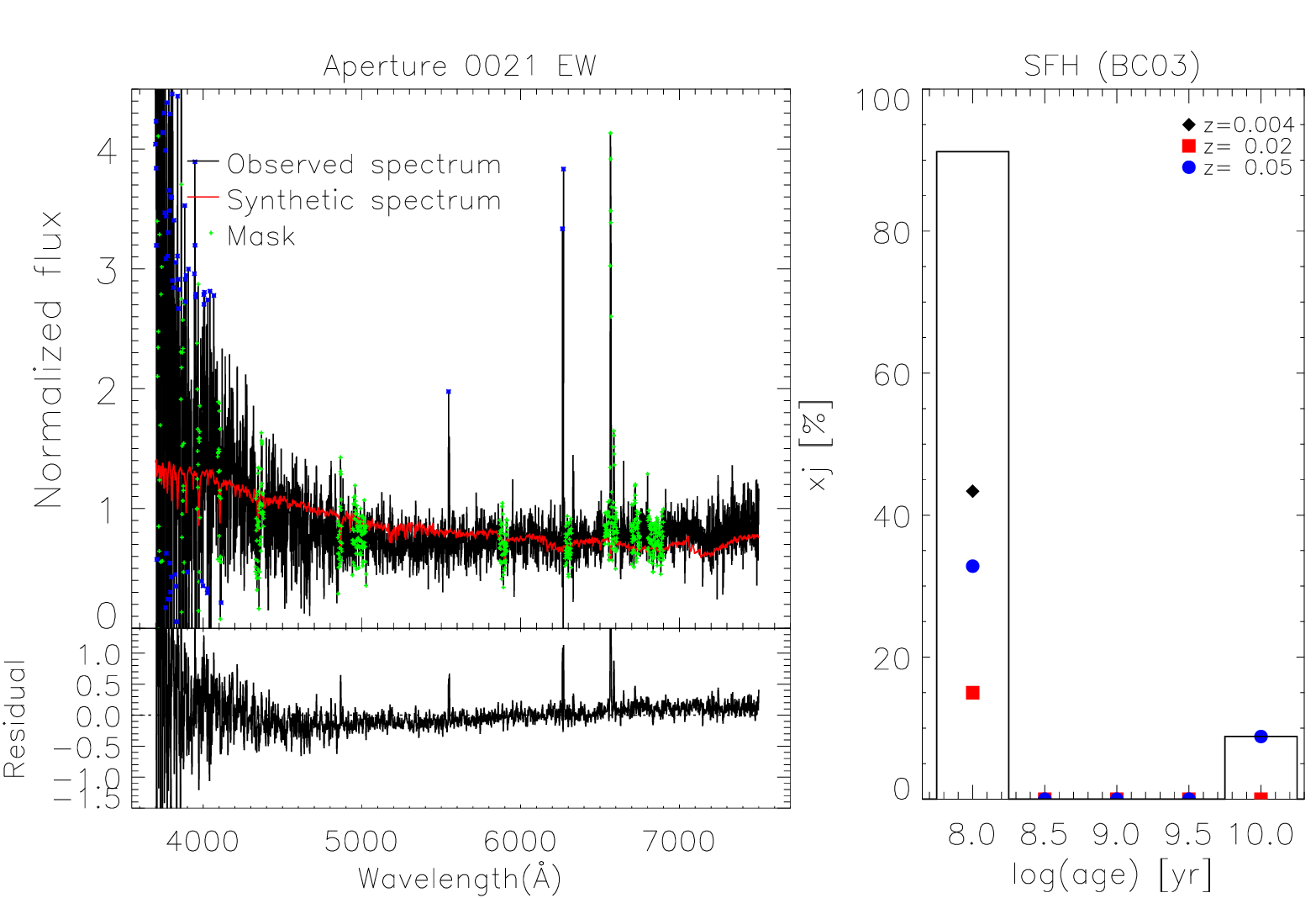}
}
\subfloat[]{
\includegraphics[scale=0.38]{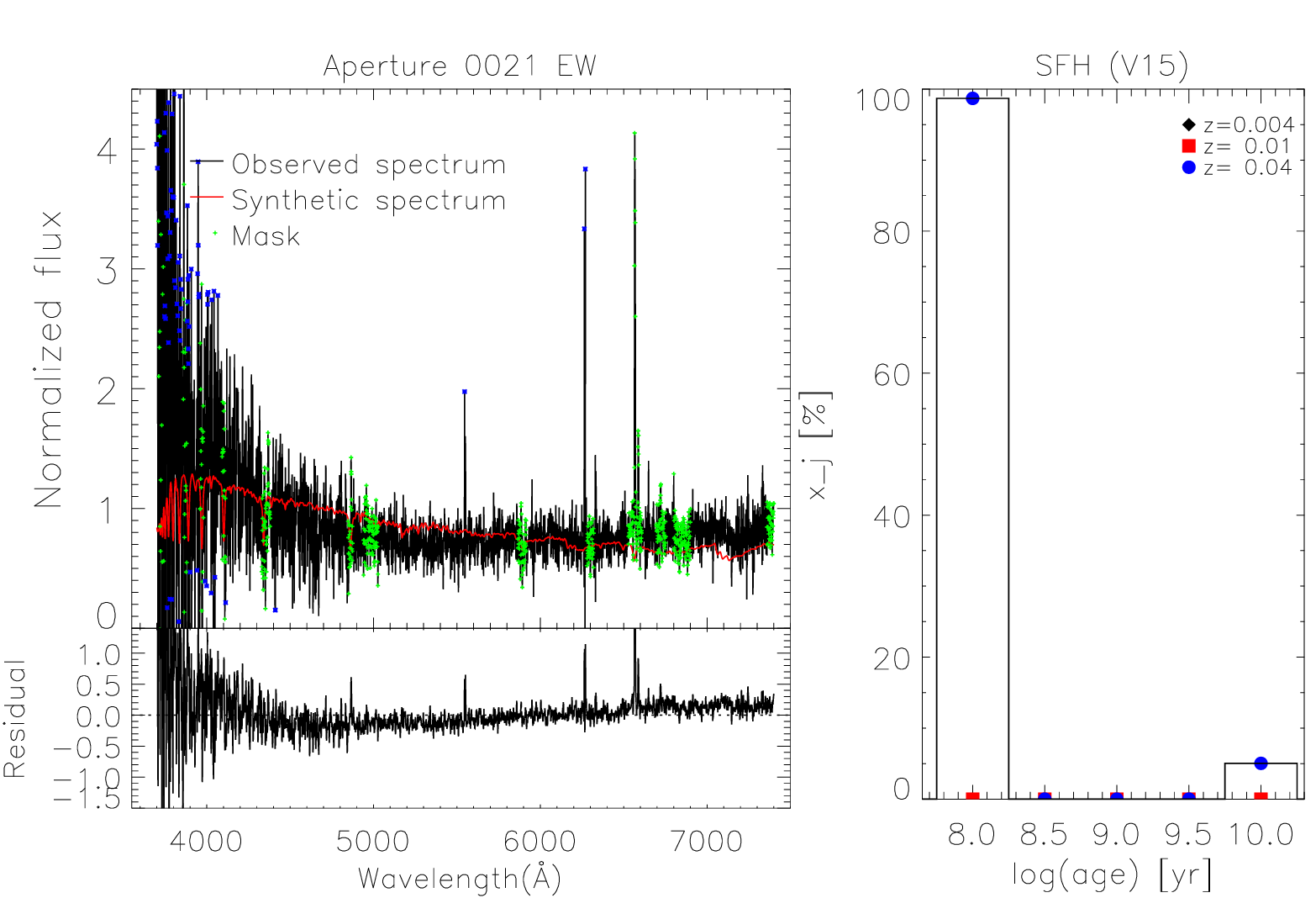}
}
\\
\subfloat[]{
\includegraphics[scale=0.38]{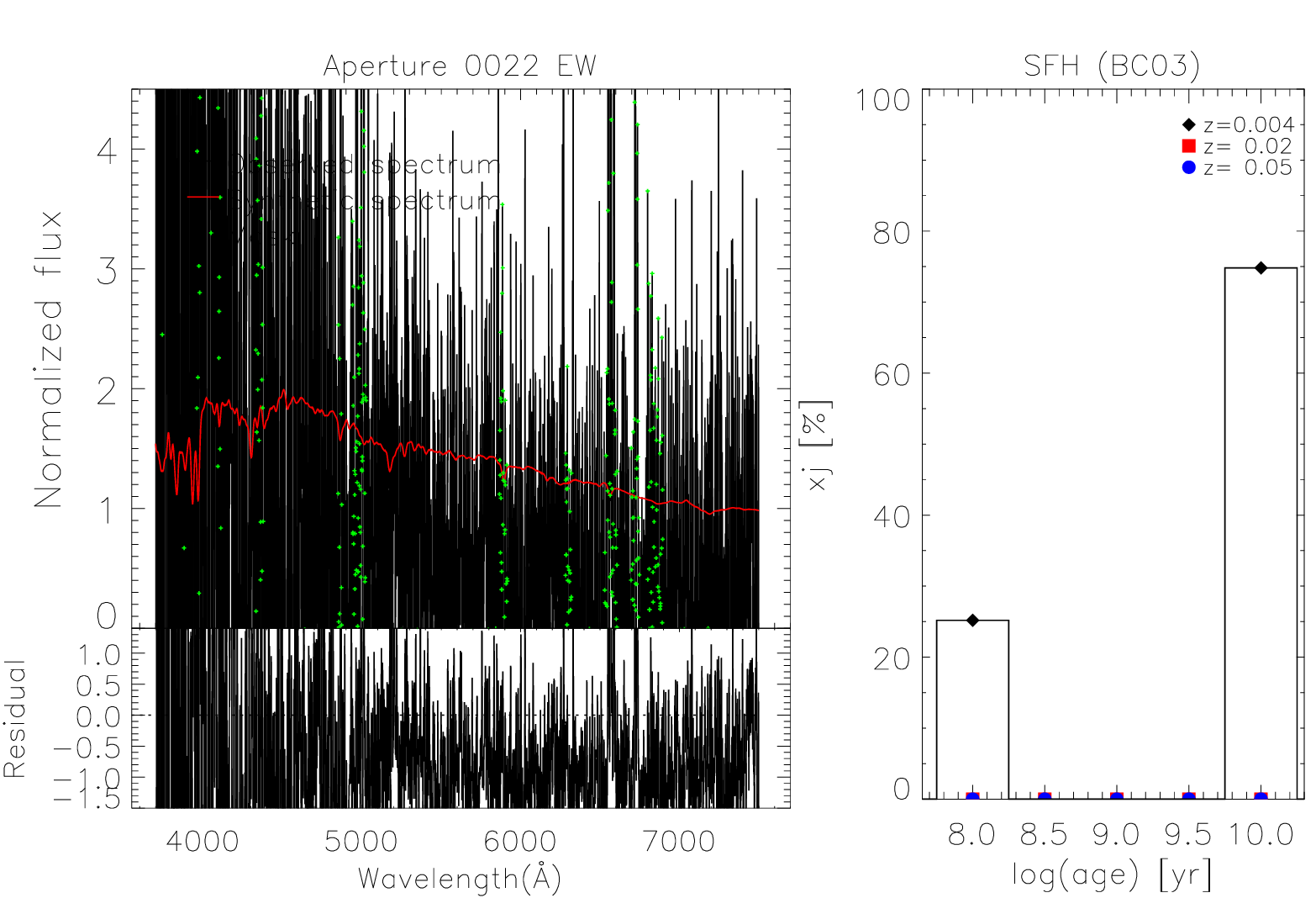}
}
\subfloat[]{
\includegraphics[scale=0.38]{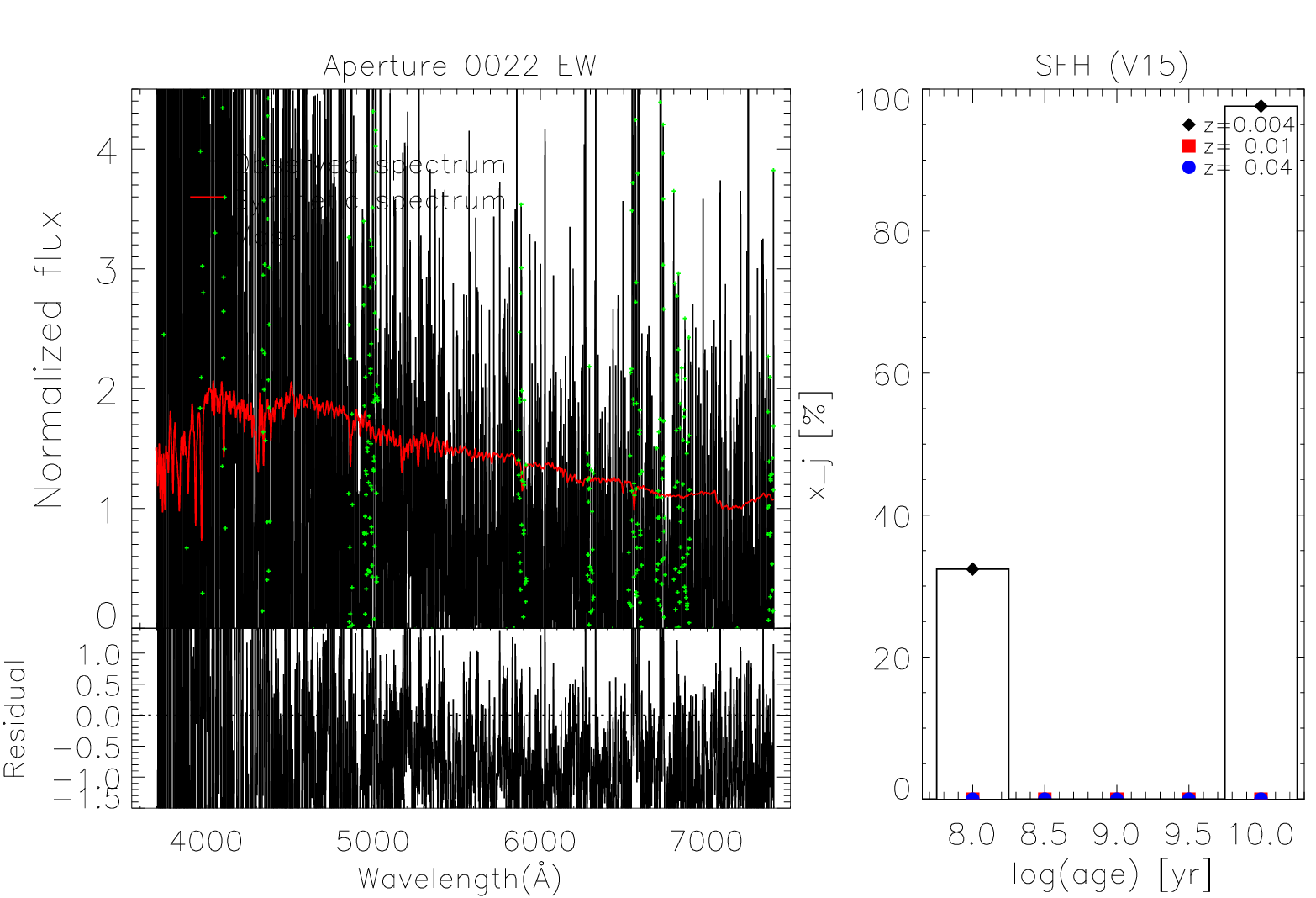}
}

\caption {Same as Figure~\ref{popstellarA1} but for apertures 20, 21,
 and 22 of the E-W  slit.
} 
\label{popstellarA13}
\end{figure*}

\end{appendix}

\end{document}